\documentclass[aps, prd, reprint, superscriptaddress, nofootinbib,
]{revtex4-2}

\usepackage{times}
\usepackage{graphicx}
\usepackage[,dvipsnames]{xcolor}
\usepackage{xspace}
\usepackage{amsmath,amsfonts,amssymb,amsthm,bm}
\usepackage{mathtools}
\usepackage[utf8]{inputenc}
\usepackage[normalem]{ulem}

\usepackage{setspace}
\usepackage[hang]{footmisc} 

\xdefinecolor{mylinkcolor}{rgb}{0,0,0.8}
\usepackage[
	bookmarksnumbered, bookmarksopen, bookmarksopenlevel=2,
	breaklinks=true,
	colorlinks=true, filecolor=mylinkcolor, citecolor=mylinkcolor,
	linkcolor=mylinkcolor, urlcolor=mylinkcolor, menucolor=mylinkcolor,
]{hyperref}
\usepackage{bookmark}

\usepackage{setspace}
\usepackage[hang]{footmisc} 
\allowdisplaybreaks

\def\mr{\mathrm}
\def\di{\mr d}
\def\cross{\times}
\DeclareMathOperator{\Order}{\mathcal{O}}
\def\mF{\mathcal{F}}
\def\e{\mathrm{e}}
\def\s{\hspace{0.5pt}}

\graphicspath{ {figs/} }

\newcommand{\AEI}{\affiliation{Max Planck Institute for Gravitational Physics (Albert Einstein Institute), Am M\"uhlenberg 1, Potsdam 14476, Germany}}
\newcommand{\Maryland}{\affiliation{Department of Physics, University of Maryland, College Park, MD 20742, USA}}
\newcommand{\PI}{\affiliation{Perimeter Institute for Theoretical Physics, 31 Caroline Street North, Waterloo, ON N2L 2Y5, Canada}}

\begin{document}

\title{Third post-Newtonian dynamics for eccentric orbits and aligned spins\\
in the effective-one-body waveform model \texttt{SEOBNRv5EHM}}

\author{Aldo Gamboa}\email{aldo.gamboa@aei.mpg.de}\AEI
\author{Mohammed Khalil}\email{mkhalil@perimeterinstitute.ca}\PI
\author{Alessandra Buonanno}\email{alessandra.buonanno@aei.mpg.de}\AEI\Maryland

\begin{abstract}
Accurate waveform models for coalescing binaries on eccentric orbits are crucial for avoiding biases in the analysis of eccentric gravitational-wave signals.
The effective-one-body (EOB) formalism combines various analytical approximation methods with information derived from numerical-relativity simulations, and it has proven reliable in modeling the inspiral-merger-ringdown waveform of binaries on generic orbits.
In this work, we derive new analytical results within the EOB formalism, specifically addressing eccentric, aligned-spin binaries.
For the first time, we obtain EOB results for eccentric orbits that are accurate up to the third post-Newtonian (PN) order for the complete far-zone energy and angular-momentum fluxes, the radiation-reaction (RR) force, and waveform modes (including memory contributions).
We obtain these results by deriving transformations from the harmonic to EOB coordinates and from the quasi-Keplerian to the Keplerian parametrization, that properly account for the spin-supplementary condition, postadiabatic contributions, and the gauge freedom in the RR force.
We employ these findings to build the inspiral part of the \texttt{SEOBNRv5EHM} waveform model, which includes the 3PN eccentricity contributions in the RR force and waveform modes using the Keplerian parametrization, and augments the EOB equations of motion with PN evolution equations for the Keplerian parameters.
The new \texttt{SEOBNR} model achieves much better accuracy than its predecessor \texttt{SEOBNRv4EHM} and other eccentric models, even for eccentricities as high as $\lesssim 0.5$.
\end{abstract}


\maketitle


\section{Introduction}

Since the first detection of gravitational waves (GWs) in 2015~\cite{LIGOScientific:2016aoc}, over a hundred GW events have been observed by the LIGO-Virgo-KAGRA (LVK) Collaboration~\cite{LIGOScientific:2018mvr,LIGOScientific:2020ibl,LIGOScientific:2021usb,LIGOScientific:2021djp,LIGOScientific:2019lzm} and by independent analyses of public data~\cite{Venumadhav:2019lyq,Nitz:2021uxj,Nitz:2021zwj,Olsen:2022pin,Wadekar:2023gea,Mehta:2023zlk}.
Such events have mostly been consistent with compact binaries on quasicircular (QC) inspirals~\cite{LIGOScientific:2023lpe,Salemi:2019owp,Romero-Shaw:2019itr,Nitz:2019spj,Romero-Shaw:2020thy,Lenon:2020oza,Wu:2020zwr,Yun:2020aow,Pal:2023dyg,Iglesias:2022xfc,Ramos-Buades:2023yhy}, which can be attributed to the circularization of coalescing binaries, as they lose energy and angular momentum due to the emission of GWs~\cite{Peters:1964zz,Hinder:2007qu}.

However, different studies have suggested the presence of orbital eccentricity signatures in some GW events detected during the first three observing runs of the LVK detectors.
For example, a recent analysis of 57 GW events~\cite{Gupte:2024jfe}, using an eccentric, aligned-spin waveform model~\cite{Ramos-Buades:2023yhy}, found that the probability of at least one event being eccentric is greater than 99.5\%, and that three events: GW200129, GW190701, and GW200208\_22 show some evidence of eccentricity.
Independently, Ref.~\cite{Romero-Shaw:2022xko} analyzed 62 events with another eccentric, aligned-spin waveform model~\cite{Cao:2017ndf} and found evidence for eccentricity in four of them:
GW190521, GW190620, GW191109, and GW200208\_22.
In addition, Refs.~\cite{Romero-Shaw:2020thy,Gayathri:2020coq} showed that the event GW190521~\cite{LIGOScientific:2020iuh} is consistent with an aligned-spin binary moving on an eccentric orbit, while Ref.~\cite{Gamba:2021gap} found that it could be interpreted as a dynamical capture of two nonspinning black holes (BHs).
Other references~\cite{Gupte:2024jfe,Iglesias:2022xfc,Bonino:2022hkj,Ramos-Buades:2023yhy} did not find evidence of eccentricity in GW190521; such a discrepancy is likely due to the short signal of GW190521, which makes it difficult to distinguish eccentricity from spin precession~\cite{CalderonBustillo:2020xms,Romero-Shaw:2022fbf}.

At least a small fraction of binaries observed by ground-based GW detectors are expected to be eccentric, depending on their astrophysical formation channels.
Binaries formed through an isolated binary evolution will have a negligible imprint of eccentricity in their gravitational signal by the time they are observable with current GW detectors \cite{Bethe:1998bn,Belczynski:2001uc,Dominik:2013tma,Belczynski:2014iua,mennekens2014massive,spera2015mass,Belczynski:2016obo,Eldridge:2016ymr,Marchant:2016wow,Mapelli:2017hqk,Mapelli:2018wys,Stevenson:2017tfq,Giacobbo:2018etu,Kruckow:2018slo,Kruckow:2018slo}.
In contrast, in dense stellar environments, such as globular clusters or galactic nuclei, dynamical captures of compact objects and three-body interactions can lead to eccentric binary inspirals observable with the LVK detectors~\cite{heggie1975binary,sigurdsson1993primordial,Kulkarni:1993fr,Miller:2001ez,Gultekin:2004pm,Sadowski:2007dz,OLeary:2008myb,downing2010compact,downing2011compact,Antonini:2012ad,Tsang:2013mca,Rodriguez:2015oxa,Askar:2016jwt,Rodriguez:2016kxx,Rodriguez:2016avt,Stone:2016wzz,Stone:2016ryd,Petrovich:2017otm,Gondan:2017wzd,Takatsy:2018euo,Rodriguez:2018rmd,Fragione:2018vty,Rasskazov:2019gjw,Fragione:2019vgr,Chattopadhyay:2023pil,Sedda:2023qlx,1910AN....183..345V,Sigurdsson:1993tui,Sigurdsson:1994ju,kozai1962secular,lidov1962evolution,giacaglia1964notes,Wen:2002km,Kimpson:2016dgk,VanLandingham:2016ccd,Hoang:2017fvh,PortegiesZwart:1999nm,Miller:2002pg,OLeary:2005vqo,Gultekin:2005fd,Samsing:2013kua,Antonini:2015zsa,Silsbee:2016djf,Antonini:2016gqe,Antonini:2017ash,Samsing:2017rat,Samsing:2017xmd,Rodriguez:2017pec,Zevin:2018kzq,Arca-Sedda:2018qgq,Fragione:2020nib,Gondan:2020svr,Vigna-Gomez:2020fvw,Tagawa:2020jnc,Britt:2021dtg}.
Therefore, GWs from eccentric binaries can be useful in distinguishing between binary formation channels~\cite{Sesana:2010qb,Breivik:2016ddj,Nishizawa:2016eza,Samsing:2018isx,Mandel:2018hfr,Cardoso:2020iji,Zevin:2021rtf,Fumagalli:2023hde,Samsing:2024syt,Zeeshan:2024ovp,Fumagalli:2024gko}.

To accurately measure eccentricity from GW signals, it is crucial to employ accurate waveform models that incorporate eccentricity effects~\cite{Ramos-Buades:2020eju,Lower:2018seu,DePorzio:2024cet,Gadre:2024ndy,Bhaumik:2024cec}; otherwise one might encounter systematic biases in parameter estimation \cite{Favata:2013rwa,Ramos-Buades:2019uvh,Cho:2022cdy,Guo:2022ehk,GilChoi:2022waq,Divyajyoti:2023rht,Gupte:2024jfe,Das:2024zib} and tests of general relativity \cite{Saini:2022igm,Saini:2023rto,Narayan:2023vhm,Gupta:2024gun,Shaikh:2024wyn,Bhat:2022amc,Bhat:2024hyb}.
Several waveform models for eccentric binaries exist in the literature based on post-Newtonian (PN) results \cite{Junker:1992kle,Gopakumar:1997bs,Gopakumar:2001dy,Memmesheimer:2004cv,Damour:2004bz,Konigsdorffer:2006zt,Arun:2007rg,Arun:2007sg,Arun:2009mc,Mishra:2015bqa,Boetzel:2019nfw,Ebersold:2019kdc,Henry:2023tka,Boetzel:2017zza,Paul:2022xfy}, such as the inspiral-only models presented in 
Refs.~\cite{Yunes:2009yz,Cornish:2010cd,ShapiroKey:2010cnz,Huerta:2014eca,Loutrel:2017fgu,Tanay:2016zog,Tanay:2019knc,Tiwari:2020hsu,Tiwari:2019jtz,Moore:2018kvz,Moore:2019xkm,Klein:2018ybm, Klein:2021jtd,Sridhar:2024zms}, or the hybrid inspiral-merger-ringdown (IMR) models that use PN results for the inspiral combined with QC numerical relativity (NR) simulations for the merger-ringdown part of the waveform~\cite{Huerta:2016rwp,Huerta:2017kez,Hinder:2017sxy,Islam:2024tcs,Ramos-Buades:2019uvh,Chattaraj:2022tay,Manna:2024ycx,Paul:2024ujx}.
In addition, recent NR simulations of eccentric binary black holes (BBHs)~\cite{Huerta:2019oxn,Boyle:2019kee,Ramos-Buades:2019uvh,Ramos-Buades:2022lgf, Nee:2025zdy} made it possible to develop an eccentric NR surrogate model~\cite{Islam:2021mha} for equal-mass, nonspinning BHs, and also to construct IMR eccentric waveform models employing phenomenological relations observed in eccentric NR simulations \cite{Setyawati:2021gom,Wang:2023ueg,Islam:2024rhm,Islam:2024zqo, Islam:2024bza}.

Another approach for developing accurate waveform models is the effective-one-body (EOB) formalism~\cite{Buonanno:1998gg,Buonanno:2000ef,Damour:2000we,Buonanno:2005xu}, which combines analytical information (from PN, post-Minkowskian, and gravitational self-force approximations) with NR results, while recovering the strong-field test-body dynamics.
EOB models for eccentric orbits started with Ref.~\cite{Bini:2012ji}, which derived 2PN results in EOB coordinates for the instantaneous energy and angular momentum fluxes, and for the radiation-reaction (RR) force.
Then, Ref.~\cite{Hinderer:2017jcs} developed a model with 1.5PN information for nonspinning BHs, including tail effects, using phase variables that evolve only due to the RR.
In addition, Refs.~\cite{Chiaramello:2020ehz,Nagar:2020xsk,Nagar:2021gss,Albanesi:2021rby,Nagar:2021xnh,Placidi:2021rkh,Albanesi:2022ywx,Albanesi:2022xge,Nagar:2023zxh,Placidi:2023ofj,Albanesi:2023bgi,Nagar:2024dzj,Gamba:2024cvy,Nagar:2024oyk,Grilli:2024lfh} have studied several prescriptions to extend the \texttt{TEOBResumS} family of EOB waveform models \cite{Nagar:2018zoe,Nagar:2019wds,Nagar:2020pcj,Gamba:2021ydi,Riemenschneider:2021ppj} to eccentric orbits for spinning BHs.
The resulting eccentric, aligned-spin waveform model is generically known as \texttt{TEOBResumS-Dal\'i}.

Within the \texttt{SEOBNR} family of waveform models, Refs.~\cite{Cao:2017ndf,Liu:2019jpg,Liu:2021pkr,Liu:2023dgl} developed the models \texttt{SEOBNRE} and \texttt{SEOBNREHM}, which extended the QC models \texttt{SEOBNRv1}~\cite{Taracchini:2012ig} and \texttt{SEOBNRv4HM}~\cite{Bohe:2016gbl,Cotesta:2018fcv} by including eccentricity in the waveform modes. Independently, Ref.~\cite{Khalil:2021txt} derived 2PN results for the RR force and waveform modes for eccentric, aligned-spin BBHs, and Ref.~\cite{Ramos-Buades:2021adz} used those results for the modes to develop the eccentric, aligned-spin model \texttt{SEOBNRv4EHM}, which targets eccentricities up to $\sim 0.3$.
Furthermore, Refs.~\cite{Liu:2023ldr,Gamba:2024cvy} have started to develop waveform models with eccentricity and spin precession.

The eccentricity contributions in almost all previous EOB models were restricted to 2PN order in the RR force or waveform modes, and resulted in modifications of the underlying QC dynamics.
Our goal in this work is to extend the state-of-the-art, QC, aligned-spin, EOB waveform model \texttt{SEOBNRv5HM}~\cite{Pompiliv5,Khalilv5,RamosBuadesv5,VandeMeentv5,Mihaylovv5} to eccentric BBHs, by incorporating 3PN information for eccentric orbits and aligned spins, while recovering the exact QC dynamics of the \texttt{SEOBNRv5HM} model.
We achieve this goal by expressing the dissipative dynamics of the model in terms of the dimensionless orbit-averaged orbital frequency $x$, an eccentricity variable $e$, and the relativistic anomaly $\zeta$ defined in the Keplerian parametrization~\cite{darwin1959gravity}, such that we recover exactly the \texttt{SEOBNRv5HM} equations of motion (EOMs) in the QC limit $e \to 0$.

The waveform model for eccentric, aligned-spin BBHs constructed using the results derived in this work is known as \texttt{SEOBNRv5EHM}, and it is presented and validated in a companion paper \cite{Gamboa:2024hli}.
The overall accuracy of \texttt{SEOBNRv5EHM} is about one order of magnitude better than the previous generation \texttt{SEOBNRv4EHM} model, and also with respect to the current \texttt{TEOBResumS-Dal\'i} model~\cite{Nagar:2024dzj}.
Importantly, we find that the 3PN eccentricity contributions are crucial for modeling highly eccentric binaries, as these systems involve high velocities during each periastron passage. 
We note that the nonspinning part of the 3PN RR force derived in this work was compared to test-mass-limit results obtained by solving numerically the Teukolsky equation in Ref.~\cite{Faggioli:2024ugn}, assessing the accuracy of the 3PN RR force and showing the improvement with respect to the 2PN results.

Here, we outline the main results and structure of this work:
\begin{enumerate}

\item
An overview of the results derived in this work is presented in Sec.~\ref{sec:nutshell}, along with a summary of the inspiral part of the \texttt{SEOBNRv5EHM} waveform model \cite{Gamboa:2024hli}.

\item
Most PN results in the literature (for the BBH dynamics, energy and angular momentum fluxes, and waveform modes) are derived using harmonic coordinates.
To use such results in an EOB model, we first work out in Sec.~\ref{sec:transParams} the complete transformations from harmonic to EOB coordinates at 3PN order. 
In these transformations, we account for the spin-supplementary condition (SSC), postadiabatic (PA) contributions, and the gauge freedom in defining the RR force.

\item
Eccentric-orbit PN results for the hereditary contributions to the fluxes and waveform modes are often expressed using the quasi-Keplerian (QK) parametrization~\cite{damour1985general,Damour:1988mr,schafer1993second,wex1995second}, while we employ the Keplerian parametrization~\cite{darwin1959gravity}. 
Therefore, also in Sec.~\ref{sec:transParams}, we give an overview of both parametrizations and obtain transformations from the QK parametrization in harmonic coordinates to the Keplerian parametrization in EOB coordinates.

\item
In Sec.~\ref{sec:fluxes}, we use the harmonic-to-EOB and QK-to-Keplerian transformations to obtain the complete 3PN energy and angular momentum fluxes, by converting the results of Refs.~\cite{Arun:2007rg,Arun:2007sg,Arun:2009mc,Henry:2023tka} to EOB coordinates and to Keplerian parameters.

\item
From the fluxes of energy and angular momentum, we obtain in Sec.~\ref{sec:evoEqs} 3PN-accurate evolution equations for the Keplerian eccentricity $e$, relativistic anomaly $\zeta$, and orbit-averaged orbital frequency $ x $, in addition to a PN relation between the orbit-averaged and instantaneous orbital frequencies.

\item
Also from the fluxes, we construct in Sec.~\ref{sec:RRforce} a 3PN-accurate EOB RR force through the use of balance equations, which relate the energy and angular momentum losses by the system to the fluxes at infinity.

\item
We compute the 3PN EOB waveform modes in Sec.~\ref{sec:modes}, by converting the results of Refs.~\cite{Mishra:2015bqa,Boetzel:2019nfw,Ebersold:2019kdc,Henry:2023tka} in harmonic-coordinates and QK parameters to EOB coordinates and Keplerian parameters, and we recast the EOB modes in a factorized form for \texttt{SEOBNRv5EHM}.

\item
Finally, in Sec.~\ref{sec:ICs}, we obtain 3PN relations for the initial conditions of eccentric, aligned-spin binaries used to initialize the evolution of the EOMs.

\end{enumerate}

We conclude in Sec.~\ref{sec:conclusions} and delegate the long PN equations derived in this work to the Appendixes and the Supplemental Material.\footnote{
The Supplemental Material 
consists of the following \texttt{Mathematica} files:
1) \texttt{EOB\_Keplerian.dat.m}, which contains the 3PN transformations between harmonic and EOB coordinates, transformations between the QK and Keplerian orbital parameters, and various relations between the EOB and Keplerian variables;
2) \texttt{EOB\_fluxes.dat.m}, which contains the 3PN EOB energy and angular momentum fluxes, Schott terms, and RR force for eccentric orbits, as well as evolution equations for the Keplerian orbital parameters $ (x, e, \zeta) $;
and 3) \texttt{EOB\_modes.dat.m}, which contains the 3PN EOB waveform modes for eccentric orbits and the eccentricity corrections to the modes employed in the \texttt{SEOBNRv5EHM} model~\cite{Gamboa:2024hli}.
}
Throughout this work, we indicate which equations are employed in the construction of the eccentric, aligned-spin waveform model \texttt{SEOBNRv5EHM} presented in Ref.~\cite{Gamboa:2024hli}.


\section*{Notation}

Our equations are written in geometric units in which Newton's constant and the speed of light equal one, $G = c = 1$. 

In PN expansions, we follow the usual convention in which the $n$PN order refers to terms of order $v^{2 n}$, where $v$ denotes the binary’s relative velocity, and we use the power-counting parameter $\epsilon^{2n} = 1$ to distinguish each PN order.

We consider a binary with masses $m_1$ and $m_2$ such that $m_1 \geq \nolinebreak m_2$, and define the following combinations of masses
\begin{equation}
\begin{gathered}
M\equiv m_1 + m_2, \qquad \mu \equiv \frac{m_1m_2}{M}, \qquad \nu \equiv \frac{\mu}{M},  \\ 
\delta \equiv\frac{m_1 - m_2}{M},  \qquad q \equiv \frac{m_1}{m_2}.
\end{gathered}
\end{equation}

We consider aligned-spin binaries whose individual component spins, $\bm{S}_1$ and $\bm{S}_2$, are proportional to the binary's orbital angular momentum unit vector $ \bm{l} $, and hence perpendicular to the orbital plane.
These systems are characterized by their dimensionless spin components given by
\begin{equation}
 \label{eq:comp_spins}
\chi_i \equiv \frac{\bm{S}_i \cdot \bm{l}}{m_i^2} = \pm \frac{|\bm{S}_i|}{m_i^2} \s,  \qquad i \in \{1,2\}, 
\end{equation}
which take values in the interval $(-1,1)$, with positive spins being in the direction of the orbital angular momentum. 

We also define the following combinations of the spins
\begin{equation}
\label{spinComb}
\chi_S \equiv \frac{\chi_1 + \chi_2}{2}, \qquad \chi_A \equiv \frac{\chi_1 - \chi_2}{2},
\end{equation}
which are useful for simplifying the resulting equations.

The spin quadrupole constants are denoted $C_{\rm iES^2}$, which equal one for BHs, but are greater than one for neutron stars. 
To simplify some expressions, we define the following combinations of spins and multipole constants
\begin{equation}
\begin{aligned}
\kappa_S &\equiv \frac{1}{2} \left[\chi _1^2 (C_{\rm 1ES^2} - 1)+\chi _2^2 (C_{\rm 2ES^2} - 1)\right], \\
\kappa_A &\equiv \frac{1}{2} \left[\chi _1^2 (C_{\rm 1ES^2} - 1) - \chi _2^2 (C_{\rm 2ES^2} - 1)\right],
\end{aligned}
\end{equation}
which equal zero for BHs.

The relative position and momentum vectors in the binary's center of mass are denoted $\bm{r}$ and $\bm{p}$, with
\begin{equation}
p^2 = p_r^2 + \frac{L^2}{r^2}, \quad
p_r= \bm{n}\cdot\bm{p}, \quad
\bm{L}=\bm{r}\times\bm{p},
\end{equation}
where $\bm{n}\equiv\bm{r}/r$, and $\bm{L}$ is the orbital angular momentum with direction $\bm l$. The unit vector perpendicular to $\bm l$ and $\bm n$ is denoted $\bm \lambda \equiv \bm l \times \bm n$.
For equatorial orbits (aligned spins), we use the polar-coordinates phase-space variables $(r,\phi,p_r,p_\phi)$, where the angular momentum reduces to $L =p_\phi$.

We denote the energy and angular momentum fluxes at infinity by $\Phi_E$ and $\Phi_J$, respectively, and the RR force by $\bm{\mF}$ with radial and azimuthal components $\mF_r$ and $\mF_\phi$, such that
\begin{equation}
\bm{\mF} = \mF_r \, \bm n +  \frac{\mF_\phi}{r} \, \bm \lambda.
\end{equation}

We employ several quantities defined in the QK and Keplerian parametrizations, which we summarize in Table~\ref{tab:QKPsummary}.

In a few of the long equations in this paper, and in the Supplemental Material, we use scaled dimensionless quantities to simplify the notation.
To restore the units, one can use the following replacements:
\begin{equation}
\label{eq:dimlessVars}
\begin{gathered}
t \to \frac{t}{M},
\quad r \to \frac{r}{M},
\quad p \to \frac{p}{\mu},
\quad p_r \to \frac{p_r}{\mu},
\\
p_\phi \to \frac{p_\phi}{M\mu},
\quad E \to \frac{E}{\mu},
\quad \Omega \to M\Omega,
\\
\bm{\mF} \to \nu\s\bm{\mF},
\quad \Phi_E \to \nu \s \Phi_E,
\quad \Phi_J \to M \nu \s \Phi_J .
\end{gathered}
\end{equation}
\begin{table}[th]
\caption{Definition of the main quantities used in the QK and Keplerian parametrizations.}
\label{tab:QKPsummary}
\begin{ruledtabular}
\begin{tabular}{p{0.25\linewidth} p{0.72\linewidth}}
Symbol & Description \\
\hline
$a_r$ & semi-major axis \\
$e_r$ & radial eccentricity \\
$e_t$ & time eccentricity \\
$e_\phi$ & phase eccentricity \\
$u$ & eccentric anomaly \\
$V$ & true anomaly \\
$n \equiv 2\pi/P$ & mean motion (radial angular frequency), with $P$ being the radial period\\
$l \equiv n (t - t_\text p)$ & mean anomaly, where $t_\text p$ is time at periastron passage \\
$\Phi$ & total phase between two successive periastron passages \\
$K\equiv \Phi/(2\pi)$ & periastron advance \\
$k\equiv K - 1$ & fractional (or relativistic) periastron advance \\
$\lambda \equiv K l$ & an auxiliary angle used to express the orbital phase as $\phi = \lambda + W(l)$, with $ W $ an oscillatory function that is $2 \pi$-periodic in $l$ \\
$\langle \Omega \rangle = K n$ & orbit-averaged orbital frequency \\
$x \equiv \langle M\Omega \rangle^{2/3}$ & dimensionless frequency variable \\
\hline
$e$ & eccentricity in the Keplerian parametrization \\
$u_p$ & inverse semilatus rectum scaled by the total mass \\
$\zeta$ & relativistic anomaly \\
\end{tabular}
\end{ruledtabular}
\end{table}


\section{
Overview of this work and the SEOBNR\lowercase{v}5EHM model
}
\label{sec:nutshell}

In the EOB formalism \cite{Buonanno:1998gg,Buonanno:2000ef}, the gravitational two-body dynamics is mapped onto the effective dynamics of a test body in a deformed Schwarzschild or Kerr background, with the deformation parameter being the symmetric mass ratio $ \nu $ of the binary. The effective dynamics is characterized by an effective Hamiltonian $H_\text{eff}$ that is related to the two-body Hamiltonian $H_\text{EOB}$ by the energy map
\begin{equation} \label{eq:EOBHam}
H_\text{EOB} = M \sqrt{1 + 2 \nu \left( \frac{H_\text{eff}}{\mu} - 1 \right)}.
\end{equation}
In the \texttt{SEOBNRv5HM} waveform model~\cite{Pompiliv5}, the aligned-spin Hamiltonian $ H _{ \text{eff}} $ reduces exactly to the Hamiltonian of a test mass in a Kerr background when $\nu \to 0$, resums the full 4PN information, and includes two parameters at higher PN orders calibrated to NR simulations (see Ref.~\cite{Khalilv5} for the explicit expression of the Hamiltonian).

The dynamics of binaries moving on equatorial orbits is determined by Hamilton's EOMs,
\begin{subequations}
\label{eq:EOMs}
\begin{align}
\dot r &= \frac{\partial H_\text{EOB}}{\partial p_r} ,
\label{eq:rdot} \\
\dot \phi &= \frac{\partial H_\text{EOB}}{\partial p_\phi},
\label{eq:phidot} \\
 \dot p_r &= - \frac{\partial H_\text{EOB}}{\partial r} + \mathcal F_r,
\label{eq:prdot} \\
 \dot p_\phi &= - \frac{\partial H_\text{EOB}}{\partial \phi} + \mathcal F_\phi,
\label{eq:pphidot}
\end{align}
\end{subequations}
where we add a RR force, with radial and azimuthal components $ \mathcal F_r $ and $ \mathcal F_\phi $, to account for the energy and angular momentum losses by the system due to the emission of GWs.
We note that $\partial H_\text{EOB} / \partial \phi$ is zero since the aligned-spin Hamiltonian is independent of $\phi$ due to rotational symmetry.

The stability of the EOMs near the event horizon can be improved with the use of the tortoise-coordinate $r_*$, which is defined by $\di r / \di r_* \equiv \xi(r)$, with the conjugate momentum $p_{r_*}$ given by \cite{Damour:2007xr,Pan:2009wj}
\begin{equation}
\label{prstar}
p_{r_*} =  p_r \s \xi(r),
\end{equation}
where the function $\xi(r)$ is given by Eq.~(44) of Ref.~\cite{Khalilv5}. 
In terms of $p_{r_*}$, the EOMs~\eqref{eq:EOMs} become
\begin{subequations}
\label{eq:EOMprStr}
\begin{align}
\dot{r} &= \xi(r) \frac{\partial H_\text{EOB}}{\partial p_{r_*}}(r,p_{r_*},p_\phi),  \\
\dot{\phi} &= \frac{\partial H_\text{EOB}}{\partial p_\phi}(r,p_{r_*},p_\phi), \label{eq:phidot_2} \\
\dot{p}_{r_*} &= - \xi(r) \frac{\partial H_\text{EOB}}{\partial r}(r,p_{r_*},p_\phi) + \xi(r) \s \mF_r, 
\label{eq:dot_prstar}
\\
\dot{p}_\phi &= \mF_\phi.
\end{align}
\end{subequations}

The RR force components in the EOMs~\eqref{eq:EOMprStr} are determined by the \emph{flux-balance} equations.
These equations relate the gauge-dependent local energy and angular momentum losses by the system, $ \dot E _{ \text{sys}} $ and $ \dot J _{ \text{sys}} $, to the gauge-invariant fluxes of energy and angular momentum at infinity, $ \Phi_E $ and  $ \Phi_J $.
The flux-balance equations for generic equatorial orbits read~\cite{Bini:2012ji}
\begin{subequations}
\label{eq:fluxbalance}
\begin{align}
\dot E _{ \text{sys}} + \dot E _{ \text{Schott}} + \Phi_E &= 0,
 \\
\dot J _{ \text{sys}} + \dot J _{ \text{Schott}} + \Phi_J &= 0,
\end{align}
\end{subequations}
where the Schott terms, $ \dot E _{ \text{Schott}}  $ and $ \dot J _{ \text{Schott}} $, represent additional gauge-dependent contributions that appear due to the interaction of the system with the radiation field, as in the case of electromagnetism \cite{Schott:1915aa}. 

To relate the RR force components to the fluxes, we note that from the EOMs~\eqref{eq:EOMprStr} we have the relations $\dot E_{\text{sys}} = \nolinebreak \dot r  \mathcal F_r  + \nolinebreak\dot \phi  \mathcal F_\phi$ and $\dot J_{\text{sys}} = \mF_\phi$, leading to
\begin{subequations}
\label{eq:forces_schott}
\begin{align}
&\dot r  \mathcal F_r  + \dot \phi  \mathcal F_\phi  + \dot E _{ \text{Schott}} + \Phi_E = 0 ,
\label{eq:frschott} \\
&\mF_\phi + \dot J _{ \text{Schott}} + \Phi_J = 0.
\label{eq:fphischott}
\end{align}
\end{subequations}
Therefore, given ansatz for the Schott terms and expressions for the fluxes, we can employ Eqs.~\eqref{eq:forces_schott} to determine PN expansions of $ \mathcal F_r  $ and $ \mathcal F_\phi $ as functions of gauge coefficients that appear due to the gauge freedom in defining the Schott terms.

For QC orbits, the Schott terms can be chosen in such a way that the RR force components satisfy the following relations employed in \texttt{SEOBNR} waveform models:
\begin{subequations}
\label{eq:qcRRforce}
\begin{align}
\mF_\phi^\text{qc} &= - \Phi_J ^{ \text{qc}} =  -\frac{M^2 \Omega}{8 \pi} \sum_{\ell=2}^8 \sum_{m=1}^{\ell} m^2\left|d_L \, h_{\ell m}^{\text{F, qc}}\right|^2,
\\
\mF_r^\text{qc} &= \frac{p_r}{p_\phi} \mF_\phi^\text{qc}, \label{eq:ratioFrFphi}
\end{align}
\end{subequations}
where $\Omega \equiv \dot{\phi}$ is the instantaneous orbital frequency, $d_L$ is the luminosity distance between the binary and the observer, and $ h_{\ell m}^{\text{F, qc}} $ are the \emph{factorized} QC gravitational waveform modes that are employed to decompose the GW polarizations as in Eq.~\eqref{eq:polarizations}, and which are given by
\begin{equation}
\label{eq:hlm_qc_intro}
h_{\ell m}^\text{F, qc} = h_{\ell m}^\text{N, qc}  \, \hat{S}_\text{eff} \, T_{\ell m}^\text{qc} \, f_{\ell m}^\text{qc} \, \e^{i \delta_{\ell m}^\text{qc}},
\end{equation}
where all the factors in Eq.~\eqref{eq:hlm_qc_intro} are explained in  Sec.~\ref{sec:factorization_modes}.

Determining a prescription for a generic RR force requires expressions for the energy and angular momentum fluxes for eccentric orbits.
These fluxes, and also the waveform modes, are known in the literature for aligned-spin BBHs moving on eccentric (quasielliptic) orbits to 3PN order.
In particular, in the nonspinning case, the 3PN energy flux was derived in Refs.~\cite{Arun:2007rg,Arun:2007sg} and the angular momentum flux in Ref.~\cite{Arun:2009mc}, while the waveform modes were completed to 3PN in Refs.~\cite{Mishra:2015bqa,Boetzel:2019nfw,Ebersold:2019kdc}.
For aligned spins, the fluxes and modes were derived to 3PN order in Refs.~\cite{Cho:2021mqw,Henry:2023tka}.
All of these results were obtained using harmonic coordinates. Thus, to include them in an EOB model, we derive transformations from harmonic to EOB coordinates in Sec.~\ref{sec:harmToEOB}.
We note that the hereditary contributions to the fluxes and modes for eccentric orbits are usually expressed using the QK parametrization of the orbit in harmonic coordinates (summarized in Sec.~\ref{sec:quasiKeplerian}).
Here, we employ a Keplerian parametrization in EOB coordinates (summarized in Sec.~\ref{sec:Keplerian}), and we relate both parametrizations in Sec.~\ref{sec:relating_parametrization}.
In all these transformations, we give a proper treatment to the dissipative contributions that appear at 2.5PN order and which originate from the gauge dependence of the RR force.  

In the Keplerian parametrization, the radial motion of the binary is described as
\begin{equation}
\label{eq:Kep_param}
r = \frac{M}{u_p (1 + e \cos \zeta)},
\end{equation}
where $u_p$ is the inverse semilatus rectum (scaled by the total mass), $ e $ is the Keplerian eccentricity, and $\zeta$ is the relativistic anomaly introduced by Darwin in Ref.~\cite{darwin1959gravity}.
In this parametrization, the deformation of the orbit is measured by $ e $, and the relative position of the binary along the orbit is determined by $ \zeta $.
Specifically, $ \zeta = 0 $ represents the closest approach, known as periastron, while $ \zeta = \pi $ corresponds to the farthest separation, known as apastron.

Hence, using coordinate transformations and the Keplerian parametrization, we derive expressions for the complete 3PN EOB fluxes, RR force, and waveform modes for eccentric orbits.
All these results are employed to obtain the RR force and modes used in the eccentric, aligned-spin waveform model \texttt{SEOBNRv5EHM} \cite{Gamboa:2024hli}.
More specifically, we calculate the eccentricity corrections to the RR force, $\mF_\phi^\text{ecc}$ and $\mF_r^\text{ecc}$, and to the waveform modes, $h_{\ell m}^\text{ecc}$, in terms of $e$, $\zeta$, and the gauge-invariant orbit-averaged orbital frequency variable $x \equiv \langle M\Omega \rangle^{2/3}$.
These corrections are computed such that
\begin{subequations}
\label{eq:corrections}
\begin{align}
\mF_\phi &= \mF_\phi^\text{modes} \mF_\phi^\text{ecc}(x,e,\zeta), \\
\mF_r &= \frac{p_r}{p_\phi} \mF_\phi^\text{modes} \mF_r^\text{ecc}(x,e,\zeta),\\
\mF_\phi^\text{modes} &= -\frac{M^2 \Omega}{8 \pi} \sum_{\ell=2}^8 \sum_{m=1}^{\ell} m^2\left|d_L h_{\ell m}^{\text{F}}\right|^2, \\
h_{\ell m}^\text{F}&=h_{\ell m}^\text{F, qc} (x) \,h_{\ell m}^\text{ecc}(x,e,\zeta).
\end{align}
\end{subequations}
Thus, the \texttt{SEOBNRv5EHM} modes and RR force contain the full 3PN information for eccentric orbits (restricted to the nonspinning limit, as explained in footnote~\ref{fn:ecc_corr}), while the eccentricity corrections $h_{\ell m}^\text{F, ecc}$, $\mF_\phi^\text{ecc}$, and $\mF_r^\text{ecc}$ reduce to 1 when $e \to 0$, thereby recovering the QC relations~\eqref{eq:qcRRforce} and~\eqref{eq:hlm_qc_intro}.
The only difference in the QC limit compared to \texttt{SEOBNRv5HM} is an overall factor of $(M\Omega)^{1/3}$ in the modes, instead of the factor $v_\phi$ given by Eq.~\eqref{eq:vphi}, as explained in Sec.~\ref{sec:factorization_modes}.

The parametrization employed in Eqs.~\eqref{eq:corrections} allows us to smoothly recover the QC EOMs used in \texttt{SEOBNRv5HM}.
However, to determine the values of $ (x, e, \zeta) $ entering in the EOMs~\eqref{eq:EOMprStr}, we need to give additional prescriptions for these parameters.
Therefore, we derive in Sec.~\ref{sec:evoEqs} evolution equations for the Keplerian parameters,
\begin{subequations}
\label{eq:EOMexzeta}
\begin{align}
\dot{e} &= -\frac{\nu e x^4}{M} \left[\frac{\left(304 + 121 e^2\right)}{15 \left(1-e^2\right)^{5/2}} + \text{3PN expansion}\right], \label{eq:e}\\
\dot{x} &=  \frac{2  \nu x^5}{3 M} \left[\frac{96 +292 e^2 + 37 e^4}{5 \left(1-e^2\right)^{7/2}} + \text{3PN expansion} \right],\label{eq:dotx}\\
\dot{\zeta} &= \frac{x^{3/2}}{M}\left[\frac{ (1+e \cos\zeta)^2}{\left(1-e^2\right)^{3/2}} + \text{3PN expansion}\right], \label{eq:zeta}
\end{align}
\end{subequations}
along with a PN relation between the orbit-averaged frequency variable $ x $ and the instantaneous EOB frequency $ \Omega $,
\begin{equation}
x = (M \Omega)^{2/3} \left[\frac{\left(1-e^2\right) }{(1+e \cos\zeta)^{4/3}} + \text{3PN expansion}\right]. \label{eq:xOmega}
\end{equation}

Thus, the \texttt{SEOBNRv5EHM} model simultaneously evolves the six differential equations~\eqref{eq:EOMprStr}, \eqref{eq:e}, and \eqref{eq:zeta}, and uses Eq.~\eqref{eq:xOmega} to compute the value of $x$ at each time step.
The model takes as inputs the initial values for $e$, $\zeta$, and the orbit-averaged orbital frequency $\langle\Omega \rangle$, and then uses the procedure described in Sec.~\ref{sec:ICs} to initialize the other dynamical variables $r$, $p_\phi$, and $p_{r_*}$.
In the $e \to 0$ limit, we get
\begin{equation}
\left.\dot{e}\right|_{e\to0} = 0,  \qquad
\left.x\right|_{e\to0} = (M \Omega)^{2/3},
\end{equation}
and no dependence in the EOMs on $\zeta$, which allows us to recover the EOMs of the QC model \texttt{SEOBNRv5HM}.
We note that Eq.~\eqref{eq:dotx} is not employed in the evolution of the binary but, instead, is used to get consistent initial conditions for binaries characterized by high frequencies and high eccentricities, as described in Sec.~\ref{sec:ICs}.


\section{Coordinate transformations and parametrizations of the dynamics}
\label{sec:transParams}

In this section, we derive a set of transformations from the harmonic coordinate system (typically employed in the PN formalism) to the EOB coordinate system.
Additionally, we derive the relations needed to transform the orbital parameters defined within the QK parametrization of the orbit (also commonly employed in PN literature) into the parameters used in the Keplerian parametrization.
In all these transformations, we properly treat the dissipative contributions that appear at 2.5PN order, and which are related to the gauge-dependence of the RR force acting on the binary.


\subsection{Harmonic to EOB coordinate transformation}
\label{sec:harmToEOB}

Several physical quantities of interest to the two-body problem, such as the waveform modes and fluxes, have been computed in \emph{harmonic} coordinates, which is the preferred system in the PN formalism for the radiative dynamics \cite{Blanchet:2008je,Blanchet:2013haa,Faye:2012we}. 
At 3PN order, the harmonic EOMs in the center-of-mass frame have some terms proportional to $\ln r_0'$, where $r_0'$ is a length scale that depends on the infrared cutoff used to regularize the integrals appearing in the PN computations~\cite{Blanchet:2000ub,Jaranowski:1999ye,Damour:2000ni}. These EOMs correspond to the so-called \emph{standard} harmonic (SH) coordinates. 
However, the logarithmic terms can be removed by applying a gauge transformation within the class of harmonic coordinates; the resulting system is known as the \emph{modified} harmonic (MH) coordinates~\cite{Mora:2003wt,Arun:2007sg}, which differ from the SH coordinates by log terms at 3PN order.

To transform a physical quantity expressed in harmonic coordinates to EOB coordinates, we need an appropriate coordinate transformation. Typically, one looks for a canonical transformation that maps the \emph{conservative} dynamics of both coordinate systems. However, since we work at 3PN order, we also need to include information about the RR gauge freedom appearing at 2.5PN order \cite{Iyer:1993xi,Iyer:1995rn,Gopakumar:1997ng}, which corresponds to the leading-order \emph{dissipative} part of the dynamics.

To find the complete (conservative + dissipative) 3PN accurate mapping between harmonic (either SH or MH) coordinates $(\bm r_\text{h}, \bm v_\text{h})$ and EOB coordinates $(\bm r,\bm p)$, we follow the method employed in Ref.~\cite{Damour:2000ni}, which was used to obtain the transformation between harmonic and Arnowitt–Deser–Misner (ADM) coordinates at 3PN order for the conservative dynamics. 
The method starts with a PN-expanded ansatz for the transformation $\bm r_\text{h} = \bm r_\text{h}( \bm{r}, \bm{p})$ in terms of \emph{undetermined coefficients}. 
Then, one computes the acceleration employing Poisson brackets and adding a generic RR force to account for the gauge freedom.
Afterward, one matches this acceleration with the known 3PN expression of the relative acceleration in either SH or MH coordinates \cite{Blanchet:2002mb,Marsat:2012fn,Bohe:2015ana}, whose 2.5PN part is the RR force in a specific gauge.
This results in a system of linear equations that can be solved for the undetermined coefficients.
The solution at 2.5PN order is given in a parametrized form depending on two arbitrary constants, $\alpha$ and $\beta$, representing the coordinate dependence of the leading order RR force~\cite{Iyer:1993xi,Iyer:1995rn}.

Following this procedure, we start by proposing the ansatz
\begin{widetext}
\begin{align} \label{eq:xharm}
\bm r_\text{h} ( \bm{r}, \bm{p}) = & \;  \bm{r}
+  \epsilon^2 \left[\left(c_n p^2+\frac{c_n}{r}\right) \bm{r}+c_n \, r \,  p_r \, \bm{p}\right] 
+ \epsilon^4
   \left[\left(c_n \, p^4+ c_n \frac{p^2}{r}+c_n \frac{p_r^2}{r}+\frac{c_n}{r^2}\right) \bm{r}+\left(c_n \, p^2+\frac{c_n}{r}\right) r \, p_r \, 
   \bm p\right]
\nonumber \\
& + \epsilon^4 \left(c_n S_1^2 + c_n S_2 S_1 + c_n S_2^2 \right) \frac{\bm n}{r}
+ \epsilon^5 \left[(c_n S_1 + c_n S_2) p_\phi \frac{\bm{n}}{r^2}
	+ \left(c_n \bm n \cross \bm S_1 + c_n \bm n \cross \bm S_2\right) \frac{p_r}{r} \right]  \nonumber \\
&
+ \epsilon^5 \left(\gamma_1\frac{p_r \, \bm r}{r}+\gamma_2 \,\bm p\right)  \frac{1}{r}
+ \epsilon^6 \bigg\{
\left[c_n \,p^6+c_n \frac{p^4}{r}+ c_n \frac{p_r^4}{r}+c_n\frac{p_r^2 \, p^2}{r}+c_n\frac{p^2}{r^2}+ c_n\frac{p_r^2}{r^2}+\frac{c_n}{r^3}
+ \,\frac{\gamma_\text{ln}}{r^3}  \ln \left(\frac{r}{r_0'}\right) \right]  \bm{r}  \nonumber \\ 
&\qquad
+\left(c_n \, p^4+ c_n\frac{p^2}{r}+ c_n\frac{p_r^2}{r}+\frac{c_n}{r^2}\right) r \, p_r \, \bm p  \bigg\}  
+ \frac{\epsilon^6}{r} \bigg[
\left(c_n S_1^2 + c_n S_2 S_1 + c_n S_2^2 \right) \bm{n} \frac{p_\phi^2}{r^2}\nonumber\\
&\qquad
+ \left(c_n S_1^2 + c_n S_2 S_1 + c_n S_2^2 \right) \frac{\bm{n}}{r}
+ \left(c_n S_1^2 + c_n S_2 S_1 + c_n S_2^2 \right) p_r \bm{p} 
+ \left(c_n S_1^2 + c_n S_2 S_1 + c_n S_2^2 \right) p_r^2 \bm{n}
\bigg],
\end{align}
\end{widetext}
where each $c_n$ represents an independent unknown coefficient, but we used the same symbol to simplify the notation. 
The spin-orbit contribution only appears at 2.5PN order, because the leading order (1.5PN) is the same between harmonic and EOB coordinates.
All terms in this ansatz correspond to the conservative dynamics, except for the 2.5PN nonspinning contribution, which we parametrize by the two constants $\gamma_1$ and $\gamma_2$.
The constant $\gamma_\text{ln}$ is only needed when transforming from SH to EOB coordinates, but is zero for the transformation from MH to EOB coordinates. 

Next, we use Poisson brackets and the EOB Hamiltonian to compute the velocity and acceleration from our ansatz \eqref{eq:xharm}
\begin{subequations}
\begin{align}
\bm{v}_\text{h}(\bm{r},\bm{p}) &= \{ \bm r_\text{h}( \bm{r}, \bm{p}), H_\text{EOB}\}  ,
\label{eq:vsh} \\
\bm{a}_\text{h}(\bm{r},\bm{p}) &= \{ \bm{v}_\text{h}(\bm{r},\bm{p}), H_\text{EOB} \}  + \bm{\mathcal{F}}^\text{LO} ( \bm{r}, \bm{p}) / \mu,
\label{eq:a_harmtoEOB}
\end{align}
\end{subequations}
where all functions in the above equations are expressed in terms of the EOB variables $(\bm{r}, \bm{p})$.
Here, $\bm{\mathcal{F}}^\text{LO}$ is the leading PN order of the RR force, whose components appear in the EOMs~\eqref{eq:EOMs}, and can be written as
\begin{equation} \label{eq:FLO}
\bm{\mathcal{F}}^\text{LO} =  \mathcal{F}^\text{LO}_r \, \bm{n} + \frac{\mathcal{F}^\text{LO}_\phi}{r} \, \bm{\lambda} ,
\end{equation}
where we recall that $\bm{n}$ and $\bm{\lambda}$ are unit vectors in the radial and azimuthal directions, respectively. 

At leading PN order, the RR force components are given by
\begin{subequations}
\label{eq:FrFphiLO}
\begin{align}
\mathcal{F}^\text{LO}_r  \!&= \frac{8 \nu^2 \epsilon ^5 M^3\!}{15 \, r^3}  \frac{p_r}{\mu} \! \bigg[(3 -3 \alpha +9 \beta) \frac{p^2}{\mu^2}\! + \!(9 \alpha -15 \beta +9)  \frac{p_r^2}{\mu^2} \nonumber\\
&\quad\qquad 
+ \frac{M\left(9 \alpha -9 \beta +17\right)}{r}\bigg],
\label{eq:FrLO} \\
 \mathcal{F}^\text{LO}_\phi  \!&= \frac{8 \nu^2 \epsilon ^5 M^4}{15 \, r^3}  \, \frac{ p_\phi}{M \mu} \!
\bigg[9 (\alpha +1) \, \frac{p_r^2}{\mu^2} -3 (2 + \alpha) \, \frac{p^2}{\mu^2} \nonumber\\
&\quad\qquad
+ \frac{3 M(\alpha -2)}{r}\bigg].
\label{eq:FphiLO}
\end{align}
\end{subequations}
To obtain these expressions, one can either compute them by projecting the RR force derived in Refs.~\cite{Iyer:1993xi,Iyer:1995rn} in the radial and azimuthal directions (that is, $\mathcal{F}^\text{LO}_r = \mu \, \bm a_\text{RR} \cdot \bm{n}$ and $\mathcal{F}^\text{LO}_\phi = \mu \s r \, \bm a_\text{RR} \cdot \bm{\lambda}$, where $\bm a_\text{RR}$ is the acceleration due to the RR force), or one can compute them using the flux-balance equations with ansatz for the Schott energy and angular momentum, as was done in Ref.~\cite{Bini:2012ji}. 
Using either method, the undetermined constants $\alpha$ and $\beta$ parametrize the RR gauge freedom, and they are defined as in Refs.~\cite{Iyer:1993xi,Iyer:1995rn}. 

For spinning binaries, in addition to the transformations above, we also need to perform a transformation for the SSC, since the spin contributions in Refs.~\cite{Henry:2023tka,Cho:2022syn} used the covariant Tulczyjew-Dixon SSC~\cite{Tulczyjew:1959,Dixon:1979} but the EOB formalism uses the canonical Newton-Wigner (NW) SSC~\cite{pryce1948mass,newton1949localized}.
The SSC reflects an arbitrariness in the definition of the center of mass for spinning bodies in general relativity. 
Hence, to convert from the covariant to NW SSC, we perform a center-of-mass shift in harmonic coordinates given by the ansatz
\begin{align}
\label{eq:SSCtrans}
\bm{r}_\text{h} &= \bm{r}_\text{h, NW} + \epsilon^3 \left(c_n \bm{v}_\text{h, NW} \cross \bm{S}_1 + c_n \bm{v}_\text{h, NW} \cross \bm{S}_2\right) \nonumber\\
&\quad
+ \epsilon^5 \Big[ (c_n / r_\text{h, NW} + c_n v_\text{h, NW}^2) \bm{v}_\text{h, NW} \cross \bm{S}_1 \nonumber\\
&\qquad
+ (c_n / r_\text{h, NW} + c_n v_\text{h, NW}^2) \bm{v}_\text{h, NW} \cross \bm{S}_2\Big].
\end{align}

To summarize, starting from the harmonic-coordinate acceleration $\bm{a}_\text{h}(\bm{r}_\text{h},\bm{v}_\text{h})$ given by Refs.~\cite{Blanchet:2002mb,Marsat:2012fn,Bohe:2015ana}, we perform the SSC transformation given in Eq.~\eqref{eq:SSCtrans}, and then the coordinate transformation in Eqs.~\eqref{eq:xharm} and \eqref{eq:vsh}, i.e.,
\begin{align}
\bm{a}_\text{h}(\bm{r}_\text{h},\bm{v}_\text{h}) \xrightarrow[\text{trans.}]{\text{SSC}}
\bm{a}_\text{h}(\bm{r}_\text{h, NW},\bm{v}_\text{h, NW}) \xrightarrow[\text{trans.}]{\text{coord.}}
\bm{a}_\text{h}(\bm{r},\bm{p}).
\end{align}
Finally, we match this $\bm{a}_\text{h}(\bm{r},\bm{p})$ with the acceleration obtained from the EOB Hamiltonian in Eq.~\eqref{eq:a_harmtoEOB} and solve for the unknowns.
The final results for the harmonic-to-EOB transformation are presented in Appendix~\ref{app:harmToEOB}.
Note that the same procedure can be applied for precessing spins, but one also needs to transform the spins, since the spin vectors can differ by a rotation between different coordinates~\cite{Damour:2007nc}. However, our results here are restricted to aligned spins.


\subsubsection*{Choosing an appropriate RR gauge}

After matching the acceleration, we obtain the following relations between the gauge constants $(\alpha, \beta)$ in the LO RR force components \eqref{eq:FrFphiLO} and the constants $(\gamma_1, \gamma_2)$ in Eq.~\eqref{eq:xharm}:
\begin{subequations}
\begin{align} \label{eq:ansatz_ab}
 \gamma_1 &= \frac{8}{15} \nu \beta,
 \\
  \gamma_2 &= \frac{8}{15} \nu (2 \beta - 3 \alpha - 3). 
\end{align}
\end{subequations}
The choice $\alpha = -1$, $\beta = 0$ corresponds to the \emph{harmonic} RR gauge~\cite{Damour:1981ntn}.
For this choice, the coefficients $\gamma_1$ and $\gamma_2$ in our ansatz \eqref{eq:xharm} are zero, as the RR gauge is not being changed with our coordinate transformations.

An alternative option, employed in Ref.~\cite{Bini:2012ji} for a ``minimal decomposition'' of the energy and angular momentum fluxes, corresponds to $ \alpha = 0 $, $ \beta = 2 $.
These values are such that the LO Schott contribution to the angular momentum is zero, $ J _{ \text{Schott}} ^{ \text{LO}}= 0 $, and make the LO RR force components \eqref{eq:FrFphiLO} agree with Eqs.~(3.66) and (3.68) of Ref.~\cite{Bini:2012ji}, respectively.
Alternatively, the values $\alpha = 5/3$, $\beta = 3 $ are associated to the ADM RR gauge~\cite{Schafer:1982, Arun:2009mc}.
However, as we will see below in Eq.~\eqref{eq:phasediff}, these choices would introduce a modification to the QC harmonic orbital phase.

Another possible gauge choice is the one used by the \texttt{SEOBNR} waveform models, which was based on Refs.~\cite{Buonanno:2000ef,Buonanno:2005xu}, but is valid for QC orbits only. 
In this gauge, the radial and tangential components of the RR force satisfy Eq.~\eqref{eq:ratioFrFphi}, which reads $p_\phi \,\mF_r^\text{qc} = p_r \, \mF_\phi^\text{qc}$.
This can be achieved by setting $\alpha = -16/3$, with an arbitrary value for $ \beta $, as can be seen from the leading order RR force in Eqs.~\eqref{eq:FrFphiLO}: if we use the QC relations $p^2 = 1/r$ and $p_r = 0$, we observe that the dependence on $\beta$ cancels out of the RR force. 
Therefore, one can freely choose the value of $\beta$ within the same EOB QC RR gauge. 
In the \texttt{SEOBNRv4EHM} model \cite{Khalil:2021txt, Ramos-Buades:2021adz}, the RR force in Eq.~(40) of Ref.~\cite{Khalil:2021txt} can be recovered with the values $\alpha = -16/3$ and $\beta = -1$ in Eqs.~\eqref{eq:FrFphiLO}. 
These values appeared naturally when generalizing the QC RR gauge choice to eccentric orbits.

However, in \texttt{SEOBNRv5EHM} \cite{Gamboa:2024hli}, we employ a different RR gauge for the following reason. If we apply the QC limit to the transformation for the azimuthal angle $\phi$ from harmonic to EOB coordinates, given by Eq.~\eqref{eq:phih}, we obtain
\begin{equation} \label{eq:phasediff}
\phi_\text h = 
\phi + 
\frac{8 \,\nu }{15} ( 2 \beta -3
   \alpha-3) \epsilon^5 x^{5/2}.
\end{equation}
Thus, depending on the RR gauge choice, we can have a modification of the QC harmonic orbital phase.\footnote{
The orbital phase for QC orbits is sometimes computed from the binding energy $E$ and energy flux $\Phi_E$ via
\begin{equation*} 
\phi = \int \Omega \, \di t =  \int \Omega \,\frac{\di E/ \di \Omega}{\Phi_E (\Omega) } \, \di \Omega  = - \frac{x^{-5/2}}{32 \nu }  + \Order(\epsilon^2).
\end{equation*}
The second step uses the balance equations~\eqref{eq:fluxbalance}, which contain Schott terms that are not necessarily zero, even for QC orbits.
Thus, the transformation in Eq.~\eqref{eq:phasediff} accurately accounts for the LO RR gauge freedom.
}
Since the expressions of the waveform modes depend on the orbital phase, this phase difference would change the expressions of the QC modes. 
To avoid this, we select the values
\begin{equation} \label{eq:abnewgauge}
\alpha = -16/3 \quad \text{and} \quad \beta = -13/2,
\end{equation}
such that the harmonic and EOB orbital phases in Eq.~\eqref{eq:phasediff} agree in the QC orbit limit, while still satisfying the QC RR gauge in Eq.~\eqref{eq:ratioFrFphi} employed in \texttt{SEOBNR} waveform models.


\subsection{The quasi-Keplerian parametrization}
\label{sec:quasiKeplerian}

The instantaneous state of a gravitationally bound system in an eccentric orbit can be parametrized in terms of an angle, which describes the motion along the orbit, and two secularly evolving quantities. The latter can be chosen to be gauge-invariant quantities such as the energy and angular momentum of the system, or gauge-dependent quantities such as the semi-major axis and some definition of the eccentricity.

A useful parametrization of the \emph{conservative} dynamics that is widely employed in PN calculations is the \emph{quasi-Keplerian} (QK) parametrization~\cite{damour1985general,Damour:1988mr,schafer1993second,wex1995second}, which was extended to 3PN order for nonspinning binaries in harmonic and ADM coordinates in Ref.~\cite{Memmesheimer:2004cv}, and to 4PN in ADM coordinates in Ref.~\cite{Cho:2021oai}. 
The 3PN spin contributions were derived in Refs.~\cite{Tessmer:2010hp,Tessmer:2012xr} in ADM coordinates using the NW SSC, and in Ref.~\cite{Henry:2023tka} in harmonic coordinates using the covariant SSC.
The QK parametrization has been used for computing the 3PN energy and angular momentum fluxes, and the waveform modes for eccentric orbits in Refs.~\cite{Ebersold:2019kdc,Boetzel:2019nfw,Mishra:2015bqa,Arun:2007sg,Arun:2007rg,Arun:2009mc,Henry:2023tka}. 
Therefore, since we use the results of these papers, we briefly summarize the QK parametrization, and then relate it to the Keplerian parametrization, which we employ in \texttt{SEOBNRv5EHM}.

In the QK parametrization, the conservative binary dynamics is parametrized as
\begin{subequations}
\label{eq:dynamical_variables_QK}
\begin{align}
r_\text h &= a_r (1 - e_r \cos u), 
\label{eq:rQK} \\
 l &\equiv n (t - t_\text p) \nonumber\\
&= u - e_t \sin u + f_t \sin V + g_t (V - u) \nonumber\\
&\qquad + i_t \sin 2 V + h_t \sin 3V ,
\label{eq:lQK} \\
\frac{\phi_\text h - \phi_\text p}{K} &= V + f_\phi \sin 2V + g_\phi \sin 3V \nonumber\\
&\qquad
+ i_\phi \sin 4 V + h_\phi \sin 5V,
\label{eq:phiQK} \\
V &\equiv 2 \arctan  \left[  \left( \frac{1 + e_\phi}{1 - e_\phi} \right)^{1/2} \tan \frac{u}{2} \right].
\label{eq:VQK}
\end{align}
\end{subequations}
Here, $r_\text h$ is the relative separation in harmonic coordinates, $a_r$ is the semi-major axis of the orbit, $e_r$ is the radial eccentricity, $u$ is the eccentric anomaly, $l$ is the mean anomaly, $e_t$ is the time eccentricity, $n = 2 \pi / P$ is the mean motion, with $P$ the orbital period, $\phi_\text h$ is the harmonic orbital phase at time $t$, $\phi_\text p$ is the orbital phase at the instant $t_\text p$ of periastron passage, $K$ is the periastron precession, $e_\phi $ is the azimuthal phase eccentricity, $V$ is the true anomaly, and the functions $f_t, g_t, i_t, h_t, f_\phi, g_\phi, i_\phi, h_\phi$ parametrize the 2PN and 3PN relativistic corrections to the motion. 
Additionally, Eq.~\eqref{eq:lQK} is known as \emph{Kepler's equation} and it can be inverted order by order in an eccentricity expansion to give $u$ as a function of $l$~\cite{Boetzel:2017zza}. References~\cite{Memmesheimer:2004cv,Henry:2023tka} provide a complete description of the QK parametrization and the explicit 3PN expressions of all the previous orbital elements.
Additionally, Fig.~1 in Ref.~\cite{Memmesheimer:2004cv} explains geometrically the difference between the eccentric and true anomalies.

Using this parametrization, we can write all the dynamical variables $ ( r _{ \text{h}}, \dot r _{ \text{h}}, \phi_{ \text{h}}, \dot \phi _{ \text{h}} ) $ in terms of two secularly varying quantities, and either the mean anomaly $l$ with an eccentricity expansion, or the eccentric anomaly $u$ without an eccentricity expansion. 
A common choice for the two secular quantities is the gauge-invariant orbit-averaged orbital frequency $x = \langle M \Omega \rangle ^{2/3}$ and the time eccentricity $e_t$.
This is a useful parametrization to recover known results for QC orbits in the limit $ e_t \to 0 $.

However, a complete parametrization of the dynamics at 3PN order requires the inclusion of dissipative effects at 2.5PN order, which lead to a secular variation of the conserved quantities, and are hence known as \emph{postadiabatic} (PA) contributions.
Importantly, the PA contributions are needed to get the complete eccentric waveform modes at 3PN order in the QK parametrization~\cite{Mishra:2015bqa, Boetzel:2019nfw}.
These contributions have been calculated using the method of variation of constants in Refs.~\cite{Damour:1983tz, Damour:2004bz, Konigsdorffer:2006zt, Moore:2016qxz, Boetzel:2019nfw}. 

Schematically, the method of variation of constants consists in the separation of the orbital elements into a secularly evolving part (RR timescale) and a fast periodic oscillatory contribution (orbital timescale). 
For this purpose, Refs.~\cite{Boetzel:2019nfw,Ebersold:2019kdc} employ the parameters $ \left( x, e_t, l, \lambda \right) $, where $\lambda$ is an auxiliary angle linear in $l$. 
Then, one performs the separation
\begin{subequations}
\label{eq:QK_PA_contributions}
\begin{align}
x(t) &= \bar x(t) + \tilde x(t),
\label{eq:xPA} \\
e_t(t) &= \bar e_t(t) + \tilde e_t(t) ,
\label{eq:ePA} \\
l(t) &= \bar l(t) + \tilde l(t),
\label{eq:lPA}\\
\lambda(t) &= \bar \lambda(t) + \tilde \lambda(t) ,
\label{eq:lambdaPA}
\end{align}
\end{subequations} 
where the bar $\, \bar{} \,$ denotes the secularly evolving (or orbit-averaged) part, and the tilde $\, \tilde{} \,$ denotes the fast periodic oscillations that start at 2.5PN order (i.e.,~$ \tilde x, \,\tilde e_t,\, \tilde l, \,\tilde \lambda \propto \mathcal O(\epsilon^5) $), namely the PA contributions. These contributions are functions of the secular variables $\left(\bar x, \bar e_t, \bar l\right)$, with the explicit expressions given in Ref.~\cite{Boetzel:2019nfw}.\footnote{
We are grateful to Yannick Boetzel and Michael Ebersold for providing us with \emph{Mathematica} notebooks with complete expressions for $ \tilde x, \tilde e_t, \tilde l $ and $ \tilde \lambda $. We provide these expressions in the Supplemental Material.
}

The harmonic orbital phase can also be decomposed into a secularly evolving part and a fast periodic oscillation as
\begin{align} \label{eq:philW}
\phi_\text h &= \lambda + W(l)\,, \nonumber\\
&= \bar \lambda + \bar W\left(\bar l\right) + \tilde \lambda + (\tilde v- \tilde l )\,, \nonumber\\
&\equiv  \bar \phi_\text h + \tilde \phi_\text h,
\end{align}
with $\bar \phi_\text h \left(\bar x , \bar e_t, \bar l\right)= \bar \lambda + \bar W\left(\bar l\right)$, where $ W(l) $ is an oscillatory function that is $2 \pi$-periodic in $l$, and $\tilde \phi_\text h \left(\bar x , \bar e_t, \bar l\right) = \tilde \lambda + (\tilde v- \tilde l )$ is the PA contribution to the phase (see Sec.~III of Ref.~\cite{Boetzel:2019nfw} for the definition of all these quantities).
Also, note that the variables appearing in Eqs.~\eqref{eq:dynamical_variables_QK} correspond to their secularly evolving contribution, since these equations are derived for the conservative part of the dynamics.

Therefore, any dynamical quantity written in terms of $ (x, e_t, l) $ receives PA contributions at 2.5PN order. To compute these contributions, one simply substitutes $ x \to \bar x + \tilde x $, $ e_t \to \bar e_t + \tilde e_t $, and $ l \to \bar l + \tilde l $, then performs a PN expansion.


\subsection{The Keplerian parametrization}
\label{sec:Keplerian}

In \texttt{SEOBNRv5EHM}, we parametrize the eccentric dynamics in terms of the gauge-invariant dimensionless orbit-averaged frequency parameter $x  = \langle M\Omega \rangle^{2/3}$, and a definition of eccentricity $ e $ coming from the \emph{Keplerian parametrization} of the dynamics~\cite{darwin1959gravity}. 
We use the Keplerian parametrization instead of the QK one because of its simplicity, as evident from the number of variables in Table~\ref{tab:QKPsummary}, and since we do not need to reparametrize the orbital phase $\phi$.
We also use $x$ and $e$ as variables, so we can easily recover the QC waveform model \texttt{SEOBNRv5HM} in the zero-eccentricity limit, since $ \langle \Omega \rangle$ reduces to the instantaneous angular frequency $\Omega$ when $e\to 0$.

In the Keplerian parametrization for the conservative dynamics, the relative separation is given by
\begin{equation} \label{eq:rupechi}
r = \frac{M}{u_p (1 + e \cos \zeta)},
\end{equation}
where $\zeta$ is the relativistic anomaly, $e$ is the Keplerian eccentricity, and $u_p$ is the inverse semilatus rectum scaled by the total mass, i.e., $u_p\equiv M/p$ with $p$ being the semilatus rectum, but we do not use this notation to avoid confusion with the total momentum. Note that the information of the relativistic precession of the periastron is contained in the evolution of the angle $ \zeta $, since in the polar plane $ \left( r, \zeta \right) $ the orbit corresponds to a nonprecessing ellipse, which slowly decreases in size due to the emission of GWs. 

Using the Keplerian parametrization, we can express the eccentric-orbit dynamics in terms of $ (x , e, \zeta) $, with no dependence on $ u_p $. 
The process to obtain the PN expressions needed for such parametrization is described in Appendix~\ref{app:Keplerian}. 
As a result of this procedure, we obtain 3PN expressions for the conserved binding energy $E = H_\text{EOB} - M$ and EOB variables $ \left( r, p_r, p_\phi \right) $ in terms of $ \left( x, e, \zeta \right) $, which at leading orders read
\begin{subequations}
\label{eq:ErprL_xez}
\begin{align}
E &= -\mu \frac{x}{2}+ \frac{\epsilon^2 \mu x^2 }{24 \left(1-e^2\right)}\left[9+\nu-e^2 (-15+\nu ) \right] + \dots,
\label{eq:Exe} \\
p_\phi &= M \mu \sqrt{ \frac{1 - e^2}{x}} + \dots ,
\label{eq:Lxe} \\
r &= M \frac{1 - e^2}{x \, (1 + e \cos \zeta)} + \dots,
 \label{eq:rxechi} \\
p_r &= \mu \sqrt{ \frac{x}{1 - e^2}}  \, e \sin \zeta + \dots.
\label{eq:prxechi} 
\end{align}
\end{subequations}
The full 3PN relations are given in Eqs.~\eqref{eq:app:ErprL_xez} of Appendix~\ref{app:Keplerian}.
However, we note that these relations do not include information about the dissipative contributions to the dynamics.
This is discussed in the following subsection.


\subsection{Relating the quasi-Keplerian and Keplerian parametrizations}
\label{sec:relating_parametrization}

The QK parametrization has been used to compute the complete waveform modes \cite{Mishra:2015bqa, Boetzel:2019nfw,Ebersold:2019kdc,Henry:2023tka}, and energy and angular momentum fluxes \cite{Arun:2007sg,Arun:2007rg,Arun:2009mc,Henry:2023tka} for eccentric orbits to 3PN order.
Therefore, to translate those results into the EOB formalism, we need to relate the QK parameters to the Keplerian parameters, which we employ in this work. 
In this subsection, we explicitly write quantities with bars and tildes, which denote the secularly evolving and fast oscillatory parts, respectively, of the different orbital elements.

We start by relating the conservative part of both parametrizations. Since the results in the literature for the modes and fluxes are given in terms of $(\bar x, \bar e_t, \bar l)$, we only need the transformation rules for $ \bar e_t $ and $ \bar l $.
No transformation is needed for $ \bar x $ since it is gauge invariant. 

The transformation for $ \bar e_t $  is easily computed by relating the gauge-invariant binding energy in both parametrizations [i.e., $E(x,e)$ from Eq.~\eqref{eq:Exe} and $E(x,e_t)$ from Eq.~(7.11c) of Ref.~\cite{Arun:2007sg}].\footnote{In Eqs.~\eqref{eq:ErprL_xez}, $ x $ and $ e $ correspond to their secularly evolving part $ \bar x $ and $ \bar e $, since these expressions are derived from the conservative dynamics.}
Combining these equations, we get 
%
\begin{equation}
\label{eq:et}
\bar e_t = \bar e \left[1+ \epsilon ^2  x (\nu -3) + \dots \right].
\end{equation}
The full 3PN expression is given by Eq.~\eqref{eq:app:et}.

The relation between the mean anomaly $\bar l$ and the relativistic anomaly $\bar \zeta$ can be derived from the transformation for $r$ from harmonic to EOB coordinates, which is given by Eq.~\eqref{eq:rh} after removing the 2.5PN RR contribution.
In that equation, we replace the EOB-coordinates $r$ by its expression in the Keplerian parametrization from Eq.~\eqref{eq:rxechi}, and replace the harmonic-coordinates $r_\text{h}$ by its expression in the QK parametrization in terms of $(\bar{x},\bar{e}_t,\bar{l})$.
The latter is obtained from Eq.~\eqref{eq:rQK} after replacing $a_r(\bar{x},\bar{e}_t)$ and the eccentricity expansion of $u(\bar{x},\bar{e}_t,\bar{l})$ from the results of Refs.~\cite{Memmesheimer:2004cv,Boetzel:2017zza,Henry:2023tka}.
The final result, at leading PN order and expanded in eccentricity to $\Order(e^6)$, is given by
%
\begin{align}
\label{eq:mean_anomaly}
\bar l &= \bar \zeta -2 \bar{e} \sin\bar{\zeta} 
+\frac{\bar{e}^2}{64} \left(3 \bar{e}^4+8 \bar{e}^2+48\right) \sin (2 \bar{\zeta }) \nonumber\\
&\quad
-\frac{\bar{e}^3}{24} \left(3 \bar{e}^2+8\right) \sin (3 \bar{\zeta })
+\frac{\bar{e}^4}{32} \left(3 \bar{e}^2+5\right) \sin (4 \bar{\zeta })\nonumber\\
&\quad
- \frac{3}{40} \bar{e}^5 \sin (5 \bar{\zeta })
+\frac{7}{192} \bar{e}^6 \sin (6 \bar{\zeta }) + \Order(\bar{e}^7) + \dots.
\end{align}
The 3PN expression expanded to $\Order(e^2)$ is in Eq.~\eqref{eq:app:mean_anomaly}, and the expansion to $\Order(e^6)$ is in the Supplemental Material.

Now, we move on to the task of relating the dissipative part of both parametrizations.
Our objective is to compute the PA contributions to the parameters $ (x, e, \zeta)$, such that
\begin{subequations}
\label{eq:ansatz_PA_Kep}
\begin{align}
x &= \bar x + \tilde x,
\\
e &= \bar e + \tilde e,
\\
\zeta &= \bar \zeta + \tilde \zeta,
\end{align}
\end{subequations}
with $ \tilde x $, $ \tilde e $, and $ \tilde \zeta $ being functions of $ (\bar x, \bar e, \bar \zeta) $ and of order $ \mathcal O (\epsilon^5) $.
It is important to note that these PA contributions depend on the chosen RR gauge.
Thus, when transforming the QK PA contributions to the Keplerian PA contributions, we need to include information about the RR gauge. 

For this procedure, we require the PN expressions for the conserved energy and angular momentum of the system in harmonic coordinates and in the QK parametrization~\cite{Memmesheimer:2004cv,Henry:2023tka}, which we denote as $\bar{E}_\text{h}(\bar{x},\bar{e}_t)$ and  $\bar{p}_{\phi, \s\text h}(\bar{x},\bar{e}_t)$.
To these expressions, we can add PA contributions from Eqs.~\eqref{eq:QK_PA_contributions}, leading to equations for $E_\text{h}\left(x, e_t,\bar{l}\right)$ and $p_{\phi, \s \text h}\left(x, e_t,\bar{l}\right)$, where $\bar{l}$ only appears at 2.5PN order.
Since the PA contributions $(\tilde{x},\tilde{e}_t)$ are known in terms of $(\bar{x},\bar{e}_t,\bar{l})$ from Ref.~\cite{Boetzel:2019nfw}, we can express the energy and angular momentum in terms of the Keplerian parameters $ \left( \bar x, \bar e, \bar \zeta \right) $ using the transformations in Eqs.~\eqref{eq:et} and \eqref{eq:mean_anomaly}. 
Thus, we get the functions $E_\text{h}\left(\bar x, \bar e, \bar \zeta \right) $ and $p_{\phi, \s\text h} \left(\bar x, \bar e, \bar \zeta \right) $ that include the PA contributions.

On the other hand, we introduce the ansatz~\eqref{eq:ansatz_PA_Kep} into the energy and angular momentum in the Keplerian parametrization, Eqs.~\eqref{eq:Exe} and \eqref{eq:Lxe}, to obtain ${E(\bar{x}+\tilde{x}, \bar{e}+\tilde{e})}$ and $ {p_\phi(\bar{x}+\tilde{x}, \bar{e}+\tilde{e})}$. 
Since the energy and angular momentum are gauge-invariant quantities, we have that $ E_\text{h}(\bar x, \bar e, \bar \zeta ) = E(\bar{x}+\tilde{x}, \bar{e}+\tilde{e}) $ and $ p_{\phi, \s \text h}(\bar x, \bar e, \bar \zeta) = p_\phi(\bar{x}+\tilde{x}, \bar{e}+\tilde{e}) $. 
This creates a system of two equations for $ \tilde x $ and $ \tilde e $, whose solution, expanded in eccentricity, 
is given by
%
\begin{subequations}
\label{eq:tilde_x_e}
\begin{align}
\tilde x  \left(\bar x, \bar e, \bar \zeta \right) &= \nu \epsilon^5 \bar{x}^{7/2} \left[\left(\frac{16 \alpha }{5}+\frac{416}{5}\right) \bar{e} \sin \bar{\zeta}
+ \Order(\bar{e}^2)\right],
\\
\tilde e  \left(\bar x, \bar e, \bar \zeta \right) &= -\nu \epsilon^5 \bar{x}^{5/2} \bigg[
\frac{64}{5} \sin\bar{\zeta}
+ \left(\frac{8 \alpha }{5}+\frac{184}{15} \right) \bar{e} \sin \left(2 \bar{\zeta }\right) \nonumber\\
&\qquad\qquad
 + \Order(\bar{e}^2)
\bigg],
\end{align}
\end{subequations}
where the $\Order(\bar{e}^6)$ expansions are provided in Eqs.~\eqref{eq:app:tilde_x_e}.

Next, to obtain $ \tilde \zeta $ we employ Eq.~\eqref{eq:rh} for $r_\text{h}(r,p)$ (or, alternatively, the equations for $\dot{r}_\text{h}$ or $\dot{\phi}_\text{h}$). 
On the left-hand side of this equation, we substitute the expression for $ r _{ \text{h}} $ in terms of $ \left( \bar x, \bar e_t, \bar l \right) $, 
and we apply the PA contributions \eqref{eq:QK_PA_contributions}; then, we substitute the transformations \eqref{eq:et} and \eqref{eq:mean_anomaly} to write $ r _{ \text{h}} $ in terms of $ \left(\bar x, \bar e, \bar \zeta \right) $. On the right-hand side of Eq.~\eqref{eq:rh}, we substitute Eqs.~\eqref{eq:rxechi}--\eqref{eq:Lxe}, the ansatz \eqref{eq:ansatz_PA_Kep} and the solutions \eqref{eq:tilde_x_e}. After solving for $ \tilde \zeta $ and expanding up to $ \mathcal O(e^6) $,  
we get
%
\begin{align} \label{eq:tildez}
\tilde \zeta  \left(\bar x, \bar e, \bar \zeta \right)  &= 
-\nu \epsilon^5 \bar{x}^{5/2} \bigg\{\frac{64 \cos (\bar{\zeta} )}{5 \bar{e}}
-\frac{16 \alpha }{5}+\frac{8 \beta }{5}+\frac{176}{5} \nonumber\\
&\qquad
+\left(\frac{8 \alpha }{5}+\frac{184}{15}\right) \cos \left(2 \bar{\zeta }\right)
+ \Order(\bar{e}) \bigg\},
\end{align}
and the $\Order(\bar{e}^5)$ expansion is provided in Eq.~\eqref{eq:app:tildez}.
Thus, Eqs.~\eqref{eq:tilde_x_e} and \eqref{eq:tildez} constitute the PA contributions to the Keplerian parameters $ (x, e, \zeta) $.

In this way, any quantity given in harmonic coordinates can now be written in terms of $ \left(\bar x, \bar e, \bar \zeta \right) $ with the transformations \eqref{eq:et} and \eqref{eq:mean_anomaly}, and the PA contributions \eqref{eq:tilde_x_e} and \eqref{eq:tildez}. For the rest of this work, we will write $ x \equiv \bar x $, $ e \equiv \bar e $, and $ \zeta \equiv \bar \zeta $ for ease of notation. The equations in Sec.~\ref{sec:Keplerian} already follow this notation, since the variables appearing there correspond to their secularly evolving part. 
In particular, including the PA contributions into Eqs.~\eqref{eq:ErprL_xez}, gives the following 2.5PN pieces:
%
\begin{widetext}
\begin{subequations}
\label{eq:dyn_vars_PA}
\begin{align}
\tilde{E} &= \mu \nu \bar{x}^{7/2} \epsilon ^5 \left[
-\frac{8}{5} (\alpha +26) \bar{e} \sin (\bar{\zeta} ) 
-\frac{2}{15} (24 \alpha +83) \bar{e}^2 \sin (2 \bar{\zeta} ) 
+ \Order(\bar{e}^3)\right], \\
\tilde{p}_\phi &= M\mu \nu  \epsilon ^5 \bar{x}^2 \left[
-\frac{8}{5}  (\alpha +18) \bar{e} \sin \left(\bar{\zeta }\right)
+ \frac{2}{5} (3-4 \alpha ) \bar{e}^2 \sin \left(2 \bar{\zeta }\right)
+ \Order(\bar{e}^3)
\right],  \label{eq:pphi_PA} \\
\tilde{r} &= M \nu  \epsilon ^5 \bar{x}^{3/2} \left[
\frac{8}{15} \bar{e} (3 \alpha -3 \beta -151) \sin \left(\bar{\zeta }\right)
+ \frac{4}{45} \bar{e}^2 (9 \alpha -9 \beta +80) \sin \left(2 \bar{\zeta }\right)
+ \Order(\bar{e}^3)\right], \label{eq:r_PA} \\
\tilde{p}_r &= \mu \nu  \epsilon ^5 x^3 \left[
-\frac{64}{5} + \frac{8}{15} \bar{e} (3 \alpha -3 \beta -89) \cos \left(\bar{\zeta }\right)
+\frac{8}{45} \bar{e}^2 \left[(18 \alpha -18 \beta -409) \cos \left(2 \bar{\zeta }\right)+9 \alpha -9 \beta -888\right]
+ \Order(\bar{e}^3) \right],
\end{align}
\end{subequations}
\end{widetext}
with the $\Order(\bar{e}^6)$ expansions provided in the Supplemental Material.
Thus, combining the conservative and PA contributions, we obtain the complete relations,
\begin{subequations}
\label{eq:ErprL_xez_PA}
\begin{align}
E &= \text{Eq.~\eqref{eq:app:Exe}} + \tilde{E},
\label{eq:Exe_PA} \\
p_\phi &= \text{Eq.~\eqref{eq:app:Lxe}} +  \tilde{p}_\phi,
\label{eq:Lxe_PA} \\
r &= \text{Eq.~\eqref{eq:app:rxechi}} + \tilde{r},
 \label{eq:rxechi_PA} \\
p_r &=  \text{Eq.~\eqref{eq:app:prxechi}} + \tilde{p}_r.
\label{eq:prxechi_PA}
\end{align}
\end{subequations}


\section{Energy and angular momentum fluxes}
\label{sec:fluxes}


\subsection{Instantaneous contributions}
\label{sec:inst_fluxes}

The 3PN instantaneous contributions to the energy and angular momentum fluxes for general orbits were derived in Refs.~\cite{Arun:2007sg,Arun:2009mc} for nonspinning binaries and in Refs.~\cite{Henry:2023tka,Cho:2021mqw} with spin contributions.
Those references performed the derivation within the PN multipolar post-Minkowskian (PN-MPM) formalism~\cite{Blanchet:2013haa}, which gives PN-expanded formulas for the fluxes (and waveform modes) in terms of radiative multipole moments, and their relations to another set of multipole moments in the source frame.
The results were expressed in harmonic (both SH and MH) coordinates, with their associated RR gauge, and using the covariant SSC. 
It is particularly important to note that the 3PN EOMs are used to compute the required time derivatives of the source moments.
Therefore, the resulting general-orbit explicit expressions of the instantaneous fluxes depend on the chosen RR gauge, but the orbit-averaged fluxes do not depend on this gauge. 

The fluxes are gauge-invariant functions (i.e., they are independent of coordinate changes that leave the spacetime asymptotically flat).
Therefore, to obtain the fluxes in EOB coordinates we can simply take the known fluxes in harmonic coordinates and transform them into the EOB variables via the relations:
\begin{subequations}
\begin{align}
\Phi_E (\bm r, \bm p) &= \Phi_E^{\text{h}} [\bm r_{\text h}(\bm r, \bm p), \bm v_{\text h}(\bm r, \bm p)],
\label{eq:Efluxeobtransf} \\
\Phi_J (\bm r, \bm p)   &= \Phi_J^{\text{h}} [\bm r_{\text h}(\bm r, \bm p), \bm v_{\text h}(\bm r, \bm p)],
\label{eq:Jfluxeobtransf}
\end{align}
\end{subequations}
where the superscript ``h'' denotes that the fluxes are written either in SH or MH coordinates.
This was the method followed in Ref.~\cite{Bini:2012ji} to obtain the 2PN instantaneous EOB fluxes. 
However, starting at 2.5PN order, we need to include the RR gauge freedom in the equations of motion in order for Eqs.~\eqref{eq:Efluxeobtransf} and \eqref{eq:Jfluxeobtransf} to be correct at 3PN order. 
This is precisely accomplished with our complete (conservative + dissipative) transformation given by Eqs.~\eqref{eq:scalars_HarmToEOB}. 
The leading PN order for the EOB instantaneous fluxes is given by
%
\begin{subequations}
\label{eq:instfluxes_LO}
\begin{align}
\Phi_E^\text{inst} &= \frac{32 \nu^2 M^4}{5 r^4} \left(\frac{p^2}{\mu^2} - \frac{11p_r^2}{12\mu^2}\right) + \Order(\epsilon^2),
\label{eq:EfluxEOB} \\
\Phi_J^\text{inst} &= \frac{8 \nu M^2 p_\phi}{5 r^3} \left(\frac{2M}{r} + \frac{2 p^2}{\mu^2} - \frac{3 p_r^2}{\mu^2}\right) + \Order(\epsilon^2).
\label{eq:JfluxEOB}
\end{align}
\end{subequations}
We include the full 3PN expressions in Eqs.~\eqref{eq:EOBenergyflux} and \eqref{eq:EOBangMtmflux}, and in the Supplemental Material.

Our 3PN EOB fluxes are perfectly consistent with the 2PN results of Ref.~\cite{Bini:2012ji}. 
We also recover the same 2.5PN part of the fluxes shown in Eqs.~(3.14a) and (3.14b) of Ref.~\cite{Arun:2009mc}, which are written in a generic RR gauge. 
Finally, we note the appearance of the constant $ r_0 $ in the 3PN part of the fluxes, which comes from the MPM formalism, and it physically represents the \emph{arbitrary} difference in origin of time in the far zone (radiative coordinates) and in the near zone. However, it was noted in Ref.~\cite{Pan:2010hz} that, within the EOB formalism, one requires the fixed value
\begin{equation} \label{eq:r0EOB}
r_0 = \frac{2 M}{\sqrt{\e}},
\end{equation}
where $\e$ is Euler's number, to make PN results consistent with waveforms computed in the test body limit using the self-force formalism.
Thus, we use the value \eqref{eq:r0EOB} for $r_0$ in this work. 

With the expressions of the EOB fluxes in hand, we can compute their orbit average using the relations given in Sec.~\ref{sec:Keplerian} and Appendix \ref{app:Keplerian}, and evaluating the following integral for the energy flux,
\begin{align} 
\langle  \Phi_E^\text{inst} \rangle 
&= \frac{1}{P} \! \oint  \Phi_E^\text{inst} \, \di t = \frac{1}{P} \! \int_{0}^{2 \pi}  \Phi_E^\text{inst}  \left( \frac{\partial H}{\partial p_r} \right)^{-1} \! \frac{\di r}{\di \zeta} \, \di\zeta, 
\end{align}
and an analogous expression for $\langle \Phi_J^\text{inst} \rangle$.
Thus, the leading PN order for the orbit-averaged EOB fluxes reads
%
\begin{subequations}
\label{eq:fluxesAvgLO}
\begin{align}
\langle \Phi_E^{\text{inst}} \rangle &= \frac{32 \nu ^2 x^5}{5 \left(1-e^2\right)^{7/2}}  \left(1+\frac{73 e^2}{24}+\frac{37 e^4}{96}\right) + \Order(\epsilon^2), \\
\langle \Phi_J^{\text{inst}} \rangle &= \frac{4 \nu ^2 M x^{7/2}}{5 \left(1-e^2\right)^2} \left(8+7 e^2\right) + \Order(\epsilon^2),
\end{align}
\end{subequations}
with the 3PN results presented in Appendix~\ref{app:fluxesKepAvg}, where we add the tail contributions computed in the next subsection.
We checked that our results for the orbit-averaged fluxes agree with the fluxes derived in Refs.~\cite{Arun:2007sg,Arun:2007rg,Arun:2009mc} after transforming the Keplerian eccentricity $ e $ to the time eccentricity $ e_t $.


\subsection{Tail contributions}
\label{sec:hered_flux}

The tail part of the fluxes corresponds to integrals that depend on the entire past history of the source.
Due to the complexity of those integrals, getting closed-form analytic solutions requires specifying the binary's orbit and treating the integrands in a small-eccentricity expansion for bound orbits, or a large-eccentricity expansion for unbound orbits.

The 3PN nonspinning tail contributions for bound orbits were derived in Refs.~\cite{Arun:2007rg,Arun:2009mc,Ebersold:2019kdc} numerically and analytically to $\Order(e_t^6)$ using an orbit average, while Ref.~\cite{Henry:2023tka} obtained the spin and 1.5PN nonspinning parts with and without an orbit average to $\Order(e_t^8)$.
The orbit-averaged tail contributions in these references were given in terms of $x$ and the time eccentricity $ e_t $, so we employ our transformation \eqref{eq:et} to write the tail contributions in the Keplerian parametrization, yielding
%
\begin{subequations}
\label{eq:tailFluxesExp}
\begin{align}
\langle \Phi_E^{\text{tail}} \rangle &= \frac{128}{5} \pi  \nu ^2 x^{13/2} \epsilon ^3 \bigg[
1+\frac{2335 e^2}{192}+\frac{42955 e^4}{768} \nonumber\\
&\qquad\qquad
+\frac{6204647 e^6}{36864} + \Order(e^8) \bigg] + \Order(\epsilon^5),
\label{eq:phiE_tail} \\
\langle \Phi_J^{\text{tail}} \rangle &= \frac{128}{5} \pi  \nu ^2 M x^5 \epsilon ^3 \bigg[
1+\frac{209 e^2}{32}+\frac{2415 e^4}{128} \nonumber\\
&\qquad\qquad
+\frac{730751 e^6}{18432} + \Order(e^8) \bigg] + \Order(\epsilon^5).
\label{eq:phiJ_tail}
\end{align}
\end{subequations}

A simple resummation of the eccentricity expansion in the tail contributions can make them accurate for high eccentricities, by simply factoring out the divergent factor $1/(1-e^2)^n$, for some power $n$ depending on the PN order.
Such a factor can be seen in the leading order fluxes in Eqs.~\eqref{eq:fluxesAvgLO}, which were obtained without an eccentricity expansion.
In Ref.~\cite{Forseth:2015oua}, the authors used analytical and numerical gravitational self-force calculations to confirm the resummation's accuracy for high eccentricities, and Ref.~\cite{Henry:2023tka} compared the resummation to the numerical PN results of Refs.~\cite{Arun:2007rg,Arun:2009mc}.\footnote{
An alternative resummation for hereditary effects was introduced in Ref.~\cite{Loutrel:2016cdw} based on superasymptotic and hyperasymptotic series that are accurate to $\sim 10^{-8}$ compared to numerical PN results, but we use the simpler and physically motivated resummation of Ref.~\cite{Henry:2023tka} since we are not targeting eccentricities near unity. 
}

The resummed tail contributions at leading order read
\begin{subequations}
\begin{align}
\langle \Phi_E^{\text{tail}} \rangle &= \frac{128 \pi \nu^2  x^{13/2}}{5 \left(1-e^2\right)^5} \epsilon^3 \bigg[1+\frac{1375 e^2}{192} +\frac{3935 e^4}{768} \nonumber\\
&\qquad\qquad
+\frac{10007 e^6}{36864} + \Order(e^8)\bigg] + \Order(\epsilon^5), \\
\langle \Phi_J^{\text{tail}} \rangle &= \frac{128 \pi \nu^2 M x^5}{5 \left(1-e^2\right)^{7/2}} \epsilon^3 \bigg[
1+\frac{97 e^2}{32}+\frac{49 e^4}{128}-\frac{49 e^6}{18432}  \nonumber\\
&\qquad\qquad
+ \Order(e^8)\bigg] + \Order(\epsilon^5).
\end{align}
\end{subequations}

The total orbit-averaged EOB fluxes are then the sum of the instantaneous and tail contributions, i.e.,
\begin{subequations}
\label{eq:EJfluxes_KepOrbAv} 
\begin{align}
\langle \Phi_E  \rangle &= \langle \Phi_E^{\text{inst}} \rangle + \langle \Phi_E^{\text{tail}} \rangle,
\label{eq:EfluxKepOrbAv} \\
\langle \Phi_J \rangle &= \langle \Phi_J^{\text{inst}} \rangle + \langle \Phi_J^{\text{tail}} \rangle.
\label{eq:JfluxKepOrbAv}
\end{align}
\end{subequations}
The full 3PN expressions are provided in Eqs.~\eqref{eq:EfluxEOBOrbAv} and \eqref{eq:JfluxEOBOrbAv}.




Finally, it is possible to write the non-orbit-averaged tail contributions to the fluxes in terms of EOB variables, by using an expansion in $p_r$, which is proportional to $e$, since at leading order it is given by $p_r = e \sqrt{u_p} \sin \zeta + \Order(\epsilon^2)$.
Thus, we choose the following ansatz for the tail part of the fluxes:
%
\begin{subequations}
\label{eq:ansatz_fluxes_tails}
\begin{align}
\Phi_E^{\text{tail},\, p_r}
&=
\epsilon^3 \nu^2 \sum_{n=0}^{3}  \Big[c_n \, p_\phi p_r^{2n} r^{n-7} + \epsilon^2 c_n \, p_\phi p_r^{2n} r^{n-8}
\nonumber\\
&\qquad\qquad\quad
+ \epsilon^3 \left(c_n + c_n \ln r\right) p_r^{2n} r^{n-8} 
\nonumber\\
&\qquad\qquad\quad
+ \epsilon^3 \left( c_n \chi_S + c_n \chi_A \delta \right) p_r^{2n} r^{n-8} 
\Big]
\nonumber\\
&\qquad\qquad + \Phi_E^{\text{tail},\s r_0} + \Order(p_r^8),
\\
\Phi_J^{\text{tail},\, p_r}
&=
\epsilon^3 \nu^2 \sum_{n=0}^{3}  \Big[c_n \, p_r^{2n} r^{n-5} + \epsilon^2 c_n \, p_r^{2n} r^{n-6} 
\nonumber\\
&\qquad\qquad\quad
+ \epsilon^3 \left(c_n + c_n \ln r\right) p_\phi  p_r^{2n} r^{n-7}
\nonumber\\
&\qquad\qquad\quad
+ \epsilon^3 \left( c_n \chi_S + c_n \chi_A \delta \right) p_\phi p_r^{2n} r^{n-7}
\Big]
\nonumber\\
&\qquad\qquad
+ \Phi_J^{\text{tail},\s r_0} + \Order(p_r^8),
\end{align}
\end{subequations}
where $ c_n $ are unknown coefficients and the sum goes to $\Order(p_r^6)$ corresponding to the $\Order(e^6)$ expansion in the tail.
The terms $ \Phi_E^{\text{tail},\s r_0} $ and $ \Phi_J^{\text{tail},\s r_0} $ are determined by ensuring that the complete fluxes (instantaneous + tails) are independent of the arbitrary constant $ r_0 $.
Afterward, we use Eqs.~\eqref{eq:Lxe_PA}-\eqref{eq:prxechi_PA} to express the above equations in terms of the Keplerian parameters $ (x, e, \zeta) $, and then we compute the orbit average of the resulting equations and match them to Eqs.~\eqref{eq:tailFluxesExp} to solve for the unknown coefficients.
This procedure leads to
\begin{subequations}
\label{eq:tailfluxes_LO}
\begin{align}
\Phi_E^{\text{tail},\,p_r} &= \frac{\epsilon^3 \pi \nu^2 M^3 p_\phi}{\mu r^4} \bigg[\frac{128 M^3}{5 r^3}+\frac{332 M^2 p_r^2}{3 \mu^2 r^2}+\frac{4 M p_r^4}{9 \mu^4r} \nonumber\\
&\qquad
-\frac{73p_r^6}{450\mu^6} + \Order(p_r^8)\bigg] + \Order(\epsilon^5), \\
\Phi_J^{\text{tail},\,p_r} &= \frac{\epsilon^3 \pi \nu^2 M^3}{r^2} \bigg[\frac{128M^3}{5 r^3} + \frac{392 M^2 p_r^2}{5 \mu^2 r^2}-\frac{49 p_r^6}{225 \mu^6} \nonumber\\
&\qquad
+ \Order(p_r^8)\bigg] + \Order(\epsilon^5),
\end{align}
\end{subequations}
with the full expressions given in the Supplemental Material.


\section{Evolution equations for the Keplerian parameters}
\label{sec:evoEqs}

The time evolution of aligned-spin binaries in EOB waveform models is governed by the EOMs \eqref{eq:EOMprStr} for the variables $ (r, \phi, p_{r_*}\!, p_\phi) $ that correspond to the four degrees of freedom of the system. 
However, in this work, we express the eccentricity corrections to the RR force and waveform modes in terms of the Keplerian parameters $ (x, e, \zeta) $, mainly to recover the QC EOMs when $e \to 0$, but also because the hereditary contributions to fluxes and modes are only known analytically in an eccentricity expansion.

It is possible to replace the eccentricity $ e $ and the other orbital parameters by expressions in terms of the canonical EOB variables and their time derivatives, as was done in Refs.~\cite{Khalil:2021txt,Placidi:2021rkh,Placidi:2023ofj}.
However, these replacements can lead to inaccuracies for moderate and large eccentricities. 
In particular, for the \texttt{SEOBNRv4EHM} model~\cite{Ramos-Buades:2021adz}, it was found that using $p_r$ and $\dot{p}_r$ in the eccentricity corrections to the RR force did not give good agreement with NR waveforms, and it was therefore decided to include eccentricity corrections in the waveform modes only.
Moreover, it was noted that the use of $ p_r $ and $ \dot p_r $ introduced some irregularities in the modes near the end of the inspiral, which were manually fixed by introducing a sigmoid function with certain tuned parameters.

Thus, for \texttt{SEOBNRv5EHM}, we use the Keplerian parameters $ (x, e, \zeta) $ for the eccentricity corrections, and we supplement the EOMs \eqref{eq:EOMprStr} with certain prescriptions for computing those parameters.
Namely, we numerically solve evolution equations for $ e $ and $ \zeta $, and use a PN formula for computing the orbit-averaged frequency $ x $.
With these prescriptions, we have a well-defined QC limit when $ e \to 0 $, and we do not have irregularities in the eccentricity corrections at the end of the inspiral. 
Furthermore, the parametrization $ (x, e, \zeta) $ allows us to write the 3PN eccentricity corrections to the RR force and modes in a relatively compact form --- for example, our final expressions at 2PN order are significantly shorter than the ones obtained in Ref.~\cite{Khalil:2021txt} with the $ (r, p_{r}, \dot p_r) $ variables.
This is particularly relevant for the computational speed of the \texttt{SEOBNRv5EHM} model, since the expressions for the RR force and modes are evaluated several times to resolve the eccentricity modulations of the dynamics and the waveform.


\subsection{Derivation of the evolution equations for $ e $ and $ x $}
\label{sec:evo_eqs_e_x}

The \emph{secular} evolution equations of the orbital elements $ e $ and $ x $ can be obtained with the energy and angular momentum flux-balance equations. First, we note that we can invert $E(x,e)$ and $p_\phi(x,e)$ from Eqs.~\eqref{eq:Exe} and \eqref{eq:Lxe} to obtain conservative 3PN accurate equations for $e = e (E, p_\phi) $  and $ x = x (E, p_\phi)$, which are given at leading order by
%
\begin{subequations}
\begin{align}
e(E, p_\phi) &= \sqrt{1 + 2 p_\phi^2 E/ (M\mu^2)} + \Order(\epsilon^2),
\label{eq:esqEL} \\ 
x(E, p_\phi) &= - 2 E/\mu  + \Order(\epsilon^2).
\label{eq:xEL} 
\end{align}
\end{subequations}

Then, we can use the chain rule to relate the secular evolution of $ e $ and $ x $ to the orbit-averaged energy and angular momentum losses, and hence to the orbit-averaged fluxes, through the balance equations~\eqref{eq:fluxbalance}, where the Schott terms vanish after an orbit average.
That is
\begin{subequations}
\label{eq:evo_eqs_xe}
\begin{align}
\frac{\di e}{\di t} &= \frac{\partial e}{\partial E}  \left \langle \frac{\di E}{\di t } \right \rangle 
	+ \frac{\partial e}{\partial p_\phi} \left \langle \frac{\di p_\phi}{\di t } \right \rangle \nonumber\\
	&= -\frac{\partial e}{\partial E} \langle \Phi_E \rangle - \frac{\partial e}{\partial p_\phi } \langle \Phi_J \rangle ,
\label{eq:dedt_original}
\\
\frac{\di x}{\di t} &= \frac{\partial x}{\partial E}  \left \langle \frac{\di E}{\di t } \right \rangle 
	+ \frac{\partial x}{\partial p_\phi} \left \langle \frac{\di p_\phi}{\di t} \right \rangle\nonumber\\
	&= -\frac{\partial x}{\partial E} \langle \Phi_E \rangle  - \frac{\partial x}{\partial p_\phi }\langle \Phi_J \rangle.
\label{eq:dxdt_original}
\end{align}
\end{subequations}
For these formulas, we have assumed that the spins are parallel to the orbital angular momentum of the binary, implying that the spins are constant throughout the entire evolution to the considered PN order. 

Thus, we need expressions of the far-zone orbit-averaged fluxes to compute the secular evolution equations of $ e $ and $ x $.
The fluxes are discussed in Sec.~\ref{sec:fluxes}, and the orbit-averaged expressions are given by Eqs.~\eqref{eq:EfluxEOBOrbAv} and~\eqref{eq:JfluxEOBOrbAv}.
After substitution, we obtain the following expressions:
\begin{subequations}
\label{eq:evo_eqs_ex}
\begin{align}
\frac{\di e}{\di t} &= - \frac{\nu e x^4}{M} \left[\frac{\left(304 + 121 e^2\right)}{15 \left(1-e^2\right)^{5/2}} + \Order(\epsilon^2) \right],
\label{eq:edotsecPNexp}
\\
\frac{\di x}{\di t}  &=  \frac{2  \nu x^5}{3 M} \left[\frac{96 +292 e^2 + 37 e^4}{5 \left(1-e^2\right)^{7/2}} + \Order(\epsilon^2) \right],
\label{eq:xdotsecPNexp}
\end{align}
\end{subequations}
where the full 3PN expression for $\di e/\di t$ is given by Eq.~\eqref{eq:edotFull}, and the one for $\di x/\di t$ is given by Eq.~\eqref{eq:xdotFull}.
Note that, in these expressions, $ e $ and $ x $ correspond to their secularly evolving part $ \bar e $ and $ \bar x $, as discussed in Sec.~\ref{sec:relating_parametrization}.
However, we omit the bars $ \,\s \bar{} \,\s$ for ease of notation.

The system of equations \eqref{eq:evo_eqs_ex} can be solved numerically given some initial values $ (e_0, x_0) $. 
However, for the solution of the EOMs in \texttt{SEOBNRv5EHM}, we only employ Eq.~\eqref{eq:edotsecPNexp}, and we use a PN formula to compute $ x $ [see Eq.~\eqref{eq:xxinst} below], so that we can exactly recover the QC expressions used in \texttt{SEOBNRv5HM} \cite{Pompiliv5}.
Equation~\eqref{eq:xdotsecPNexp} is employed only to extend the available parameter space of \texttt{SEOBNRv5EHM}, as discussed at the end of Sec.~\ref{sec:ICs_ecc_orb}.
We note that Eq.~\eqref{eq:edotsecPNexp} becomes trivial when the initial eccentricity is zero, which is an advantage for the QC limit of the model since $ e $ will be fixed to zero throughout the entire evolution of the system.


\subsection{Derivation of the evolution equation for $ \zeta $}

For the conservative dynamics, without PA contributions, the 3PN evolution equation for the relativistic anomaly $ \zeta $ can be obtained by taking a time derivative of the Keplerian relation for $r$, Eq.~\eqref{eq:rupechi}, and solving for $\dot{\zeta}$, yielding
\begin{equation}
\dot{\zeta} = \frac{\dot{r}}{u_p \, e \,r^2\sin\zeta}.
\end{equation}
Then, we substitute the EOM for $\dot{r}$ \eqref{eq:rdot} and express all variables in terms of $(x,e,\zeta)$ by using Eqs.~\eqref{eq:ErprL_xez}.

Including the 2.5PN PA contributions requires additional steps. First, we express Eq.~\eqref{eq:rupechi} in terms of the dimensionless orbit-averaged frequency $ x $, and we add the corresponding PA contributions given by Eqs.~\eqref{eq:tilde_x_e} and \eqref{eq:tildez}.
This results in Eq.~\eqref{eq:rxechi_PA} for $r(x,e,\zeta)$, which we denote by $r_{ x e \zeta}$.
Next, we multiply both sides of that equation by $ (1 + e \cos \zeta) $, and solve for $ \zeta $, yielding
\begin{equation} 
\cos \zeta = \frac{1}{e} \left[-1 + \frac{r _{ x e \zeta}}{r} (1 + e \cos \zeta) \right].
\end{equation}
Now, we take a time derivative of the previous equation and solve for $ \dot \zeta $ to obtain
\begin{equation} 
\dot \zeta = - \frac{1}{\sin \zeta} \frac{\di}{\di t}\left\{ \frac{1}{e} \left[-1 + \frac{r _{ x e \zeta}}{r} (1 + e \cos \zeta) \right] \right\}.
\end{equation}
For the differentiation, we employ Eqs.~\eqref{eq:evo_eqs_ex} for $ \di e/\di t $ and $ \di x/\di t $, and Eq.~\eqref{eq:rdot} for $ \dot r $. Then, we substitute Eqs.~\eqref{eq:ErprL_xez_PA} and expand to 3PN order.
The resulting expression is a recursive relation for $ \dot \zeta $, so we solve this equation by iterating and performing a PN expansion. Hence, we obtain
%
\begin{equation} \label{eq:dotz_LO}
\dot \zeta = \frac{x^{3/2} (1 + e \cos\zeta)^2}{M\left(1-e^2\right)^{3/2}} + \Order(\epsilon^2),
\end{equation}
where the full 3PN relation is given by Eq.~\eqref{eq:chidotFull}.


\subsection{Prescription for computing $ x $}
\label{sec:prescription_x}

One of the main variables employed by the QC model \texttt{SEOBNRv5HM} is the \emph{instantaneous} orbital angular frequency $ \Omega = \dot \phi $.
This poses a problem since our expressions for the eccentricity corrections to the modes and fluxes are expressed in terms of the \emph{orbit-averaged} frequency $ x  = \langle M \Omega \rangle^{2/3}$. 
Thus, to exactly recover the expressions used in \texttt{SEOBNRv5HM}, we decided to use a PN formula for $ x $ as a function of $ (\Omega, e, \zeta) $, such that we get $ x = (M\Omega)^{2/3} $ in the limit $ e \to 0 $. 

To obtain such a formula for $ x $, we start by writing the EOM~\eqref{eq:phidot} for $\dot{\phi}$ in terms of $ \left( x, e, \zeta \right) $ using the relations between the EOB dynamical variables and Keplerian parameters from Eqs.~\eqref{eq:Lxe}--\eqref{eq:prxechi}, yielding
%
\begin{equation} \label{eq:phidotxezeta}
\dot \phi =  \Omega= \frac{x^{3/2} (1 + e \cos\zeta)^2}{M\left(1-e^2\right)^{3/2}} + \mathcal O (\epsilon^2) .
\end{equation}
Then, we PN-invert this equation to get $ x(\Omega,e,\zeta) $.
To leading order, the result reads
%
\begin{equation} \label{eq:xxinst}
x = \frac{(M \Omega)^{2/3} \left(1-e^2\right)}{(1 + e \cos \zeta)^{4/3}} +  \mathcal O (\epsilon^2),
\end{equation}
where the full 3PN expression is given by Eq.~\eqref{eq:xxinstFull}.
Eq.~\eqref{eq:xxinst} has the advantage of exactly recovering the relation $ x = (M\Omega)^{2/3} $ when $ e \to 0 $, and allowing any quantity expressed in terms of $ \left( x, e, \zeta \right) $ to be treated as a function of $ \left( \Omega, e, \zeta \right) $.
Note that we do not include PA contributions in Eqs.~\eqref{eq:phidotxezeta} and \eqref{eq:xxinst}. This is because we require a formula for the \emph{secularly evolving} variable $ \bar x $, since this variable enters in our expressions for the modes and RR force (the PA contributions are already taken into account in these expressions; we recall that we denote $ e = \bar e $, $ x = \bar x $, and $ \zeta = \bar \zeta $, for ease of notation).

From Eq.~\eqref{eq:xxinst}, we note that one needs a synchronization between $ \Omega $ and $ \zeta $ to get a secular (nonoscillating) evolution for $ x $.
The value of $ \Omega $ comes directly from the EOB equation \eqref{eq:phidot_2}, whereas $ \zeta $ comes from the Keplerian (PN-expanded) equation~\eqref{eq:dotz_LO}.
In practice, since Eqs.~\eqref{eq:dotz_LO} and \eqref{eq:xxinst} are truncated at 3PN order, we observe residual oscillations in $ x $.
In general, these oscillations do not significantly alter the modes, as testified by the high accuracy of the \texttt{SEOBNRv5EHM} model when compared against NR waveforms~\cite{Gamboa:2024hli}.
However, \emph{some} binary systems with high eccentricities and/or high frequencies rupture the synchronization between the EOB frequency $ \Omega $ and the Keplerian relativistic anomaly $ \zeta $, due to the high velocities involved.
As a consequence of this \emph{desynchronization}, the waveform modes of these challenging systems acquire unphysical modifications.
We refer the reader to the companion paper~\cite{Gamboa:2024hli} for a more detailed discussion about the robustness and accuracy of the \texttt{SEOBNRv5EHM} model.
Future developments will require revisiting the parametrization of the eccentricity corrections in terms of $ (x,e, \zeta)$ to avoid as much as possible unphysical features in the waveforms.

In this way, the system of equations used in the \texttt{SEOBNRv5EHM} model~\cite{Gamboa:2024hli} to solve the dynamics of eccentric binary systems is given by Eqs.~\eqref{eq:EOMprStr} for $(r,\phi,p_{r_*}\!,p_\phi)$, in addition to Eq.~\eqref{eq:edotsecPNexp} for $\dot{e}(x,e)$, Eq.~\eqref{eq:dotz_LO} for $\dot{\zeta}(x,e,\zeta)$, and Eq.~\eqref{eq:xxinst} for $x(\Omega,e,\zeta)$.


\section{Radiation-reaction force for eccentric orbits}
\label{sec:RRforce}

In the QC model \texttt{SEOBNRv5HM}, the RR force is chosen to be in a gauge that satisfies Eqs.~\eqref{eq:qcRRforce}, which can be written as
\begin{equation}
\label{eq:RRforceQCgauge}
\mF_\phi^\text{qc} = - \frac{\Phi_E^\text{qc}}{\Omega}, \qquad
\mF_r^\text{qc} = \frac{p_{r}}{p_\phi} \mF_\phi^\text{qc},
\end{equation}
where $\Phi_E^\text{qc}$ is the energy flux for QC orbits, expressed in terms of the waveform modes as
\begin{equation}
\label{eq:qcEflux}
\Phi_E^\text{qc} = \frac{M^2 \Omega^2}{8 \pi} \sum_{\ell=2}^8 \sum_{m=1}^{\ell} m^2\left|d_L h_{\ell m}^{\text{F, qc}}\right|^2.
\end{equation}
This way of writing the RR force improves the results coming from the EOMs when compared against NR simulations, because of the resummation of the waveform modes and their calibration to gravitational self-force and NR results. 

For general orbits, as discussed in Sec.~\ref{sec:nutshell}, one should use the \emph{flux-balance} equations~\eqref{eq:fluxbalance} to compute the RR force by solving the following equations:
\begin{subequations}
\label{eq:fluxbalanceII}
\begin{align}
&\dot r  \mathcal F_r  + \dot \phi  \mathcal F_\phi  + \dot E _{ \text{Schott}} + \Phi_E = 0 , \\
&\mF_\phi + \dot J _{ \text{Schott}} + \Phi_J = 0.
\end{align}
\end{subequations}
Therefore, given ansatz for the Schott terms and expressions for the fluxes in terms of EOB variables, we employ Eqs.~\eqref{eq:fluxbalanceII} to determine PN expansions of $ \mathcal F_r  $ and $ \mathcal F_\phi $ as functions of $ (r, p_r, p_\phi) $.
These expansions include arbitrary gauge constants that appear due to the ansatz for the Schott terms. 
In this work, we conveniently select the gauge constants in such a way that, in the QC orbit limit, we recover the RR force used in the \texttt{SEOBNRv5HM} waveform model.


\subsection{Ansatz for the Schott terms}

To solve the balance equations~\eqref{eq:fluxbalanceII} for the RR force components, it is convenient to work with the EOB variables $(r,p_r,p_\phi)$ instead of the Keplerian parameters. 
Thus, we employ the instantaneous~\eqref{eq:instfluxes_LO} and tail~\eqref{eq:tailfluxes_LO} contributions to the fluxes in terms of EOB variables, and we write appropriate ansatz for the Schott terms.
These terms are divided into nonspinning (S$^0$), spin-orbit (SO), and spin-spin (SS) parts, as
\begin{subequations}
\begin{align}
E _{ \text{Schott}}
&=
E _{ \text{Schott}} ^{ \text{S}^0}
+ E _{ \text{Schott}} ^{ \text{SO}}
+ E _{ \text{Schott}} ^{ \text{SS}} \, ,
\\
J _{ \text{Schott}}
&=
J _{ \text{Schott}} ^{ \text{S}^0}
+ J _{ \text{Schott}} ^{ \text{SO}}
+ J _{ \text{Schott}} ^{ \text{SS}} \, ,
\end{align}
\end{subequations}
where
\begin{subequations}
\begin{widetext}
\label{eq:SchottAnsatz}
\begin{align}
E_\text{Schott} ^{ \text{S}^0} &= \frac{\nu p_r}{r^2} \bigg\{
c_n  \, p_r^2 + c_n  \, p^2 + \frac{c_n}{r} 
+ \epsilon^2 \left[c_n  \, p^4+ \frac{c_n  \, p^2}{r} + \frac{c_n}{r^2} \right]
+ \epsilon^4 \left[c_n  \, p^6 + \frac{c_n  \, p^4}{r} + \frac{c_n  \, p^2}{r^2} + \frac{c_n}{r^3}\right]\nonumber\\
&\qquad
+ \epsilon^6 \left[c_n  \, p^8 + \frac{c_n  \, p^6}{r} + \frac{(c_n + c_n \ln r) p^4}{r^2} + \frac{(c_n + c_n \ln r) p^2}{r^3} + \frac{c_n + c_n \ln r}{r^4}\right] 
\bigg\} \nonumber\\
&\quad
+ \epsilon^3 \frac{\nu p_r}{r^5} \left\{c_n  \, p_\phi + \epsilon^2 p_\phi \left(c_n  \, p^2 + \frac{c_n}{r}\right) 
+ \epsilon^3 \left[(c_n + c_n \ln r) p^2 + \frac{c_n + c_n \ln r}{r}\right] \right\} \nonumber\\
&\quad
+ \epsilon^5 \nu \frac{p_r^2}{r^3} \left[c_n  \, p^4 + \frac{c_n  \, p^2}{r} + \frac{c_n}{r^2}\right],
\\
E _{ \text{Schott}} ^{ \text{SO}}
&=
\frac{\epsilon^3 \nu p_r p_\phi}{r^4}
\left[
\chi_S \left(c_n p^2 + \frac{c_n}{r}\right)
+ \chi_A \delta \left(c_n p^2 + \frac{c_n}{r}\right)
\right]
\nonumber\\
&\quad
+ \frac{\epsilon^5 \nu p_r p_\phi}{r^4}
\left[
\chi_S \left(c_n p^4 + \frac{c_n}{r^2} + \frac{c_n p^2}{r} \right)
+ \chi_A \delta \left(c_n p^4 + \frac{c_n}{r^2} + \frac{c_n p^2}{r} \right)
\right]
\nonumber\\
&\quad
+ \frac{\epsilon^6 \nu p_r^2 p_\phi}{r^4}
\left[
\chi_S \left(c_n p^4 + \frac{c_n}{r^2} + \frac{c_n p^2}{r} \right)
+ \chi_A \delta \left(c_n p^4 + \frac{c_n}{r^2} + \frac{c_n p^2}{r} \right)
\right]
\nonumber\\
&\quad
+ \epsilon^6 \nu p_r
\left[
\chi_S \left(\frac{c_n p^6}{r^3}  + \frac{c_n}{r^6}\right)
+ \chi_A \delta \left(\frac{c_n p^6}{r^3}  + \frac{c_n}{r^6}\right)
\right],
\\
E _{ \text{Schott}} ^{ \text{SS}}
&=
\frac{\epsilon^4 \nu p_r p^2}{r^4}
\left(
c_n \chi_S^2
+ c_n \chi_S \chi_A \delta
+ c_n \chi_A^2
+ c_n \kappa_S
+ c_n \kappa_A \delta
\right)
\nonumber\\
&\quad
+ \frac{\epsilon^6 \nu p_r}{r^4}
\bigg[
\chi_S^2 \left(\frac{c_n p^2}{r}  + \frac{c_n}{r^2}\right)
+ \chi_S \chi_A \delta \left(\frac{c_n p^2}{r}  + \frac{c_n}{r^2}\right)
+ \chi_A^2 \left(\frac{c_n p^2}{r}  + \frac{c_n}{r^2}\right)
+ \kappa_S \left(\frac{c_n p^2}{r}  + \frac{c_n}{r^2}\right)
\nonumber\\
&\qquad
+ \kappa_A \delta \left(\frac{c_n p^2}{r}  + \frac{c_n}{r^2}\right)
\bigg],
\\
J_\text{Schott} ^{ \text{S}^0}
&=
\frac{\nu p_r p_\phi}{r^2} \left[c_n + \epsilon^2 \frac{c_n}{r} + \epsilon^4 \frac{c_n}{r^2} + \epsilon^6 \frac{(c_n + c_n \ln r)}{r^3}\right] 
+ \frac{\nu \epsilon^3 p_r}{r^3} \left[
c_n + \epsilon^2 \frac{c_n}{r}
+ \epsilon^3 \frac{p_\phi}{r^2} (c_n + c_n \ln r)
\right]
+ \epsilon^5 c_n \frac{\nu p_\phi}{r^5},
\\
J _{ \text{Schott}} ^{ \text{SO}}
&=
\frac{\epsilon^3 \nu p_r}{r^3}
\left(
c_n \chi_S
+ c_n \chi_A \delta
\right)
+ \frac{\epsilon^5 \nu p_r p^2}{r^3}
\left(
c_n \chi_S
+ c_n \chi_A \delta
\right)
+ \frac{\epsilon^6 \nu p^4}{r^3}
\left(
c_n \chi_S
+ c_n \chi_A \delta
\right)
\nonumber\\
&\quad
+ \frac{\epsilon^6 \nu p_r p_\phi p^4}{r^3}
\left(
c_n \chi_S
+ c_n \chi_A \delta
\right),
\\
J _{ \text{Schott}} ^{ \text{SS}}
&=
\frac{\epsilon^4 \nu p_r p_\phi}{r^4}
(
c_n \chi_S^2
+ c_n \chi_S \chi_A \delta
+ c_n \chi_A^2
+ c_n \kappa_S
+ c_n \kappa_A \delta
)
\nonumber\\
&\quad
+ \frac{\epsilon^6 \nu p_r p_\phi}{r^4}
\bigg[
\chi_S^2 \left( c_n p^2 + \frac{c_n}{r} \right)
+ \chi_S \chi_A \delta \left( c_n p^2 + \frac{c_n}{r} \right)
+ \chi_A^2 \left( c_n p^2 + \frac{c_n}{r} \right)
+ \kappa_S \left( c_n p^2 + \frac{c_n}{r} \right)
\nonumber\\
&\qquad
+ \kappa_A \delta \left( c_n p^2 + \frac{c_n}{r} \right)
\bigg].
\end{align}
\end{widetext}
\end{subequations}

Each of the terms in the equations above is associated with certain contributions to the fluxes.
Namely,
\begin{itemize}
\item
In the nonspinning terms, $ E _{ \text{Schott}} ^{ \text{S}^0}$ and $ J _{ \text{Schott}} ^{ \text{S}^0}$, the first bracket corresponds to the instantaneous contributions of the fluxes, the second bracket to the tail part, and the last terms (at 2.5PN order) to the PA contributions.
\item
In the SO terms, $ E _{ \text{Schott}} ^{ \text{SO}}$ and $ J _{ \text{Schott}} ^{ \text{SO}}$, all the terms correspond to instantaneous contributions, except for the last terms (in the respective last line), which are associated with the tail contributions.
\item
In the SS terms, $ E _{ \text{Schott}} ^{ \text{SS}}$ and $ J _{ \text{Schott}} ^{ \text{SS}}$, all the terms correspond to instantaneous contributions.
\end{itemize}



%
%

\subsection{PN-expanded radiation-reaction force}
\label{sec:PN_RR}

Inserting the ansatz for the Schott terms \eqref{eq:SchottAnsatz} in the balance equations~\eqref{eq:fluxbalanceII}, we solve for the RR force components while enforcing the gauge choice in Eq.~\eqref{eq:RRforceQCgauge} for the circular-orbit limit.\footnote{
At 3PN order, the circular-orbit limit is obtained by setting $ p_r = - 64 \s \nu \s \epsilon^5 /(5 \s r^3) $ and $p_\phi = \sqrt{r} + \epsilon^2 3 / (2 \sqrt{r}) + \dots$, which is the solution of $\dot{p}_r = -\partial H_\text{EOB}/\partial r = 0$ for $p_\phi(r)$.
}
Additionally, for the leading PN order, we choose the unknowns to give agreement with the QC orbital phase in harmonic coordinates, as discussed in Sec.~\ref{sec:harmToEOB}.
This procedure gives a unique solution for the unknowns in the ansatz \eqref{eq:SchottAnsatz}, which was chosen to contain the minimum number of unknowns needed to solve the flux-balance equations.

The final expressions for the components of the PN-expanded RR force obtained with this method are\footnote{
The RR force components in Eqs.~\eqref{eq:RRforces_EOB} have recently been used in Ref.~\cite{Faggioli:2024ugn} to study the impact of different types of factorizations of the RR force in the test-mass limit.
There, the eccentricity contributions are written in terms of the parametrization $ (r, p_{r_*}, \dot p_{r_*}) $, which was used in the \texttt{SEOBNRv4EHM} model \cite{Ramos-Buades:2021adz}, as opposed to the parametrization $ (x, e, \zeta) $ that we employ in this work.
}
%
\begin{widetext}
\begin{subequations} \label{eq:RRforces_EOB}
\begin{align}
\mF_\phi ^{ \text{PN}}( r, p_r, p_\phi )  &= \frac{8\nu p_\phi  }{15 r^3} \left[ \frac{10 p_\phi ^2}{r^2}-\frac{22}{r}-29 p_r^2\right] + \frac{\epsilon ^2 \nu p_\phi}{r^3} \bigg[
p_r^4 \left(\frac{44 \nu }{7}+\frac{116}{35}\right)
-\frac{p_{\phi }^2}{r^3}\left(\frac{484 \nu }{105}+\frac{3833}{105}\right) +\frac{p_r^2 p_{\phi }^2}{r^2}\left(\frac{332 \nu }{35}+\frac{24}{7}\right) \nonumber\\
&\qquad
-\frac{p_{\phi }^4}{r^4}\left(\frac{8 \nu }{35}+\frac{186}{35}\right)
+\frac{p_r^2}{r}\left(\frac{8324 \nu }{105}+\frac{15563}{105}\right) +\frac{1}{r^2} \left(\frac{1684 \nu }{105}+\frac{2071}{35}\right)
\bigg] + \Order(\epsilon^3), \\
\mF_r ^{ \text{PN}} ( r, p_r, p_\phi ) &= \frac{4 \nu p_r }{15 r^3}\left[-\frac{79 p_\phi^2}{r^2}+\frac{55}{r}+38 p_r^2\right]  + \frac{\epsilon^2 \nu p_r}{r^3} \bigg[
p_r^4\left(\frac{16 \nu }{35}+\frac{372}{35}\right)
-\frac{p_{\phi }^2}{r^3}\left(\frac{3032 \nu }{105}+\frac{3317}{105}\right) +\frac{p_r^2 p_{\phi }^2}{r^2}\left(\frac{442}{35}-\frac{28 \nu }{5}\right) \nonumber\\
&\qquad
-\frac{p_{\phi }^4}{r^4}\left(\frac{332 \nu }{35}+\frac{24}{7}\right)
+\frac{p_r^2}{r}\left(\frac{1160 \nu }{21}+\frac{14057}{105}\right) +\frac{1}{r^2} \left(\frac{5204 \nu }{105}+\frac{921}{7}\right)
\bigg] + \Order(\epsilon^3),
\end{align}
\end{subequations}
where, in these equations, we used dimensionless variables scaled as in Eq.~\eqref{eq:dimlessVars}.

Next, we write these components in terms of the Keplerian parameters $ \left( x, e, \zeta \right) $ by employing Eqs.~\eqref{eq:Lxe_PA}--\eqref{eq:prxechi_PA}, yielding
%
\begin{subequations} \label{eq:RRforces_Kep}
\begin{align}
\mF_\phi ^{ \text{PN}} ( x, e, \zeta )  &= - \frac{\nu  x^{7/2} (e \cos\zeta+1)^3}{\left(1-e^2\right)^{7/2}} \left[\frac{52}{5} e^2 \cos (2 \zeta )-\frac{76 e^2}{15}-\frac{16}{15} e \cos\zeta-\frac{32}{5}\right]\nonumber\\
&\quad
+\epsilon ^2 \frac{\nu  x^{9/2} (e \cos\zeta+1)^3}{\left(1-e^2\right)^{9/2}} \bigg\{
e^4 \left(\frac{11569}{420}-\frac{773 \nu }{315}\right)
-e^4 \left(\frac{3 \nu }{7}+\frac{19}{28}\right) \cos (4 \zeta )
-e^3 \left(\frac{908 \nu }{35}+\frac{955}{21}\right) \cos (3 \zeta ) \nonumber\\
&\qquad
+\left[e^4 \left(\frac{932 \nu }{105}-\frac{2153}{35}\right)+e^2 \left(-\frac{1178 \nu }{21}-\frac{17977}{210}\right)\right] \cos (2 \zeta )
+e^2 \left(\frac{1946 \nu }{45}+\frac{3121}{42}\right)
+\frac{56 \nu }{3}+\frac{2494}{105} \nonumber\\
&\qquad
+\left[e^3 \left(\frac{1208 \nu }{63}+\frac{186}{35}\right)+e \left(\frac{836 \nu }{45}+\frac{629}{35}\right)\right] \cos\zeta
\bigg\}  + \Order(\epsilon^3),
\label{eq:Fphi_PN_Kep} \\
\mF_r  ^{ \text{PN}} ( x, e, \zeta )  &= -\frac{e \nu  x^{9/2} (e \cos\zeta+1)^3 \sin\zeta}{\left(1-e^2\right)^{9/2}} \left[\frac{78}{5} e^2 \cos (2 \zeta )+\frac{82 e^2}{15}+\frac{412}{15} e \cos\zeta+\frac{32}{5}\right] \nonumber\\
&\quad
+ \epsilon ^2 \frac{e \nu  x^{11/2} (e \cos\zeta+1)^3 \sin\zeta}{\left(1-e^2\right)^{11/2}} \bigg\{
e^4 \left(\frac{6339}{140}-\frac{86 \nu }{7}\right)
-e^4 \left(\frac{3 \nu }{7}+\frac{19}{28}\right) \cos (4 \zeta )
-e^3 \left(\frac{194 \nu }{7}+\frac{2687}{42}\right) \cos (3 \zeta ) \nonumber\\
&\qquad
+\left[e^4 \left(\frac{3849}{35}-\frac{993 \nu }{35}\right)+e^2 \left(-\frac{5083 \nu }{105}-\frac{25901}{210}\right)\right] \cos (2 \zeta ) 
+e^2 \left(\frac{717}{10}-\frac{165 \nu }{7}\right)
+\frac{104 \nu }{5}+\frac{3166}{105} \nonumber\\
&\qquad
+\left[e^3 \left(\frac{39677}{210}-\frac{8432 \nu }{105}\right)+e \left(\frac{1654 \nu }{105}+\frac{853}{105}\right)\right] \cos\zeta
\bigg\} + \Order(\epsilon^3).
\label{eq:Fr_PN_Kep}
\end{align}
\end{subequations}
The complete 3PN expressions for the components of the RR force are provided in the Supplemental Material.
\end{widetext}
%


\subsection{Factorization into a QC part and an eccentric correction}
\label{sec:fact_QCEcc}

In EOB models, the QC RR force is expressed as a sum over waveform modes, which are factorized and resummed to improve agreement with NR results.
Therefore, to include the eccentricity contributions in the RR force, we use the following straightforward generalization of the RR force from the QC model \texttt{SEOBNRv5HM}~\cite{Pompiliv5}:
\begin{subequations}
\label{eq:fact_RRforces_QcEcc}
\begin{align}
\mF_\phi ^{ \text{F}}&= \mF_\phi^\text{modes} \mF_\phi^\text{ecc}(x, e, \zeta), 
 \\
\mF_r ^{ \text{F}}&= \frac{p_{r}}{p_\phi} \mF_\phi^\text{modes} \mF_r^\text{ecc}(x, e, \zeta),
\\
\mF_\phi^\text{modes}  &= -\frac{M^2 \Omega}{8 \pi} \sum_{\ell=2}^8 \sum_{m=1}^{\ell} m^2\left|d_L h_{\ell m}^{\text{F}}\right|^2, 
\end{align}
\end{subequations}
where $h_{\ell m}^{\text{F}}$ are the factorized modes valid for eccentric orbits, which we discuss in Sec.~\ref{sec:modes} and are given by Eq.~\eqref{eq:fact_QCecc}, while $ \mF_\phi^\text{ecc} $ and $ \mF_r^\text{ecc} $ are constructed such that one recovers the PN-expanded RR force to 3PN order, as we discuss below.
In the following calculations, we restrict to nonspinning binaries;
the resulting 3PN eccentric corrections to the RR force are the ones employed in the eccentric model \texttt{SEOBNRv5EHM} \cite{Gamboa:2024hli}.\footnote{
The calculations in this work were carried out in parallel with the development of the \texttt{SEOBNRv5EHM} model \cite{Gamboa:2024hli}.
In particular, the spinning parts of the RR force \eqref{eq:RRforces_EOB} and of the waveform modes \eqref{eq:modes_EOB_final} were derived in a late stage of the \texttt{SEOBNRv5EHM} model, and hence they were not included in the eccentricity corrections employed in the model.
\label{fn:ecc_corr}
}

In Ref.~\cite{Khalil:2021txt}, it was proposed to write the eccentric modes in terms of the EOB variables and, hence, to have the eccentricity corrections $ \mF_\phi^\text{ecc} $ and $ \mF_r^\text{ecc} $ also as functions of the EOB variables. However, we are interested in having the eccentricity corrections to the RR force in terms of the Keplerian parameters $ \left( x, e, \zeta \right) $ for the reasons discussed in Sec.~\ref{sec:evoEqs}.
Thus, to obtain these corrections we employ the PN-expanded force given in Eqs.~\eqref{eq:RRforces_Kep}, and we rewrite all the elements entering in $ \mF_\phi^\text{modes} $ in terms of the Keplerian parameters by employing Eqs.~\eqref{eq:Lxe_PA}--\eqref{eq:prxechi_PA}.
Then, the eccentricity corrections are obtained by performing a PN expansion of $ \mF_\phi^\text{ecc} = \mF_\phi ^{ \text{PN}}/ \mF_\phi^\text{modes} $ and $ \mF_r^\text{ecc} = p_\phi \mF_r ^{ \text{PN}}/ (p_r \mF_\phi^\text{modes}) $, with $ p_\phi $ and $ p_r $ also expressed in terms of the Keplerian parameters.
This procedure leads to the following nonspinning expressions:
%
\begin{widetext}
\begin{subequations}
\begin{align}
\mF_\phi^\text{ecc} &= 1+2 e^2+\frac{5 e^4}{4}+ \frac{181 e^6}{192} -\frac{389 e^6 \cos (6 \zeta )}{1024}+\frac{365}{512} e^5 \cos (5 \zeta )+\left(\frac{2503 e^6}{3072}+\frac{47 e^4}{96}-\frac{5 e^2}{12}\right) \cos (2 \zeta )\nonumber\\
&\quad
-\left(\frac{427 e^6}{384}+\frac{97 e^4}{96}\right) \cos (4 \zeta ) 
-\left(\frac{451 e^5}{256}+\frac{209 e^3}{96}+\frac{11 e}{6}\right) \cos (\zeta )+\left(\frac{923 e^5}{1536}+\frac{101 e^3}{96}\right) \cos (3 \zeta ) + \Order(e^7) \nonumber\\
&\quad
+ x \epsilon ^2 \bigg\{
e^6 \left(\frac{132443 \nu }{258048}-\frac{1265}{344064}\right) \cos (6 \zeta )
-e^6 \left(\frac{299273 \nu }{64512}+\frac{3603457}{258048}\right)
-e^5 \left(\frac{15391 \nu }{21504}+\frac{506803}{258048}\right) \cos (5 \zeta ) \nonumber\\
&\qquad
-e^4 \left(\frac{39737 \nu }{8064}+\frac{48601}{4608}\right)
+\left[e^6 \left(\frac{8327 \nu }{7168}+\frac{2809507}{258048}\right)+e^4 \left(\frac{1235 \nu }{1152}+\frac{132871}{32256}\right)\right] \cos (4 \zeta )\nonumber\\
&\qquad
-e^2 \left(\frac{2623 \nu }{504}+\frac{13775}{2016}\right)+\left[e^5 \left(-\frac{88937 \nu }{64512}-\frac{2683735}{258048}\right)+e^3 \left(-\frac{14603 \nu }{8064}-\frac{165845}{32256}\right)\right] \cos (3 \zeta )\nonumber\\
&\qquad
+\left[e^5 \left(\frac{14083 \nu }{3584}+\frac{2063917}{129024}\right)+e^3 \left(\frac{34439 \nu }{8064}+\frac{112499}{10752}\right)+e \left(\frac{32 \nu }{21}+\frac{304}{63}\right)\right] \cos (\zeta ) \nonumber\\
&\qquad
+\left[e^6 \left(\frac{36565 \nu }{28672}+\frac{3087451}{1032192}\right)+e^4 \left(\frac{2651 \nu }{2016}+\frac{19115}{4032}\right)+e^2 \left(\frac{2623 \nu }{1008}+\frac{16153}{4032}\right)\right] \cos (2 \zeta )
 + \Order(e^7) \bigg\} + \Order(\epsilon^3), 
 \\
\mF_r^\text{ecc} &= 1-\frac{33 e^2}{16} -\frac{37 e^4}{32}-\frac{5089 e^6}{6144} + \frac{997 e^6 \cos (6 \zeta )}{4096}-\frac{707 e^5 \cos (5 \zeta )}{2048}+\left(-\frac{8543 e^6}{12288}-\frac{13 e^4}{24}-\frac{23 e^2}{48}\right) \cos (2 \zeta ) \nonumber\\
&\quad
+\left(\frac{4745 e^6}{6144}+\frac{35 e^4}{96}\right) \cos (4 \zeta )
+\left(\frac{1601 e^5}{1024}+\frac{823 e^3}{384}+\frac{55 e}{24}\right) \cos (\zeta )
-\left(\frac{2317 e^5}{6144}+\frac{67 e^3}{384}\right) \cos (3 \zeta )  + \Order(e^7) \nonumber\\
&\quad
+ x \epsilon ^2 \bigg\{
e^6 \left(\frac{30175 \nu }{516096}-\frac{23521}{28672}\right) \cos (6 \zeta )
+e^6 \left(-\frac{1057199 \nu }{258048}-\frac{5459537}{516096}\right)
+e^5 \left(-\frac{176951 \nu }{172032}-\frac{1077113}{2064384}\right) \cos (5 \zeta ) \nonumber\\
&\qquad
-e^4 \left(\frac{93665 \nu }{32256}+\frac{1529785}{129024}\right)
+\left[e^6 \left(\frac{66323 \nu }{12288}+\frac{2900741}{516096}\right)+e^4 \left(\frac{76829 \nu }{32256}+\frac{65011}{18432}\right)\right] \cos (4 \zeta ) \nonumber\\
&\qquad
-e^2 \left(\frac{205 \nu }{504}+\frac{89963}{8064}\right)+\left[e^5 \left(-\frac{3960025 \nu }{516096}-\frac{24115325}{2064384}\right)+e^3 \left(-\frac{32489 \nu }{8064}-\frac{596539}{64512}\right)\right] \cos (3 \zeta ) \nonumber\\
&\qquad
+\left[e^5 \left(\frac{1433245 \nu }{86016}+\frac{25572275}{1032192}\right)+e^3 \left(\frac{109385 \nu }{8064}+\frac{173647}{7168}\right)+e \left(\frac{1481 \nu }{96}+\frac{201317}{8064}\right)\right] \cos (\zeta ) \nonumber\\
&\qquad
+\left[e^6 \left(\frac{1779787 \nu }{172032}+\frac{5310979}{258048}\right)+e^4 \left(\frac{96431 \nu }{8064}+\frac{729403}{32256}\right)+e^2 \left(\frac{8353 \nu }{1008}+\frac{158051}{8064}\right)\right] \cos (2 \zeta )
 + \Order(e^7)
\bigg\} \nonumber\\
&\quad + \Order(\epsilon^3).
\end{align}
\end{subequations}
These expressions are obtained in an eccentricity expansion up to $\Order(e^6)$. The complete nonspinning 3PN expressions are provided in the Supplemental Material.
\end{widetext}


\section{Waveform modes}
\label{sec:modes}

The gravitational waveform modes $h_{\ell m}$ are the decomposition of the complex polarization waveform $h =\nolinebreak h_+ -\nolinebreak i h_\times$ into spin-weighted $s = -2$ spherical harmonics $_{-2} Y_{\ell,\s m}(\iota, \varphi)$ such that
\begin{equation}
\label{eq:polarizations}
h = 
\sum^{\infty}_{\ell =2} \sum_{m=-\ell }^{\ell }{} \!_{-2} Y_{\ell,\s m}(\iota, \varphi) \, h_{\ell m},
\end{equation}
where $ (\iota, \varphi) $ are the inclination and azimuthal angles of the line of sight measured in the source frame.
The modes $h_{\ell m}$ can be computed from the waveform $h$ by inverting the above relation, or directly from the radiative multipole moments, as explained in detail in, e.g., Refs.~\cite{Kidder:2007rt,Blanchet:2008je,Faye:2012we,Blanchet:2013haa}.

To the 3PN order, the waveform modes are composed of the following contributions:
\begin{enumerate}

\item Instantaneous contributions, which were derived in Ref.~\cite{Mishra:2015bqa} in harmonic coordinates. These also include the PA contributions derived in Ref.~\cite{Boetzel:2019nfw} using the QK parametrization in an eccentricity expansion to $\Order(e_t^6)$.

\item Tail contributions, which start at 1.5PN order relative to the leading order of the $(2,2)$ mode, and were derived in Ref.~\cite{Boetzel:2019nfw} in harmonic coordinates using the QK parametrization in an eccentricity expansion to $\Order(e_t^6)$.

\item Oscillatory memory contributions, which appear in the even $\ell+m$ modes and start in the $(2,2)$ mode at 1.5PN for eccentric orbits, but at 2.5PN for circular orbits, and were derived in Ref.~\cite{Ebersold:2019kdc} using the QK parametrization in an eccentricity expansion to $\Order(e_t^6)$.

\item Direct current (DC) memory contributions, which start at Newtonian order and appear in the $(\text{even } \ell, m= 0)$ modes, but their effect is negligible for bound orbits at the current sensitivity of GW detectors~\cite{Grant:2022bla}.

\end{enumerate}

The full 3PN spin contributions to the modes (instantaneous, tail, and memory) were derived in Ref.~\cite{Henry:2023tka} using harmonic coordinates and the covariant SSC. 
The tail and memory contributions in that reference were also expressed using the QK parametrization in an eccentricity expansion to $\Order(e_t^6)$.

Therefore, to obtain the 3PN EOB waveform modes, we start from the known harmonic-coordinates expressions and transform them to EOB coordinates and the Keplerian parametrization.
This is allowed since the modes are gauge invariant once the orientation of the source frame is fixed as in Eq.~\eqref{eq:polarizations}. 
For this purpose, we employ
\begin{enumerate}

\item The complete (conservative + dissipative) transformation from harmonic $(r_\text{h},\dot{r}_\text{h},\phi_\text{h},\dot{\phi}_\text{h})$ to EOB coordinates $ (r, p_r, \phi, p_\phi) $ given by Eqs.~\eqref{eq:scalars_HarmToEOB}.

\item The transformations from EOB variables $ (r, p_r, p_\phi) $ to the Keplerian parameters $ (x, e, \zeta) $ given by Eqs.~\eqref{eq:ErprL_xez_PA}. 

\item The transformations from the harmonic-coordinates QK parametrization to the EOB Keplerian parametrization given by Eqs.~\eqref{eq:et} and \eqref{eq:mean_anomaly}, with the PA contributions \eqref{eq:tilde_x_e} and \eqref{eq:tildez}.
\end{enumerate}

We specialize our results to bound orbits, and present the modes in the general form 
\begin{equation} \label{eq:modes_EOB}
 h_{\ell m}  = \frac{8 M \nu}{d_L} \sqrt{ \frac{\pi}{5} } \, x \, \e^{- i m \phi }   H_{\ell m} (x, e, \zeta),
\end{equation}
with $H_{\ell m}$ a complex amplitude split into instantaneous, tail, and memory contributions,
\begin{equation}
\label{eq:H_lm}
H_{\ell m} (x, e, \zeta) \equiv H_{\ell m}^\text{inst} +  H_{\ell m}^\text{tail} + H_{\ell m}^\text{mem}
\end{equation}
where $ x = \langle M \Omega \rangle^{2/3} $ is the dimensionless orbit-averaged orbital frequency, $ e $ is the Keplerian eccentricity, and $ \zeta $ is the relativistic anomaly (see Sec.~\ref{sec:Keplerian}). 

The instantaneous, tail, and memory parts of the EOB modes are presented in Sec.~\ref{sec:inst_modes}, \ref{sec:tail_modes}, and \ref{sec:memory_modes}, respectively. The complete modes are in Sec.~\ref{sec:complete_modes}, and the factorization of these modes employed in the \texttt{SEOBNRv5EHM} waveform model \cite{Gamboa:2024hli} is presented in Sec.~\ref{sec:factorization_modes}.


\subsection{Instantaneous and postadiabatic contributions}
\label{sec:inst_modes}

The 3PN instantaneous contributions to the modes for generic orbits were computed in Refs.~\cite{Mishra:2015bqa,Henry:2023tka} using the MH coordinate system $ (r_\text h, \dot r_\text h, \phi _{ \text{h}}, \dot \phi_\text h) $ and the covariant SSC.
We write their modes in the general form,
\begin{equation} \label{eq:modes_inst_MH}
h^\text{h,\,inst} _{\ell m} = \frac{8 M \nu}{d_L} \sqrt{ \frac{\pi}{5} } \e^{- i m \phi_\text h } H^\text{h,\,inst}_{\ell m}(r_\text h, \dot r_\text h, \dot \phi_\text h), 
\end{equation}
where the $H^\text{h,\,inst} _{\ell m}$ function defined here differs from the one employed in Ref.~\cite{Mishra:2015bqa} by a factor of 2. Then, the EOB instantaneous modes are computed by substituting our transformation rules \eqref{eq:scalars_HarmToEOB} in Eq.~\eqref{eq:modes_inst_MH}, and performing a PN expansion. Our results take the general form 
\begin{equation} \label{eq:modes_inst_EOB}
h^\text{inst} _{\ell m}  = \frac{8 M \nu}{d_L} \sqrt{ \frac{\pi}{5} } \e^{- i m \phi } H^\text{inst}_{\ell m}(r, p_r, p_\phi).
\end{equation}
The explicit expressions for all the amplitudes $H^\text{inst} _{\ell m}$, up to $\ell = 8$, are given in the Supplemental Material. 
The leading order of the instantaneous part of the (2,2) mode reads 
\begin{equation}
\label{eq:22mode_inst}
H^\text{inst}_{22} = \frac{\mu ^2 M r-r^2 p_r^2+p_{\phi }^2+2 i r p_r p_{\phi }}{2 \mu ^2 r^2} + \Order(\epsilon^2).
\end{equation}
The 3PN part of the modes contains the constant $r_0$, which also appears in the computation of the fluxes (see Sec.~\ref{sec:inst_fluxes}). Hence, for \texttt{SEOBNRv5EHM}, we use the value $r_0 = 2M/\sqrt{\e}$, where $\e$ is Euler's number.

Next, we specialize our EOB modes to the case of bound eccentric orbits using the Keplerian parametrization with the parameters $ \left( x, e, \zeta \right) $. 
Schematically, we simply substitute the relations \eqref{eq:ErprL_xez_PA} in the modes \eqref{eq:modes_inst_EOB}, and then perform a PN expansion. This procedure automatically accounts for the PA contributions which depend on the chosen RR gauge. 
We note that, to 3PN accuracy, only the Newtonian and 0.5PN parts of the instantaneous $(2,0)$, $(2,1)$, $(2,2)$, $(3,1)$, and $(3,3)$ modes are affected by the PA corrections included in Eqs.~\eqref{eq:ErprL_xez_PA}. 
In contrast, the hereditary modes at the 3PN order are not affected by the PA contributions. 

Thus, we obtain the complete instantaneous part of the EOB modes at 3PN order for bound eccentric orbits in the Keplerian parametrization. We write the results in the form \eqref{eq:modes_EOB}, and include the expressions for the $ H ^{ \text{inst}}  _{ \ell m}$ in the Supplemental Material.   
The leading order of the $(2,2)$ mode reads
\begin{equation}
\label{eq:22mode_inst_Kep}
 H^\text{inst}_{22}(x, e, \zeta) =  \frac{4+ (\text{e}^{-i \zeta } +5 \text{e}^{i \zeta }) e+2 \text{e}^{2 i \zeta } e^2}{4 (1- e^2)} + \Order(\epsilon^2),
\end{equation}
with the 3PN expression given by Eq.~\eqref{eq:app:instModes}.
The constant $x_0$ appearing in the 3PN part of the modes is defined by $x_0 \equiv M/r_0$.
For \texttt{SEOBNRv5EHM}, we employ a fixed value for $ r_0$ given by Eq.~\eqref{eq:r0EOB}, leading to
\begin{equation} \label{eq:x0_EOB}
x_{0,\s\text{EOB}} = \frac{\sqrt{\e}}{2},
\end{equation}
where $\e$ is Euler's number.

We have checked that we recover the QC modes obtained in Refs.~\cite{Blanchet:2008je, Henry:2022dzx} in the limit $e \to 0$ and when choosing either the harmonic RR gauge ($\alpha = -1$, $\beta = 0$) or our new RR gauge ($ \alpha = -16/3 $, $ \beta = -13/2 $). 
The modes employed in \texttt{SEOBNRv5EHM} make use of this latter choice for $ \alpha $ and $ \beta $.


\subsection{Tail contributions}
\label{sec:tail_modes}

In contrast to the instantaneous contributions, the tail and memory contributions depend on the entire history of the binary, and thus it is not possible to write them in an explicit form valid for generic orbits. 
Instead, one employs a parametrization of the orbit to compute the relevant integrals coming from the PN-MPM formalism in a small-eccentricity expansion for bound orbits, or a large-eccentricity expansion for scattering trajectories. 

The nonspinning tail contributions to the modes at 3PN order were computed in Ref.~\cite{Boetzel:2019nfw} within the MH coordinate system and employing the QK parametrization.
The spinning tail contributions at 3PN order were derived in Ref.~\cite{Henry:2023tka} employing the same coordinate system and parametrization of the orbit, and the covariant SSC.
The results are given as eccentricity expansions up to $ \mathcal O (e_t^6) $, where $ e_t $ is the time eccentricity, and they take the form
\begin{equation} \label{eq:modes_tail_QK}
 h^{\text{QK,\,tail}} _{\ell m}   = \frac{8 M \nu}{d_L} \sqrt{ \frac{\pi}{5} } \, x \, \e^{- i m \phi_\text h }  H^{\text{QK,\,tail}} _{\ell m}(x, e_t, l),
\end{equation}
where $ l $ is the mean anomaly, and the subscript `h' in the phase emphasizes that these expressions were computed within the MH coordinate system.

To obtain the tail contributions to the EOB modes, we simply substitute the transformations from Eqs.~\eqref{eq:phih}, \eqref{eq:et} and \eqref{eq:mean_anomaly} into the results of Refs.~\cite{Boetzel:2019nfw,Henry:2023tka}, and then we perform PN and eccentricity expansions to $ \mathcal O (e^6) $ at 3PN order. 
The resulting tail contributions to the EOB modes are written in the form \eqref{eq:modes_EOB} in terms of the Keplerian parameters, and the amplitudes $H^\text{tail,\,exp}_{\ell m}$ are given in the Supplemental Material expanded to $\Order(e^6)$. 
Here, we write the leading order tail contribution to the $(2,2)$ mode expanded to $ \mathcal{O}(e) $ as
\begin{widetext}
\begin{subequations}
\label{eq:22mode_tail_exp}
\begin{align}
 H^\text{tail,\,exp}_{22}(x, e, \zeta) &=    
\frac{\epsilon^3 \text{e}^{-6 i \zeta } x^{3/2}}{11520}  \bigg \{23040 \text{e}^{6 i \zeta } \pi +2880 \text{e}^{5 i \zeta } e \Big[6 i \left(-9 \ln 2 +\text{e}^{2 i \zeta } \ln 2+9 \ln 3\right)+\left(11+13 \text{e}^{2 i \zeta }\right) \pi \Big] \nonumber \\ 
&\qquad  
+1080 i \text{e}^{3 i \zeta } \Big[64 \text{e}^{3 i \zeta }+8 \text{e}^{2 i \zeta } \left(11+13 \text{e}^{2 i \zeta }\right) e\Big] \ln \left(\frac{x}{x_0'}\right) \bigg\} + \Order(\epsilon^5),
\end{align}
\end{subequations}
\end{widetext}
where the superscript ``exp'' in the above equation reflects the fact that the tail contributions are computed in an eccentricity expansion; we used this notation to differentiate these amplitudes from the \emph{resummed} amplitudes, discussed below.

The constant $x_0'$ that appears in these expressions is related to the constant $r_0$ by [see, e.g.,~Eq.~(49) of Ref.~\cite{Boetzel:2019nfw}]
\begin{equation} \label{eq:x0p}
x_0' = \left( \frac{M}{4 r_0} \e^{11/12 - \gamma_\text E} \right)^{2/3},
\end{equation}
where $\gamma_\text E \approx 0.577$ is Euler's constant and $\e$ is Euler's number. Substituting $r_0$ from Eq.~\eqref{eq:r0EOB} into Eq.~\eqref{eq:x0p} we obtain
\begin{equation} \label{eq:x0p_EOB}
x_{0,\s \text{EOB}}' = \frac{1}{4} \e^{17/18 - 2 \gamma_\text E /3 }.
\end{equation}

Our results agree with the expressions given in Refs.~\cite{Blanchet:2008je, Henry:2022dzx} after taking the QC orbit limit $e \to 0$.


\subsubsection*{Resummation of the tail contributions}
\label{sec:tail_modes_resummation}

The instantaneous modes expressed in the Keplerian or QK parametrizations have a special pattern that appears at each PN order.
This pattern can be seen in Eq.~\eqref{eq:22mode_inst_Kep}, where we note that there is a prefactor of $ 1/(1 - e^2)^{n+1} $ at each $ n $-PN order.
This prefactor originates from the types of integrals that appear in the computation of the multipole moments. However, the tail contributions to the modes do not have this prefactor since they were computed in an eccentricity expansion.

For \texttt{SEOBNRv5EHM}, we propose an \emph{eccentricity resummation} of the tail contributions that follows the same pattern as the instantaneous part of the modes. That is, we introduce a factor of $ 1/(1 - e^2)^{n+1} $ at each $ n $-PN order, such that the eccentricity expansion of the resummed modes coincides with the original expression.

The resulting eccentricity-resummed PN-expanded amplitudes $ H^\text{tail}_{\ell m} $ are the ones employed when computing the eccentricity corrections to the \emph{factorized} EOB modes (see Sec.~\ref{sec:factorization_modes}). The complete expressions of $ H^\text{tail}_{\ell m} $ are given in the Supplemental Material,
and here we present the leading order contribution to the $(2,2)$ mode as
\begin{align}
\label{eq:22mode_tail}
&H^\text{tail}_{22}(x, e, \zeta) =\frac{\epsilon^3 x^{3/2}}{(1 - e^2)^{5/2}} \bigg[
\pi  \left(2+\frac{11}{4} e \e^{-i \zeta}+\frac{13}{4} e \e^{i \zeta}\right) \nonumber\\
&\quad\qquad
+i\left(\frac{33}{4} e \e^{-i \zeta}+\frac{39}{4} e \e^{i \zeta}+6 \right) \ln \left(\frac{x}{x_0'}\right) \nonumber\\
&\quad\qquad
+\frac{3}{2} i e \e^{i \zeta } \ln 2 
+\frac{27}{2} i e \e^{-i \zeta } \left(\ln 3-\ln 2\right) + \Order(e^2)
\bigg] \nonumber\\
&\quad\qquad + \Order(\epsilon^5),
\end{align}
with the 3PN expression given by Eq.~\eqref{eq:app:tailModes}.


\subsection{Memory contributions}
\label{sec:memory_modes}
 
Similar to the tail contributions, computing the memory part of the modes also requires a parametrization of the orbit to perform eccentricity expansions.
The memory contributions appear in the modes with even $\ell+m$, and consist of a DC contribution that appears only in $m=0$ modes and an oscillatory contribution.
The memory contributions to the modes at 3PN order were computed in Refs.~\cite{Ebersold:2019kdc,Henry:2023tka} within the harmonic-coordinates QK parametrization and with the covariant SSC.
Their results take the following form:
\begin{equation}
\label{eq:modes_memory_QK}
 h^{\text{QK,\,mem}} _{\ell m}   = \frac{8 M \nu}{d_L} \sqrt{ \frac{\pi}{5} } \, x \, \e^{- i m \psi }  H^{\text{QK,\,mem}} _{\ell m}(x, e_t, \xi; e_i),
\end{equation}
where the amplitudes $H^{\text{QK,\,mem}} _{\ell m}$ can be split into oscillatory and DC memory contributions
\begin{align} 
H^{\text{QK,\,mem}} _{\ell m} &= 
 H^{\text{QK,\,osc}} _{\ell m}(x, e_t, \xi) + H^{\text{QK,\,DC}} _{\ell m}(x, e_t; e_i),
\end{align}
with a constant initial eccentricity $e_i$. 
The phase angles $\xi$ and $\psi$ are associated with a shift of the time coordinate  \cite{Blanchet:1996pi, Arun:2004ff}, and were introduced to eliminate the dependence of the modes on the freely specifiable constant $x_0'$ that appears in the hereditary contributions to the modes. 

In our case, we are interested in the expressions of the modes in terms of the orbital phase $\phi$ and the relativistic anomaly $\zeta$.
Thus, we need to transform the phase angles $\xi$ and $\psi$ to $\phi$ and $\zeta$. 
To compute the transformation of the phases, we first write here the definitions of the angles $\xi$ and $\psi$
\begin{subequations}
\begin{align}
\xi &= \bar l - 3 \epsilon^3 M_\text{ADM} \, \bar n \ln \left( \frac{\bar x}{x_0'} \right),  
\label{eq:xi} \\
 \psi &= \bar \lambda_\xi + \bar W_\xi + \tilde \lambda_\xi + (\tilde v_\xi - \tilde l _\xi),
\label{eq:psi}
\end{align}
\end{subequations}
where we have used bars and tildes to denote the secularly evolving and rapidly oscillating (postadiabatic) parts, respectively, of each orbital parameter (see Sec.~\ref{sec:quasiKeplerian}),
$M_\text{ADM} \equiv M [1 - \nu \epsilon^2 \bar x /2 + \Order(\epsilon^4)] $ is the ADM mass of the system, $\bar n$ is the (gauge-invariant) mean motion, and 
\begin{align} \label{eq:lambdaxi}
\bar \lambda_\xi &= \bar \lambda - 3 \epsilon^3 M _{ \text{ADM}}  \left( 1 + \bar k \right) \bar n \ln  \left( \frac{\bar x}{x_0'} \right) \nonumber\\
&= \bar \lambda - 3 \epsilon^3 \left( 1 - \epsilon^2 \frac{\nu\bar x}{2} \right) \bar x^{3/2} \ln  \left( \frac{\bar x}{x_0'} \right) ,
\end{align}
with $\bar k$ the relativistic periastron precession. The definitions of the other quantities, $\bar W_\xi$,  $\tilde \lambda_\xi$, $\tilde v_\xi$, and $\tilde l _\xi$, are given in Appendix~B of Ref.~\cite{Boetzel:2019nfw}. Using these definitions, we can write the angles $ \xi $ and $ \psi $ in terms of the parameters $\bar x$, $\bar l$ and $\bar e_t$.

We note that the \emph{functional form} (i.e.,~the dependence on $\bar x$, $\bar e_t$ and $\xi$) of Eq.~\eqref{eq:psi} is the same as the functional form of Eq.~\eqref{eq:philW}, but with $\bar l$ relabeled as $\xi$ in the latter equation.
This is why the definition of $\psi$ in Eq.~\eqref{eq:psi} has subscripts denoting the $\xi$ dependence. 
Therefore, if one knows the complete expression of $ \bar W_\xi + \tilde \lambda_\xi + (\tilde v_\xi - \tilde l _\xi)$ (as in Eq.~(B1c) of Ref.~\cite{Boetzel:2019nfw}\footnote{The expression given in Eq.~(B1c) of Ref.~\cite{Boetzel:2019nfw} is accurate to $\mathcal O (e_t^2)$, but the authors of  Ref.~\cite{Ebersold:2019kdc} kindly sent us the complete expression accurate to $\mathcal O (e_t^6)$ in private communications.}), then one can automatically obtain the form of $\bar W\left(\bar l\right) + \tilde \lambda + (\tilde v- \tilde l )$, just by interchanging the labels from $ \xi $ to $ \bar l $. 
Hence, by using Eqs.~\eqref{eq:philW}, \eqref{eq:psi}  and \eqref{eq:lambdaxi} we obtain the following relation between $\psi$ and the orbital phase $\phi_\text h$ 
\begin{align} \label{eq:psiphi}
\psi &= \phi_\text h - 3 \left( 1 - \frac{\nu \bar x}{2} \right) \bar x^{3/2} \ln  \left( \frac{\bar x}{x_0'} \right) \nonumber\\
&\qquad
+ \bar W_\xi + \tilde \lambda_\xi + (\tilde v_\xi - \tilde l _\xi) \nonumber\\
&\qquad
- \bar W\left(\bar l\right) - \tilde \lambda - (\tilde v- \tilde l ).
\end{align}
In this way, Eqs.~\eqref{eq:xi} and \eqref{eq:psiphi} give us the transformation from ($\xi$, $\psi$) to $\left( \bar l, \phi_\text h \right)$. 

Finally, using the previous equations and our relations \eqref{eq:phih}, \eqref{eq:et}, and \eqref{eq:mean_anomaly}, we translate the memory modes \eqref{eq:modes_memory_QK} into the EOB modes, and express them in the form \eqref{eq:modes_EOB}.
The associated amplitudes are given as
\begin{equation} \label{eq:H_mem_exp}
H^{\text{mem,\,exp}} _{\ell m}(x, e, \zeta; e_i) =  H^{\text{osc,\,exp}} _{\ell m}(x, e, \zeta) + H^{\text{DC}} _{\ell m}(x, e; e_i),
\end{equation}
where $H^{\text{DC}} _{\ell m} = 0$ for $ m \neq 0$, and the subscript ``exp'' helps us to distinguish between the eccentricity-expanded oscillatory amplitudes and the eccentricity-resummed oscillatory amplitudes discussed below (we do not perform a resummation of the DC contributions; hence, we do not put the subscript ``exp'' in $ H^{\text{DC}} _{ \ell m}$).
The amplitudes $H^{\text{osc,\,exp}} _{\ell m}$ and $H^{\text{DC}}_{\ell m}$ are given in the Supplemental Material, and here we present the leading order of the oscillatory contributions to the $(2,2)$ mode as
%
\begin{subequations}
\label{eq:22mode_osc_exp}
\begin{align}
H^\text{osc,\,exp}_{22}(x, e, \zeta) &=  -i\nu \e^{2 i \zeta } \epsilon^3 x^{3/2} \left[\frac{55 e^6}{224}+\frac{23 e^4}{168}+\frac{13 e^2}{252}\right] \nonumber\\
&\qquad 
+ \Order(e^8) + \Order(\epsilon^5) .
\end{align}
\end{subequations}

Our results for the memory contributions agree with those from Refs.~\cite{Ebersold:2019kdc,Henry:2023tka} in the QC orbit limit $ e \to 0 $.

We note that the \texttt{SEOBNRv5EHM} model \cite{Gamboa:2024hli} does not include the DC memory contributions to the modes, because they have a negligible effect for bound orbits, at least for the sensitivity of current GW detectors.
This could be an improvement for future waveform models.


\subsubsection*{Resummation of the oscillatory memory contributions}
\label{sec:memory_modes_resummation}

As with the tail contributions to the modes, we perform an eccentricity resummation of the amplitudes $ H^\text{osc,\,exp}_{\ell m} $ for \texttt{SEOBNRv5EHM}.
Hence, we introduce a factor of $ 1/(1 - \nolinebreak e^2)^{n+1} $ at each $ n $-PN order, such that the eccentricity expansion of the resummed modes coincides with the original expression. 

Thus, we obtain the eccentricity-resummed PN-expanded oscillatory memory contributions to the EOB modes, $ H^\text{osc}_{\ell m} $, which are used to compute the eccentricity corrections to the \emph{factorized} modes (see Sec.~\ref{sec:factorization_modes}).
The complete expressions are given in the Supplemental Material, and we show here the leading order contribution to the $(2,2)$ mode as
%
\begin{subequations}
\label{eq:22mode_osc}
\begin{align}
H^\text{osc}_{22}(x, e, \zeta) &= -\frac{ i\nu \epsilon ^3 x^{3/2} \e^{2 i \zeta }}{252 \left(1-e^2\right)^{5/2}} \left[13e^2 +2 e^4 + \Order(e^8)\right] \nonumber\\
&\qquad
+ \Order(\epsilon^5),
\end{align}
\end{subequations}
with the 3PN expression given by Eq.~\eqref{eq:app:oscMemModes}.


\subsection{Complete modes for bound orbits}
\label{sec:complete_modes}

The complete EOB waveform modes for bound eccentric orbits written in the Keplerian parametrization are obtained by adding all the contributions from the previous subsections. 
Thus, the final modes are given by
\begin{subequations}
\label{eq:modes_EOB_final}
\begin{align}
h_{\ell m} &= \frac{8 M \nu}{d_L} \sqrt{ \frac{\pi}{5} } \, x \, \e^{- i m \phi }   H_{\ell m}(x, e, \zeta),
\\
H_{\ell m}(x, e, \zeta) &=   H^{\text{inst}} _{\ell m}(x, e, \zeta) +  H^{\text{tail}} _{\ell m}(x, e, \zeta) \nonumber\\
&\quad + H^{\text{osc}} _{\ell m}(x, e, \zeta) + H^{\text{DC}} _{\ell m}(x, e; e_i),
\label{eq:modes_EOB_amplitudes}
\end{align}
\end{subequations}
and all the Keplerian parameters $(x, e, \zeta)$ here correspond to their secularly evolving part $(\bar x, \bar e, \bar \zeta)$, but we omitted the bar for ease of notation. The complete modes are explicitly given in the Supplemental Material.

We performed two checks of our results.
In the first one, we took the limit $e \to 0$, and we were able to recover the same QC modes as in Ref.~\cite{Henry:2022ccf} at 3PN order.
For the second check, we compared the expressions of the PN-expanded eccentric EOB modes obtained in two different ways.
More specifically, we applied all our transformations \eqref{eq:phih}, \eqref{eq:et}, \eqref{eq:mean_anomaly}, \eqref{eq:xi} and \eqref{eq:psiphi}, \emph{directly} to the full 3PN eccentric modes given in Refs.~\cite{Ebersold:2019kdc,Henry:2023tka} written in the QK parametrization, and then we compared the resulting expressions against our modes in Eq.~\eqref{eq:modes_EOB_final}, which were computed by transforming and adding each contribution \emph{separately};
in the process of comparison, we expanded in eccentricity to $ \mathcal O (e^6)$.
After doing this expansion, we find perfect consistency, except for some terms proportional to $e^6$ at the 2.5PN order.\footnote{
The inconsistency appears at 2.5PN order and at $ \mathcal O(e^6) $ for the modes $\{(2,2), (2,0), (2,1), (3,1), (3,3)\}$, and it originates from the PA contributions to the instantaneous part of the modes.
Since the inconsistency only occurs for terms proportional to $e^6$, this suggests that the issue could be related to an incomplete cut in eccentricity.
Specifically, the PA contribution to the radial anomaly (either mean anomaly $ \tilde l $ or relativistic anomaly $ \tilde \zeta $) contains a term proportional to $ 1/e $, as seen in Eq.~\eqref{eq:tildez} or Eq.~(39c) in Ref.~\cite{Boetzel:2019nfw}.
Hence, in some parts of the calculations, one needs to cut at an order higher than $ e^6 $ to achieve the complete final expressions of the modes accurate up to $ \mathcal O(e^6) $.
}
This is a confirmation that all our transformations are mutually consistent.

Finally, we note that the QK instantaneous contributions to the modes of Ref.~\cite{Ebersold:2019kdc} are computed \emph{with} eccentricity expansions and in terms of the mean anomaly $ l $,\footnote{
The instantaneous contributions to the QK modes were first obtained in Ref.~\cite{Mishra:2015bqa} without eccentricity expansions, and in terms of the eccentric anomaly $u$.
Then, Ref.~\cite{Boetzel:2019nfw} used the general 3PN solution of the Kepler's equation \cite{Boetzel:2017zza} to write the QK instantaneous modes in terms of the mean anomaly $l$.
In the process, they performed eccentricity expansions.
}
whereas our instantaneous modes are computed \emph{without} such expansions and in terms of the relativistic anomaly $ \zeta $.
More precisely, we took the general-orbit expressions for the instantaneous modes \eqref{eq:modes_inst_EOB}, and then we expressed them in terms of the Keplerian parameters without an expansion in eccentricity.\footnote{
Only the PA contributions are computed in an eccentricity expansion.
This is the case for both the QK and Keplerian parametrizations.
}
Because of this, our instantaneous modes have eccentricity contributions in the form of square roots and logarithms, which capture a nontrivial behavior with respect to eccentricity.
This is especially relevant for high values of eccentricity.


\subsection{Factorization of the EOB eccentric modes}
\label{sec:factorization_modes}

The agreement between the PN-expanded modes and the modes extracted from NR simulations can be highly improved with a suitable \emph{factorization}~\cite{Damour:2007xr,Damour:2007yf,Damour:2008gu,Pan:2010hz}.
This process consists in expressing the modes $h _{ \ell m}$ as a product of certain factors which, in principle, capture some relevant physical properties of the modes, and control their numerical behavior. 
This is especially important towards the end of the inspiral phase, when the standard PN expressions start to break down. 

In the family of waveform models \texttt{SEOBNRv5}, the QC modes are factorized as~\cite{Pompiliv5}
\begin{equation}
\label{eq:qcFactModes}
h_{\ell m}^\text{F, qc} = h_{\ell m}^\text{N, qc}  \, \hat{S}_\text{eff} \, T_{\ell m}^\text{qc} \, f_{\ell m}^\text{qc} \, \e^{i \delta_{\ell m}^\text{qc}}.
\end{equation}
The first factor $h_{\ell m}^\text{N, qc}$ is the leading PN order for QC orbits, which is given for all modes by~\cite{Damour:2008gu,Pan:2010hz}
\begin{equation}\label{eq:hnewt}
h_{\ell m}^\text{N, qc}=\frac{\nu M}{d_L} n_{\ell m} c_{\ell+\epsilon_{\ell m}}(\nu)v_\phi^{\ell+\epsilon_{\ell m}}Y_{\ell-\epsilon_{\ell m},-m}\left(\frac{\pi}{2},\phi\right),
\end{equation}
where $Y_{\ell m}$ is the scalar spherical harmonic, 
$\epsilon_{\ell m}$ is the parity of the mode, such that $\epsilon_{\ell m} = 0$ if $\ell+m$ is even, and $\epsilon_{\ell m} = 1$ if $\ell+m$ is odd,
the factors $n_{\ell m}$ and $c_{k}(\nu)$ are given by Eqs.~(5)--(7) of Ref.~\cite{Damour:2008gu}, 
while $v_\phi$ is defined as
\begin{equation}
\label{eq:vphi}
	v_\phi \equiv M \Omega r_\Omega,
\end{equation}
with
\begin{equation}
\label{eq:rOmega}
	r_\Omega \equiv \left.\left(M\frac{\partial H_{\rm EOB}}{\partial p_\phi} \right)^{-2/3}\right|_{p_r=0} \!.
\end{equation}

The second factor in Eq.~\eqref{eq:qcFactModes} is the dimensionless effective source term $\hat{S}_\text{eff}$, which is given by
\begin{equation} \label{eq:Seff}
\hat{S}_\text{eff}= \left\{
        \begin{array}{ll}
            H_\text{eff} / \mu, & \quad \ell + m \text{ even} \\
          ( M \Omega ) ^{1/3} \, p_\phi/(M \mu), & \quad \ell + m \text{ odd}
        \end{array}
    \right. \!,
\end{equation}
and the other factors are detailed in Ref.~\cite{Pompiliv5}. These factors depend directly on the instantaneous angular frequency $ \dot \phi = \Omega $, or equivalently, on the dimensionless quantity $ ( M \Omega)^{2/3} $.

For \texttt{SEOBNRv5EHM}, we propose the following factorization
\begin{equation}
\label{eq:fact_QCecc} 
h_{\ell m}^\text{F}  = h_{\ell m}^\text{F, qc}(x) \,  h_{\ell m}^\text{ecc} ( x, e, \zeta),
\end{equation}
where $ x = \langle M \Omega\rangle^{2/3} $ represents the \emph{orbit-averaged} orbital frequency, instead of the \emph{instantaneous} one.
We proceed now to explain the different factors entering this expression.

The first factor $ h_{\ell m}^\text{F, qc} $ has the same form as in Eq.~\eqref{eq:qcFactModes}, except that instead of using $ v_\phi = M \Omega \, r_\Omega $ in the Newtonian part,\footnote{
Historically, the factor $ v_\phi = M \Omega \, r_\Omega $ was introduced in Ref.~\cite{Pan:2009wj} to improve the agreement with QC NR waveforms. Now that EOB models are more sophisticated, this factor is not needed anymore and will be removed in future \texttt{SEOBNR} models.
}
we replace it by
\begin{equation}
v_\Omega \equiv \langle M \Omega \rangle^{1/3} = x^{1/2}.
\end{equation}
This is because the generalization of $ r_\Omega $ from Eq.~\eqref{eq:rOmega} to eccentric orbits is not straightforward due to the $ p_r = 0 $ requirement, which would change the PN expansion of the eccentric modes unless a correction factor is applied.
Thus, we decided to use $ v_\Omega $ after checking that this substitution does not produce a significant change in the zero-eccentricity limit of \texttt{SEOBNRv5EHM} with respect to the underlying QC model \texttt{SEOBNRv5HM} (e.g., see Figure 2 of the companion paper~\cite{Gamboa:2024hli}).
Additionally, in Eqs.~\eqref{eq:Seff} and \eqref{eq:fact_QCecc}, we replace the original dependence of the different factors entering $ h_{\ell m}^\text{F, qc} $ on the instantaneous frequency $ M \Omega $ by a dependence on the orbit-averaged frequency $ x $.
As we discuss in Sec.~\ref{sec:prescription_x}, in \texttt{SEOBNRv5EHM} we employ a prescription for computing $ x $ such that we are guaranteed to recover the original dependence on $ M \Omega $ in the QC limit $ e \to 0 $ [see Eq.~\eqref{eq:xxinst}].

The second factor, $ h_{\ell m}^\text{ecc} $, entering in Eq.~\eqref{eq:fact_QCecc} is a multiplicative eccentricity correction that receives contributions at all PN orders.
These eccentricity corrections are computed by requiring that the PN expansion of Eq.~\eqref{eq:fact_QCecc} is consistent with Eq.~\eqref{eq:modes_EOB_final} in the nonspinning limit (see footnote~\ref{fn:ecc_corr}).
We remark that, in these calculations, we employ the eccentricity-resummed tail and oscillatory memory contributions.\footnote{
In the computation of $ h_{\ell m}^\text{ecc} $, we have to take into account the fact that the factors in  Eq.~\eqref{eq:qcFactModes} used in \texttt{SEOBNRv5HM} \cite{Pompiliv5} are not complete at 3PN order.
These factors are not always equivalent to the full 3PN expressions derived in Ref.~\cite{Henry:2022ccf};
the missing terms were not known at the time of \texttt{SEOBNRv5HM} development.
Since \texttt{SEOBNRv5EHM} reduces to \texttt{SEOBNRv5HM} in the QC limit, we had to keep track of the missing terms.
The full known PN information will be used in the future to build the next generation of waveform models.
}
The resulting corrections depend only on the Keplerian parameters $ (x, e, \zeta) $ and satisfy $ h_{\ell m}^\text{ecc} (x, e = 0, \zeta) = 1 $. In particular, if we initialize a system with $ e = 0 $, the eccentricity contributions will remain equal to one throughout the entire evolution. This is important for recovering the underlying QC model.

The different nonspinning 3PN expressions for $ h_{\ell m}^\text{ecc}  $ are given in the Supplemental Material, 
and here we present the eccentricity correction to the $ (2,2) $ mode to 1PN order,
\begin{widetext}
\begin{align}
\label{eq:fact_QCecc_22} 
h_{22}^\text{ecc} ( x, e, \zeta) &= \frac{4+2 e^2 \e^{2 i \zeta }+e \e^{-i \zeta }+5 e \e^{i \zeta }}{4(1- e^2)} 
+ \frac{x e \epsilon^2}{(1-e^2)^2} \bigg\{
e \left(\frac{26 \nu }{7}-\frac{559}{84}\right)
+e^2 \e^{-3 i \zeta } \left(\frac{9 \nu }{56}+\frac{1}{112}\right) 
+e^2 \e^{3 i \zeta } \left(\frac{\nu }{8}-\frac{49}{48}\right)\nonumber\\
&\quad
+e \e^{-2 i \zeta } \left(\frac{15 \nu }{14}-\frac{95}{168}\right)
+\e^{2 i \zeta } \left[e^3 \left(\frac{6 \nu }{7}-\frac{41}{21}\right)+e \left(\frac{\nu }{14}-\frac{153}{56}\right)\right]
+\e^{-i \zeta } \left[e^2 \left(\frac{7 \nu }{8}-\frac{59}{48}\right)+\frac{27 \nu }{14}-\frac{23}{14}\right]\nonumber\\
&\quad
+\e^{i \zeta } \left[e^2 \left(\frac{143 \nu }{56}-\frac{2071}{336}\right)+\frac{\nu }{14}-\frac{13}{7}\right]
\bigg\} + \Order(\epsilon^3).
\end{align}
\end{widetext}

As an additional comment, we note that the factors $ h_{\ell m}^\text{ecc} $ are not the only eccentricity corrections to the modes. All the factors in Eq.~\eqref{eq:qcFactModes} have a dependence on the Hamiltonian or the orbit-averaged frequency $ x $. These variables contain eccentricity contributions \emph{per se}, which will also affect the numerical value of the modes.

In conclusion, in \texttt{SEOBNRv5EHM} \cite{Gamboa:2024hli}, we employ a factorization of
the modes that depends on the Keplerian parameters $ ( x, e, \zeta )
$, and such that we can exactly recover the \texttt{SEOBNRv5HM}
factorized QC modes in the limit of zero eccentricity (except for the
factor $ v_\phi $, as we discussed above). This should be contrasted
with the \texttt{SEOBNRv4EHM} model \cite{Ramos-Buades:2021adz}, in which the eccentricity corrections
contain residual contributions in the QC limit.\footnote{
The modes employed in \texttt{SEOBNRv4EHM} \cite{Ramos-Buades:2021adz} are
  factorized as $h_{\ell m} ^{ \text{F, v4}} =\nolinebreak h_{\ell m}^\text{N, qc} \, \hat S_\text{eff}\, (T_{\ell m}^\text{qc} + T_{\ell m}^\text{ecc})\, (f_{\ell m} ^{ \text{qc}} + f_{\ell m} ^{\text{ecc}} )\, \e^{i \delta_{\ell m}^\text{qc}} $, where the superscript ``qc'' denotes that the corresponding quantity is given by the associated QC expression, $ \hat S _{ \text{eff}} $ is the same as in Eq.~\eqref{eq:Seff}, and the factors $ T_{\ell m}^\text{ecc} $ and $ f_{\ell m} ^{ \text{ecc}} $ account for the rest of the eccentricity corrections.
The eccentricity contributions are written in terms of the variables $ \left( r, p_r, \dot p_r \right) $, such that the QC limit is recovered by setting $ p_r = 0 = \dot p_r $.
However, the values of $ p_r $ and $ \dot p_r $ are not identically zero throughout the entire evolution of a QC system.
Thus, there is always a residual contribution from the eccentricity corrections.
This is not the case for \texttt{SEOBNRv5EHM}.
}
However, in Ref.~\cite{Ramos-Buades:2021adz}, it was checked that the impact on the QC limit was negligible when comparing the accuracy of the model to NR waveforms.


\section{Initial conditions}
\label{sec:ICs}

To compute the evolution of an eccentric, aligned-spin binary system, we numerically solve a set of time-domain ordinary differential equations. Hence, we require a prescription for calculating the initial conditions that correspond to the physical system of interest.

In general, we require 7 input parameters to compute the initial conditions and the dynamics of an aligned-spin eccentric BBH system: the component masses $ m_1 $ and $ m_2 $ (or, equivalently, the mass-ratio $ q $ and the total mass $ M $), the dimensionless spin components $ \chi_1 $ and $ \chi_2 $ in the direction of the orbital angular momentum, the eccentricity of the orbit, a radial phase parameter, and a starting frequency. 
The masses and spins are intrinsic properties of the compact objects, whereas the eccentricity, radial phase parameter, and starting frequency are properties of their orbit.
More specifically, the eccentricity measures the orbit's deformation, and the radial phase parameter determines the relative position of the BHs in their orbit. Together, these input parameters determine a unique physical system.

Our definitions for the eccentricity and radial phase parameter correspond to the Keplerian eccentricity $ e $ and relativistic anomaly $ \zeta $, respectively, defined within the Keplerian parametrization \eqref{eq:Kep_param}. This parametrization is particularly useful since it allows us to employ the \emph{orbit-averaged} orbital frequency $ \langle M \Omega \rangle $ as the starting frequency of the system. 

Therefore, the main challenge is to find the starting values $ (r_0, \phi_0, p_{r_* 0}, p_{\phi 0}) $ that physically correspond to the given input parameters $ (q, \chi_1, \chi_2, e, \zeta, \langle M \Omega \rangle) $, where the total mass $ M $ is just a scaling parameter.
In the following subsections, we discuss a general procedure for translating the input parameters into initial conditions for bound eccentric orbits, under the PA approximation. For the starting value of the orbital phase, we set $ \phi_0 = 0 $, by convention.


\subsection{Postadiabatic approximation for the initial conditions}

In the PA approximation, the dissipative effects due to the emission of GWs represent a small modification to the conservative dynamics~\cite{Buonanno:2000ef,Damour:2002vi,Buonanno:2005xu,Damour:2012ky}.
Thus, the inspiral dynamics can be approximated as a sequence of conservative orbits with decreasing energy and angular momentum.

Under this approximation, we can write the initial values of the dynamical variables as a conservative plus a dissipative contribution.
At first-order in the PA approximation, we have
\begin{subequations}
\label{eq:PA_approx}
\begin{align}
r_0
&= r_{ 0, \s \text{cons}} + r_{ 0, \s \text{diss}} ,
\\
p _{\phi 0}
&= p_{ \phi 0, \s \text{cons}} + p_{\phi 0, \s \text{diss}} ,
\\
p _{r_* 0}
&= p_{ r_* 0, \s \text{cons}} + p_{r_* 0, \s \text{diss}} ,
\end{align}
\end{subequations}
where the terms $ \left( r_{ 0, \s \text{cons}} ,  \, p_{ \phi 0, \s \text{cons}}, \, p_{ r_* 0, \s \text{cons}} \right) $ satisfy conservative EOMs, and $ \left( r_{ 0, \s \text{diss}},  \, p_{ \phi 0, \s \text{diss}}, \, p_{ r_* 0, \s \text{diss}} \right) $ are the first-order PA contributions to the dynamics, which start at 2.5PN order.

To compute the PA contributions, we use the EOM for $ \dot r $ and we relate it to the loss of energy of the system. 
To derive this equation, we first note that by taking a time derivative of Eq.~\eqref{eq:dot_prstar}, we get
\begin{align}
\label{eq:ddprdt}
\frac{\di \dot p_{r_*}}{\di t} =& 
- \frac{\partial}{\partial r} \left( \xi \,\frac{\partial H_{ \text{EOB}}}{\partial r} \right) \dot r
- \xi \, \frac{\partial^2 H_{ \text{EOB}}}{\partial p_{ r_*} \partial r} \, \dot p_{ r_*}
\nonumber
\\
&- \xi \, \frac{\partial^2 H_{ \text{EOB}}}{\partial p_{ \phi} \partial r} \, \dot p_{ \phi}
+ \mathcal{\dot F}_r,
\end{align}
where we have used the chain rule.

Next, we consider the time derivative of the aligned-spin EOB Hamiltonian
\begin{equation}
\label{eq:dHdt}
\frac{\di H_{ \text{EOB}}}{\di t} = 
\frac{\partial H_{ \text{EOB}}}{\partial r} \dot r
+ \frac{\partial H_{ \text{EOB}}}{\partial p_{r_*}} \dot p_{r_*}
+ \frac{\partial H_{ \text{EOB}}}{\partial p_\phi} \dot p_\phi.
\end{equation}

From Eqs.~\eqref{eq:ddprdt} and \eqref{eq:dHdt} we eliminate $ \dot p_\phi $ and solve for $ \dot r $.
After some algebraic manipulations, we get
\begin{widetext}
\begin{equation}
\label{eq:rdot_diss_ecc}
\dot r = 
 \frac{ \displaystyle
-  \dot H _{ \text{EOB}}\frac{\partial^2 H_{ \text{EOB}}}{\partial p_\phi \partial r}
+ \dot p_{r_*} \left( 
	\frac{\partial^2 H_{ \text{EOB}}}{\partial p_\phi \partial r} 
	- \frac{\partial^2 H_{ \text{EOB}}}{\partial p_{r_*} \partial r} 
	\frac{\partial H_{ \text{EOB}}}{\partial p_\phi}
 \right)
+ \frac{ \mathcal{  \dot  F}_r  - \ddot p_{r_*}}{\xi} \frac{\partial H_{ \text{EOB}}}{\partial p_\phi}
}
{
\displaystyle
\frac{\partial H_{ \text{EOB}}}{\partial p_\phi} \frac{\partial^2 H_{ \text{EOB}}}{\partial r^2} 
+ \frac{\partial H_{ \text{EOB}}}{\partial r} 
\left( 
	\frac{\xi'}{\xi} \frac{\partial H_{ \text{EOB}}}{\partial p_\phi}
	- \frac{\partial^2 H_{ \text{EOB}}}{\partial p_\phi \partial r}
\right) 
} .
\end{equation}
\end{widetext}

The next step is to connect Eq.~\eqref{eq:rdot_diss_ecc} with the dissipation caused by the emission of GWs.
This is established by using the following formula for $ \dot H _{ \text{EOB}} $ in terms of the RR force
\begin{align}
\label{eq:dotH}
\dot H _{ \text{EOB}}
&= \dot r \, \mathcal{F}_r + \dot \phi \, \mathcal{F}_\phi,
\nonumber
\\
&= \left( 
\frac{\partial H_{ \text{EOB}}}{\partial p_{r_*}} \frac{p_{r_*}}{p_\phi} \mathcal{F}_r ^{ \text{ecc}}
+ \frac{\partial H_{ \text{EOB}}}{\partial p_\phi} \mathcal{F}_\phi ^{ \text{ecc}}
 \right)
 \mathcal{F} ^{ \text{modes}} _\phi,
\end{align}
where $ \mathcal{F}_r ^{ \text{ecc}} $, $ \mathcal{F}_\phi ^{ \text{ecc}} $, and $ \mathcal{F} ^{ \text{modes}} _\phi$ are defined as in Eqs.~\eqref{eq:fact_RRforces_QcEcc}, and we repeat here the definition of $ \mathcal{F} ^{ \text{modes}} _\phi$ for clarity, 
\begin{equation}
\mF^\text{modes}_\phi = -\frac{M^2 \Omega}{8 \pi} \sum_{\ell=2}^8 \sum_{m=1}^{\ell} m^2\left|d_L h_{\ell m}^{\text{F}}\right|^2,
\end{equation}
with $ h_{\ell m} ^{ \text{F}} $ the factorized modes for eccentric orbits.

Now, we discuss how these results are applied to the computation of the initial conditions for QC and eccentric orbits.


\subsection{Initial conditions for quasicircular orbits}

Conservative QC orbits satisfy the simple relations,
\begin{equation}
p_{r_*, \s \text{cons}}
= 0,
\qquad
\dot p_{r_*, \s \text{cons}}
= 0.
\end{equation}
Thus, from the EOMs \eqref{eq:EOMprStr}, we get
\begin{subequations}
\label{eq:EOM_cons}
\begin{align}
\Omega &= \frac{\partial H_\text{EOB}}{\partial p_\phi}(r,p_{r_*}=0,p_\phi), \\
0 &= \frac{\partial H_\text{EOB}}{\partial r}(r,p_{r_*}=0,p_\phi) ,
\end{align}
\end{subequations}
where $ \Omega = \dot{\phi} $.
Hence, by specifying the initial value of the angular velocity $ \Omega_0 $, we can compute numerically the roots of Eqs.~\eqref{eq:EOM_cons} to get the initial values $ (r_{0, \s \text{cons}}, \, p_{\phi 0, \s \text{cons}}) $.
This determines the conservative part of the initial conditions.

The next step is to determine the dissipative contributions to the initial values. In the QC case, only the radial momentum becomes modified at leading order in RR (see, e.g., Refs.~\cite{Buonanno:2000ef,Buonanno:2005xu}).
Consequently, the initial values for QC orbits satisfy
\begin{equation}
\label{eq:diss_QC_0}
r_{0, \s \text{diss}}
= 0,
\qquad
p_{\phi 0, \s \text{diss}}
= 0,
\qquad
p _{r_* 0}
= p_{r_* 0, \s \text{diss}} \,,
\end{equation}
at leading order in dissipative contributions. 

To determine the radial momentum $p_{r_* 0, \s \text{diss}} $, we apply Eqs.~\eqref{eq:rdot_diss_ecc} and \eqref{eq:dotH} for QC orbits. In this way, we obtain
\begin{equation}
\label{eq:rdot_diss_QC}
\xi \frac{\partial H _{ \text{EOB}}}{\partial p_{r_*}}
= \dot r
= 
- \frac{\Phi_E ^{ \text{qc}}}{\partial  H_{ \text{EOB}}/ \partial p_\phi} \frac{\partial ^2  H_{ \text{EOB}} / \partial p_\phi \partial r }{ \partial ^2  H_{ \text{EOB}} / \partial^2 r}.
\end{equation}
By substituting the initial values of the separation $ r_0 = r_{0, \s \text{cons}} $ and angular momentum $ p_{\phi 0} = p_{\phi 0, \s \text{cons}} $ in Eq.~\eqref{eq:rdot_diss_QC}, we obtain an equation which we solve numerically to find the root $ p_{r_*\s 0} $.

This is the procedure that is employed in the QC \texttt{SEOBNRv5HM} model \cite{Pompiliv5} to find the starting values $ \left( r_0, \,p_{r_* 0}, \,p_{\phi 0} \right) $ that correspond to given input parameters.


\subsection{Initial conditions for eccentric orbits}
\label{sec:ICs_ecc_orb}

The easiest way to find the initial conditions for eccentric orbits given the input parameters $ (q, \chi_1, \chi_2, e, \zeta, \langle M \Omega \rangle) $ is to employ the PN formulas for $ r $, $ p_\phi $ and $ p_{r_*} $ given in Eqs.~\eqref{eq:ErprL_xez_PA}, with $ p_{r_*} = p_r \, \xi(r) $. However, we would like to have a prescription that recovers the same initial conditions as the \texttt{SEOBNRv5HM} model in the QC limit \cite{Pompiliv5}. 
Our starting point is the procedure used in \texttt{SEOBNRv4EHM} \cite{Khalil:2021txt, Ramos-Buades:2021adz, Ramos-Buades:2023yhy}, also based on the PA approximation \eqref{eq:PA_approx}, where we split the problem into conservative and dissipative parts.

We start with the conservative part. The values of $ r_{0, \s \text{cons}} $ and $  p_{\phi 0, \s \text{cons}} $ are determined by numerically finding the roots of the equations
\begin{subequations}
\label{eq:root_cons}
\begin{align}
\Omega(\langle\Omega\rangle, e, \zeta)  &= 
  \frac{\partial  H_ \text{EOB}[r, p_{r_* } (\langle\Omega\rangle, e, \zeta) , p_\phi] }{\partial p_\phi}  ,
\label{eq:root_cons_b}
\\
\dot p_{r_*} (\langle\Omega\rangle, e, \zeta)  
&= 
- \xi (r) \, \frac{\partial  H_ \text{EOB}[r, p_{r_* } (\langle\Omega\rangle, e, \zeta) , p_\phi] }{\partial r} ,
\label{eq:root_cons_a} 
\end{align}
\end{subequations}
where $ \Omega(\langle\Omega\rangle, e, \zeta) $, $ p_{r_* } (\langle\Omega\rangle, e, \zeta) $, and $ \dot p_{r_*} (\langle\Omega\rangle, e, \zeta) $ are calculated with 3PN-accurate formulas without dissipative contributions, given in Eqs.~\eqref{eq:omega_kep}, \eqref{eq:prstar_xez}, and \eqref{eq:dot_pr_xez}. 
Additionally, Eqs.~\eqref{eq:Lxe} and \eqref{eq:rxechi} are employed for the initial guesses of the root solver. In the QC limit, we have $ p_{r_* } = 0 $ and $ \dot p_{r_*} = 0 $, and hence we recover the same equations employed in \texttt{SEOBNRv5HM} \cite{Pompiliv5, Buonanno:2005xu}.

In \texttt{SEOBNRv4EHM}, the formulas for $ p_{r_* } $ and $ \dot p_{r_* } $ entering in Eqs.~\eqref{eq:root_cons} are written in terms of $ p_\phi $ instead of $ \langle \Omega \rangle $. However, we decided to change this because we observed that a dependence on $ p_\phi $ causes unphysical discontinuities in the generated dynamics when varying the relativistic anomaly from $ 0 $ to $ 2 \pi $. The origin of these discontinuities comes from the solutions to Eqs.~\eqref{eq:root_cons}:
if one uses PN formulas in terms of $ p_\phi $, then the root solver has more freedom to roam around in the parameter space, and hence it is likely that it will find another solution for $ \left( r_{0, \s \text{cons}} , p_{\phi 0, \s \text{cons}} \right) $ that does not physically correspond to the input parameters.
Thus, we solve this problem which affects \texttt{SEOBNRv4EHM} by using PN formulas that depend on $ \langle \Omega \rangle $.

As opposed to the QC case \eqref{eq:diss_QC_0}, not only the radial momentum but also the separation and angular momentum receive PA contributions. To calculate their dissipative part, we simply employ the relations \eqref{eq:pphi_PA} and \eqref{eq:r_PA}, such that
\begin{equation}
r_{0, \s \text{diss}} = \tilde{r},
\qquad
p_{\phi 0, \s \text{diss}} = \tilde{p}_{\phi}.
\end{equation}
In this way, the initial values for the separation and angular momentum are given by
\begin{equation}
\label{eq:initial_values_r_pphi}
r_{0} = r_{0, \s \text{cons}} + \tilde{r},
\qquad
p_{\phi 0} = p_{\phi 0, \s \text{cons}} + \tilde{p}_{\phi}.
\end{equation}
Hence, we do not need to add extra equations for the root-solving procedure, and we recover the same prescription used in \texttt{SEOBNRv5HM} since the PA contributions vanish for $ e = 0 $.

The dissipative value of the radial momentum is calculated using Eq.~\eqref{eq:rdot_diss_ecc}. More precisely, we split this equation into conservative and dissipative parts. The conservative part is obtained by expanding the equation $ \dot r   = \xi \, \partial H _{ \text{EOB}} / \partial p_{r_*}$ in terms of the parameters $ ( x, e, \zeta ) $ using Eqs.~\eqref{eq:ErprL_xez}, without any PA contribution. Then, using Eq.~\eqref{eq:x_pphie} we change the dependence on $ x $ to a dependence on $ p_{\phi, \s \text{cons}} $.\footnote{
We find this change convenient since we directly employ the value $ p_{\phi, \s \text{cons}}$ obtained from the solution to the conservative equations \eqref{eq:root_cons}.
Additionally, we remember that the parameters $ \left( x, e, \zeta \right) $ correspond to their secularly evolving parts $ \left( \bar x, \bar e, \bar \zeta \right) $. Equation \eqref{eq:x_pphie} which changes $ x $ into $ p_\phi $ is a \emph{conservative} transformation. Hence, in these equations we have $ p_\phi = \bar p_\phi  = p_{\phi, \s \text{cons}}$.}
Thus, we get
%
\begin{align}
\dot r _{ \text{cons}} (p_{\phi, \s\text{cons}}, e, \zeta) &= \frac{e \sin (\zeta )}{p_{\phi }/(M\mu)} + \frac{\epsilon ^2 e \sin\zeta}{2 p_{\phi }^3/(M\mu)^3} \Big[e^2 (1-\nu) \nonumber\\
&\qquad
-6 e \cos\zeta+\nu -3\Big] + \Order(\epsilon^3),
\end{align}
such that $ \dot r _{ \text{cons}}|_{e=0} = 0 $. The dissipative part is obtained by expanding Eq.~\eqref{eq:rdot_diss_ecc} in terms of the parameters $ \left( x, e, \zeta \right) $, and including their PA contributions via the transformations \eqref{eq:ansatz_PA_Kep}. After that, we change to a dependence on $ p_{\phi, \s \text{cons}} $ using Eq.~\eqref{eq:x_pphie}. Thus, we have
\begin{align}
\label{eq:dot_r_diss}
\dot r _{ \text{diss}} =&
\; \dot r _{ \text{diss}, \s \text{ecc}} (p_{\phi, \s \text{cons}}, e, \zeta)
\nonumber 
\\ 
&- \frac{
\dot H _{ \text{EOB}}\frac{\partial^2 H_{ \text{EOB}}}{\partial p_\phi \partial r}
}
{
\frac{\partial H_{ \text{EOB}}}{\partial p_\phi} \frac{\partial^2 H_{ \text{EOB}}}{\partial r^2} 
\! + \! \frac{\partial H_{ \text{EOB}}}{\partial r} \!
\left( 
	\frac{\xi'}{\xi} \frac{\partial H_{ \text{EOB}}}{\partial p_\phi}
	- \frac{\partial^2 H_{ \text{EOB}}}{\partial p_\phi \partial r}
\right) 
} ,
\end{align}
where $ \dot H _{ \text{EOB}} $ is given by Eq.~\eqref{eq:dotH}, $ \xi ' = d\xi/dr $, and
%
\begin{align}
&\dot r _{ \text{diss}, \s \text{ecc}} (p_{\phi, \s \text{cons}}, e, \zeta)
= \frac{e \nu  \epsilon ^5}{p_\phi^6/(M\mu)^6} \bigg\{
\!\!\left(\!\frac{24 \alpha }{5}-\frac{8 \beta }{5}-\frac{8}{3}\!\right) \! \cos \zeta  \nonumber\\
&\qquad
+ e \left[\left(\frac{72 \alpha }{5}-\frac{16 \beta }{5}-\frac{1424}{45}\right) \! \cos (2 \zeta )-\frac{8 \beta }{5}-\frac{1544}{15}\right] \nonumber\\
&\qquad+ \Order(e^2)
\bigg\} ,
\end{align}
such that $ \dot r _{ \text{diss}, \s \text{ecc}} |_{e=0} = 0$. We remark that these dissipative contributions start at 2.5PN order.

In this way, the initial value of the radial momentum $ p_{r_* 0} $ is obtained by numerically finding the root of
\begin{equation} 
\label{eq:dissipative_equation}
\left(  \dot r _{ \text{cons}} + \dot r _{ \text{diss}} \right)|_0 
= \xi(r) \, \frac{\partial H _{ \text{EOB}}(r_0, p_{r_*} , p_{\phi 0}) }{\partial p_{r_*}}.
\end{equation}
In this equation, the EOB Hamiltonian is always evaluated at the initial values $ (r_0, p_{\phi 0}) $ from \eqref{eq:initial_values_r_pphi}, whereas the PN-expanded formulas are evaluated at $ (p_{\phi, \s \text{cons}}, e, \zeta) $.

This prescription for computing the initial conditions is highly convenient since it allows us to recover the QC initial conditions of \texttt{SEOBNRv5HM} when $ e = 0 $. The only differences come from the term $ \partial H _{ \text{EOB}}/\partial r $ in the denominator of the second term in Eq.~\eqref{eq:dot_r_diss}, and from the first term of Eq.~\eqref{eq:dotH}, which is proportional to $p_{r_*} \partial H / \partial p_{r_*} $. However, we note that the terms $ \partial H _{ \text{EOB}}/\partial r $ and $p_{r_*} \partial H / \partial p_{r_*}$ are zero at Newtonian order for QC orbits, and only start to contribute at 5PN order.
Hence, the difference between the QC expression \eqref{eq:rdot_diss_QC} and the generalized version in Eq.~\eqref{eq:dissipative_equation} becomes negligible for QC orbits and, in practice, both expressions are equivalent in the $ e = 0 $ limit under the first-order PA approximation.

Additionally, we note that the prescription for the initial conditions in \texttt{SEOBNRv4EHM} only uses Eq.~\eqref{eq:rdot_diss_QC} as a first approximation to get the dissipative part of the radial momentum. However, this prescription has two disadvantages: 1) it does not include the dissipative contributions to the separation and angular momentum, and 2) it has a nonrobust and problematic behavior because the factor $ \partial^2 H _{ \text{EOB}} / \partial r^2 $ in the denominator of Eq.~\eqref{eq:rdot_diss_QC} becomes zero or very close to zero for certain values of the relativistic anomaly for moderate eccentricities. Our new prescription is constructed in such a way to solve all these problems.

A final advantage of the Keplerian parametrization in terms of $ (x, e, \zeta) $ is that it allows us to increase the available parameter space of the model.
Certain combinations of the input parameters $ (\langle M \Omega \rangle_0, e_0, \zeta_0) $ are associated to binaries with a very small initial separation ($ r_0 < 10 \, M $).
Naturally, this occurs for systems with high eccentricities, high frequencies, and/or initialized close to periastron $ \zeta = 0 $.
These configurations break down the PN approximations due to the high velocities involved; in particular, the prescription given above for the initial conditions starts to have problems for these challenging systems.
Therefore, in \texttt{SEOBNRv5EHM} \cite{Gamboa:2024hli}, we extend the parameter space by evolving backward in time the two secular evolution equations \eqref{eq:evo_eqs_ex} along with Eq.~\eqref{eq:dotz_LO}, in case the starting separation is smaller than $ 10M $, to get the values of $ (x, e, \zeta) $ in a past time when the initial separation is such that $ r_0 = 10 \, M $.
These new values of the Keplerian parameters are then used with the prescription for the initial conditions given above.
This procedure leads to a binary system whose evolution eventually passes through the original input parameters to a good approximation, depending on how extreme the values of eccentricity and frequency are.


\section{Conclusions}
\label{sec:conclusions}

Developing accurate waveform models for eccentric binaries is important for current and future gravitational-wave detectors to avoid biases in parameter estimation and tests of general relativity, and to distinguish between different binary formation channels.
In this work, we derived new analytical results for inspiraling eccentric-orbit aligned-spin binaries in the EOB formalism, which we used, with collaborators, to develop the inspiral-merger-ringdown waveform model \texttt{SEOBNRv5EHM}, presented in Ref.~\cite{Gamboa:2024hli}.

We first derived 3PN transformations from harmonic to EOB coordinates, and used them to obtain the complete 3PN energy and angular momentum fluxes and waveform modes for eccentric, aligned-spin binaries, based on the harmonic-coordinates results of Refs.~\cite{Arun:2007sg,Arun:2007rg,Arun:2009mc,Mishra:2015bqa,Boetzel:2019nfw,Ebersold:2019kdc,Henry:2023tka}, which also employ the quasi-Keplerian parametrization of the orbit~\cite{damour1985general,Damour:1988mr,schafer1993second,wex1995second}.
We expressed the EOB fluxes and modes in a suitable factorized form using instead a Keplerian parametrization of the orbit in terms of the orbit-averaged frequency $x$, eccentricity $e$, and relativistic anomaly $\zeta$.
From the fluxes, we constructed an EOB RR force in a gauge compatible with the QC \texttt{SEOBNR} models, and we obtained evolution equations for the Keplerian parameters $ (x, e, \zeta) $.
We also derived 3PN relations for computing the initial conditions used in evolving the equations of motion.
As shown in the companion paper~\cite{Gamboa:2024hli}, the new contributions to the EOB RR force and waveform modes at 3PN order are important for modeling highly eccentric binaries, due to the large velocities involved during their periastron passages.
 
Some of the results obtained in this work are valid for generic planar orbits (i.e., bound or unbound orbits), such as the instantaneous contributions to the fluxes and modes expressed in terms of the EOB variables $(r,\phi,p_r,p_\phi)$.
However, our final results are specialized to bound orbits, and expressed in terms of the Keplerian parameters $ \left( x, e, \zeta \right) $, to easily recover the \texttt{SEOBNRv5HM} QC model \cite{Pompiliv5} when $e \to 0$, and because the hereditary contributions to the fluxes and modes are only known analytically in an eccentricity expansion.
Still, the approach followed in this work can be extended to hyperbolic orbits; 
for example, the quasi-Keplerian parametrization is known for hyperbolic orbits to 3PN order~\cite{Cho:2018upo}, and the Keplerian parametrization can also be extended to hyperbolic orbits via analytic continuation.
Some results exist in the literature for the fluxes and modes for hyperbolic orbits, such as Refs.~\cite{Cho:2021onr,Bini:2021qvf,DeVittori:2014psa,Bini:2023mdz}, though the hereditary contributions are generally known at lower orders, in PN and eccentricity expansions, than for bound orbits.

One of our next important goals is the derivation of an accurate and consistent EOB framework that can be employed in waveform models for eccentric binaries with generic spin orientations.
This task has recently started with the works of Refs.~\cite{Liu:2023ldr, Gamba:2024cvy}.
Such a model for eccentric, spin-precessing binaries will benefit from revisiting the prescription for initial conditions, as well as the parametrization and factorization of the eccentricity corrections to the RR force and modes employed in the \texttt{SEOBNRv5EHM} model \cite{Gamboa:2024hli}, and it will be a primary focus of our work in the immediate future.


\section*{Acknowledgments}

We thank Raffi Enficiaud, Guglielmo Faggioli, Quentin Henry, Benjamin Leather, Oliver Long, Philip Lynch, Antoni Ramos-Buades, and Maarten van de Meent for useful discussions.
We also thank K. G. Arun, Yannick Boetzel and Michael Ebersold for providing us with \emph{Mathematica} files containing some of the results of Refs.~\cite{Mishra:2015bqa,Boetzel:2019nfw,Ebersold:2019kdc,Arun:2007sg,Arun:2009mc}, and thank Jan Steinhoff for providing us with a \emph{Mathematica} package for performing canonical transformations.

M.K.'s work is supported by Perimeter Institute for Theoretical Physics. Research at Perimeter Institute is supported in part by the Government of Canada through the Department of Innovation, Science and Economic Development and by the Province of Ontario through the Ministry of Colleges and Universities.

\vspace{0.5cm}
\appendix

\section{Harmonic to EOB coordinate transformations}
\label{app:harmToEOB}

In this appendix, we provide the transformations from harmonic coordinates using the covariant SSC to EOB coordinates using the canonical NW SSC. 
To simplify the notation, we use the scaled dimensionless variables defined in Eq.~\eqref{eq:dimlessVars}.

As described in Sec.~\ref{sec:harmToEOB}, we obtain solutions for the unknown coefficients in the ansatz~\eqref{eq:xharm} for the transformation $\bm r_\text{h} ( \bm{r}, \bm{p})$ from harmonic to EOB coordinates, and subsequently for $\bm{v}_\text{h}(\bm{r},\bm{p})$ from Eq.~\eqref{eq:vsh}, which are included in the Supplemental Material.
With this transformation, we compute the scalars $(r_\text h, \phi_\text h, \dot r_\text h, \dot \phi_\text h)$ for aligned spins, and combine them with the covariant to NW SSC transformation to obtain
%
\begin{widetext}
\small
\begin{subequations}
\label{eq:scalars_HarmToEOB}
\begin{align}
r_\text{h} &= \sqrt{\bm r_\text{h} \cdot \bm r_\text{h}} \nonumber\\
&= r 
+ \epsilon^2 \left[-\frac{\nu }{2}+\frac{\nu  p_{\phi }^2}{2 r}+\frac{3}{2} \nu  r p_r^2-1\right] -\frac{\nu  \epsilon ^3 p_{\phi } \chi _S}{r}
+ \epsilon^4 \bigg[
\frac{ p_{\phi }^4}{r^3}\left(-\frac{\nu^2 }{8}-\frac{\nu}{8}\right) 
+\frac{p_{\phi }^2}{r^2}\left(\frac{3 \nu^2 }{8}-\frac{\nu}{8}\right) 
+\frac{p_r^2 p_{\phi }^2}{r}\left(\frac{3 \nu^2 }{4}-\frac{3\nu}{4}\right)\nonumber \\
&\qquad 
+\left(\frac{3 \nu^2 }{8}-\frac{5\nu}{8}\right) r p_r^4-7 \nu p_r^2
+\frac{1}{r}\left(\frac{19\nu}{4}-\frac{\nu^2 }{4}\right)\bigg]
+ \frac{\epsilon ^4}{r} \left[\left(\frac{1}{2}-2 \nu \right) \chi _A^2+\delta  \chi _A \chi _S+\frac{\chi _S^2}{2}\right]
- \epsilon^5 \frac{8 \nu  (1+\alpha -\beta ) \,  p_r}{5  \,   r}
\nonumber \\
&\quad
+ \epsilon^5 \left\{
\frac{p_{\phi }^3 }{r^3} \left[\left(\frac{\nu ^2}{4}+\frac{\nu }{8}\right) \chi _S-\frac{\delta  \nu  \chi _A}{8} \right]
+\frac{p_\phi}{r^2}\left[\frac{33}{16} \delta  \nu  \chi _A+\left(\frac{59 \nu }{16}-\frac{11 \nu ^2}{8}\right) \chi _S\right]
+\frac{p_\phi p_r^2}{r} \left[\left(\frac{5 \nu ^2}{4}+\frac{\nu }{8}\right) \chi _S-\frac{\delta  \nu  \chi _A}{8} \right] \right\}
\nonumber \\
&\quad 
+ \epsilon^6 \Bigg\{
\frac{p_{\phi }^6}{r^5} \!\left(\frac{\nu ^3}{16}+\frac{\nu^2 }{16}+\frac{\nu}{16}\right)  
+\frac{p_r^2 p_{\phi }^4}{r^3} \!\left(\frac{5 \nu ^3}{16}-\frac{15 \nu^2 }{16}+\frac{9\nu}{16}\right)\! 
+\frac{p_{\phi }^2}{r^3}\left(\frac{\nu ^3}{2}-\frac{11 \nu^2 }{4}+\frac{91\nu}{12}\right)\! 
+\frac{p_r^2 p_{\phi }^2}{r^2} \!\left(\frac{7 \nu ^3}{16}-\frac{135 \nu^2 }{16}+\frac{37\nu}{16}\right)\!  
\nonumber\\
&\qquad
+\frac{p_{\phi }^4}{r^4}\left(\frac{\nu}{16}-\frac{5 \nu ^3}{16}+\frac{\nu^2 }{16}\right)
+r p_r^6 \left(\frac{7\nu}{16}-\frac{\nu ^3}{16}-\frac{9 \nu^2 }{16}\right)
+\frac{p_r^2}{r}\left(\frac{275\nu}{24}-\frac{\nu ^3}{8}+\frac{37 \nu^2 }{8}\right)  
+p_r^4 \left(\frac{5 \nu ^3}{48} - \frac{133 \nu^2 }{48}+\frac{179\nu}{48}\right) \nonumber\\
&\qquad 
+\frac{p_r^4 p_{\phi }^2}{r}\left(\frac{15\nu}{16}-\frac{5 \nu ^3}{16}-\frac{25 \nu^2 }{16}\right)
+ \left[ \frac{137 \nu }{105}-\frac{41 \pi ^2 \nu }{64}+\frac{7 \nu ^2}{4}-\frac{\nu    ^3}{4}+\gamma_\text{ln} \ln \left(\frac{r}{r_0'}\right) \right]\frac{1}{r^2} \Bigg\} \nonumber\\
&\quad
+\epsilon^6 \Bigg\{
\frac{1}{r^2} \!\left[\!
\left(\frac{1}{2}-\nu ^2+\frac{3 \nu }{4}\right)\! \chi _S^2 
+ \left(\frac{3 \nu }{2}+1\right)\! \delta\chi _A \chi _S
+\left(\frac{1}{2}-\nu ^2-\frac{5 \nu }{4}\right)\! \chi _A^2
- \left(\frac{\nu }{4}+\frac{1}{2}\right)\! \delta\kappa _A
-\left(\frac{\nu ^2}{2}-\frac{3 \nu }{4}+\frac{1}{2}\right)\! \kappa _S\right] \nonumber\\
&\qquad
+ \frac{p_{\phi }^2}{r^3} \left[\left(\nu ^2-\frac{\nu }{4}\right) \chi _A^2-\frac{\delta}{2}  \nu  \chi _A \chi _S-\frac{\nu  \chi _S^2}{4}\right]
+ \frac{p_r^2}{r} \left[\left(\frac{9 \nu }{4}-9 \nu ^2\right) \chi _A^2+\frac{9}{2} \delta  \nu  \chi _A \chi _S+\left(\frac{\nu ^2}{2}+\frac{9 \nu }{4}\right) \chi _S^2\right]
\Bigg\},
\label{eq:rh} 
\\ 
\dot r_\text{h} &= \frac{\bm v_\text{h} \cdot \bm r_\text{h}}{r_\text{h}} \nonumber\\
&= p_r 
+\epsilon^2 \left[\left(-\frac{1}{2}+\nu \right) p_r^3+\frac{p_\phi^2p_r}{r^2} \left(-\frac{1}{2}+2    \nu \right) +\frac{ p_r}{r}(-3-2 \nu )\right]
+\frac{\nu  \epsilon ^3 p_r p_{\phi } \chi _S}{r^2}
+\epsilon^4 \Bigg[
\frac{p_\phi^2 p_r^3}{r^2}\left(\frac{3}{4}-3 \nu -\frac{\nu    ^2}{2}\right) 
\nonumber \\
&\qquad 
+\left(\frac{3}{8}-\nu \right) p_r^5
+\frac{ p_\phi^4    p_r}{r^4}\left(\frac{3}{8}-2 \nu +\nu ^2\right)
+\frac{p_\phi^2 p_r}{4 r^3}\left(6-55 \nu +\nu    ^2\right) 
+\frac{\left(\frac{5}{2}-4 \nu \right) p_r^3}{r}
+\frac{\left(6+39 \nu -5 \nu ^2\right) p_r}{4    r^2}\Bigg]\nonumber \\
&\quad 
+\frac{\epsilon ^4 p_r }{r^2}\left[\left(\frac{1}{2}-2 \nu \right) \chi _A^2+\delta  \chi _A \chi _S+\frac{\chi _S^2}{2}\right] 
+\epsilon^5 \left[
\frac{8 (1+\alpha -\beta ) \nu  p_r^2}{5 r^2}-\frac{8    \nu  (1+\alpha -\beta ) p_\phi^2}{5 r^4}
+\frac{8 \nu  (1+\alpha -\beta )}{5    r^3}\right]\nonumber\\
&\quad
+\epsilon^5 \Bigg\{
\frac{p_r p_{\phi }^3}{r^4} \left[\frac{1}{8} \delta  \nu  \chi _A+\left(\frac{5 \nu ^2}{4}-\frac{5 \nu }{8}\right) \chi _S\right]
+\frac{p_r p_{\phi }}{r^3} \left[\frac{97}{8} \delta  \nu  \chi _A+\left(\frac{43 \nu }{8}-\frac{27 \nu ^2}{4}\right) \chi _S\right]\nonumber\\
&\qquad
+\frac{p_r^3 p_{\phi }}{r^2} \left[\frac{1}{8} \delta  \nu  \chi _A+\left(-\frac{7 \nu ^2}{4}-\frac{5 \nu }{8}\right) \chi _S\right]
\Bigg\}
+\epsilon^6 \Bigg\{
\frac{p_r p_{\phi }^6}{r^6}\left(-2 \nu ^2+2 \nu -\frac{5}{16}\right) 
+\left(-\frac{\nu ^2}{4}+\nu -\frac{5}{16}\right) p_r^7
\nonumber\\
&\qquad
+\frac{p_r^3 p_{\phi }^4}{r^4}\left(-2 \nu ^3-2 \nu ^2+5 \nu -\frac{15}{16}\right) 
+\frac{p_r p_{\phi }^2}{r^4}\left(-\frac{\nu ^3}{2}+\frac{119 \nu ^2}{8}+\frac{73 \nu }{6}-\frac{3}{4}\right) 
+\frac{p_r^5}{r}\left(\frac{3 \nu ^2}{2}+6 \nu -\frac{21}{8}\right)  
 \nonumber\\
&\qquad
+\frac{p_r p_{\phi }^4}{r^5}\left(\frac{3 \nu ^3}{2}-\frac{103 \nu ^2}{8}+\frac{45 \nu }{4}-\frac{9}{8}\right) 
+\frac{ p_r^5 p_{\phi }^2}{r^2}\left(\frac{\nu ^3}{2}-\frac{\nu ^2}{4}+4 \nu -\frac{15}{16}\right)
+\frac{ p_r^3}{r^2}\left(\frac{13 \nu ^3}{12}+\frac{191 \nu ^2}{24}+\frac{25 \nu }{2}-\frac{15}{4}\right)\nonumber\\
&\qquad
+\frac{p_r^3 p_{\phi }^2}{r^3}\left(-\frac{13 \nu ^3}{12}+\frac{433 \nu ^2}{24}+\frac{127 \nu }{6}-\frac{15}{4}\right)
+\frac{p_r}{r^3}\left(-\nu ^3-\frac{15 \nu ^2}{4}+\frac{41 \pi ^2 \nu }{32}+\frac{3751 \nu }{105}+\frac{1}{2}\right) 
\Bigg\} \nonumber\\
&\quad
+ \epsilon^6 \Bigg\{
\frac{p_r^3}{r^2} \left[\left(-6 \nu ^2+\frac{9 \nu }{2}-\frac{3}{4}\right) \chi _A^2+  \left(3 \nu -\frac{3}{2}\right) \delta\chi _A \chi _S+\left(-\frac{\nu ^2}{2}+\frac{3 \nu }{2}-\frac{3}{4}\right) \chi _S^2\right] \nonumber\\
&\qquad
+\frac{p_r }{r^3}\left[\left(6 \nu ^2-\frac{\nu }{2}-\frac{1}{2}\right)\! \chi _A^2
+  (7 \nu -1) \delta \chi _A \chi _S 
-\left(2 \nu ^2-\frac{19 \nu }{2}+\frac{1}{2}\right)\! \chi _S^2
+\left(3 \nu ^2+\frac{9 \nu }{2}-\frac{5}{2}\right)\! \kappa _S
- \left(\frac{\nu }{2}+\frac{5}{2}\right)\! \delta \kappa _A\right] \nonumber\\
&\qquad
+\frac{p_r p_{\phi }^2 }{r^4}\left[\left(4 \nu ^2-2 \nu +\frac{1}{4}\right) \chi _A^2+  \left(\frac{1}{2}-2 \nu \right) \delta\chi _A \chi _S+\left(\nu ^2-\nu +\frac{1}{4}\right) \chi _S^2\right]
\Bigg\},
\label{eq:rdh}
\\
\dot \phi_\text{h} &= \frac{1}{r_\text{h}}\sqrt{\bm v_\text{h} \cdot \bm v_\text{h} - \dot r_\text{h}^2} \nonumber\\
&= \frac{p_\phi}{r^2}
+\epsilon^2 \left[\frac{ p_{\phi }^3}{r^4}\left(\frac{\nu }{2}-\frac{1}{2}\right)
-\frac{p_{\phi }}{r^3}+\frac{ p_r^2 p_{\phi }}{r^2}\left(-\frac{3 \nu }{2}-\frac{1}{2}\right)\right]
+\epsilon ^3 \left[\frac{2 \delta  \chi _A+(2-2 \nu ) \chi _S}{r^3}+\frac{\nu  p_{\phi }^2 \chi _S}{r^4}-\frac{\nu  p_r^2 \chi _S}{r^2}\right]
\nonumber \\
&\quad 
+\epsilon^4 \Bigg[\frac{p_{\phi }^5}{r^6}\left(-\frac{\nu ^2}{8}-\frac{5 \nu }{8}+\frac{3}{8}\right) 
+\frac{p_{\phi }^3}{r^5}\left(\frac{3 \nu ^2}{4}-\frac{17 \nu }{4}+\frac{1}{2}\right) 
+\frac{p_r^2 p_{\phi }^3}{r^4}\left(-\frac{9 \nu ^2}{4}+\frac{3 \nu }{4}+\frac{3}{4}\right) 
+\frac{ p_{\phi }}{r^4}\left(-\frac{\nu ^2}{4}+\frac{9 \nu }{4}-\frac{1}{2}\right)\nonumber \\
&\qquad 
+\frac{p_r^2 p_{\phi }}{r^3}\left(-\frac{\nu ^2}{2}+12 \nu +\frac{3}{2}\right) 
+\frac{p_r^4 p_{\phi }}{r^2}\left(\frac{15 \nu ^2}{8}+\frac{11 \nu }{8}+\frac{3}{8}\right)  \Bigg]
+ \frac{\epsilon ^4 p_{\phi }}{r^4} \left[(4 \nu -1) \chi _A^2-2 \delta  \chi _A \chi _S-\chi _S^2\right]\nonumber \\
&\quad
+\epsilon^5 \frac{ 8 \nu  (3+3 \alpha -2 \beta )    p_\phi p_r}{5 r^4}
+\epsilon^5 \Bigg\{
\frac{p_{\phi }^4}{r^6} \left[\frac{\delta \nu  \chi _A}{8}-\left(\frac{\nu ^2}{4}+\frac{5 \nu }{8}\right)\! \chi _S\right]
+\frac{p_{\phi }^2}{r^5} \left[\frac{3\delta  \nu  \chi _A}{8}+\left(\frac{11 \nu ^2}{4}-\frac{55 \nu }{8}\right)\! \chi _S\right]
-\frac{6 \nu ^2 p_r^2 p_{\phi }^2 \chi _S}{r^4}\nonumber \\
&\qquad
+\frac{p_r^2}{r^3} \left[\left(6 \nu ^2-\frac{7 \nu }{2}\right) \chi _S-\frac{13\delta  \nu  \chi _A}{2} \right]
+\frac{p_r^4 }{r^2}\left[\left(\frac{9 \nu ^2}{4}+\frac{5 \nu }{8}\right) \chi _S-\frac{\delta  \nu  \chi _A}{8} \right]
+\frac{1}{r^4}\left[\frac{\delta  \nu  \chi _A}{2} +\left(-3 \nu ^2-\frac{\nu }{2}\right) \chi _S\right]
\Bigg\}\nonumber \\
&\quad
+\epsilon^6 \Bigg\{
\frac{p_{\phi }^7}{r^8}\left(\frac{\nu ^3}{16}+\frac{11 \nu }{16}-\frac{5}{16}\right) 
+\frac{p_r^4 p_{\phi }}{r^3}\left(\frac{17 \nu ^3}{12}-\frac{653 \nu ^2}{24}-\frac{317 \nu }{24}-\frac{15}{8}\right)
+\frac{p_r^2 p_{\phi }^5}{r^6}\left(-\frac{\nu ^3}{16}+\frac{15 \nu ^2}{4}+\frac{\nu }{16}-\frac{15}{16}\right) \nonumber \\
&\qquad
+\frac{ p_{\phi }^3}{r^6}\left(\frac{5 \nu ^3}{4}-\frac{37 \nu ^2}{8}+\frac{13 \nu }{24}+\frac{1}{4}\right)
+\frac{ p_r^2 p_{\phi }^3}{r^5}\left(-\frac{35 \nu ^3}{8}+\frac{139 \nu ^2}{4}-\frac{15 \nu }{4}-\frac{9}{4}\right)
+\frac{ p_{\phi }}{r^5}\left(-\frac{\nu ^3}{2}+\frac{17 \nu ^2}{4}+\frac{161 \nu }{24}-\frac{1}{2}\right)\nonumber \\
&\qquad
+\frac{ p_r^2 p_{\phi }}{r^4}\left(\frac{5 \nu ^3}{8}-\frac{41 \nu ^2}{4}-\frac{113 \nu }{8}-\frac{3}{4}\right)
+\frac{p_{\phi }^5}{r^7}\left(-\frac{\nu ^3}{2}-\frac{\nu ^2}{4}+\frac{27 \nu }{8}-\frac{3}{8}\right) 
+\frac{ p_r^4 p_{\phi }^3}{r^4}\left(\frac{91 \nu ^3}{16}+\frac{3 \nu ^2}{2}-\frac{31 \nu }{16}-\frac{15}{16}\right) \nonumber \\
&\qquad
+\frac{ p_r^6 p_{\phi }}{r^2}\left(-\frac{35 \nu ^3}{16}-\frac{9 \nu ^2}{4}-\frac{21 \nu }{16}-\frac{5}{16}\right)
\Bigg\}
+ \epsilon^6 \Bigg\{
\frac{p_{\phi }^3 }{r^6}\left[\left(4 \nu ^2-5 \nu +1\right) \chi _A^2+  (2-2 \nu ) \delta \chi _A \chi _S+\left(\nu ^2-\nu +1\right) \chi _S^2\right]\nonumber \\
&\qquad
+\frac{p_{\phi }}{r^5} \left[\delta  \left(-\nu -\frac{1}{2}\right) \kappa _A+\left(4 \nu ^2+3 \nu -1\right) \chi _A^2+\delta  (4 \nu -2) \chi _A \chi _S+\left(2 \nu ^2-\frac{1}{2}\right) \kappa _S+\left(-4 \nu ^2+5 \nu -1\right) \chi _S^2\right]\nonumber \\
&\qquad
-\frac{3 \nu ^2 p_r^2 p_{\phi } \chi _S^2}{r^4}
\Bigg\},
\label{eq:phidh} \\
\phi_\text{h} &= \arctan  \left( \frac{y_\text{h}}{x_\text{h}} \right) \nonumber\\
&= \phi + \epsilon^2 \frac{\nu  p_\phi p_r}{ r}
+ \frac{\nu  \epsilon ^3 p_r \chi _S}{r}
+ \epsilon^4 \left[
\frac{ p_r p_{\phi }}{r^2}\left(\frac{3 \nu ^2}{4}-\frac{15 \nu }{4}\right)
-\frac{\nu  p_r p_{\phi }^3}{2 r^3}
+\frac{p_r^3 p_{\phi }}{r}\left(-\nu ^2-\frac{\nu }{2}\right) \right]
- \epsilon^5\frac{8 \nu  (3+3 \alpha -2    \beta ) p_\phi}{15  r^3}\nonumber \\
&\quad
+ \epsilon^5 \left\{
\frac{p_r p_{\phi }^2}{r^3} \left[\frac{\delta}{8} \nu  \chi _A+\left(\frac{\nu ^2}{4}-\frac{\nu }{8}\right) \chi _S\right]
+\frac{p_r}{r^2} \left[\frac{41}{16} \delta  \nu  \chi _A+\left(\frac{35 \nu }{16}-\frac{3 \nu ^2}{8}\right) \chi _S\right]
+\frac{p_r^3}{r} \left[\frac{\delta}{8} \nu  \chi _A-\left(\frac{7 \nu ^2}{4}+\frac{\nu }{8}\right) \chi _S\right]\!
\right\}\nonumber \\
&\quad
+ \epsilon^6 \Bigg\{
\frac{p_r p_{\phi }^5}{r^5}\left(\frac{3 \nu }{8}-\frac{\nu ^2}{4}\right) 
+\frac{p_r p_{\phi }^3}{r^4}\left(-\frac{\nu ^3}{8}-\frac{15 \nu ^2}{8}+\frac{7 \nu }{8}\right) 
+\frac{ p_r^3 p_{\phi }^3}{r^3}\left(-\frac{\nu ^3}{3}+\frac{\nu ^2}{2}+\frac{3 \nu }{4}\right)
+\frac{ p_r^5 p_{\phi }}{r}\left(\nu ^3+\frac{3 \nu ^2}{4}+\frac{3 \nu }{8}\right) \nonumber \\
&\qquad
+\frac{p_r p_{\phi }}{r^3}\left(\frac{3 \nu ^3}{4}-\frac{3 \nu ^2}{4}-\frac{95 \nu }{24}\right)
+\frac{p_r^3 p_{\phi }}{r^2}\left(-\frac{19 \nu ^3}{12}+\frac{55 \nu ^2}{6}+\frac{55 \nu }{24}\right) 
\Bigg\}
+ \epsilon ^6 \frac{\nu ^2 p_r p_{\phi } \chi _S^2}{r^3},
\label{eq:phih} 
\end{align}
\end{subequations}
\normalsize
\end{widetext}
where $x_\text{h}$  and $y_\text{h}$ in the last equation are the harmonic-coordinates Cartesian components in the orbital plane.
The constant $\gamma_\text{ln}$ that appears at 3PN order is zero for modified-harmonic (MH) coordinates, but is given by $\gamma_\text{ln} = 22 \nu /3 $ for standard-harmonic (SH) coordinates.
Additionally, the magnitude of the velocity can be easily obtained from the above equations using $v _{ \text{h}}^2 = \bm{v}_\text{h}\cdot\bm{v}_\text{h} = \dot r _{ \text{h}} ^2 + r _{ \text{h}}^2 \, \dot \phi _{ \text{h}}^2$.

Our results are consistent with previous derivations at 2PN order in Refs.~\cite{Bini:2012ji,Khalil:2021txt}. Furthermore, we can identify the 2.5PN nonspinning part of our transformations as the gauge freedom in defining the RR force, as discussed in Refs.~\cite{Iyer:1993xi,Iyer:1995rn,Gopakumar:1997ng}, which is allowed by the flux-balance equations. 

The 3PN transformation from the QK parametrization $(e_t,l)$ in harmonic coordinates to the Keplerian parametrization $(e,\zeta)$ in EOB coordinates is given by
\begin{widetext}
\small
\begin{subequations}
\begin{align}
e_t &= e \Bigg\{
1+ \epsilon ^2  x (\nu -3)
+ \frac{2 x^{3/2} \epsilon ^3 (\delta \chi_A+\chi_S)}{\sqrt{1-e^2}}
+ \frac{x^2 \epsilon ^4 }{6 \left(1-e^2\right)} \left[e^2 \left(3 \nu-4 \nu ^2 -18\right)+9 \sqrt{1-e^2} (2 \nu -5)+4 \nu ^2-45 \nu +30\right] \nonumber\\
&\qquad
- \frac{x^2 \epsilon ^4}{2 \left(1-e^2\right)}  \left[\delta  \kappa _A+(3-12 \nu ) \chi _A^2+6 \delta  \chi _A \chi _S-2 \nu  \kappa _S+\kappa _S+3 \chi _S^2\right]
-\frac{x^{5/2} \epsilon ^5}{48 \left(1-e^2\right)^{3/2}} \bigg[\delta  \chi _A \left[\left(123 e^2-47\right) \nu +96\right] \nonumber\\
&\quad\qquad
+\left[-2 \left(9 e^2-5\right) \nu ^2+\left(201 e^2+475\right) \nu +96\right] \chi _S
+\sqrt{1-e^2} \left[192 \delta  (\nu -3) \chi _A-96 \left(\nu ^2-8 \nu +6\right) \chi _S\right]\bigg] \nonumber\\
&\qquad
+\frac{x^3 \epsilon ^6}{6720 \left(1-e^2\right)^2} \bigg[280 e^4 \nu  \left(8 \nu ^2+8 \nu -21\right)-560 e^2 \left(8 \nu ^3-120 \nu ^2-3 \nu -66\right)+2240 \nu ^3+24640 \nu ^2 \nonumber\\
&\qquad\quad
+47144 \nu -26880 -35 \sqrt{1-e^2} \left[96 \left(7 e^2+8\right) \nu ^2+\left(-672 e^2+123 \pi ^2-9824\right) \nu +1440\right]\bigg] \nonumber\\
&\qquad
+ \frac{x^3 \epsilon ^6}{6 \left(1-e^2\right)^2} \bigg[
\chi _S^2 \left[3 e^2 \left(2 \nu ^2-7 \nu +9\right)-6 \nu ^2+30 \nu -3\right]
-3 (4 \nu -1) \chi _A^2 \left[e^2 (\nu +9)-6 \nu -1\right] \nonumber\\
&\qquad\quad
-6 \delta  \chi _A \chi _S \left[3 e^2 (\nu -3)-2 \nu +1\right]
+ \delta  \kappa _A \left[e^2 (\nu +3)+2 \nu -3\right]
+ \kappa _S\left[e^2 \left(-2 \nu ^2-5 \nu +3\right)+32 \nu ^2+8 \nu -3\right] \nonumber\\
&\qquad\quad
+ \sqrt{1-e^2} \Big[
12 \delta  (9 \nu -11) \chi _A \chi _S+\left(-24 \nu ^2+96 \nu -66\right) \chi _S^2
-6 \left(6 \nu ^2-46 \nu +11\right) \chi _A^2
+3 \delta \kappa _A  (5 \nu -14)\nonumber\\
&\qquad\qquad +3 \kappa _S \left(-6 \nu ^2+33 \nu -14\right)\Big]
\bigg]
\Bigg\}. 
\label{eq:app:et}\\
l &= \zeta - 2 e \sin\zeta + \frac{3}{4} e^2 \sin (2 \zeta )
+ x \epsilon ^2 \left[3 e \sin\zeta-3 e^2 \sin\zeta \cos\zeta\right]
+ e x^{3/2} \epsilon ^3 \sin\zeta (e \cos\zeta-1) \left[2 \delta  \chi _A-(\nu -2) \chi _S\right] \nonumber\\
&\quad
-\frac{e}{2} x^2 \epsilon ^4 \sin\zeta (e \cos\zeta-1) \left[\delta  \kappa _A+2 (1-4 \nu ) \chi _A^2+2 \chi _S \left(2 \delta  \chi _A+\chi _S\right)+(1-2 \nu ) \kappa _S\right] \nonumber\\
&\quad
+ x^2 \epsilon ^4 \left[\frac{e^2}{8} (26 \nu -25) \sin (2 \zeta )+2 e (5-4 \nu ) \sin\zeta\right]
+ \frac{x^{5/2} \epsilon ^5}{32} \Big\{e^2 \sin (2 \zeta ) \left[\delta  (256-159 \nu ) \chi _A+\left(66 \nu ^2-533 \nu +256\right) \chi _S\right] \nonumber\\
&\qquad
+e \sin\zeta \left[\delta  (436 \nu -768) \chi _A+\left(-200 \nu ^2+1468 \nu -768\right) \chi _S\right]\Big\} \nonumber\\
&\quad
+ x^3 \epsilon ^6 \left\{
e \left[\frac{41 \nu ^2}{3}+\left(\frac{41 \pi ^2}{16}-\frac{2023}{12}\right) \nu +44\right] \sin\zeta
+e^2 \left[-\frac{19 \nu ^2}{3}+\left(\frac{433}{8}-\frac{287 \pi ^2}{256}\right) \nu -\frac{19}{2}\right] \sin (2 \zeta )
\right\} \nonumber\\
&\quad
+ x^3 \epsilon ^6 \bigg\{
\frac{e}{6}  \sin\zeta \Big[
\left(75 \nu ^2-295 \nu +138\right) \chi _S^2
+\left(124 \nu ^2-595 \nu +138\right) \chi _A^2+\delta  (276-338 \nu ) \chi _A \chi _S+\delta  (108-53 \nu ) \kappa _A \nonumber\\
&\qquad
+\left(58 \nu ^2-269 \nu +108\right) \kappa _S\Big]
+ \frac{e^2}{24}  \sin (2 \zeta ) \Big[
\left(-108 \nu ^2+404 \nu -210\right) \chi _S^2
+\left(-140 \nu ^2+896 \nu -210\right) \chi _A^2\nonumber\\
&\qquad
+\delta  (460 \nu -420) \chi _A \chi _S
+\delta  (73 \nu -144) \kappa _A +\left(-62 \nu ^2+361 \nu -144\right) \kappa _S
\Big]
\bigg\} + \Order(e^3).
 \label{eq:app:mean_anomaly}
\end{align}
\end{subequations}
\normalsize
The eccentricity expansion in $l(e,\zeta)$ is provided in the Supplemental Material to $\Order(e^6)$.
\end{widetext}
%


\section{PN expressions in the Keplerian parametrization}
\label{app:Keplerian}

\subsection{Conservative dynamics}
The Keplerian parametrization is defined by
\begin{equation}
\label{eq:app:rKepler}
r = \frac{M}{u_p (1 + e \cos \zeta)},
\end{equation} 
where $u_p$ is the inverse semilatus rectum scaled by the total mass, $ e $ is the Keplerian eccentricity, and $\zeta$ is the relativistic anomaly.
At the turning points of an eccentric orbit (periastron and apastron), we have $\zeta = 0$ or $\zeta=\pi$, and thus, $r_\pm = M / (u_p(1 \pm e))$. 
Evaluating the Hamiltonian at the turning points, at which $p_r = 0$, and solving for the energy and angular momentum yields
\begin{subequations}
\begin{align}
\frac{E}{\mu} &= \frac{1}{2} \left(e^2-1\right) u_p - \epsilon^2 \frac{u_p^2}{8} \left(e^2-1\right)^2 (\nu -3) + \Order(\epsilon^3),
\\
\frac{p_\phi}{M\mu} &= \frac{1}{\sqrt{u_p}} + \epsilon^2\frac{\sqrt{u_p}}{2} \left(e^2+3\right) + \Order(\epsilon^3),
\end{align}
\end{subequations}
which can be inverted to obtain $e(E,p_\phi)$ and $u_p(E,p_\phi)$.
In addition, the Hamiltonian can be inverted to obtain $p_r(E,p_\phi,r)$, leading to
\begin{align}
\label{eq:app:prKepler}
\frac{p_r}{\mu} &=  e \sqrt{u_p} \sin\zeta 
\nonumber
\\
& \quad +\epsilon^2\frac{e u_p^{3/2}}{2} \!\left(1+e^2+2 e \cos\zeta\right) \sin\zeta + \Order(\epsilon^3).
\end{align}

The radial and azimuthal periods are given, respectively, by
\begin{subequations}
\begin{align}
P &= \oint dt = \oint \left(\frac{\partial H}{\partial p_r}\right)^{-1} dr = 2\int_{0}^{\pi}  \left(\frac{\partial H}{\partial p_r}\right)^{-1} \frac{dr}{d\zeta} \, d\zeta \nonumber\\
& = \frac{2 \pi M}{\left(u_p-e^2 u_p\right)^{3/2}} -\epsilon^2\frac{\pi M (\nu -6)}{\sqrt{u_p-e^2 u_p}} + \Order(\epsilon^3), \\
P_\phi &=  \oint \dot{\phi} \, dt = \oint \frac{\partial H}{\partial p_\phi} dt \nonumber\\
&= 2 \pi + 6 \pi \epsilon^2 u_p + \Order(\epsilon^3).
\end{align}
\end{subequations}

The associated orbit-averaged frequencies are
\begin{equation}
n = \frac{2\pi}{P}, \qquad
\langle \Omega \rangle = \frac{P_\phi}{P}.
\end{equation}

The dimensionless frequency $x \equiv \langle M \Omega \rangle^{2/3}$ is given by
\begin{align}
x &= u_p(1-e^2) + \epsilon^2\frac{u_p^2}{3} \left(e^2-1\right) \left[e^2 (\nu -6)-\nu \right] + \Order(\epsilon^3),
\end{align}
which can be inverted to obtain $u_p(x,e)$,
\begin{align}
\label{eq:app:upxe}
u_p &= -\frac{x}{e^2-1} + \epsilon^2 \frac{x^2 \left[e^2 (\nu -6)-\nu \right]}{3 \left(e^2-1\right)^2} + \Order(\epsilon^3).
\end{align} 

Plugging $u_p(x,e)$ into Eqs.~\eqref{eq:app:rKepler}-\eqref{eq:app:prKepler}, we obtain $E$, $p_\phi$, $r$, and $p_r$ as functions of $(x,e,\zeta)$, which are given by
\begin{widetext}
\small
\begin{subequations}
\label{eq:app:ErprL_xez}
\begin{align}
E &= -\frac{\mu x}{2}\Bigg\{
1- \frac{\epsilon^2 x }{12 \left(1-e^2\right)}\left[9+\nu -e^2 (-15+\nu ) \right]
-\frac{4 \epsilon^3 x^{3/2}}{3 \left(1-e^2\right)^{3/2}}  \left[(\nu -2) \chi _S-2 \delta  \chi _A\right]
\nonumber  \\
&\quad
- \frac{\epsilon^4 x^2}{24 \left(1-e^2\right)^2}\bigg[201-105 \nu +\nu ^2 -2 e^2 \left(87+15 \nu +\nu ^2\right)+e^4 \left(-15+15 \nu +\nu ^2\right)-24 \left(1-e^2\right)^{3/2} (5-2 \nu )\bigg]
\nonumber  \\
&\quad
-\frac{\epsilon ^4 x^2 }{\left(1-e^2\right)^2} \left[\chi _S^2+(1-4 \nu ) \chi _A^2+2 \delta  \chi _A \chi _S+ (1-2 \nu)  \kappa _S + \delta  \kappa _A\right]
-\frac{\epsilon ^5 x^{5/2} }{9 \left(1-e^2\right)^{5/2}} \bigg\{
\delta  \chi _A \left[8 e^2 (\nu +21)+55 \nu -144\right]
 \nonumber\\
&\qquad
+ \left[-4 e^2 \left(\nu ^2+10 \nu -42\right)-14 \nu ^2+217 \nu -144\right] \chi _S
+\left(1-e^2\right)^{3/2} \left[12 \left(\nu ^2-8 \nu +6\right) \chi _S-24 \delta  (\nu -3) \chi _A\right]\bigg\}
 \nonumber\\
&\quad
- \frac{\epsilon^6 x^3 }{5184 \left(1-e^2\right)^3}  \bigg\{67635+27 \left(492 \pi ^2-18319\right) \nu +27378 \nu ^2+35 \nu ^3
+3 e^4 \left[33507-261 \nu +66 \nu ^2+35 \nu ^3\right]
\nonumber  \\
&\qquad
+3 e^2 \left[-55539+9 \left(123 \pi ^2-1651\right) \nu +6078 \nu ^2-35 \nu ^3\right]
-5 e^6 \left[999+1215 \nu +90 \nu ^2+7 \nu ^3\right]
\nonumber  \\
&\qquad
-18 \left(1-e^2\right)^{3/2} \left[720-\left(10256-123 \pi ^2\right) \nu +1056 \nu ^2-48 e^2 \left(105-43 \nu -8 \nu ^2\right)\right] \bigg\} \nonumber\\
&\quad
-\frac{x^3 \epsilon ^6}{18 \left(1-e^2\right)^3} \bigg\{
\chi _A^2 \left[3 e^2 \left(16 \nu ^2+461 \nu -117\right)+132 \nu ^2-857 \nu +197\right]
-2 \delta  \chi _A \chi _S \left[3 e^2 (\nu +117)+253 \nu -197\right]\nonumber\\
&\qquad
+  \chi _S^2 \left[3 e^2 (5 \nu -117)+128 \nu ^2-437 \nu +197\right]
+ 3  \kappa _S \left[e^2 \left(8 \nu ^2+32 \nu -21\right)+22 \nu ^2-161 \nu +63\right] \nonumber\\
&\qquad
-3 \delta  \kappa _A \left[e^2 (10 \nu +21)+7 (5 \nu -9)\right]
+\left(1-e^2\right)^{3/2} \Big[
24 \delta \chi _A \chi _S  (9 \nu -11) -12 \chi _A^2 \left(6 \nu ^2-46 \nu +11\right) \nonumber\\
&\quad\qquad
-12 \left(4 \nu ^2-16 \nu +11\right) \chi _S^2
+6 \delta \kappa _A  (5 \nu -14)+6  \kappa _S\left(-6 \nu ^2+33 \nu -14\right)
\Big]
\bigg\}
\Bigg\},
\label{eq:app:Exe} \\
p_\phi &=  M\mu\sqrt{ \frac{1 - e^2}{x}} \Bigg\{
1 + \frac{x \epsilon ^2}{1-e^2} \left[e^2 \left(\frac{3}{2}-\frac{\nu }{6}\right)+\frac{\nu }{6}+\frac{3}{2}\right]
+ \frac{x^{3/2} \epsilon ^3}{\left(1-e^2\right)^{3/2}} \left[\left(e^2 (\nu -2)+\frac{5 \nu }{3}-\frac{10}{3}\right) \chi _S - \delta \chi _A \left(2 e^2+\frac{10}{3}\right) \right] \nonumber\\
&\quad
+\frac{x^2 \epsilon ^4}{\left(1-e^2\right)^2} \left[e^4 \left(\frac{\nu ^2}{24}+\frac{11 \nu }{24}-\frac{1}{8}\right)+e^2 \left(-\frac{\nu ^2}{12}-\frac{19 \nu }{12}\right)+\frac{\nu ^2}{24}-\frac{27 \nu }{8}+\frac{47}{8}+\sqrt{1-e^2} \left(e^2 \left(\frac{5}{2}-\nu \right)+\nu -\frac{5}{2}\right)\right] \nonumber\\
&\quad
+ \frac{x^2 \epsilon ^4}{\left(1-e^2\right)^2} \left\{\!\left(\frac{3 e^2}{2}+1\right)\! \chi _S^2+ \!\left(\frac{e^2}{2}+1\right)\! \delta \kappa _A+\chi _A^2 \!\left[e^2 \!\left(\frac{3}{2}-6 \nu \right)-4 \nu +1\right]\! + \left(3 e^2+2\right) \delta \chi _A \chi _S+ \!\left[e^2 \left(\frac{1}{2}-\nu \right)-2 \nu +1\right]\! \kappa _S \!\right\} \nonumber\\
&\quad
+ \frac{x^{5/2} \epsilon ^5}{\left(1-e^2\right)^{5/2}} \bigg\{
\left[e^4 \left(-\frac{19 \nu ^2}{24}+\frac{67 \nu }{48}+3\right)+e^2 \left(\frac{3133 \nu }{144}-\frac{205 \nu ^2}{72}\right)-\frac{31 \nu ^2}{36}+\frac{1231 \nu }{72}-11\right] \chi _S
+\delta  \chi _A \bigg[e^4 \left(\frac{49 \nu }{48}+3\right)\nonumber\\
&\qquad
+\frac{1063 e^2 \nu }{144}+\frac{313 \nu }{72}-11\bigg] 
+\sqrt{1-e^2} \left[\delta  \chi _A \left(e^2 \!\left(\frac{4 \nu }{3}-4\right)-\frac{4 \nu }{3}+4\right)+\left(e^2 \!\left(-\frac{2 \nu ^2}{3}+\frac{16 \nu }{3}-4\right)+\frac{2 \nu ^2}{3}-\frac{16 \nu }{3}+4\right) \chi _S\right]\!
\bigg\} \nonumber\\
&\quad
+ \frac{x^3 \epsilon ^6}{\left(1-e^2\right)^3} \bigg\{
e^6 \left(-\frac{7 \nu ^3}{1296}-\frac{\nu ^2}{18}-\frac{23 \nu }{48}-\frac{31}{48}\right)+e^4 \left(\frac{7 \nu ^3}{432}-\frac{17 \nu ^2}{36}+\frac{169 \nu }{48}-\frac{53}{16}\right)\nonumber\\
&\qquad
+e^2 \left[\frac{16 \nu ^2}{9}-\frac{7 \nu ^3}{432}+\left(\frac{287 \pi ^2}{128}-\frac{2939}{48}\right) \nu -\frac{37}{16}\right]
+\frac{7 \nu ^3}{1296}+\frac{25 \nu ^2}{8}+\left(\frac{123 \pi ^2}{64}-\frac{3151}{48}\right) \nu +\frac{155}{16}\nonumber\\
&\qquad
+\sqrt{1-e^2} \left[e^4 \left(\frac{2 \nu ^2}{3}+\frac{31 \nu }{12}-\frac{25}{4}\right)+e^2 \left(\frac{7 \nu ^2}{6}+\left(\frac{41 \pi ^2}{192}-\frac{367}{18}\right) \nu +\frac{15}{2}\right)-\frac{11 \nu ^2}{6}+\left(\frac{641}{36}-\frac{41 \pi ^2}{192}\right) \nu -\frac{5}{4}\right]
\bigg\} \nonumber\\
&\quad
+ \frac{x^3 \epsilon ^6}{\left(1-e^2\right)^3} \bigg\{
\chi _A^2 \left[e^4 \left(-3 \nu ^2+\frac{111 \nu }{4}-\frac{27}{4}\right)+e^2 \left(\frac{10 \nu ^2}{3}+\frac{79 \nu }{4}-\frac{67}{12}\right)+\frac{14 \nu ^2}{3}-\frac{260 \nu }{9}+\frac{59}{9}\right]
+\delta  \chi _A\chi _S \bigg[e^4 \left(\frac{11 \nu }{2}-\frac{27}{2}\right)\nonumber\\
&\qquad
+e^2 \left(-\frac{77 \nu }{6}-\frac{67}{6}\right)-\frac{170 \nu }{9}+\frac{118}{9}\bigg]
+ \chi _S^2\left[e^4 \left(-\nu ^2+\frac{19 \nu }{4}-\frac{27}{4}\right)+e^2 \left(\frac{11 \nu ^2}{3}-\frac{41 \nu }{4}-\frac{67}{12}\right)+\frac{44 \nu ^2}{9}-\frac{146 \nu }{9}+\frac{59}{9}\right]\nonumber\\
&\qquad
+\delta  \kappa _A \left[e^4 \!\left(\frac{\nu }{4}-\frac{5}{4}\right)+e^2 \!\left(\frac{21}{4}-\frac{23 \nu }{6}\right)-\frac{25 \nu }{6}+7\right]
+ \kappa _S \left[e^4 \!\left(\frac{11 \nu }{4}-\frac{\nu ^2}{2}-\frac{5}{4}\right)+e^2 \!\left(\frac{2 \nu ^2}{3}-\frac{43 \nu }{3}+\frac{21}{4}\right)+\frac{7 \nu ^2}{3}-\frac{109 \nu }{6}+7\right] \nonumber\\
&\qquad
+ \sqrt{1-e^2} \bigg[
\chi _S^2\left(e^2 \left(\frac{4 \nu ^2}{3}-\frac{16 \nu }{3}+\frac{11}{3}\right)-\frac{4 \nu ^2}{3}+\frac{16 \nu }{3}-\frac{11}{3}\right)
+\chi _A^2 \left(e^2 \left(2 \nu ^2-\frac{46 \nu }{3}+\frac{11}{3}\right)-2 \nu ^2+\frac{46 \nu }{3}-\frac{11}{3}\right)\nonumber\\
&\qquad\quad
+\delta  \chi _A \chi _S\left(e^2 \left(\frac{22}{3}-6 \nu \right)+6 \nu -\frac{22}{3}\right) 
+ \kappa _S\left(e^2 \left(\nu ^2-\frac{11 \nu }{2}+\frac{7}{3}\right)-\nu ^2+\frac{11 \nu }{2}-\frac{7}{3}\right)\nonumber\\
&\qquad\quad
+\delta  \kappa _A \left(e^2 \left(\frac{7}{3}-\frac{5 \nu }{6}\right)+\frac{5 \nu }{6}-\frac{7}{3}\right)
\bigg]
\bigg\}
\Bigg\},
\label{eq:app:Lxe} \\
r &= \frac{M(1 - e^2)}{x \, (1 + e \cos \zeta)} \Bigg\{
1 + \frac{x \epsilon ^2}{1-e^2} \left[e^2 \left(2-\frac{\nu }{3}\right)+\frac{\nu }{3}\right]
+ \frac{x^{3/2} \epsilon ^3}{\left(1-e^2\right)^{3/2}} \left[\delta  \left(-2 e^2-\frac{2}{3}\right) \chi _A+\left(e^2 (\nu -2)+\frac{\nu }{3}-\frac{2}{3}\right) \chi _S\right] \nonumber\\
&\quad
+ \frac{x^2 \epsilon ^4}{\left(1-e^2\right)^2} \left[e^4 \left(\frac{\nu ^2}{9}+\frac{5 \nu }{12}+1\right)+e^2 \left(-\frac{2 \nu ^2}{9}-\frac{13 \nu }{6}-\frac{3}{2}\right)+\sqrt{1-e^2} \left(e^2 (5-2 \nu )+2 \nu -5\right)+\frac{\nu ^2}{9}-\frac{13 \nu }{4}+5\right] \nonumber\\
&\quad
+ \frac{x^2 \epsilon ^4}{\left(1-e^2\right)^2} \left[  \left(\frac{e^2}{2}+\frac{1}{2}\right)\delta \kappa _A+e^2 (1-4 \nu ) \chi _A^2+2 e^2 \delta \chi _A \chi _S+\left(e^2 \left(\frac{1}{2}-\nu \right)-\nu +\frac{1}{2}\right) \kappa _S+e^2 \chi _S^2\right]\nonumber\\
&\quad
+ \frac{x^{5/2} \epsilon ^5}{\left(1-e^2\right)^{5/2}} \bigg\{
\chi _S \!\left[e^4 \!\left(\frac{107 \nu }{48}-\frac{23 \nu ^2}{24}+2\right)+e^2 \!\left(\frac{51 \nu }{4}-\frac{\nu ^2}{2}-2\right)-\frac{13 \nu ^2}{24}+\frac{481 \nu }{48}-8\right]\!
+\delta  \chi _A \bigg[e^4 \!\left(\frac{65 \nu }{48}+2\right)+e^2 \!\left(\frac{13 \nu }{4}-2\right) \nonumber\\
&\qquad
+\frac{115 \nu }{48}-8\bigg]
+\sqrt{1-e^2} \left[\delta  \chi _A \left(e^2 \left(\frac{8 \nu }{3}-8\right)-\frac{8 \nu }{3}+8\right)+\left(e^2 \left(-\frac{4 \nu ^2}{3}+\frac{32 \nu }{3}-8\right)+\frac{4 \nu ^2}{3}-\frac{32 \nu }{3}+8\right) \chi _S\right]
\bigg\}\nonumber\\
&\quad
+ \frac{x^3 \epsilon ^6}{\left(1-e^2\right)^3} \bigg\{
e^6 \left[-\frac{2 \nu ^3}{81}-\frac{5 \nu ^2}{36}+\frac{\nu }{8}-\frac{2}{3}\right]
+e^4 \left[\frac{2 \nu ^3}{27}-\frac{\nu ^2}{18}+\frac{47 \nu }{24}-4\right]
+e^2 \left[-\frac{2 \nu ^3}{27}+\frac{169 \nu ^2}{36}+\left(\frac{123 \pi ^2}{64}-\frac{541}{8}\right) \nu +4\right]\nonumber\\
&\qquad
+\frac{2 \nu ^3}{81}+\frac{17 \nu ^2}{4}+\left(\frac{41 \pi ^2}{32}-\frac{1435}{24}\right) \nu +10
+\sqrt{1-e^2} \bigg[e^4 \left(\frac{5 \nu ^2}{3}+\frac{4 \nu }{3}-5\right)+e^2 \left(\frac{5 \nu ^2}{3}+\left(\frac{41 \pi ^2}{96}-\frac{352}{9}\right) \nu +15\right)\nonumber\\
&\qquad\quad
-\frac{10 \nu ^2}{3}+\left(\frac{340}{9}-\frac{41 \pi ^2}{96}\right) \nu -10\bigg]
\bigg\} \nonumber\\
&\quad
+ \frac{x^3 \epsilon ^6}{\left(1-e^2\right)^3} \bigg\{
\chi _S^2 \left[e^4 \left(-\nu ^2+\frac{13 \nu }{3}-5\right)+e^2 \left(\frac{16 \nu ^2}{3}-\frac{55 \nu }{3}+\frac{1}{3}\right)+\frac{41 \nu ^2}{9}-\frac{149 \nu }{9}+\frac{65}{9}\right]
+\delta  \chi _A \chi _S \bigg[e^4 \left(\frac{14 \nu }{3}-10\right)\nonumber\\
&\qquad\quad
+e^2 \left(\frac{2}{3}-\frac{64 \nu }{3}\right)-\frac{170 \nu }{9}+\frac{130}{9}\bigg]
+\chi _A^2 \left[e^4 \left(-\frac{4 \nu ^2}{3}+\frac{61 \nu }{3}-5\right)+e^2 \left(6 \nu ^2-\frac{13 \nu }{3}+\frac{1}{3}\right)+\frac{16 \nu ^2}{3}-\frac{281 \nu }{9}+\frac{65}{9}\right]\nonumber\\
&\qquad
+ \kappa _S\left[e^4 \left(-\frac{\nu ^2}{3}+\frac{13 \nu }{6}-1\right)+e^2 \left(\frac{7 \nu ^2}{3}-\frac{97 \nu }{6}+6\right)+3 \nu ^2-\frac{31 \nu }{2}+6\right]
+\delta  \kappa _A \bigg[e^4 \left(\frac{\nu }{6}-1\right)
+e^2 \left(6-\frac{25 \nu }{6}\right)-\frac{7 \nu }{2}+6\bigg]\nonumber\\
&\qquad
+ \sqrt{1-e^2} \bigg[
+ \chi _S^2\left(e^2 \left(\frac{8 \nu ^2}{3}-\frac{32 \nu }{3}+\frac{22}{3}\right)-\frac{8 \nu ^2}{3}+\frac{32 \nu }{3}-\frac{22}{3}\right)
+\chi _A^2 \left(e^2 \left(4 \nu ^2-\frac{92 \nu }{3}+\frac{22}{3}\right)-4 \nu ^2+\frac{92 \nu }{3}-\frac{22}{3}\right)\nonumber\\
&\qquad\quad
+\delta  \chi _A \chi _S \left(e^2 \left(\frac{44}{3}-12 \nu \right)+12 \nu -\frac{44}{3}\right)
+ \kappa _S\left(e^2 \left(2 \nu ^2-11 \nu +\frac{14}{3}\right)-2 \nu ^2+11 \nu -\frac{14}{3}\right)\nonumber\\
&\qquad\quad
+\delta  \kappa _A \left(e^2 \left(\frac{14}{3}-\frac{5 \nu }{3}\right)+\frac{5 \nu }{3}-\frac{14}{3}\right)
\bigg] \bigg\}
\Bigg\},
 \label{eq:app:rxechi} \\
p_r &=  \mu\sqrt{\frac{x}{1 - e^2}}  \, e \sin \zeta \Bigg\{
1 + \frac{x \epsilon ^2}{1-e^2} \left[e^2 \left(\frac{\nu }{6}-\frac{1}{2}\right)+e \cos\zeta-\frac{\nu }{6}+\frac{1}{2}\right]
+ \frac{x^{3/2} \epsilon ^3}{\left(1-e^2\right)^{3/2}} \bigg[\chi _S \left(e (2-\nu ) \cos\zeta-\frac{5 \nu }{3}+\frac{10}{3}\right) \nonumber\\
&\qquad
+\delta  \chi _A \left(2 e \cos\zeta+\frac{10}{3}\right)\bigg]
+ \frac{x^2 \epsilon ^4}{\left(1-e^2\right)^2} \bigg\{
e^4 \left(-\frac{\nu ^2}{72}-\frac{11 \nu }{24}-\frac{1}{8}\right)+\left[e^3 \left(\frac{\nu }{2}-\frac{5}{2}\right)+e \left(-\frac{11 \nu }{2}-\frac{3}{2}\right)\right] \cos\zeta \nonumber\\
&\qquad
+e^2 \left(\frac{3}{4}-\frac{3 \nu }{2}\right) \cos (2 \zeta )+e^2 \left(\frac{\nu ^2}{36}-\frac{5 \nu }{12}+\frac{7}{4}\right)+\sqrt{1-e^2} \left(e^2 \left(\nu -\frac{5}{2}\right)-\nu +\frac{5}{2}\right)-\frac{\nu ^2}{72}-\frac{\nu }{8}-\frac{57}{8}
\bigg\} \nonumber\\
&\quad
+ \frac{x^2 \epsilon ^4}{\left(1-e^2\right)^2} \bigg\{
\chi _A^2 \!\left[e^2 \left(2 \nu -\frac{1}{2}\right) \cos (2 \zeta )+e^2 \left(2 \nu -\frac{1}{2}\right) +e (12 \nu -3) \cos\zeta+10 \nu -\frac{5}{2}\right]
+\kappa _S \left[e \left(\nu -\frac{1}{2}\right) \cos\zeta+2 \nu -1\right]\nonumber\\
&\qquad
+\chi _S^2 \left[-\frac{1}{2} e^2 \cos (2 \zeta )-\frac{e^2}{2}-3 e \cos\zeta-\frac{5}{2}\right]
+\delta  \chi _A \chi _S \left[-e^2 \cos (2 \zeta )-e^2-6 e \cos\zeta-5\right]
+\delta  \kappa _A \left(-\frac{1}{2} e \cos\zeta-1\right)
\bigg\} \nonumber\\
&\quad
+ \frac{x^{5/2} \epsilon ^5}{\left(1-e^2\right)^{5/2}} \bigg\{
\sqrt{1-e^2} \left[\delta  \chi _A \left(e^2 \left(4-\frac{4 \nu }{3}\right)+\frac{4 \nu }{3}-4\right)+\left(e^2 \left(\frac{2 \nu ^2}{3}-\frac{16 \nu }{3}+4\right)-\frac{2 \nu ^2}{3}+\frac{16 \nu }{3}-4\right) \chi _S\right]\nonumber\\
&\qquad
+\delta  \chi _A \bigg[\left( e \left(25-\frac{335 \nu }{24}\right)-e^3 \left(\frac{311 \nu }{192}+5\right)\right) \cos\zeta-\frac{27}{64} e^3 \nu  \cos (3 \zeta )+e^2 \left(3-\frac{59 \nu }{16}\right) \cos (2 \zeta )-e^2 \left(\frac{97 \nu }{24}+10\right)\nonumber\\
&\qquad\quad
-\frac{565 \nu }{48}+27\bigg]
+\chi _S \bigg[e^3 \left(\frac{9 \nu ^2}{32}-\frac{57 \nu }{64}\right) \cos (3 \zeta )+\left(e^3 \left(\frac{125 \nu ^2}{96}-\frac{461 \nu }{192}-5\right)+e \left(\frac{77 \nu ^2}{12}-\frac{1121 \nu }{24}+25\right)\right) \cos\zeta\nonumber\\
&\qquad\quad
+e^2 \left(\frac{17 \nu ^2}{8}-\frac{161 \nu }{16}+3\right) \cos (2 \zeta )
+e^2 \left(\frac{31 \nu ^2}{12}-\frac{187 \nu }{24}-10\right)+\frac{103 \nu ^2}{24}-\frac{2047 \nu }{48}+27\bigg]
\bigg\} \nonumber\\
&\quad
+ \frac{x^3 \epsilon ^6}{\left(1-e^2\right)^3} \bigg\{
e^6 \left(-\frac{5 \nu ^3}{1296}-\frac{\nu ^2}{18}+\frac{5 \nu }{16}+\frac{37}{48}\right)+e^4 \left(\frac{\nu }{2}-\frac{3 \nu ^2}{8}\right) \cos (4 \zeta )+e^4 \left(\frac{5 \nu ^3}{432}-\frac{5 \nu ^2}{9}+\frac{299 \nu }{48}-\frac{93}{16}\right) \nonumber\\
&\qquad
+e^3 \left(-\frac{3 \nu ^2}{4}-\frac{11 \nu }{2}+\frac{5}{8}\right) \cos (3 \zeta )+e^2 \left[-\frac{5 \nu ^3}{432}+\frac{77 \nu ^2}{18}+\left(-\frac{1027}{48}-\frac{41 \pi ^2}{64}\right) \nu +\frac{421}{16}\right]
+\frac{5 \nu ^3}{1296} +\left(\frac{4217}{48}-\frac{205 \pi ^2}{64}\right) \nu \nonumber\\
&\qquad
+\left[e^5 \left(\frac{\nu ^2}{24}-\frac{65 \nu }{24}+\frac{31}{8}\right)+e^3 \left(-\frac{\nu ^2}{2}+\frac{19 \nu }{12}+\frac{99}{8}\right)+e \left(\frac{317 \nu ^2}{24}+\left(-\frac{5}{24}-\frac{41 \pi ^2}{16}\right) \nu -\frac{329}{8}\right)\right] \cos\zeta +\frac{4 \nu ^2}{3} -\frac{639}{16}\nonumber\\
&\qquad
+\left[e^4 \left(-\frac{5 \nu ^2}{4}+\frac{59 \nu }{8}-\frac{27}{8}\right)+e^2 \left(\frac{17 \nu ^2}{4}+\left(-\frac{781}{24}-\frac{41 \pi ^2}{128}\right) \nu -\frac{21}{8}\right)\right] \cos (2 \zeta ) 
+\sqrt{1-e^2} \bigg[e^4 \left(-\frac{\nu ^2}{3}-\frac{41 \nu }{12}+\frac{25}{4}\right)\nonumber\\
&\qquad\quad
+\left(e^3 \left(3 \nu -\frac{15}{2}\right)+e \left(\frac{15}{2}-3 \nu \right)\right) \cos\zeta+e^2 \left(\left(\frac{451}{18}-\frac{41 \pi ^2}{192}\right) \nu-\frac{11 \nu ^2}{6} -15\right)
+\frac{13 \nu ^2}{6}+\left(\frac{41 \pi ^2}{192}-\frac{779}{36}\right) \nu +\frac{35}{4}\bigg]
\bigg\}\nonumber\\
&\quad
+ \frac{x^3 \epsilon ^6}{\left(1-e^2\right)^3} \bigg\{
\chi _A^2\bigg[\left(\frac{5 \nu ^2}{3}-\frac{113 \nu }{12}+\frac{9}{4}\right) e^4+\left(\frac{3 \nu ^2}{2}+\frac{21 \nu }{8}-\frac{3}{4}\right) \cos (3 \zeta ) e^3+\left(\frac{40 \nu ^2}{3}-\frac{437 \nu }{6}+\frac{35}{2}\right) e^2-5 \nu ^2+\frac{3331 \nu }{36}-\frac{799}{36}\nonumber\\
&\qquad\quad
+\left(\!\left(\frac{29 \nu ^2}{2}-\frac{581 \nu }{8}+\frac{69}{4}\right)\! e^3+\!\left(\! 8 \nu ^2+\frac{316 \nu }{3}-\frac{79}{3}\right)\! e\right) \cos\zeta+ \!\left(\!\left(\frac{5 \nu ^2}{3}-\frac{113 \nu }{12}+\frac{9}{4}\right)\! e^4+\!\left(\frac{22 \nu ^2}{3}+\frac{305 \nu }{12}-\frac{27}{4}\right) e^2\!\right) \cos (2 \zeta )\bigg]\nonumber\\
&\qquad
+\delta  \chi _S \chi _A \bigg[\left(\frac{9}{2}-\frac{5 \nu }{6}\right) e^4+\left(-\frac{9 \nu }{4}-\frac{3}{2}\right) \cos (3 \zeta ) e^3+\left(35-\frac{65 \nu }{3}\right) e^2 +\left(\left(\frac{69}{2}-\frac{79 \nu }{4}\right) e^3+\left(-\frac{2 \nu }{3}-\frac{158}{3}\right) e\right) \cos\zeta \nonumber\\
&\qquad\quad
+\left(\left(\frac{9}{2}-\frac{5 \nu }{6}\right) e^4+\left(-\frac{61 \nu }{6}-\frac{27}{2}\right) e^2\right) \cos (2 \zeta )-\frac{799}{18}+\frac{467 \nu }{18}\bigg]
+ \chi _S^2\bigg[\left(\frac{9}{4}-\frac{5 \nu }{12}\right) e^4+\left(\frac{3 \nu ^2}{8}-\frac{15 \nu }{8}-\frac{3}{4}\right) \cos (3 \zeta ) e^3\nonumber\\
&\qquad\quad
+\left(\frac{7 \nu ^2}{2}-\frac{113 \nu }{6}+\frac{35}{2}\right) e^2+\frac{799 \nu }{36}
+\left(\left(\frac{25 \nu ^2}{8}-\frac{129 \nu }{8}+\frac{69}{4}\right) e^3+\left(-\frac{11 \nu ^2}{6}-\frac{2 \nu }{3}-\frac{79}{3}\right) e\right) \cos\zeta
-\frac{64 \nu ^2}{9}-\frac{799}{36} \nonumber\\
&\qquad\quad
+\left(\left(\frac{9}{4}-\frac{5 \nu }{12}\right) e^4+\left(\frac{3 \nu ^2}{2}-\frac{103 \nu }{12}-\frac{27}{4}\right) e^2\right) \cos (2 \zeta )\bigg]
+\delta \kappa _A \bigg[\left(\frac{3}{8}-\frac{3 \nu }{8}\right) \cos (3 \zeta ) e^3+\left(\frac{7}{2}-\frac{3 \nu }{2}\right) e^2+5 \nu -14\nonumber\\
&\qquad\quad
+\left(\frac{1}{2}-\frac{7 \nu }{4}\right) \cos (2 \zeta ) e^2
+\left(\left(\frac{23}{8}-\frac{37 \nu }{24}\right) e^3+\left(-\frac{\nu }{12}-\frac{37}{4}\right) e\right) \cos\zeta\bigg]
+ \kappa _S\bigg[\left(\frac{3 \nu ^2}{4}-\frac{9 \nu }{8}+\frac{3}{8}\right) \cos (3 \zeta ) e^3+33 \nu-14\nonumber\\
&\qquad\quad
+\left(5 \nu ^2-\frac{17 \nu }{2}+\frac{7}{2}\right) e^2+\left(\frac{9 \nu ^2}{2}-\frac{11 \nu }{4}+\frac{1}{2}\right) \cos (2 \zeta ) e^2 
+\left(\left(\frac{37 \nu ^2}{12}-\frac{175 \nu }{24}+\frac{23}{8}\right) e^3+\left(\frac{49 \nu ^2}{6}+\frac{221 \nu }{12}-\frac{37}{4}\right) e\right) \cos\zeta\bigg]\nonumber\\
&\qquad
+\sqrt{1-e^2} \bigg[\left(\left(-2 \nu ^2+\frac{46 \nu }{3}-\frac{11}{3}\right) e^2+2 \nu ^2-\frac{46 \nu }{3}+\frac{11}{3}\right) \chi _A^2+\delta  \left(\left(6 \nu -\frac{22}{3}\right) e^2-6 \nu +\frac{22}{3}\right) \chi _S \chi _A\nonumber\\
&\qquad\quad
+\left(\left(-\frac{4 \nu ^2}{3}+\frac{16 \nu }{3}-\frac{11}{3}\right) e^2+\frac{4 \nu ^2}{3}-\frac{16 \nu }{3}+\frac{11}{3}\right) \chi _S^2+\left(\left(-\nu ^2+\frac{11 \nu }{2}-\frac{7}{3}\right) e^2+\nu ^2-\frac{11 \nu }{2}+\frac{7}{3}\right) \kappa _S\nonumber\\
&\qquad\quad
+\left(\left(\frac{5 \nu }{6}-\frac{7}{3}\right) e^2+\frac{7}{3}-\frac{5 \nu }{6}\right) \delta  \kappa _A\bigg]
\bigg\}
\Bigg\}.
\label{eq:app:prxechi} 
\end{align}
\end{subequations}
\normalsize

\subsection{Postadiabatic contributions}

The PA contributions to the Keplerian parameters $ (x, e, \zeta) $ are discussed in Sec.~\ref{sec:relating_parametrization}, and they depend on the radiation-reaction gauge constants $ \alpha $ and $ \beta $ defined in Eq.~\eqref{eq:FrFphiLO}.
These PA contributions enter at 2.5PN order and are given by
\small
\begin{subequations}
\label{eq:app:tilde_x_e}
\begin{align}
\tilde x &= \nu \epsilon^5 \bar{x}^{7/2} \bigg\{\left(\frac{16 \alpha }{5}+\frac{416}{5}\right) \bar{e} \sin (\bar{\zeta} ) 
+ \left(\frac{32 \alpha }{5}+\frac{332}{15}\right) \bar{e}^2 \sin \left(2 \bar{\zeta }\right) 
+ \bar{e}^3 \sin \left(\bar{\zeta }\right) \bigg[\left(8 \alpha +\frac{8 \beta }{5}+\frac{1472}{45}\right) \cos \left(2 \bar{\zeta }\right) \nonumber\\
&\quad\qquad
+\frac{112 \alpha }{5}-\frac{8 \beta }{5}+\frac{19744}{45}\bigg]
+ \bar{e}^4 \sin \left(2 \bar{\zeta }\right) \left[\left(\frac{8 \alpha }{5}+\frac{8 \beta }{5}-\frac{1}{15}\right) \cos \left(2 \bar{\zeta }\right)+\frac{136 \alpha }{5}-\frac{8 \beta }{5}+\frac{892}{15}\right] \nonumber\\
&\qquad
+\bar{e}^5 \left[\left(\frac{256 \alpha }{5}-\frac{44 \beta }{5}+\frac{16888}{15}\right) \sin \left(\bar{\zeta }\right)+\left(\frac{74 \alpha }{5}+\frac{13 \beta }{5}+\frac{3304}{45}\right) \sin \left(3 \bar{\zeta }\right)+\left(\frac{\beta }{5}+\frac{24}{25}\right) \sin \left(5 \bar{\zeta }\right)\right] \nonumber\\
&\qquad
+ \bar{e}^6 \sin \left(2 \bar{\zeta }\right) \left[\left(\frac{28 \alpha }{5}+\frac{28 \beta }{5}-\frac{74}{5}\right) \cos \left(2 \bar{\zeta }\right)-\frac{14}{15} \cos \left(4 \bar{\zeta }\right)+\frac{336 \alpha }{5}-\frac{28 \beta }{5}+\frac{305}{3}\right] \nonumber\\
&\qquad
+ \bar{e}^7 \bigg[\left(\frac{1057 \alpha }{10}-\frac{203 \beta }{10}+\frac{34118}{15}\right) \sin \left(\bar{\zeta }\right)+\left(\frac{343 \alpha }{10}+\frac{28 \beta }{5}+\frac{2891}{15}\right) \sin \left(3 \bar{\zeta }\right)+\left(\frac{7 \beta }{10}+\frac{177}{25}\right) \sin \left(5 \bar{\zeta }\right)+\frac{8}{35} \sin \left(7 \bar{\zeta }\right)\bigg] \nonumber\\
&\qquad
+ \Order(\bar{e}^8)\bigg\},
\\
\tilde e &= -\nu \epsilon^5 \bar{x}^{5/2} \bigg\lbrace
\frac{64}{5} \sin\bar{\zeta}
+ \left(\frac{8 \alpha }{5}+\frac{184}{15}\right) \bar{e} \sin \left(2 \bar{\zeta }\right)
+ \bar{e}^2 \sin \bar{\zeta } \left[\left(\frac{16 \alpha }{5}+\frac{4 \beta }{5}+\frac{508}{45}\right) \cos \left(2 \bar{\zeta }\right)+\frac{32 \alpha }{5}-\frac{4 \beta }{5}+\frac{5348}{45}\right] \nonumber\\
&\qquad
+ \bar{e}^3 \sin \left(2 \bar{\zeta }\right) \left[\left(\frac{4 \alpha }{5}+\frac{4 \beta }{5}+\frac{59}{30}\right) \cos \left(2 \bar{\zeta }\right)+8 \alpha -\frac{4 \beta }{5}+\frac{382}{15}\right]
+ \bar{e}^4 \sin \left(\bar{\zeta }\right) \bigg[\left(\frac{48 \alpha }{5}+2 \beta +\frac{662}{15}\right) \cos \left(2 \bar{\zeta }\right)\nonumber\\
&\quad\qquad
+\frac{\beta}{5} \cos \left(4 \bar{\zeta }\right)+\frac{88 \alpha }{5}-\frac{11 \beta }{5}+300\bigg]
+ \bar{e}^5 \sin \left(2 \bar{\zeta }\right) \left[\left(2 \alpha +2 \beta -\frac{17}{12}\right) \cos \left(2 \bar{\zeta }\right)+17 \alpha -2 \beta +\frac{385}{12}\right] \nonumber\\
&\qquad
+\bar{e}^6 \left[\left(23 \alpha -\frac{23 \beta }{4}+\frac{5839}{12}\right) \sin \left(\bar{\zeta }\right)+\left(9 \alpha +\frac{3 \beta }{2}+\frac{1793}{36}\right) \sin \left(3 \bar{\zeta }\right)+\left(\frac{\beta }{4}+\frac{38}{25}\right) \sin \left(5 \bar{\zeta }\right)\right]
+ \Order(\bar{e}^7)
\bigg\rbrace. \\
\tilde \zeta  &= 
-\nu \epsilon^5 \bar{x}^{5/2} \bigg\{\frac{64 \cos (\bar{\zeta} )}{5 \bar{e}}
-\frac{16 \alpha }{5}+\frac{8 \beta }{5}+\frac{176}{5} 
+\left(\frac{8 \alpha }{5}+\frac{184}{15}\right) \cos \left(2 \bar{\zeta }\right)
+ \bar{e} \cos \left(\bar{\zeta }\right) \bigg[\left(\frac{16 \alpha }{5}+\frac{4 \beta }{5}+\frac{508}{45}\right) \cos \left(2 \bar{\zeta }\right) \nonumber\\
&\qquad\quad
-8 \alpha +4 \beta +\frac{1916}{9}\bigg]
+ \bar{e}^2 \left[\left(\frac{8 \alpha }{5}+\frac{12 \beta }{5}+\frac{482}{9}\right) \cos \left(2 \bar{\zeta }\right)+\left(\frac{2 \alpha }{5}+\frac{2 \beta }{5}+\frac{59}{60}\right) \cos \left(4 \bar{\zeta }\right)-\frac{54 \alpha }{5}+6 \beta +\frac{6668}{15}\right] \nonumber\\
&\qquad
+ \bar{e}^3 \cos \left(\bar{\zeta }\right) \left[\left(\frac{36 \alpha }{5}+\frac{14 \beta }{5}+\frac{1859}{45}\right) \cos \left(2 \bar{\zeta }\right)+\frac{1}{5} \beta  \cos \left(4 \bar{\zeta }\right)-\frac{104 \alpha }{5}+\frac{53 \beta }{5}+\frac{36553}{30}\right] \nonumber\\
&\qquad
+ \bar{e}^4 \left[\left(\alpha +6 \beta +\frac{97681}{360}\right) \cos \left(2 \bar{\zeta }\right)+\left(\alpha +\beta -\frac{91}{144}\right) \cos \left(4 \bar{\zeta }\right)-21 \alpha +12 \beta +\frac{35935}{18}\right] \nonumber\\
&\qquad
+\bar{e}^5 \cos \left(\bar{\zeta }\right) \left[\left(12 \alpha +\frac{11 \beta }{2}+\frac{212989}{2250}\right) \cos \left(2 \bar{\zeta }\right)+\left(\frac{\beta }{2}+\frac{26869}{9000}\right) \cos \left(4 \bar{\zeta }\right)-37 \alpha +19 \beta +\frac{6323737}{1500}\right]
+ \Order(\bar{e}^6) \bigg\}.
\label{eq:app:tildez}
\end{align}
\end{subequations}
\normalsize
\end{widetext}

\subsection{Relations for the initial conditions}

The relations in this section can be used to derive the equations needed for computing the initial conditions in \texttt{SEOBNRv5EHM}, as explained in Sec.~\ref{sec:ICs}.

The first expression that we need is the conservative \emph{instantaneous} angular frequency $ \Omega $ as a function of the \emph{orbit-averaged} angular frequency $ \langle \Omega \rangle $, eccentricity $ e $, and relativistic anomaly $ \zeta $. To obtain this expression, first we compute $ \partial H_\text{EOB} / \partial p_\phi $ as a function of $ (r, p_r, p_\phi) $ in a PN expansion. Then, we substitute Eqs.~\eqref{eq:app:ErprL_xez}  to obtain a dependence on $ (\langle \Omega \rangle, e, \zeta) $. The resulting equation is
\begin{align}
\label{eq:omega_kep}
&\Omega (\langle \Omega \rangle, e, \zeta) = \frac{ \langle \Omega \rangle (1 + e \cos \zeta)^2}{(1 - e^2)^{3/2}} \nonumber\\
&\qquad - \frac{ \epsilon ^2 e \langle \Omega \rangle^{5/3} (3 e+2 \cos\zeta) (e \cos\zeta+1)^2}{\left(1-e^2\right)^{5/2}} + \Order(\epsilon^3).
\end{align}
%

Next, we compute conservative expressions for $ p_{r_*} $ and $ \dot p_{r_*} $ as functions of $ (\langle \Omega \rangle, e, \zeta) $. The equation for $ p_{r_*} $ is simply derived by multiplying Eq.~\eqref{eq:prxechi} by the tortoise-coordinates function $ \xi(r) $ expressed in terms of $ (\langle \Omega \rangle, e, \zeta) $ [which is obtained by substituting Eq.~\eqref{eq:app:rxechi} in the formula for $\xi(r)$ given by Eq.~(44) of Ref.~\cite{Khalilv5}]. We obtain
\begin{align}
\label{eq:prstar_xez}
p_{r_*}
&= \frac{e \mu \langle M \Omega \rangle^{1/3} \sin (\zeta )}{\sqrt{1-e^2}} 
-\frac{ \epsilon ^2e \mu \langle M \Omega \rangle \sin\zeta}{6 \left(1-e^2\right)^{3/2}}  \nonumber\\
&\quad \times \left(\nu +9+6 e \cos\zeta-e^2 (\nu -3)\right) + \Order(\epsilon^3),
\end{align}

The conservative equation for $ \dot p_{r_*} $ can be computed from
\begin{align}
\dot p_{r_*} &= \frac{\di}{\di t}\xi(r)p_r = \xi(r) \dot{p}_r + \xi'(r) p_r \dot{r} \nonumber\\
&= -\xi(r) \frac{\partial H_\text{EOB}}{\partial \phi} + \xi'(r) p_r \frac{\partial H_\text{EOB}}{\partial p_r}.
\end{align}
We PN-expand this expression and write the result in terms of $(\langle \Omega \rangle ,e,\zeta)$, leading to
\begin{align}
\label{eq:dot_pr_xez}
\dot p_{r_*} &= \frac{e \mu\langle M \Omega \rangle^{4/3} \cos\zeta (e \cos\zeta +1)^2}{M\left(1-e^2\right)^2} \nonumber\\
&\quad
+ \frac{\epsilon ^2 e \mu \langle M \Omega \rangle^2 (e \cos\zeta +1)^2}{6 M \left(1-e^2\right)^3}  \Big[-3 e (5 \cos (2 \zeta )+3) \nonumber\\
&\qquad + \left(e^2 (\nu -21)-\nu -27\right) \cos\zeta\Big] + \Order(\epsilon^3).
\end{align}

Finally, for the dissipative part of the initial conditions, we require to change the dependence on $ x = \langle M \Omega \rangle $ to a dependence on $ p_\phi $, since we use the value of $ p_\phi $ coming from the solution to the conservative equations, as explained in Sec.~\ref{sec:ICs}. To obtain an expression for $ x $ in terms of $ p_\phi $, we simply invert Eq.~\eqref{eq:app:Lxe}, leading to
\begin{equation}
\label{eq:x_pphie}
x(p_\phi, e) = \frac{1-e^2}{p_{\phi }^2/(M\mu)^2} +\epsilon ^2  \frac{e^4 (\nu -9)-2 e^2 \nu +\nu +9}{3 p_{\phi }^4/(M\mu)^4} + \Order(\epsilon^3).
\end{equation}
The full 3PN expressions for the relations in this appendix are provided in the Supplemental Material.


\section{Energy and angular momentum fluxes in EOB coordinates}
\label{app:fluxesKepAvg}

\subsection{Instantaneous contributions to the fluxes for generic orbits}

In this appendix, we use the scaled dimensionless variables defined in Eq.~\eqref{eq:dimlessVars} to simplify the notation.

The 3PN instantaneous contribution to the energy flux in EOB coordinates is given by
\begin{widetext}
\small
\begin{align}
\label{eq:EOBenergyflux}
\Phi_E^\text{inst} &= \frac{\nu ^2}{r^4} \left[\frac{32 p_{\phi }^2}{5 r^2}+\frac{8 p_r^2}{15}\right]
+ \frac{\nu ^2 \epsilon ^2}{r^4} \Bigg[\frac{p_{\phi }^4}{r^4}\left(\frac{898}{105}-\frac{568 \nu }{35}\right)+\frac{ p_{\phi }^2}{r^3}\left(\frac{992 \nu }{105}-\frac{1088}{21}\right)+\frac{p_r^2 p_{\phi }^2}{r^2}\left(-\frac{352 \nu }{21}-\frac{3536}{105}\right) +p_r^4 \left(-\frac{16 \nu }{21}-\frac{104}{35}\right) \nonumber\\
&\qquad
+\frac{ p_r^2}{r}\left(\frac{64}{21}-\frac{32 \nu }{105}\right) +\frac{1}{r^2} \left(\frac{32}{105}-\frac{128 \nu }{105}\right)\Bigg]\nonumber\\
&\quad
+ \frac{\nu ^2 \epsilon ^3 p_\phi}{r^6} \Bigg\{
\chi _S \left[\frac{ p_{\phi }^2}{r^2}\left(\frac{96 \nu }{5}-\frac{296}{15}\right)+p_r^2\left(\frac{48 \nu }{5}-\frac{16}{3}\right) +\frac{1}{r}\left(\frac{96}{5}-\frac{128 \nu }{15}\right)\right]
+\delta  \chi _A \left[\frac{96}{5 r}-\frac{296 p_{\phi }^2}{15 r^2}-\frac{16 p_r^2}{3}\right] \Bigg\} \nonumber\\
&\quad
+\frac{\nu ^2 \epsilon ^4}{r^6} \Bigg\{
\chi _A^2 \left[\frac{2 p_{\phi }^2}{5 r^2}+\left(\frac{128 \nu }{15}-\frac{8}{15}\right) p_r^2\right]
+\chi _S^2 \left[\frac{ p_{\phi }^2}{r^2}\left(\frac{2}{5}-\frac{8 \nu }{5}\right)+\left(-\frac{32 \nu }{5}-\frac{8}{15}\right) p_r^2\right]
+\delta  \chi _A \chi _S \left(\frac{4 p_{\phi }^2}{5 r^2}-\frac{16 p_r^2}{15}\right)  \nonumber\\
&\qquad
+\kappa _S \left[\frac{ p_{\phi }^2}{r^2}\left(\frac{96}{5}-\frac{192 \nu }{5}\right)+\left(\frac{16 \nu }{5}-\frac{8}{5}\right) p_r^2\right]
+\delta  \kappa _A \left(\frac{96 p_{\phi }^2}{5 r^2}-\frac{8 p_r^2}{5}\right)
\Bigg\} \nonumber\\
&\quad
+ \frac{\nu ^2 \epsilon ^4}{r^4} \Bigg\{
\frac{ p_{\phi }^6}{r^6}\left(\frac{7496 \nu ^2}{315}-\frac{3352 \nu }{315}-\frac{212}{105}\right)
+\frac{ p_{\phi }^4}{r^5}\left(-\frac{2224 \nu ^2}{63}+\frac{50896 \nu }{315}-\frac{4232}{35}\right) 
+\frac{p_r^2 p_{\phi }^4}{r^4}\left(\frac{3152 \nu ^2}{35}+\frac{548 \nu }{21}+\frac{1256}{35}\right) \nonumber\\
&\qquad
+\frac{ p_{\phi }^2}{r^4}\left(\frac{15224 \nu ^2}{945}-\frac{13688 \nu }{315}+\frac{56846}{405}\right)
+\frac{ p_r^2 p_{\phi }^2}{r^3}\left(-\frac{13408 \nu ^2}{315}+\frac{47984 \nu }{315}+\frac{39496}{105}\right)
+\frac{p_r^4 p_{\phi }^2}{r^2}\left(\frac{656 \nu ^2}{21}+\frac{688 \nu }{7}+\frac{1600}{21}\right)  \nonumber\\
&\qquad
+\frac{ p_r^2}{r^2}\left(-\frac{64 \nu ^2}{945}+\frac{512 \nu }{21}-\frac{12808}{405}\right)
+\frac{1}{r^3} \left(-\frac{4352 \nu ^2}{945}+\frac{384 \nu }{35}-\frac{464}{189}\right)
+\frac{ p_r^4}{r}\left(\frac{928 \nu ^2}{945}-\frac{4016 \nu }{945}+\frac{8}{9}\right) \nonumber\\
&\qquad
+p_r^6\left(\frac{8 \nu ^2}{9}+\frac{1396 \nu }{315}+\frac{2056}{315}\right) 
\Bigg\} 
+ \frac{\nu ^3 \epsilon ^5 p_r}{r^5} \Bigg[
\frac{p_{\phi }^2}{r^2}\left(\frac{6016 \alpha }{75}-\frac{896 \beta }{15}-\frac{93904}{1575}\right) 
+\left(\frac{128 \alpha }{25}-\frac{128 \beta }{25}+\frac{256}{35}\right) p_r^2\nonumber\\
&\qquad
-\frac{197584 p_{\phi }^4}{525 r^3} +\frac{39136 p_r^2 p_{\phi }^2}{525 r}+\frac{64 r p_r^4}{105}
+\frac{1}{r}\left(\frac{128 \alpha }{75}-\frac{128 \beta }{75}+\frac{2656}{1575}\right)
\Bigg] \nonumber\\
&\quad
+\frac{\nu ^2 \epsilon ^5 p_\phi}{r^6} \Bigg\{
\delta  \chi _A \Bigg[\frac{p_{\phi }^4}{r^4}\left(52 \nu -\frac{1342}{105}\right) +\frac{p_{\phi }^2}{r^3}\left(\frac{22516}{105}-\frac{15788 \nu }{105}\right) +\frac{p_r^2 p_{\phi }^2}{r^2}\left(\frac{5202 \nu }{35}+\frac{4288}{105}\right)
+\frac{p_r^2}{r}\left(\frac{1856}{105}-\frac{2496 \nu }{35}\right)  \nonumber\\
&\qquad
+p_r^4\left(\frac{598 \nu }{35}+\frac{712}{21}\right)
+\frac{1}{r^2} \left(\frac{1392 \nu }{35}-\frac{13112}{105}\right)\Bigg]
+\chi _S \Bigg[\frac{p_{\phi }^4}{r^4}\left(\frac{7876 \nu }{105}-\frac{1616 \nu ^2}{35}-\frac{1342}{105}\right) 
+p_r^4\left(\frac{712}{21}-\frac{172 \nu ^2}{5}-\frac{6194 \nu }{105}\right)  \nonumber\\
&\qquad
+\frac{ p_{\phi }^2}{r^3}\left(\frac{424 \nu ^2}{5}-\frac{38356 \nu }{105}+\frac{22516}{105}\right)
+\frac{ p_r^2 p_{\phi }^2}{r^2}\left(-\frac{1052 \nu ^2}{5}+\frac{4682 \nu }{105}+\frac{4288}{105}\right)
+\frac{ p_r^2}{r}\left(\frac{2368 \nu ^2}{35}-\frac{26848 \nu }{105}+\frac{1856}{105}\right)\nonumber\\
&\qquad
+\frac{1}{r^2}\left(-\frac{416 \nu ^2}{15}+\frac{9776 \nu }{105}-\frac{13112}{105}\right)\Bigg]
\Bigg\} \nonumber\\
&\quad
+ \frac{\nu ^2 \epsilon ^6}{r^4} \Bigg\{
\frac{p_{\phi }^8}{r^8}\left(-\frac{91832 \nu ^3}{3465}+\frac{2432 \nu ^2}{495}-\frac{102506 \nu }{3465}+\frac{4051}{462}\right) 
+\frac{p_{\phi }^6}{r^7}\left(\frac{29176 \nu ^3}{495}-\frac{827912 \nu ^2}{3465}+\frac{94756 \nu }{315}+\frac{5762}{385}\right) \nonumber\\
&\qquad
+\frac{p_r^2 p_{\phi }^6}{r^6}\left(-\frac{249296 \nu ^3}{1155}+\frac{196552 \nu ^2}{3465}+\frac{1207736 \nu }{3465}-\frac{459248}{3465}\right)
+\frac{1}{r^4} \left(-\frac{387808 \nu ^3}{31185}+\frac{9752 \nu ^2}{297}-\frac{896552 \nu }{31185}+\frac{11080}{2079}\right) \nonumber\\
&\qquad
+\frac{p_{\phi }^4 }{r^6}\left[-\frac{220912 \nu ^3}{4455}+\frac{36944 \nu ^2}{231}+\left(-\frac{2225572}{31185}-\frac{369 \pi ^2}{10}\right) \nu -\frac{27392}{175} \ln \left(\frac{r}{r_0}\right)+\frac{466174244}{259875}\right] \nonumber\\
&\qquad
+\frac{p_r^2 p_{\phi }^4}{r^5}\left(\frac{1272856 \nu ^3}{3465}-\frac{5055784 \nu ^2}{3465}-\frac{49976 \nu }{105}+\frac{57676}{1155}\right) 
+\frac{p_r^4 p_{\phi }^4}{r^4}\left(-\frac{103784 \nu ^3}{385}-\frac{36184 \nu ^2}{105}-\frac{222892 \nu }{385}-\frac{37316}{1155}\right) \nonumber\\
&\qquad
+\frac{p_r^4 p_{\phi }^2}{r^3}\left(\frac{222560 \nu ^3}{2079}-\frac{2109416 \nu ^2}{10395}-\frac{4329188 \nu }{3465}-\frac{3735964}{3465}\right) 
+\frac{p_r^6 p_{\phi }^2}{r^2}\left(-\frac{11392 \nu ^3}{231}-\frac{679576 \nu ^2}{3465}-\frac{868708 \nu }{3465}-\frac{58064}{495}\right) \nonumber\\
&\qquad
+\frac{p_{\phi }^2}{r^5} \left[\frac{83872 \nu ^3}{2835}+\frac{12272 \nu ^2}{10395}+\left(\frac{533 \pi ^2}{10}-\frac{836672}{945}\right) \nu +\frac{27392}{525} \ln \left(\frac{r}{r_0}\right)-\frac{69690092}{111375}\right] \nonumber\\
&\qquad
+\frac{p_r^2 p_{\phi }^2}{r^4} \left[-\frac{832432 \nu ^3}{6237}+\frac{45464 \nu ^2}{77}+\left(\frac{492 \pi ^2}{5}-\frac{89433188}{31185}\right) \nu +\frac{54784}{175} \ln \left(\frac{r}{r_0}\right)-\frac{1101433658}{259875}\right] \nonumber\\
&\qquad
+\frac{p_r^2}{r^3} \left[\frac{13312 \nu ^3}{2079}+\frac{4864 \nu ^2}{165}-\frac{4}{405} \left(1107 \pi ^2-3592\right) \nu +\frac{27392 }{1575}\ln \left(\frac{r}{r_0}\right)-\frac{105332728}{779625}\right] \nonumber\\
&\qquad
+\frac{p_r^4}{r^2} \left[\frac{29632 \nu ^3}{6237}-\frac{41392 \nu ^2}{1155}+\left(-\frac{292616}{6237}-\frac{41 \pi ^2}{10}\right) \nu +\frac{6848}{525} \ln \left(\frac{r}{r_0}\right)+\frac{1254788}{23625}\right] \nonumber\\
&\qquad
+ p_r^8\left(-\frac{32 \nu ^3}{33}-\frac{2668 \nu ^2}{495}-\frac{36238 \nu }{3465}-\frac{3352}{385}\right)
+\frac{p_r^6}{r}\left(-\frac{20288 \nu ^3}{10395}+\frac{6728 \nu ^2}{2079}+\frac{19268 \nu }{3465}-\frac{81704}{2475}\right) 
\Bigg\} \nonumber\\
&\quad
+ \frac{\nu ^2 \epsilon ^6}{r^6} \Bigg\{
\chi _A^2\Bigg[p_r^4\left(-\frac{544 \nu ^2}{35}-\frac{800 \nu }{21}-\frac{128}{35}\right) +\frac{ p_{\phi }^2 p_r^2}{r^2}\left(\frac{24128 \nu ^2}{105}-\frac{15566 \nu }{105}+\frac{320}{21}\right)+\frac{p_r^2}{r}\left(\frac{2432 \nu ^2}{105}-\frac{2384 \nu }{35}-\frac{1616}{105}\right) \nonumber\\
&\qquad\quad
+\frac{p_{\phi }^4}{r^4}\left(-\frac{288 \nu ^2}{35}-\frac{706 \nu }{35}+\frac{202}{21}\right) +\frac{p_{\phi }^2}{r^3}\left(\frac{320 \nu ^2}{21}+\frac{32476 \nu }{105}-\frac{3044}{35}\right) +\frac{1}{r^2}\left(\frac{256 \nu ^2}{35}-\frac{928 \nu }{15}+\frac{1576}{105}\right)\Bigg] \nonumber\\
&\qquad
+ \chi _S^2\Bigg[p_r^4\left(\frac{200 \nu ^2}{7}+\frac{656 \nu }{15}-\frac{128}{35}\right) +\frac{p_{\phi }^2 p_r^2}{r^2}\left(\frac{4496 \nu ^2}{35}-\frac{20854 \nu }{105}+\frac{320}{21}\right) +\frac{p_r^2}{r} \left(-\frac{2944 \nu ^2}{105}+\frac{15664 \nu }{105}-\frac{1616}{105}\right) \nonumber\\
&\qquad\quad
+\frac{p_{\phi }^4}{r^4}\left(\frac{232 \nu ^2}{7}-\frac{1186 \nu }{15}+\frac{202}{21}\right) +\frac{p_{\phi }^2}{r^3}\left(\frac{848 \nu ^2}{105}-\frac{348 \nu }{35}-\frac{3044}{35}\right) +\frac{1}{r^2} \left(\frac{64 \nu ^2}{21}-\frac{320 \nu }{21}+\frac{1576}{105}\right)\Bigg]\nonumber\\
&\qquad
+ \delta  \chi _S \chi _A\Bigg[p_r^4\left(-\frac{944 \nu }{105}-\frac{256}{35}\right) +\frac{p_{\phi }^2 p_r^2}{r^2}\left(\frac{640}{21}-\frac{6004 \nu }{21}\right) +\frac{p_r^2}{r} \left(\frac{2048 \nu }{105}-\frac{3232}{105}\right)+\frac{p_{\phi }^4}{r^4}\left(\frac{404}{21}-\frac{1276 \nu }{21}\right) \nonumber\\
&\qquad\quad
+\frac{p_{\phi }^2}{r^3}\left(-\frac{728 \nu }{15}-\frac{6088}{35}\right) +\frac{1}{r^2}\left(\frac{3152}{105}-\frac{256 \nu }{15}\right)\Bigg]
+ \delta  \kappa _A \Bigg[p_r^4\left(\frac{104 \nu }{35}+\frac{208}{35}\right) +\frac{ p_{\phi }^2 p_r^2}{r^2}\left(-\frac{1944 \nu }{35}-\frac{3728}{35}\right)\nonumber\\
&\qquad\quad
+\frac{p_r^2}{r} \left(\frac{816}{35}-\frac{32 \nu }{105}\right) 
+\frac{p_{\phi }^4}{r^4}\left(\frac{1128}{35}-\frac{2444 \nu }{35}\right) +\frac{p_{\phi }^2}{r^3}\left(-\frac{40 \nu }{3}-\frac{4808}{35}\right) +\frac{1}{r^2} \left(-\frac{40 \nu }{3}-\frac{4808}{35}\right) \Bigg]\nonumber\\
&\qquad
+ \kappa _S\Bigg[p_r^4\left(-\frac{272 \nu ^2}{35}-\frac{312 \nu }{35}+\frac{208}{35}\right) +\frac{p_{\phi }^2 p_r^2}{r^2}\left(\frac{6368 \nu ^2}{35}+\frac{5512 \nu }{35}-\frac{3728}{35}\right) +\frac{p_r^2}{r} \left(\frac{1312 \nu ^2}{105}-\frac{704 \nu }{15}+\frac{816}{35}\right) \nonumber\\
&\qquad\quad
+\frac{p_{\phi }^4}{r^4}\left(\frac{880 \nu ^2}{7}-\frac{940 \nu }{7}+\frac{1128}{35}\right) +\frac{p_{\phi }^2}{r^3}\left(-\frac{2048 \nu ^2}{35}+\frac{27448 \nu }{105}-\frac{4808}{35}\right) +\frac{1}{r^2}\left(\frac{384 \nu ^2}{35}-\frac{928 \nu }{105}+\frac{32}{21}\right)\Bigg]
\Bigg\},
\end{align}
\normalsize

The 3PN instantaneous contribution to the angular momentum flux in EOB coordinates is given by
\small
\begin{align}
\label{eq:EOBangMtmflux}
\frac{\Phi_J^\text{inst}}{M} &= \frac{\nu ^2 p_\phi}{r^3} \left(\frac{16 p_{\phi }^2}{5 r^2}-\frac{1}{5} 8 p_r^2+\frac{16}{5 r}\right) 
+ \frac{\epsilon^2 \nu ^2 p_\phi}{r^3} \Bigg[
\frac{p_{\phi }^4}{r^4}\left(\frac{22}{21}-\frac{424 \nu }{105}\right) 
+\frac{p_{\phi }^2}{r^3}\left(-\frac{1096 \nu }{105}-\frac{1976}{105}\right) 
+\left(\frac{236 \nu }{105}+\frac{548}{105}\right) p_r^4 \nonumber\\
&\qquad
+\frac{p_r^2 p_{\phi }^2}{r^2}\left(\frac{88}{105}-\frac{548 \nu }{105}\right)
+\frac{p_r^2}{r}\left(\frac{1864}{105}-\frac{128 \nu }{15}\right) 
+\frac{1}{r^2} \left(\frac{344 \nu }{105}-\frac{2644}{105}\right) \Bigg] \nonumber\\
&\quad
+ \frac{\epsilon ^3 \nu ^2}{r^3} \Bigg\{
\chi _S \left[\frac{p_{\phi }^4}{r^4}\left(\frac{64 \nu }{15}-\frac{24}{5}\right) 
+\frac{ p_{\phi }^2}{r^3}\left(\frac{16 \nu }{3}+\frac{88}{15}\right)
-\frac{p_r^2 p_{\phi }^2}{r^2}\left(\frac{64 \nu }{15}+\frac{24}{5}\right) 
-\frac{8}{15}  \nu  p_r^4
+\frac{ p_r^2}{r}\left(\frac{32 \nu }{15}-\frac{16}{5}\right)
+\frac{1}{r^2} \left(\frac{24}{5}-\frac{32 \nu }{15}\right)\right] \nonumber\\
&\qquad
+\delta  \chi _A \left[-\frac{24 p_{\phi }^4}{5 r^4}+\frac{88 p_{\phi }^2}{15 r^3}-\frac{24 p_r^2 p_{\phi }^2}{5 r^2}-\frac{16 p_r^2}{5 r}+\frac{24}{5 r^2}\right]
\Bigg\}
+ \frac{ \epsilon ^4\nu ^2 p_\phi}{r^5} \Bigg\{
\chi _S^2 \left[-\frac{16 p_{\phi }^2}{5 r^2}-\frac{32}{5}  p_r^2+\frac{1}{r}\left(\frac{18}{5}-\frac{8 \nu }{5}\right)\right] \nonumber\\
&\qquad
+\chi _A^2 \left[\frac{p_{\phi }^2}{r^2}\left(\frac{64 \nu }{5}-\frac{16}{5}\right) +\left(\frac{128 \nu }{5}-\frac{32}{5}\right) p_r^2+\frac{1}{r}\left(\frac{18}{5}-\frac{64 \nu }{5}\right)\right]
+\delta  \chi _A \chi _S \left(\frac{36}{5 r}-\frac{32 p_{\phi }^2}{5 r^2}-\frac{1}{5} 64 p_r^2\right) \nonumber\\
&\qquad
+\kappa _S \left[\frac{p_{\phi }^2}{r^2}\left(\frac{24}{5}-\frac{48 \nu }{5}\right) +\left(\frac{72 \nu }{5}-\frac{36}{5}\right) p_r^2+\frac{1}{r}\left(\frac{48}{5}-\frac{96 \nu }{5}\right)\right]
+\delta  \kappa _A \left(\frac{24 p_{\phi }^2}{5 r^2}-\frac{36}{5}  p_r^2+\frac{48}{5 r}\right)\!
\Bigg\}\nonumber\\
&\quad
+ \frac{ \epsilon ^4 p_\phi \nu ^2}{r^3} \Bigg\{
\frac{p_{\phi }^6}{r^6}\left(\frac{50 \nu ^2}{21}-\frac{1051 \nu }{315}-\frac{10}{63}\right) 
+\frac{p_{\phi }^4}{r^5}\left(\frac{7802 \nu ^2}{315}+\frac{6238 \nu }{315}-\frac{673}{35}\right)
+\frac{p_r^2 p_{\phi }^4}{r^4}\left(\frac{2801 \nu ^2}{105}+\frac{328 \nu }{15}-\frac{1021}{105}\right)  \nonumber\\
&\qquad
+\frac{p_{\phi }^2}{r^4}\left(-\frac{8896 \nu ^2}{315}+\frac{4854 \nu }{35}-\frac{1232}{45}\right)
+\frac{p_r^2 p_{\phi }^2}{r^3}\left(\frac{10088 \nu ^2}{315}+\frac{22868 \nu }{315}+\frac{2951}{63}\right)
+\frac{p_r^4 p_{\phi }^2}{r^2}\left(\frac{164 \nu ^2}{35}-\frac{1802 \nu }{105}+\frac{158}{105}\right)  \nonumber\\
&\qquad
+\frac{p_r^2}{r^2}\left(-\frac{4504 \nu ^2}{315}+\frac{17422 \nu }{315}-\frac{20612}{315}\right)
+\left(-\frac{857 \nu ^2}{315}-\frac{499 \nu }{63}-\frac{1973}{315}\right) p_r^6
+\frac{p_r^4}{r}\left(\frac{4462 \nu ^2}{315}-\frac{1034 \nu }{63}-\frac{7076}{105}\right) \nonumber\\
&\qquad
+\frac{1}{r^3} \left(\frac{1352 \nu ^2}{315}-\frac{428 \nu }{45}+\frac{36868}{567}\right)
\Bigg\}
+ \frac{ \epsilon ^5 \nu ^3 p_r p_\phi}{r^6} \Bigg[
\frac{512 \alpha }{25}-\frac{384 \beta }{25}-\frac{12112}{525}
-\frac{27744 p_{\phi }^4}{175 r^2} -\frac{624}{175}  r^2 p_r^4 +\frac{40232}{175} p_r^2 p_{\phi }^2 \nonumber\\
&\qquad
+\frac{p_{\phi }^2}{r}\left(\frac{1152 \alpha }{25}-\frac{768 \beta }{25}-\frac{92752}{525}\right)
+\left(-\frac{384 \alpha }{25}+\frac{64 \beta }{5}+\frac{1376}{75}\right) r p_r^2
\Bigg] \nonumber\\
&\quad
+ \frac{ \epsilon ^5 \nu ^2}{r^3} \Bigg\{
\delta \chi _A \Bigg[
\frac{p_{\phi }^6}{r^6}\left(\frac{659 \nu }{105}-\frac{12}{35}\right) 
+\frac{p_{\phi }^4}{r^5}\left(\frac{4936}{105}-\frac{164 \nu }{35}\right) 
+\frac{p_r^2 p_{\phi }^4}{r^4}\left(\frac{1458 \nu }{35}+\frac{108}{5}\right) 
+\frac{p_{\phi }^2}{r^4}\left(-\frac{6082 \nu }{105}-\frac{538}{105}\right)
-\frac{19}{105} \nu  p_r^6  \nonumber\\
&\qquad\quad
+\frac{p_r^2 p_{\phi }^2}{r^3}\left(\frac{9454}{105}-\frac{9 \nu }{10}\right) 
+\frac{p_r^4 p_{\phi }^2}{r^2}\left(\frac{51 \nu }{5}-\frac{72}{35}\right) 
+\frac{p_r^2}{r^2}\left(\frac{312}{7}-\frac{508 \nu }{35}\right) 
+\frac{p_r^4}{r}\left(\frac{278 \nu }{35}-\frac{304}{35}\right) 
+\frac{1}{r^3} \left(\frac{544 \nu }{105}-\frac{1032}{35}\right)\Bigg] \nonumber\\
&\qquad
+\chi _S \Bigg[
\frac{p_r^2 p_{\phi }^4}{r^4}\left(\frac{108}{5}-\frac{2484 \nu ^2}{35}-\frac{1188 \nu }{35}\right)
+\frac{p_{\phi }^2}{r^4}\left(\frac{3568 \nu ^2}{105}-\frac{12374 \nu }{105}-\frac{538}{105}\right)
+\frac{p_r^2 p_{\phi }^2}{r^3}\left(-\frac{3721 \nu ^2}{105}+\frac{337 \nu }{210}+\frac{9454}{105}\right) \nonumber\\
&\qquad\quad
+\frac{ p_{\phi }^6}{r^6}\left(\frac{20 \nu ^2}{21}+\frac{43 \nu }{3}-\frac{12}{35}\right)
+\frac{p_{\phi }^4}{r^5}\left(\frac{4936}{105}-\frac{722 \nu ^2}{21}-\frac{6344 \nu }{105}\right)
+\frac{p_r^4 p_{\phi }^2}{r^2}\left(\frac{106 \nu ^2}{5}+\frac{2259 \nu }{35}-\frac{72}{35}\right) 
+\left(\frac{118 \nu ^2}{105}+\frac{191 \nu }{105}\right) p_r^6 \nonumber\\
&\qquad\quad
+\frac{ p_r^2}{r^2}\left(\frac{416 \nu ^2}{35}-\frac{4376 \nu }{105}+\frac{312}{7}\right)
+\frac{ p_r^4}{r}\left(-\frac{36 \nu ^2}{5}+\frac{1774 \nu }{105}-\frac{304}{35}\right)
+\frac{1}{r^3} \left(-\frac{512 \nu ^2}{105}+\frac{1864 \nu }{105}-\frac{1032}{35}\right) \Bigg]
\Bigg\} \nonumber\\
&\quad
+ \frac{\epsilon ^6 \nu ^2 p_\phi}{r^3}  \Bigg\{
\left(\frac{21587 \nu ^3}{6930}+\frac{35639 \nu ^2}{3465}+\frac{76399 \nu }{6930}+\frac{47141}{6930}\right) p_r^8
+\frac{p_{\phi }^2 p_r^6}{r^2}\left(-\frac{221 \nu ^3}{198}+\frac{386377 \nu ^2}{6930}+\frac{24181 \nu }{693}-\frac{4405}{693}\right)  \nonumber\\
&\qquad
+\frac{p_r^6}{r}\left(-\frac{68122 \nu ^3}{3465}+\frac{9353 \nu ^2}{1155}+\frac{312014 \nu }{3465}+\frac{1849598}{17325}\right)
+\frac{p_{\phi }^4 p_r^4}{r^4}\left(-\frac{188519 \nu ^3}{2310}-\frac{132691 \nu ^2}{1540}-\frac{14753 \nu }{660}+\frac{99931}{4620}\right)  \nonumber\\
&\qquad
+\frac{p_{\phi }^2 p_r^4}{r^3}\left(-\frac{88211 \nu ^3}{1155}-\frac{460168 \nu ^2}{3465}+\frac{872321 \nu }{6930}-\frac{1038307}{6930}\right)
+\frac{p_{\phi }^6 p_r^2}{r^6}\left(-\frac{74707 \nu ^3}{2310}-\frac{13196 \nu ^2}{385}-\frac{11173 \nu }{770}+\frac{38251}{3465}\right)  \nonumber\\
&\qquad
+\frac{p_{\phi }^4 p_r^2}{r^5}\left(-\frac{223304 \nu ^3}{3465}-\frac{280351 \nu ^2}{693}-\frac{1213357 \nu }{3465}+\frac{22346}{385}\right)
+\frac{ p_{\phi }^8}{r^8}\left(-\frac{271 \nu ^3}{1155}+\frac{10349 \nu ^2}{1540}+\frac{1807 \nu }{420}-\frac{3463}{6930}\right)\nonumber\\
&\qquad
+\frac{p_{\phi }^6}{r^7}\left(-\frac{25093 \nu ^3}{495}+\frac{6653 \nu ^2}{3465}+\frac{180227 \nu }{3465}+\frac{217447}{13860}\right) 
+\frac{p_{\phi }^2 p_r^2}{r^4}\Bigg[\frac{126358 \nu ^3}{1155}-\frac{31831 \nu ^2}{105}+\left(-\frac{13276}{7}+\frac{369 \pi ^2}{5}\right) \nu \nonumber\\
&\qquad\quad
+\frac{46224}{175} \ln \left(\frac{r}{r_0} \!\right)-\frac{2219044}{875} \Bigg]\!
+\frac{ p_r^2}{r^3}\left[-\frac{16040 \nu ^3}{693}+\frac{29098 \nu ^2}{693}-\left(\frac{789710}{6237}+\frac{123 \pi ^2}{40}\right) \nu +\frac{39376}{525} \ln \left(\frac{r}{r_0} \!\right)-\frac{471863527}{779625}\right]\nonumber\\
&\qquad
+\frac{p_{\phi }^4}{r^6}\left[\frac{40349 \nu ^3}{385}-\frac{543947 \nu ^2}{1386}+\left(\frac{2726203}{6930}-\frac{369 \pi ^2}{40}\right) \nu -\frac{6848}{175} \ln \left(\frac{r}{r_0}\right)+\frac{21540643}{57750}\right]\nonumber\\
&\qquad
+\frac{ p_r^4}{r^2}\left[\frac{113411 \nu ^3}{3465}-\frac{542981 \nu ^2}{6930}+\left(\frac{2778313}{6930}-\frac{123 \pi ^2}{5}\right) \nu -\frac{6848}{175} \ln \left(\frac{r}{r_0}\right)+\frac{127177579}{173250}\right]\nonumber\\
&\qquad
+\frac{p_{\phi }^2}{r^5}\left[-\frac{2018 \nu ^3}{33}+\frac{908632 \nu ^2}{3465}+\left(-\frac{1720487}{2835}-\frac{123 \pi ^2}{20}\right) \nu -\frac{37664}{525} \ln \left(\frac{r}{r_0}\right)+\frac{1374786181}{1559250}\right] \nonumber\\
&\qquad
+\frac{1}{r^4} \left[\frac{27292 \nu ^3}{3465}+\frac{1522 \nu ^2}{315}+\left(-\frac{11839606}{31185}+\frac{943 \pi ^2}{40}\right) \nu +\frac{3424}{525} \ln \left(\frac{r}{r_0}\right)-\frac{109925608}{779625}\right]
\Bigg\} \nonumber\\
&\quad
+ \frac{ \epsilon ^6 \nu ^2 p_\phi}{r^5} \Bigg\{
\chi _A^2 \Bigg[\left(-\frac{8416 \nu ^2}{105}-\frac{2328 \nu }{35}+\frac{2824}{105}\right) p_r^4
+\frac{p_{\phi }^2 p_r^2}{r^2}\left(\frac{848 \nu ^2}{35}-\frac{324 \nu }{35}-\frac{584}{35}\right) 
+\frac{ p_r^2}{r}\left(\frac{14752 \nu ^2}{105}-\frac{1201 \nu }{5}+\frac{5323}{105}\right)\nonumber\\
&\qquad\quad
+\frac{p_{\phi }^4}{r^4}\left(-\frac{64 \nu ^2}{3}+\frac{556 \nu }{35}-\frac{8}{21}\right) 
+\frac{p_{\phi }^2}{r^3}\left(\frac{2272 \nu ^2}{35}-\frac{117 \nu }{5}+\frac{149}{105}\right) 
+\frac{1}{r^2} \left(-\frac{2656 \nu ^2}{105}+\frac{9718 \nu }{105}-\frac{2722}{105}\right)\Bigg]\nonumber\\
&\qquad
+ \chi _S^2\Bigg[\left(-4 \nu ^2+\frac{8 \nu }{21}+\frac{2824}{105}\right) p_r^4+\frac{p_{\phi }^2 p_r^2}{r^2}\left(24 \nu ^2+\frac{892 \nu }{35}-\frac{584}{35}\right) +\frac{p_r^2}{r}\left(\frac{716 \nu ^2}{105}-\frac{853 \nu }{21}+\frac{5323}{105}\right) \nonumber\\
&\qquad\quad
+\frac{p_{\phi }^4}{r^4}\left(-\frac{172 \nu }{15}-\frac{8}{21}\right) +\frac{ p_{\phi }^2}{r^3}\left(\frac{3284 \nu ^2}{105}-\frac{5989 \nu }{105}+\frac{149}{105}\right)+\frac{1}{r^2}\left(\frac{184 \nu ^2}{35}-\frac{98 \nu }{5}-\frac{2722}{105}\right)\Bigg]\nonumber\\
&\qquad
+ \delta  \chi _S \chi _A \Bigg[\left(\frac{4352 \nu }{105}+\frac{5648}{105}\right) p_r^4+\frac{p_{\phi }^2 p_r^2}{r^2}\left(-\frac{1768 \nu }{35}-\frac{1168}{35}\right) +\frac{ p_r^2}{r}\left(\frac{10646}{105}-\frac{8194 \nu }{105}\right)+\frac{ p_{\phi }^4}{r^4}\left(\frac{304 \nu }{105}-\frac{16}{21}\right) \nonumber\\
&\qquad\quad
+\frac{ p_{\phi }^2}{r^3}\left(\frac{298}{105}-\frac{1570 \nu }{21}\right)+\frac{1}{r^2}\left(-\frac{1076 \nu }{35}-\frac{5444}{105}\right)\Bigg]
+ \delta  \kappa _A \Bigg[\left(\frac{26 \nu }{5}+\frac{170}{7}\right) p_r^4+\frac{p_{\phi }^2 p_r^2}{r^2}\left(\frac{354 \nu }{35}-\frac{108}{7}\right) \nonumber\\
&\qquad\quad
+\frac{ p_r^2}{r}\left(\frac{7964}{105}-\frac{1976 \nu }{105}\right)+\frac{p_{\phi }^4}{r^4}\left(\frac{23}{7}-\frac{388 \nu }{35}\right) +\frac{p_{\phi }^2}{r^3}\left(-\frac{1276 \nu }{21}-\frac{1432}{105}\right) +\frac{1}{r^2}\left(-\frac{148 \nu }{21}-\frac{7204}{105}\right)\Bigg]\nonumber\\
&\qquad
+ \kappa _S \Bigg[\left(-\frac{1324 \nu ^2}{35}-\frac{1518 \nu }{35}+\frac{170}{7}\right) p_r^4+\frac{p_{\phi }^2 p_r^2}{r^2}\left(\frac{972 \nu ^2}{35}+\frac{1434 \nu }{35}-\frac{108}{7}\right) +\frac{ p_r^2}{r}\left(\frac{3056 \nu ^2}{35}-\frac{5968 \nu }{35}+\frac{7964}{105}\right)\nonumber\\
&\qquad\quad
+\frac{ p_{\phi }^4}{r^4}\left(\frac{88 \nu ^2}{5}-\frac{618 \nu }{35}+\frac{23}{7}\right)+\frac{p_{\phi }^2}{r^3}\left(\frac{3328 \nu ^2}{35}-\frac{1172 \nu }{35}-\frac{1432}{105}\right) +\frac{1}{r^2}\left(-\frac{1016 \nu ^2}{35}+\frac{4556 \nu }{35}-\frac{7204}{105}\right)\Bigg]
\Bigg\}.
\end{align}
\normalsize

\subsection{Complete orbit-averaged fluxes in the Keplerian parametrization}

The complete 3PN orbit-averaged energy flux in the Keplerian parametrization 
is given by
\small
\begin{subequations}
\label{eq:EfluxEOBOrbAv} 
\begin{align}
\langle \Phi_E \rangle &\equiv \frac{32 \nu ^2 x^5}{5 \left(1-e^2\right)^{7/2}}  \Big[ \mathcal{H}^{\text{N}} + \mathcal{H}^{1\text{PN}} + \mathcal{H}^{1.5\text{PN}} + \mathcal{H}^{2\text{PN}}+ \mathcal{H}^{2.5\text{PN}} + \mathcal{H}^{3\text{PN}} + \mathcal{H}^{\text{SO}} + \mathcal{H}^{\text{SS}}\Big], \\
\mathcal{H}^{\text{N}} &= 1+\frac{73 e^2}{24}+\frac{37 e^4}{96} , \\
\mathcal{H}^{1\text{PN}} &= - \frac{x \epsilon ^2}{1-e^2} \left[
e^6 \left(\frac{185 \nu }{576}+\frac{4037}{1792}\right)+e^4 \left(\frac{55 \nu }{6}+\frac{9253}{384}\right)+e^2 \left(\frac{305 \nu }{18}+\frac{15901}{672}\right)
+\frac{35 \nu }{12}+\frac{1247}{336}\right], \\
\mathcal{H}^{1.5\text{PN}}&= \frac{4 \pi  x^{3/2} \epsilon ^3}{\left(1-e^2\right)^{3/2}} \left[1+\frac{1375 e^2}{192} +\frac{3935 e^4}{768} +\frac{10007 e^6}{36864} + \Order(e^8)\right] ,\\
\mathcal{H}^{2\text{PN}}&= \frac{x^2 \epsilon ^4}{\left(1-e^2\right)^2} \bigg\lbrace
e^8 \left(\frac{1295 \nu ^2}{6912}+\frac{49283 \nu }{32256}+\frac{499451}{64512}\right)+e^6 \left(\frac{9965 \nu ^2}{864}+\frac{30395 \nu }{384}+\frac{231899}{2304}\right) \nonumber\\
&\quad
+e^4 \left(\!\frac{45665 \nu ^2}{864}+\frac{2143381 \nu }{8064}+\frac{2161337}{24192}\right)
+e^2 \left(\!\frac{8395 \nu ^2}{216}+\frac{51047 \nu }{288}-\frac{1430873}{18144}\right)
+\frac{65 \nu ^2}{18}+\frac{12799 \nu }{504}-\frac{203471}{9072} \nonumber\\
&\quad
+ \sqrt{1-e^2} \left[e^6 \left(\frac{259 \nu }{96}-\frac{1295}{192}\right)+e^4 \left(\frac{595 \nu }{32}-\frac{2975}{64}\right)+e^2 \left(\frac{1715}{48}-\frac{343 \nu }{24}\right)-7 \nu +\frac{35}{2}\right]
\bigg\rbrace,\\
\mathcal{H}^{2.5\text{PN}}&= -\frac{\pi  x^{5/2} \epsilon ^5}{\left(1-e^2\right)^{5/2}} \bigg[
\frac{8191}{672}+\frac{583 \nu }{24}
+e^2 \left(\frac{521593 \nu }{2016}+\frac{62003}{336}\right)
+e^4 \left(\frac{11179741 \nu }{32256}+\frac{20327389}{43008}\right) \nonumber\\
&\quad
+ e^6 \left(\frac{28633781 \nu }{387072}+\frac{87458089}{387072}\right)
+ \Order(e^8)
\bigg],\\
\mathcal{H}^{3\text{PN}}&=\frac{x^3 \epsilon ^6}{\left(1-e^2\right)^3} \Bigg\lbrace
-\frac{775 \nu ^3}{324}-\frac{94403 \nu ^2}{3024}+\left(\frac{8009293}{54432}-\frac{41 \pi ^2}{64}\right) \nu +\left(\frac{287 \pi ^2}{192}-\frac{165761}{1008}\right) \nu +\frac{16 \pi ^2}{3}+\frac{5599693439}{69854400} \nonumber\\
&\quad
+ e^2 \left[-\frac{95335 \nu ^3}{1944}-\frac{14879 \nu ^2}{27}+\left(\frac{324260}{1701}+\frac{4879 \pi ^2}{1536}\right) \nu +\frac{680 \pi ^2}{9}+\frac{292692719843}{139708800}+\frac{18832 \ln 2}{45}-\frac{234009 \ln 3}{560}\right] \nonumber\\
&\quad
+ e^4 \left[\frac{5171 \pi ^2}{36}+\frac{120664043497}{31046400}+\frac{2106081 \ln 3}{448}-\frac{1100515 \nu ^3}{7776}-\frac{4722371 \nu ^2}{4032}-\left(\frac{21903607}{27216}+\frac{29971 \pi ^2}{1024}\right) \nu -\frac{745897 \ln 2}{84}\right] \nonumber\\
&\quad
+ e^6 \bigg[\frac{1751 \pi ^2}{36}+\frac{66948078809}{62092800}+\frac{17845567 \ln 2}{180}-\frac{864819261 \ln 3}{35840}-\frac{5224609375 \ln 5}{193536}-\frac{448415 \nu ^3}{5184}-\frac{36009965 \nu ^2}{48384} \nonumber\\
&\qquad
-\left(\frac{62053829}{41472}+\frac{84501 \pi ^2}{4096}\right) \nu \bigg] 
+ e^8 \left[\frac{99 \pi ^2}{64}-\frac{33102576139}{141926400}-\frac{309815 \nu ^3}{31104}-\frac{18205613 \nu ^2}{193536}-\left(\frac{45857803}{129024}+\frac{4059 \pi ^2}{4096}\right) \nu \right] \nonumber\\
&\quad
+ e^{10} \left[-\frac{10175 \nu ^3}{124416}-\frac{311165 \nu ^2}{387072}-\frac{561073 \nu }{193536}-\frac{233745653}{11354112}\right] \nonumber\\
&\quad
+ \sqrt{1-e^2} \bigg[
e^8 \left(-\frac{185 \nu ^2}{36}-\frac{239455 \nu }{16128}+\frac{617515}{10752}\right)+e^6 \left(-\frac{28685 \nu ^2}{288}+\left(\frac{1983559}{48384}-\frac{10619 \pi ^2}{18432}\right) \nu +\frac{1128608203}{2419200}\right) \nonumber\\
&\qquad
+e^4 \left(-\frac{25655 \nu ^2}{288}+\left(\frac{4351741}{8064}-\frac{24395 \pi ^2}{6144}\right) \nu -\frac{210234049}{403200}\right)
+\frac{455 \nu ^2}{12}+\left(\frac{287 \pi ^2}{192}-\frac{165761}{1008}\right) \nu -\frac{14047483}{151200} \nonumber\\
&\qquad
+e^2 \left(\frac{11225 \nu ^2}{72}+\left(\frac{14063 \pi ^2}{4608}-\frac{2427389}{6048}\right) \nu -\frac{75546769}{100800}\right)
\bigg] \nonumber\\
&\quad
+ \ln \left(\frac{2 \left(1-e^2\right) \gamma_ { \text{E}}' \sqrt{x}}{\sqrt{1-e^2}+1}\right) \left[-\frac{10593 e^8}{2240}-\frac{187357 e^6}{1260}-\frac{553297 e^4}{1260}-\frac{14552 e^2}{63}-\frac{1712}{105}\right]
\Bigg\rbrace,\\
\mathcal{H}^{\text{SO}}&= -\frac{x^{3/2} \epsilon ^3}{\left(1-e^2\right)^{3/2}} \bigg\lbrace
\left[e^6 \left(\frac{\nu }{8}-\frac{83}{64}\right)+e^4 \left(-\frac{299 \nu }{36}-\frac{589}{288}\right)+e^2 \left(\frac{1019}{72}-\frac{349 \nu }{18}\right)-3 \nu +\frac{11}{4}\right] \chi _S \nonumber\\
&\qquad
+ \delta  \left[-\frac{83 e^6}{64}-\frac{589 e^4}{288}+\frac{1019 e^2}{72}+\frac{11}{4}\right] \chi _A
\bigg\rbrace \nonumber\\
&\quad
+ \frac{x^{5/2} \epsilon ^5}{\left(1-e^2\right)^{5/2}} \bigg\lbrace
\chi_S \bigg[
e^8 \left(\frac{413 \nu ^2}{768}-\frac{12731 \nu }{5376}-\frac{171947}{14336}\right)+e^6 \left(-\frac{8201 \nu ^2}{768}-\frac{1318297 \nu }{10752}-\frac{12903}{224}\right) \nonumber\\
&\qquad
+e^4 \left(-\frac{9043 \nu ^2}{72}-\frac{256847 \nu }{1008}+\frac{86255}{1344}\right)
+e^2 \left(-\frac{11939 \nu ^2}{96}-\frac{55627 \nu }{1344}+\frac{1774}{21}\right)-\frac{115 \nu ^2}{9}-\frac{109 \nu }{9}+\frac{389}{16}
\bigg] \nonumber\\
&\quad
+ \delta  \chi _A \left[\frac{365 \nu }{36}+\frac{389}{16}-e^8 \left(\!\frac{2659 \nu }{1536}+\frac{171947}{14336}\right)-e^6 \left(\!\frac{45437 \nu }{1536}+\frac{12903}{224}\right)+e^4 \left(\!\frac{1889 \nu }{72}+\frac{86255}{1344}\right)+e^2 \left(\!\frac{18425 \nu }{192}+\frac{1774}{21}\right)\right]\nonumber\\
&\quad
+ \sqrt{1-e^2} \bigg[
\chi _S \bigg[e^6 \left(\frac{259 \nu ^2}{144}-\frac{259 \nu }{18}+\frac{259}{24}\right)+e^4 \left(\frac{595 \nu ^2}{48}-\frac{595 \nu }{6}+\frac{595}{8}\right)+e^2 \left(-\frac{343 \nu ^2}{36}+\frac{686 \nu }{9}-\frac{343}{6}\right) \nonumber\\
&\qquad\quad
-\frac{14 \nu ^2}{3}+\frac{112 \nu }{3}-28\bigg] 
+ \delta  \chi _A \left[e^6 \left(\frac{259}{24}-\frac{259 \nu }{72}\right)+e^4 \left(\frac{595}{8}-\frac{595 \nu }{24}\right)+e^2 \left(\frac{343 \nu }{18}-\frac{343}{6}\right)+\frac{28 \nu }{3}-28\right]
\bigg]
\bigg\rbrace \nonumber\\
&\quad
+ \frac{\pi  x^3 \epsilon^6}{\left(1-e^2\right)^3} \bigg\lbrace
\chi _S \left[e^6 \left(\frac{237541 \nu }{13824}+\frac{12169}{128}\right)+e^4 \left(\frac{49039 \nu }{288}+\frac{12029}{576}\right)+e^2 \left(\frac{9869 \nu }{72}-\frac{1711}{18}\right)+\frac{34 \nu }{3}-\frac{65}{6}\right] \nonumber\\
&\qquad
+ \delta \chi _A  \left[\frac{12169 e^6}{128}+\frac{12029 e^4}{576}-\frac{1711 e^2}{18}-\frac{65}{6}\right]
\bigg\rbrace,\\
\mathcal{H}^{\text{SS}}&= \frac{x^2 \epsilon ^4}{\left(1-e^2\right)^2} \bigg\{
\left[e^6 \left(\frac{13 \nu }{4}-\frac{199}{256}\right)+e^4 \left(\frac{413 \nu }{24}-\frac{1445}{384}\right)+e^2 \left(\frac{199}{32}-\frac{45 \nu }{2}\right)-8 \nu +\frac{33}{16}\right] \chi _A^2 
+ \bigg[-\frac{199 e^6}{128}-\frac{1445 e^4}{192}\nonumber\\
&\qquad\quad
+\frac{199 e^2}{16}+\frac{33}{8}\bigg] \delta\chi _A \chi _S 
+\left[e^6 \left(-\frac{9 \nu }{64}-\frac{199}{256}\right)+e^4 \left(-\frac{69 \nu }{32}-\frac{1445}{384}\right)+e^2 \left(\frac{199}{32}-\frac{19 \nu }{8}\right)-\frac{\nu }{4}+\frac{33}{16}\right] \chi _S^2\nonumber\\
&\qquad
+\delta  \left(\frac{11 e^6}{32}+\frac{1195 e^4}{96}+\frac{449 e^2}{24}+2\right) \kappa _A+\left(\! e^6 \!\left(\frac{11}{32}-\frac{11 \nu }{16}\right)+e^4 \!\left(\frac{1195}{96}-\frac{1195 \nu }{48}\right)+e^2 \left(\frac{449}{24}-\frac{449 \nu }{12}\right)-4 \nu +2\!\right) \kappa _S
\!\bigg\}\nonumber\\
&\quad
+ \frac{x^3 \epsilon^6}{(1 - e^2)^3} \Bigg\{
\chi _A^2 \bigg[e^8 \left(\frac{151 \nu ^2}{48}-\frac{606029 \nu }{10752}+\frac{195537}{14336}\right)+e^6 \left(\frac{8437 \nu ^2}{288}-\frac{2034103 \nu }{8064}+\frac{303295}{5376}\right)\nonumber\\
&\qquad\quad
+e^4 \left(\frac{12817 \nu ^2}{72}+\frac{1936265 \nu }{12096}-\frac{1484029}{24192}\right)+e^2 \left(\frac{10025 \nu ^2}{36}+\frac{260423 \nu }{3024}-\frac{266911}{6048}\right)+29 \nu ^2+\frac{45751 \nu }{504}-\frac{12325}{504}\bigg]\nonumber\\
&\qquad
+\chi _S^2 \bigg[e^8 \left(\frac{169 \nu ^2}{128}-\frac{7745 \nu }{1536}+\frac{195537}{14336}\right)+e^6 \left(\frac{1145 \nu ^2}{48}-\frac{34783 \nu }{1152}+\frac{303295}{5376}\right) +e^4 \left(\frac{57781 \nu ^2}{432}-\frac{447259 \nu }{1728}-\frac{1484029}{24192}\right) \nonumber\\
&\qquad\quad
+e^2 \left(\frac{6191 \nu ^2}{54}-\frac{134389 \nu }{432}-\frac{266911}{6048}\right)+\frac{5 \nu ^2}{18}+\frac{295 \nu }{72}-\frac{12325}{504}\bigg] \nonumber\\
&\qquad
+\delta  \chi _A \chi _S \bigg[e^8 \left(\frac{195537}{7168}-\frac{10519 \nu }{1536}\right)+e^6 \left(\frac{303295}{2688}-\frac{32701 \nu }{576}\right)+e^4 \left(-\frac{297329 \nu }{864}-\frac{1484029}{12096}\right)-\frac{53 \nu }{18}-\frac{12325}{252}\nonumber\\
&\qquad\quad
+e^2 \left(-\frac{86723 \nu }{216}-\frac{266911}{3024}\right)\bigg]
+ \delta  \kappa _A \bigg[e^8 \left(-\frac{3269 \nu }{3072}-\frac{4979}{1792}\right)+e^6 \left(-\frac{69331 \nu }{1152}-\frac{5255}{48}\right)+e^4 \left(-\frac{41125 \nu }{144}-\frac{70151}{336}\right)\nonumber\\
&\qquad\quad
+e^2 \left(-\frac{9817 \nu }{48}-\frac{1935}{28}\right)-\frac{241 \nu }{24}-\frac{340}{21}\bigg]
+ \kappa _S\bigg[e^8 \left(\frac{47 \nu ^2}{24}+\frac{96613 \nu }{21504}-\frac{4979}{1792}\right)+e^6 \left(\frac{51481 \nu ^2}{576}+\frac{182909 \nu }{1152}-\frac{5255}{48}\right)\nonumber\\
&\qquad\quad
+e^4 \left(\frac{58241 \nu ^2}{144}+\frac{133031 \nu }{1008}-\frac{70151}{336}\right)+e^2 \left(\frac{6647 \nu ^2}{24}-\frac{22279 \nu }{336}-\frac{1935}{28}\right)+\frac{29 \nu ^2}{2}+\frac{1251 \nu }{56}-\frac{340}{21}\bigg] \nonumber\\
&\qquad
+ \sqrt{1-e^2} \bigg\{
\chi _A^2 \bigg[e^6 \left(-\frac{259 \nu ^2}{48}+\frac{5957 \nu }{144}-\frac{2849}{288}\right)+e^4 \left(-\frac{595 \nu ^2}{16}+\frac{13685 \nu }{48}-\frac{6545}{96}\right)+e^2 \left(\frac{343 \nu ^2}{12}-\frac{7889 \nu }{36}+\frac{3773}{72}\right)\nonumber\\
&\qquad\qquad
+14 \nu ^2-\frac{322 \nu }{3}+\frac{77}{3}\bigg]
+\chi _S^2 \bigg[e^6 \left(-\frac{259 \nu ^2}{72}+\frac{259 \nu }{18}-\frac{2849}{288}\right)+e^4 \left(-\frac{595 \nu ^2}{24}+\frac{595 \nu }{6}-\frac{6545}{96}\right)\nonumber\\
&\qquad\qquad
+e^2 \left(\frac{343 \nu ^2}{18}-\frac{686 \nu }{9}+\frac{3773}{72}\right)+\frac{28 \nu ^2}{3}-\frac{112 \nu }{3}+\frac{77}{3}\bigg] 
+ \delta  \chi _A \chi _S \bigg[e^6 \left(\frac{259 \nu }{16}-\frac{2849}{144}\right)+e^4 \left(\frac{1785 \nu }{16}-\frac{6545}{48}\right)\nonumber\\
&\qquad\qquad
+e^2 \left(\frac{3773}{36}-\frac{343 \nu }{4}\right)-42 \nu +\frac{154}{3} \bigg]
+ \kappa _S\bigg[e^6 \left(-\frac{259 \nu ^2}{96}+\frac{2849 \nu }{192}-\frac{1813}{288}\right)+e^4 \left(-\frac{595 \nu ^2}{32}+\frac{6545 \nu }{64}-\frac{4165}{96}\right)\nonumber\\
&\qquad\qquad
+e^2 \left(\frac{343 \nu ^2}{24}-\frac{3773 \nu }{48}+\frac{2401}{72}\right)+7 \nu ^2-\frac{77 \nu }{2}+\frac{49}{3}\bigg]
+\delta  \kappa _A \bigg[e^6 \left(\frac{1295 \nu }{576}-\frac{1813}{288}\right)+e^4 \left(\frac{2975 \nu }{192}-\frac{4165}{96}\right)\nonumber\\
&\qquad\qquad
+e^2 \left(\frac{2401}{72}-\frac{1715 \nu }{144}\right)-\frac{35 \nu }{6}+\frac{49}{3}\bigg]
\bigg\}
\Bigg\},
\end{align}
\end{subequations}
\normalsize
where we defined $\gamma_ { \text{E}}' \equiv \text{e}^{-11/12 + \gamma_ { \text{E}} + 2 \ln 2}$ to simplify a bit the notation.

The complete 3PN orbit-averaged angular momentum flux in the Keplerian parametrization 
is given by
\small
\begin{subequations}
\label{eq:JfluxEOBOrbAv}
\begin{align}
\langle \Phi_J \rangle &\equiv \frac{4 \nu ^2 M x^{7/2}}{5 \left(1-e^2\right)^2} \Big[ \mathcal{G}^{\text{N}} + \mathcal{G}^{1\text{PN}} + \mathcal{G}^{1.5\text{PN}} + \mathcal{G}^{2\text{PN}} + \mathcal{G}^{2.5\text{PN}} + \mathcal{G}^{3\text{PN}} + \mathcal{G}^{\text{SO}} + \mathcal{G}^{\text{SS}}\Big],\\
\mathcal{G}^{\text{N}} &= 8+7 e^2, \\
\mathcal{G}^{1\text{PN}}&= -\frac{x \epsilon ^2}{1-e^2} \left[e^4 \left(\frac{107 \nu }{12}+\frac{5713}{336}\right)+e^2 \left(\frac{197 \nu }{3}+\frac{2777}{42}\right)+\frac{70 \nu }{3}+\frac{1247}{42}\right] , \\
\mathcal{G}^{1.5\text{PN}}&= \frac{32 \pi  x^{3/2} \epsilon ^3}{\left(1-e^2\right)^{3/2}} \left[ 1+\frac{97 e^2}{32}+\frac{49 e^4}{128}-\frac{49 e^6}{18432} + \Order(e^8) \right],\\
\mathcal{G}^{2\text{PN}}&= \frac{x^2 \epsilon ^4}{\left(1-e^2\right)^2} \bigg\lbrace\!
\frac{260 \nu ^2}{9}+\frac{11287 \nu }{63}-\frac{135431}{1134}
+e^2 \left(\!\frac{617 \nu ^2}{3}+\frac{19199 \nu }{28}-\frac{190087}{756}\right)
+e^4 \left(\!\frac{436 \nu ^2}{3}+\frac{11311 \nu }{24}+\frac{192133}{3024}\right) \nonumber\\
&\quad
+ e^6 \left(\frac{289 \nu ^2}{36}+\frac{613 \nu }{21}+\frac{3499}{288}\right)
+ \sqrt{1-e^2} \left[e^4 (28 \nu -70)+e^2 (4 \nu -10)-32 \nu +80\right]
\bigg\rbrace,\\
\mathcal{G}^{2.5\text{PN}}&= -\frac{\pi  x^{5/2} \epsilon ^5}{\left(1-e^2\right)^{5/2}} \bigg[e^6 \left(\frac{273865 \nu }{12096}+\frac{6395111}{96768}\right)+e^4 \left(\frac{848429 \nu }{1344}+\frac{3557227}{5376}\right)+e^2 \left(\frac{24043 \nu }{21}+\frac{102121}{168}\right) 
+\frac{583 \nu }{3}+\frac{8191}{84}\bigg],\\
\mathcal{G}^{3\text{PN}}&= \frac{x^3 \epsilon ^6}{\left(1-e^2\right)^3} \Bigg\lbrace
-\frac{1550 \nu ^3}{81}-\frac{94403 \nu ^2}{378}+\left(\frac{41 \pi ^2}{6}-\frac{48907}{63}\right) \nu +\frac{4340155 \nu }{6804}+\frac{128 \pi ^2}{3}+\frac{5599693439}{8731800} \nonumber\\
&\quad
+ e^2 \left[-\frac{25378 \nu ^3}{81}-\frac{267515 \nu ^2}{108}+\left(\frac{1528855}{13608}+\frac{41 \pi ^2}{2}\right) \nu +\frac{916 \pi ^2}{3}+\frac{31462815089}{4365900}+\frac{78324 \ln 2}{35}-\frac{78003 \ln 3}{35}\right]  \nonumber\\
&\quad
+ e^4 \left[218 \pi ^2+\frac{73302046931}{11642400}+\frac{3042117 \ln 3}{140}-\frac{4040534 \ln 2}{105}-\frac{66433 \nu ^3}{108}-\frac{3524231 \nu ^2}{1008}-\left(\frac{36637039}{18144}+\frac{11521 \pi ^2}{256}\right) \nu \right] \nonumber\\
&\quad
+ e^6 \bigg[-\frac{64291 \nu ^3}{324}-\frac{18775 \nu ^2}{18}+\left(-\frac{28220923}{18144}-\frac{615 \pi ^2}{128}\right) \nu +\frac{23 \pi ^2}{2}+\frac{990174139}{970200}+\frac{659100847 \ln 2}{1890}\nonumber\\
&\qquad
-\frac{42667641 \ln 3}{448}-\frac{1044921875 \ln 5}{12096}\bigg] 
+ e^8 \left[-\frac{14839 \nu ^3}{2592}-\frac{27803 \nu ^2}{896}-\frac{614653 \nu }{12096}+\frac{29539919}{709632}\right] \nonumber\\
&\quad
+ \sqrt{1 - e^2} \bigg[
e^6 \left(-\frac{461 \nu ^2}{6}+\frac{2195 \nu }{168}+\frac{36405}{112}\right)+e^4 \left(-\frac{781 \nu ^2}{2}+\left(\frac{545719}{504}-\frac{287 \pi ^2}{48}\right) \nu +\frac{196073}{720}\right) \nonumber\\
&\qquad
+e^2 \left(274 \nu ^2+\left(-\frac{20131}{63}-\frac{41 \pi ^2}{48}\right) \nu -\frac{554692}{315}\right)+\frac{580 \nu ^2}{3}+\left(\frac{41 \pi ^2}{6}-\frac{48907}{63}\right) \nu -\frac{379223}{630}
\bigg] \nonumber\\
&\quad
+ \left(-\frac{2461 e^6}{70}-\frac{23326 e^4}{35}-\frac{98012 e^2}{105}-\frac{13696}{105}\right) \ln \left(\frac{2 \left(1-e^2\right) \gamma_ { \text{E}}' \sqrt{x}}{\sqrt{1-e^2}+1}\right)
\Bigg\rbrace ,\\
\mathcal{G}^{\text{SO}}&= \frac{x^{3/2} \epsilon ^3}{\left(1-e^2\right)^{3/2}} \left\{
\delta \chi_A  \left[\frac{13 e^4}{4}-\frac{140 e^2}{3}-22\right] +\chi_S \left[e^4 \left(\frac{10 \nu }{3}+\frac{13}{4}\right)+e^2 \left(\frac{188 \nu }{3}-\frac{140}{3}\right)+24 \nu -22\right]
\right\} \nonumber\\
&\quad
+ \frac{x^{5/2} \epsilon ^5}{\left(1-e^2\right)^{5/2}} \bigg\{
\delta  \chi_A \left[e^6 \left(-\frac{53 \nu }{4}-\frac{7127}{672}\right)+e^4 \left(\frac{943 \nu }{12}+\frac{87781}{336}\right)+e^2 \left(\frac{36763 \nu }{72}+\frac{5273}{28}\right)+\frac{1018 \nu }{9}+\frac{197}{2}\right]\nonumber\\
&\qquad
+\chi_S \bigg[e^6 \left(-\frac{97 \nu ^2}{36}-\frac{2655 \nu }{56}-\frac{7127}{672}\right)+e^4 \left(-\frac{567 \nu ^2}{2}-\frac{14263 \nu }{42}+\frac{87781}{336}\right)+e^2 \left(-\frac{7555 \nu ^2}{12}+\frac{131575 \nu }{504}+\frac{5273}{28}\right)\nonumber\\
&\qquad\quad
-\frac{1064 \nu ^2}{9}+\frac{280 \nu }{9}+\frac{197}{2}\bigg]
+\sqrt{1-e^2} \bigg[\delta  \chi _A \left[e^4 \left(112-\frac{112 \nu }{3}\right)+e^2 \left(16-\frac{16 \nu }{3}\right)+\frac{128 \nu }{3}-128\right]\nonumber\\
&\qquad\quad
+ \chi _S \left[e^4 \left(\frac{56 \nu ^2}{3}-\frac{448 \nu }{3}+112\right)+e^2 \left(\frac{8 \nu ^2}{3}-\frac{64 \nu }{3}+16\right)-\frac{64 \nu ^2}{3}+\frac{512 \nu }{3}-128\right]\bigg]
\bigg\}\nonumber\\
&\quad
+ \frac{\pi  x^3 \epsilon^6}{\left(1-e^2\right)^3} \bigg\{
\left[e^6 \left(\frac{8927}{216}-\frac{184 \nu }{27}\right)+e^4 \left(\frac{1345 \nu }{6}+\frac{543}{8}\right)+e^2 \left(\frac{1720 \nu }{3}-\frac{1342}{3}\right)+\frac{272 \nu }{3}-\frac{260}{3}\right] \chi _S \nonumber\\
&\qquad
+ \left[\frac{8927 e^6}{216}+\frac{543 e^4}{8}-\frac{1342 e^2}{3}-\frac{260}{3}\right] \delta \chi _A \bigg\} ,\\
\mathcal{G}^{\text{SS}}&= \frac{x^2\epsilon^4}{(1-e^2)^2} \bigg\{
\chi _A^2 \left[e^4 \left(21 \nu -\frac{81}{16}\right)+e^2 \left(\frac{39}{2}-72 \nu \right)-64 \nu +\frac{33}{2}\right]
+ \delta \chi _A \chi _S \left[33-\frac{81 e^4}{8}+39 e^2\right]
+ \delta \kappa _A \left[\frac{35 e^4}{4}+64 e^2+16\right] \nonumber\\
&\qquad
+ \chi _S^2 \left[e^4 \left(-\frac{3 \nu }{4}-\frac{81}{16}\right)+e^2 \left(\frac{39}{2}-6 \nu \right)-2 \nu +\frac{33}{2}\right]
+ \kappa _S\left[e^4 \left(\frac{35}{4}-\frac{35 \nu }{2}\right)+e^2 (64-128 \nu )-32 \nu +16\right]
\bigg\}\nonumber\\
&\quad
+ \frac{x^3\epsilon^6}{(1-e^2)^3} \Bigg\lbrace
\delta  \kappa _A \left[e^6 \left(-\frac{397 \nu }{16}-\frac{271}{6}\right)+e^4 \left(-\frac{3005 \nu }{6}-\frac{63599}{168}\right)+e^2 \left(-\frac{5417 \nu }{6}-\frac{5717}{42}\right)-\frac{301 \nu }{3}-\frac{1544}{21}\right] \nonumber\\
&\qquad
+\chi _A^2 \bigg[e^6 \left(52 \nu ^2-\frac{10023 \nu }{56}+\frac{9085}{224}\right)+e^4 \left(\frac{2579 \nu ^2}{6}+\frac{51983 \nu }{56}-\frac{44855}{168}\right)+e^2 \left(\frac{3968 \nu ^2}{3}-\frac{20954 \nu }{63}-\frac{3133}{252}\right)-\frac{6781}{63} \nonumber\\
&\qquad\quad 
+\frac{22567 \nu }{63}+280 \nu ^2\bigg]
+\delta  \chi _A \chi _S \bigg[e^6 \!\left(\frac{9085}{112}-\frac{641 \nu }{8}\right)-e^4 \!\left(\frac{4913 \nu }{9}+\frac{44855}{84}\right)-e^2 \!\left(\frac{17557 \nu }{9}+\frac{3133}{126}\right)-\frac{1508 \nu }{9}-\frac{13562}{63}\bigg]\nonumber\\
&\qquad
+ \kappa _S\bigg[e^6 \left(43 \nu ^2+\frac{3145 \nu }{48}-\frac{271}{6}\right)+e^4 \left(\frac{9227 \nu ^2}{12}+\frac{21529 \nu }{84}-\frac{63599}{168}\right)+e^2 \left(\frac{3796 \nu ^2}{3}-\frac{26485 \nu }{42}-\frac{5717}{42}\right)\nonumber\\
&\qquad\quad
+140 \nu ^2+\frac{327 \nu }{7}-\frac{1544}{21}\bigg]
+ \chi _S^2\bigg[e^6 \left(14 \nu ^2-\frac{507 \nu }{8}+\frac{9085}{224}\right)+e^4 \left(\frac{3083 \nu ^2}{18}-\frac{29245 \nu }{72}-\frac{44855}{168}\right) \nonumber\\
&\qquad\quad
+e^2 \left(\frac{1562 \nu ^2}{3}-\frac{14116 \nu }{9}-\frac{3133}{252}\right)+\frac{308 \nu ^2}{9}-\frac{857 \nu }{9}-\frac{6781}{63}\bigg] \nonumber\\
&\qquad
+ \sqrt{1-e^2} \bigg\{
\left[e^4 \left(-\frac{112 \nu ^2}{3}+\frac{448 \nu }{3}-\frac{308}{3}\right)+e^2 \left(-\frac{16 \nu ^2}{3}+\frac{64 \nu }{3}-\frac{44}{3}\right)+\frac{128 \nu ^2}{3}-\frac{512 \nu }{3}+\frac{352}{3}\right] \chi _S^2\nonumber\\
&\quad\qquad
+\chi _A^2 \left[e^4 \left(-56 \nu ^2+\frac{1288 \nu }{3}-\frac{308}{3}\right)+e^2 \left(-8 \nu ^2+\frac{184 \nu }{3}-\frac{44}{3}\right)+64 \nu ^2-\frac{1472 \nu }{3}+\frac{352}{3}\right]\nonumber\\
&\quad\qquad
+\delta  \chi _A \chi _S\left[e^4 \left(168 \nu -\frac{616}{3}\right)+e^2 \left(24 \nu -\frac{88}{3}\right)-192 \nu +\frac{704}{3}\right] \nonumber\\
&\quad\qquad
+ \kappa _S\left[e^4 \left(-28 \nu ^2+154 \nu -\frac{196}{3}\right)+e^2 \left(-4 \nu ^2+22 \nu -\frac{28}{3}\right)+32 \nu ^2-176 \nu +\frac{224}{3}\right]\nonumber\\
&\quad\qquad
+\delta  \kappa _A \left[e^4 \left(\frac{70 \nu }{3}-\frac{196}{3}\right)+e^2 \left(\frac{10 \nu }{3}-\frac{28}{3}\right)-\frac{80 \nu }{3}+\frac{224}{3}\right]
\bigg\}
\Bigg\rbrace.
\end{align}
\end{subequations}
\normalsize

\section{Equations of motion for the Keplerian parameters}
\label{app:evoEqs}

The 3PN evolution equation for the eccentricity is given by
%
\small
\begin{subequations}
\label{eq:edotFull}
\begin{align}
\frac{\di e}{\di t} &\equiv - \frac{\nu e x^4}{M\left(1-e^2\right)^{5/2}} \Big[\varepsilon^\text{N} + \varepsilon^{1\text{PN}} + \varepsilon^{1.5\text{PN}} + \varepsilon^{2\text{PN}} + \varepsilon^{2.5\text{PN}} + \varepsilon^{3\text{PN}} + \varepsilon^{\text{SO}} + \varepsilon^{\text{SS}}\Big], \\
\varepsilon^\text{N} &=\frac{121 e^2}{15}+\frac{304}{15} , \\
\varepsilon^{1\text{PN}} &= \frac{x \epsilon ^2}{1-e^2} \left[e^4 \left(-\frac{353 \nu }{45}-\frac{10543}{280}\right)+e^2 \left(-\frac{3061 \nu }{30}-\frac{19349}{105}\right)-\frac{3508 \nu }{45}-\frac{2283}{35}\right], \\
\varepsilon^{1.5\text{PN}} &= \frac{\pi  x^{3/2} \epsilon ^3}{\left(1-e^2\right)^{3/2}} \left[-\frac{49 e^6}{720}+\frac{24217 e^4}{1440}+\frac{5969 e^2}{30}+\frac{394}{3} + \Order(e^8)\right], \\
\varepsilon^{2\text{PN}} &= \frac{x^2 \epsilon ^4}{\left(1-e^2\right)^2} \bigg\lbrace
e^6 \left(\frac{269 \nu ^2}{45}+\frac{42451 \nu }{1260}+\frac{352147}{3360}\right)+e^4 \left(\frac{15101 \nu ^2}{90}+\frac{4599977 \nu }{5040}+\frac{11094859}{15120}\right)+e^2 \bigg(\frac{47027 \nu ^2}{120}+\frac{459707 \nu }{280} \nonumber\\
&\qquad
-\frac{189827}{504}\bigg) +\frac{1792 \nu ^2}{15}+\frac{2453 \nu }{6}-\frac{765197}{1890}
+ \sqrt{1-e^2} \left[e^4 \!\left(\frac{121 \nu }{3}-\frac{605}{6}\right)+e^2 \!\left(61 \nu -\frac{305}{2}\right)-\frac{304 \nu }{3}+\frac{760}{3}\right]\!\!
\bigg\rbrace, \\
\varepsilon^{2.5\text{PN}} &= \frac{\pi  x^{5/2} \epsilon ^5}{\left(1-e^2\right)^{5/2}} \bigg[
e^6 \left(-\frac{185051 \nu }{12096}-\frac{1725289}{24192}\right)+e^4 \left(-\frac{401755 \nu }{432}-\frac{16513345}{8064}\right)+e^2 \left(-\frac{6653761 \nu }{2520}-\frac{8393617}{3360}\right) \nonumber\\
&\qquad
-\frac{288944 \nu }{315}-\frac{87947}{210} + \Order(e^8)
\bigg], \\
\varepsilon^{3\text{PN}} &= \frac{x^3 \epsilon ^6}{\left(\sqrt{1-e^2}+1\right) \left(1-e^2\right)^3} \Bigg\{
e^8 \left[-\frac{908 \nu ^3}{243}+\frac{11591 \nu ^2}{189}+\frac{5329 \nu }{63}-\frac{1738243127}{1774080}\right] 
+e^6\bigg[\frac{191 \pi ^2}{10}-\frac{45530 \nu ^3}{243}-\frac{447409 \nu ^2}{630}\nonumber\\
&\qquad
+\left(-\frac{86352167}{15120}-\frac{9061 \pi ^2}{5760}\right) \nu
+\frac{263529979 \ln 2}{945}-\frac{20437 \gamma_E }{350}-\frac{42667641 \ln 3}{560}-\frac{208984375 \ln 5}{3024}-\frac{500266412287}{232848000}\bigg] \nonumber\\
&\quad
+ e^4 \bigg[-\frac{184621 \nu ^3}{216}-\frac{49874227 \nu ^2}{6720}+\left(-\frac{340057397}{36288}-\frac{790193 \pi ^2}{5760}\right) \nu +\frac{21442 \pi ^2}{45}+\frac{1520628574 \ln 2}{4725}-\frac{2294294 \gamma_E }{1575}\nonumber\\
&\qquad
-\frac{208984375 \ln 5}{2016}-\frac{340795107 \ln 3}{5600}+\frac{2864930559373}{116424000}\bigg]
+ e^2 \bigg[-\frac{2767045 \nu ^3}{3888}-\frac{90282343 \nu ^2}{15120}+\left(\frac{83899451}{11340}-\frac{8077 \pi ^2}{45}\right) \nu \nonumber\\
&\qquad
+\frac{44512 \pi ^2}{45}+\frac{546021 \ln 3}{50}-\frac{47727136 \ln 2}{1575}-\frac{4762784 \gamma_E }{1575}+\frac{1011350391713}{43659000}\bigg]
+\frac{12304 \pi ^2}{45} +\frac{213942960841}{43659000} \nonumber\\
&\quad
+\left(-\frac{65951621}{22680}+\frac{17917 \pi ^2}{180}\right) \nu -\frac{9463 \nu ^2}{40}-\frac{156006 \ln 3}{175}-\frac{176336 \ln 2}{225}-\frac{46313 \nu ^3}{486}-\frac{1316528 \gamma_E }{1575}\nonumber\\
&\quad
+\sqrt{1-e^2} \bigg\{ e^8\left[-\frac{908 \nu ^3}{243}-\frac{51397 \nu ^2}{1890}-\frac{57623 \nu }{840}-\frac{390348887}{1774080}\right]
+ e^6 \bigg[-\frac{45530 \nu ^3}{243}-\frac{937157 \nu ^2}{630}-\frac{20970012893}{77616000}\nonumber\\
&\qquad\quad
-\left(\frac{13208801}{3024}+\frac{6519 \pi ^2}{640}\right) \nu +\frac{191 \pi ^2}{10}+\frac{263529979 \ln 2}{945}-\frac{20437 \gamma_E }{350}-\frac{42667641 \ln 3}{560}-\frac{208984375 \ln 5}{3024}\bigg]\nonumber\\
&\qquad
+ e^4 \bigg[-\frac{184621 \nu ^3}{216}-\frac{29314405 \nu ^2}{4032}-\left(\frac{280844453}{36288}+\frac{865223 \pi ^2}{5760}\right) \nu +\frac{21442 \pi ^2}{45}+\frac{1520628574 \ln 2}{4725}-\frac{2294294 \gamma_E }{1575}\nonumber\\
&\qquad\quad
-\frac{208984375 \ln 5}{2016}-\frac{340795107 \ln 3}{5600}+\frac{2199381667981}{116424000}\bigg]
+ e^2 \bigg[\left(\frac{10380283}{2268}-\frac{3157 \pi ^2}{20}\right) \nu-\frac{2767045 \nu ^3}{3888}-\frac{79477927 \nu ^2}{15120}\nonumber\\
&\qquad\quad
+\frac{44512 \pi ^2}{45}+\frac{546021 \ln 3}{50}-\frac{47727136 \ln 2}{1575}-\frac{4762784 \gamma_E }{1575}+\frac{182487950377}{8731800}\bigg]
+\left(-\frac{65951621}{22680}+\frac{17917 \pi ^2}{180}\right) \nu\nonumber\\
&\qquad
+\frac{12304 \pi ^2}{45} -\frac{9463 \nu ^2}{40}-\frac{156006 \ln 3}{175}-\frac{176336 \ln 2}{225}-\frac{46313 \nu ^3}{486}-\frac{1316528 \gamma_E }{1575}+\frac{212593610377}{43659000}\bigg\}  \nonumber\\
&\quad
+\left(\sqrt{1-e^2}+1\right) \left(-\frac{20437 e^6}{350}-\frac{2294294 e^4}{1575}-\frac{4762784 e^2}{1575}-\frac{1316528}{1575}\right) \ln \left[\frac{2 \left(1-e^2\right) \sqrt{x}}{\sqrt{1-e^2}+1}\right]
\Bigg\}, \\
\varepsilon^{\text{SO}} &= \frac{x^{3/2} \epsilon ^3}{\left(1-e^2\right)^{3/2}} \left\{ \delta  \chi _A \left[\frac{623 e^4}{30}-\frac{124 e^2}{45}-\frac{8116}{45}\right]
+ \chi _S\left[e^4 \left(\frac{623}{30}-\frac{46 \nu }{15}\right)+e^2 \left(\frac{3986 \nu }{45}-\frac{124}{45}\right)+\frac{6656 \nu }{45}-\frac{8116}{45}\right] \right\} \nonumber\\
&\quad
+ \frac{x^{5/2} \epsilon ^5}{\left(1-e^2\right)^{5/2}} \Bigg\{
\chi _S \bigg[e^6 \left(\frac{4867 \nu ^2}{360}-\frac{133459 \nu }{2520}-\frac{382371}{2240}\right)+e^4 \left(-\frac{173947 \nu ^2}{1080}-\frac{16825619 \nu }{15120}-\frac{410449}{840}\right) \nonumber\\
&\qquad\quad
+e^2 \left(-\frac{215363 \nu ^2}{180}-\frac{1266857 \nu }{840}+\frac{2238367}{1260}\right)-\frac{95788 \nu ^2}{135}+\frac{197122 \nu }{189}+\frac{6409}{105}\bigg]  
+\delta  \chi _A \bigg[e^6 \left(-\frac{27473 \nu }{720}-\frac{382371}{2240}\right)\nonumber\\
&\qquad\quad
+e^4 \left(-\frac{136879 \nu }{432}-\frac{410449}{840}\right)+e^2 \left(\frac{249289 \nu }{360}+\frac{2238367}{1260}\right)+\frac{136124 \nu }{135}+\frac{6409}{105}\bigg]
\nonumber\\
&\qquad
+ \sqrt{1-e^2} \bigg\{
\chi _S\left[e^4 \left(\frac{242 \nu ^2}{9}-\frac{1936 \nu }{9}+\frac{484}{3}\right)+e^2 \left(\frac{122 \nu ^2}{3}-\frac{976 \nu }{3}+244\right)-\frac{608 \nu ^2}{9}+\frac{4864 \nu }{9}-\frac{1216}{3}\right] \nonumber\\
&\qquad\quad
+\delta  \chi _A \left[e^4 \left(\frac{484}{3}-\frac{484 \nu }{9}\right)+e^2 \left(244-\frac{244 \nu }{3}\right)+\frac{1216 \nu }{9}-\frac{1216}{3}\right]
\bigg\}
\Bigg\} \nonumber\\
&\quad
+ \frac{\pi  x^3 \epsilon ^6}{\left(1-e^2\right)^3} \bigg\{
\chi _S \left[e^6 \left(\frac{22073}{360}-\frac{42287 \nu }{2160}\right)+e^4 \left(\frac{95561 \nu }{1080}+\frac{1125617}{1080}\right)+e^2 \left(\frac{74954 \nu }{45}-\frac{39223}{45}\right)+\frac{39496 \nu }{45}-\frac{49184}{45}\right] \nonumber\\
&\qquad
+\delta \chi _A \left[\frac{22073 e^6}{360}+\frac{1125617 e^4}{1080}-\frac{39223 e^2}{45}-\frac{49184}{45}\right]
\bigg\} , \\
\varepsilon^{\text{SS}} &= \frac{x^2 \epsilon ^4}{\left(1-e^2\right)^2} \bigg\{
 \chi _S^2\left[e^4 \left(-\frac{3 \nu }{2}-\frac{321}{40}\right)+e^2 \left(-18 \nu -\frac{881}{30}\right)-12 \nu +\frac{269}{3}\right]
+\delta  \chi _A \chi _S \left[-\frac{321 e^4}{20}-\frac{881 e^2}{15}+\frac{538}{3}\right] \nonumber\\
&\qquad
+\chi _A^2 \left[e^4 \left(\frac{168 \nu }{5}-\frac{321}{40}\right)+e^2 \left(\frac{2032 \nu }{15}-\frac{881}{30}\right)-\frac{1040 \nu }{3}+\frac{269}{3}\right] +\delta \kappa _A \left[\frac{143 e^4}{15}+\frac{1814 e^2}{15}+\frac{1684}{15}\right]  \nonumber\\
&\qquad
+ \kappa _S\left[e^4 \left(\frac{143}{15}-\frac{286 \nu }{15}\right)+e^2 \left(\frac{1814}{15}-\frac{3628 \nu }{15}\right)-\frac{3368 \nu }{15}+\frac{1684}{15}\right]
\bigg\} \nonumber\\
&\quad
+ \frac{x^3 \epsilon ^6}{\left(\sqrt{1-e^2}+1\right) \left(1-e^2\right)^3} \Bigg\{
\chi _A^2 \bigg[e^6 \left(\frac{2452 \nu ^2}{15}-\frac{334597 \nu }{252}+\frac{6397597}{20160}\right)+e^4 \left(\frac{15121 \nu ^2}{45}-\frac{13837483 \nu }{5040}+\frac{3131923}{5040}\right) \nonumber\\
&\qquad\quad
+e^2 \left(\frac{7622 \nu ^2}{5}+\frac{34227913 \nu }{3780}-\frac{4553009}{1890}\right)+\frac{104648 \nu ^2}{45}-\frac{9635809 \nu }{1890}+\frac{2069461}{1890}\bigg]\nonumber\\
&\qquad
+\delta  \chi _A \chi _S \left[e^6 \left(\frac{6397597}{10080}-\frac{95329 \nu }{240}\right)+e^4 \left(\frac{3131923}{2520}-\frac{311833 \nu }{360}\right)-e^2 \left(\frac{469207 \nu }{270}+\frac{4553009}{945}\right)-\frac{569339 \nu }{135}+\frac{2069461}{945}\right]\nonumber\\
&\qquad
+ \chi _S^2\bigg[e^6 \!\left(\frac{3557 \nu ^2}{45}-\frac{121969 \nu }{360}+\frac{6397597}{20160}\right)
+e^4 \!\left(\frac{52361 \nu ^2}{180}-\frac{436553 \nu }{720}+\frac{3131923}{5040}\right)+\frac{143894 \nu ^2}{135} -\frac{944683 \nu }{270}+\frac{2069461}{1890}\nonumber\\
&\qquad\quad
+e^2 \left(\frac{111989 \nu ^2}{135}-\frac{624677 \nu }{540}-\frac{4553009}{1890}\right) \bigg]
+ \kappa _S\bigg[e^6 \left(\frac{260 \nu ^2}{3}-\frac{127601 \nu }{1120}+\frac{67987}{2520}\right) \nonumber\\
&\qquad\quad
+e^4 \left(\frac{99643 \nu ^2}{90}+\frac{1167671 \nu }{1260}-\frac{431441}{504}\right)+e^2 \left(\frac{45107 \nu ^2}{15}+\frac{25645 \nu }{21}-\frac{104320}{63}\right)+\frac{60964 \nu ^2}{45}-\frac{510122 \nu }{315}+\frac{88517}{315}\bigg]\nonumber\\
&\qquad
+\delta  \kappa _A \left[e^6 \left(\frac{67987}{2520}-\frac{86359 \nu }{1440}\right)+e^4 \left(-\frac{70681 \nu }{90}-\frac{431441}{504}\right)+e^2 \left(-\frac{18815 \nu }{9}-\frac{104320}{63}\right)-\frac{47584 \nu }{45}+\frac{88517}{315}\right] \nonumber\\
&\qquad
+ \sqrt{1-e^2} \bigg\{
\chi _A^2 \bigg[e^6 \left(\frac{414 \nu ^2}{5}-\frac{19861 \nu }{28}+\frac{379573}{2240}\right)+e^4 \left(\frac{9631 \nu ^2}{45}-\frac{9123403 \nu }{5040}+\frac{2004643}{5040}\right)+\frac{104648 \nu ^2}{45}-\frac{9635809 \nu }{1890} \nonumber\\
&\qquad\qquad
+e^2 \left(\frac{25906 \nu ^2}{15}+\frac{28354633 \nu }{3780}-\frac{3850769}{1890}\right)+\frac{2069461}{1890}\bigg]
+ \chi _S^2 \bigg[e^6 \left(\frac{379 \nu ^2}{15}-\frac{14843 \nu }{120}+\frac{379573}{2240}\right)+\frac{2069461}{1890}\nonumber\\
&\qquad\qquad
+e^4 \left(\frac{37721 \nu ^2}{180}-\frac{202313 \nu }{720}+\frac{2004643}{5040}\right)+e^2 \left(\frac{130229 \nu ^2}{135}-\frac{916517 \nu }{540}-\frac{3850769}{1890}\right)+\frac{143894 \nu ^2}{135}-\frac{944683 \nu }{270}\bigg]
\nonumber\\
&\qquad\quad
+\delta  \chi _A \chi _S \bigg[e^6 \left(\frac{379573}{1120}-\frac{37249 \nu }{240}\right)+e^4 \left(\frac{2004643}{2520}-\frac{180073 \nu }{360}\right)-e^2 \left(\frac{633367 \nu }{270}+\frac{3850769}{945}\right)-\frac{569339 \nu }{135}+\frac{2069461}{945}\bigg]\nonumber\\
&\qquad\quad
+\delta  \kappa _A \left[e^6 \left(-\frac{12653 \nu }{480}-\frac{18797}{280}\right)+e^4 \left(-\frac{33053 \nu }{45}-\frac{503177}{504}\right)+e^2 \left(-2175 \nu -\frac{9936}{7}\right)-\frac{47584 \nu }{45}+\frac{88517}{315}\right]\nonumber\\
&\qquad\quad
+ \kappa _S\bigg[e^6 \left(\frac{139 \nu ^2}{3}+\frac{362557 \nu }{3360}-\frac{18797}{280}\right)+e^4 \left(\frac{94153 \nu ^2}{90}+\frac{1590401 \nu }{1260}-\frac{503177}{504}\right)+e^2 \left(\frac{46627 \nu ^2}{15}+\frac{4647 \nu }{7}-\frac{9936}{7}\right)\nonumber\\
&\qquad\qquad
+\frac{60964 \nu ^2}{45}-\frac{510122 \nu }{315}+\frac{88517}{315}\bigg]\bigg\}
\Bigg\}.
\end{align}
\end{subequations}
\normalsize

The 3PN evolution equation for the orbit-averaged orbital frequency $ x = \langle M \Omega \rangle^{2/3} $ is given by 
%
\small
\begin{subequations}
\label{eq:xdotFull}
\begin{align}
\frac{\di x}{\di t} &\equiv  \frac{2 \nu x^5}{3 M\left(1-e^2\right)^{7/2}} \Big[\mathcal{X}^\text{N} + \mathcal{X}^{1\text{PN}} + \mathcal{X}^{1.5\text{PN}} + \mathcal{X}^{2\text{PN}} + \mathcal{X}^{2.5\text{PN}} + \mathcal{X}^{3\text{PN}} + \mathcal{X}^{\text{SO}} + \mathcal{X}^{\text{SS}}\Big], \\
\mathcal{X}^\text{N} &=\frac{37 e^4}{5} + \frac{292 e^2}{5}+\frac{96}{5} , \\
\mathcal{X}^{1\text{PN}} &= \frac{x \epsilon ^2}{1-e^2} \left[
e^6 \left(\frac{37 \nu }{5}+\frac{6931}{280}\right)
+ e^4 \left(\frac{369 \nu }{2}+\frac{7079}{20}\right)
+e^2 \left(\frac{1594 \nu }{5}+\frac{15411}{35}\right)
+ \frac{264 \nu }{5}+\frac{1486}{35}
\right], \\
\mathcal{X}^{1.5\text{PN}} &= \frac{\pi  x^{3/2} \epsilon ^3}{\left(1-e^2\right)^{3/2}} \left[\frac{10007 e^6}{480}+\frac{787 e^4}{2}+550 e^2+\frac{384}{5} + \Order(e^8)\right], \\
\mathcal{X}^{2\text{PN}} &= \frac{x^2 \epsilon ^4}{\left(1-e^2\right)^2} \bigg\lbrace
e^8 \left(\frac{259 \nu ^2}{45}+\frac{4037 \nu }{140}+\frac{244901}{3360}\right)
+e^6 \left(\frac{23077 \nu ^2}{90}+\frac{2068219 \nu }{1680}+\frac{393677}{336}\right)
\nonumber\\
&\qquad
+e^4 \left(\frac{369319 \nu ^2}{360}+\frac{646521 \nu }{140}+\frac{448457}{1260}\right)
+e^2 \bigg(\frac{12679 \nu ^2}{18}+\frac{20747 \nu }{7}-\frac{1339343}{945}\bigg)
\nonumber\\
&\qquad
 +\frac{944 \nu ^2}{15}+\frac{15677 \nu }{105}-\frac{11257}{945}
+ \sqrt{1-e^2} \left[e^6 \left(\frac{158 \nu }{5}-79\right)+e^4 \left(\frac{438 \nu }{5}-219\right)+e^2 (250-100 \nu )-\frac{96 \nu }{5}+48\right]\!\!
\bigg\rbrace, \\
\mathcal{X}^{2.5\text{PN}} &= \frac{\pi  x^{5/2} \epsilon ^5}{\left(1-e^2\right)^{5/2}} \bigg[
e^8 \left(\frac{16809}{320}-\frac{10007 \nu }{2880}\right)+e^6 \left(-\frac{2490491 \nu }{1680}-\frac{17257369}{5040}\right)+e^4 \left(-\frac{3741187 \nu }{560}-\frac{18599341}{2240}\right)
\nonumber\\
&\qquad
+e^2 \left(-\frac{171104 \nu }{35}-\frac{115991}{35}\right)-\frac{2268 \nu }{5}-\frac{4159}{35} + \Order(e^8)
\bigg], \\
\mathcal{X}^{3\text{PN}} &= \frac{x^3 \epsilon ^6}{ \left(1-e^2\right)^3} \Bigg\{
e^{10} \left(-\frac{296 \nu ^3}{81}-\frac{2666 \nu ^2}{105}-\frac{19828 \nu }{315}-\frac{38469511}{197120}\right)
+e^8 \bigg[-\frac{20672 \nu ^3}{81}-\frac{185084 \nu ^2}{105}-\frac{31793 \nu }{7}
\nonumber\\
&\qquad \quad
-\frac{297}{640} \pi ^2 (41 \nu -64)-\frac{31779 \gamma_E }{350}-\frac{4419999333}{2464000}-\frac{31779 \ln 2}{175}\bigg]
+e^6 \bigg[-\frac{384497 \nu ^3}{216}-\frac{29069589 \nu ^2}{2240}-\frac{376329641 \nu }{20160}
\nonumber\\
&\qquad \quad
+\pi ^2 \left(\frac{14008}{15}-\frac{261211 \nu }{640}\right)-\frac{1498856 \gamma_E }{525}+\frac{1123773865243}{38808000}+\frac{199270808 \ln 2}{105}-\frac{1044921875 \ln 5}{2016}-\frac{2594457783 \ln 3}{5600}\bigg]
\nonumber\\
&\qquad
+e^4 \bigg[-\frac{3391213 \nu ^3}{1296}-\frac{69436847 \nu ^2}{3360}-\frac{60059983 \nu }{18144}+\pi ^2 \left(\frac{41368}{15}-\frac{197087 \nu }{320}\right)-\frac{4426376 \gamma_E }{525}+\frac{4756012160407}{58212000}
\nonumber\\
&\qquad \quad
+\frac{6318243 \ln 3}{70}-\frac{98360392 \ln 2}{525}\bigg]
+e^2 \bigg[-\frac{279061 \nu ^3}{324}-\frac{1224481 \nu ^2}{168}+\frac{503543 \nu }{4536}+\frac{1}{48} \pi ^2 (9717 \nu +69632)-\frac{465664 \gamma_E }{105}
\nonumber\\
&\qquad \quad
+\frac{176753408437}{4851000}-\frac{1404054 \ln 3}{175}-\frac{438272 \ln 2}{525}\bigg]
-\frac{1121 \nu ^3}{27}-\frac{16073 \nu ^2}{140}+\pi ^2 \left(\frac{369 \nu }{2}+\frac{512}{5}\right)-\frac{57265081 \nu }{11340}
\nonumber\\
&\qquad
-\frac{54784 \gamma_E }{175}+\frac{5133257111}{1455300}-\frac{109568 \ln 2}{175}
+\sqrt{1-e^2} \bigg[ 
e^8 \left(-\frac{896 \nu ^2}{15}-\frac{35317 \nu }{420}+\frac{28563}{56}\right)
\nonumber\\
&\qquad \quad
+e^6 \left(-\frac{14263 \nu ^2}{15}+\frac{572071 \nu }{315}-\frac{6683 \pi ^2 \nu }{480}+\frac{17442583}{7875}\right)
+e^4 \left(-\frac{1415 \nu ^2}{3}+\frac{120059 \nu }{60}-\frac{2009 \pi ^2 \nu }{160}-\frac{163017049}{21000}\right)
\nonumber\\
&\qquad \quad
+e^2 \left(\frac{20338 \nu ^2}{15}+\frac{7339 \pi ^2 \nu }{240}-\frac{241103 \nu }{63}-\frac{25778072}{2625}\right)+\frac{632 \nu ^2}{5}+\frac{9874 \nu }{105}-\frac{41 \pi ^2 \nu }{10}-\frac{1425319}{1125}
\bigg]
\nonumber\\
&\qquad
+\left(-\frac{31779 e^8}{350}-\frac{1498856 e^6}{525}-\frac{4426376 e^4}{525}-\frac{465664 e^2}{105}-\frac{54784}{175}\right) \ln \left[\frac{2 \left(1-e^2\right) \sqrt{x}}{\sqrt{1-e^2}+1}\right]
\Bigg\}, \\
\mathcal{X}^{\text{SO}} &= \frac{x^{3/2} \epsilon ^3}{\left(1-e^2\right)^{3/2}} \Bigg\{ 
\chi_S \left[e^6 \left(\frac{249}{10}-\frac{12 \nu }{5}\right)+e^4 \left(\frac{2036 \nu }{15}+\frac{1301}{15}\right)+e^2 \left(\frac{1336 \nu }{3}-\frac{6268}{15}\right)+\frac{608 \nu }{5}-\frac{904}{5}\right]
\nonumber\\
&\qquad
+  \left(\frac{249 e^6}{10}+\frac{1301 e^4}{15}-\frac{6268 e^2}{15}-\frac{904}{5}\right) \chi_A \delta
\Bigg\} 
\nonumber\\
&\quad
+ \frac{x^{5/2} \epsilon ^5}{\left(1-e^2\right)^{5/2}} \Bigg\{
\delta  \chi_A \bigg[e^8 \left(-\frac{2991 \nu }{80}-\frac{376401}{2240}\right)+e^6 \left(-\frac{10117 \nu }{16}-\frac{612083}{420}\right)+e^4 \left(\frac{4611 \nu }{10}+\frac{40462}{15}\right)
\nonumber\\
&\qquad \quad
+e^2 \left(\frac{99563 \nu }{30}+\frac{107066}{35}\right)+\frac{4636 \nu }{5}-\frac{62638}{105}\bigg]
+\chi_S \bigg[e^8 \left(\frac{429 \nu ^2}{40}-\frac{15573 \nu }{280}-\frac{376401}{2240}\right)
\nonumber\\
&\qquad \quad
+e^6 \left(-\frac{8041 \nu ^2}{40}-\frac{3133943 \nu }{1680}-\frac{612083}{420}\right)+e^4 \left(-\frac{12258 \nu ^2}{5}-\frac{79817 \nu }{15}+\frac{40462}{15}\right)
\nonumber\\
&\qquad \quad
+e^2 \left(-\frac{45197 \nu ^2}{15}+\frac{125759 \nu }{210}+\frac{107066}{35}\right)-\frac{2528 \nu ^2}{5}+\frac{36760 \nu }{21}-\frac{62638}{105}\bigg]
+ \sqrt{1 - e^2} \bigg\{ \delta  \chi_A \bigg[e^6 \left(\frac{968}{5}-\frac{968 \nu }{15}\right)
\nonumber\\
&\qquad \qquad
+e^4 \left(\frac{1464}{5}-\frac{488 \nu }{5}\right)+e^2 \left(\frac{2432 \nu }{15}-\frac{2432}{5}\right)\bigg]
+\chi_S \bigg[e^6 \left(\frac{484 \nu ^2}{15}-\frac{3872 \nu }{15}+\frac{968}{5}\right)
\nonumber\\
&\qquad \qquad
+e^4 \left(\frac{244 \nu ^2}{5}-\frac{1952 \nu }{5}+\frac{1464}{5}\right)+e^2 \left(-\frac{1216 \nu ^2}{15}+\frac{9728 \nu }{15}-\frac{2432}{5}\right)\bigg] \bigg\}
\Bigg\}
\nonumber\\
&\quad
+ \frac{\pi  x^3 \epsilon ^6}{\left(1-e^2\right)^3} \Bigg\{
\chi_S \bigg[e^8 \left(\frac{49 \nu }{120}-\frac{49}{60}\right)+e^6 \left(\frac{42985 \nu }{144}+\frac{339871}{180}\right)+e^4 \left(\frac{50807 \nu }{15}+\frac{4957}{30}\right)+e^2 \left(\frac{55156 \nu }{15}-\frac{58736}{15}\right)
\nonumber\\
&\qquad
+\frac{2368 \nu }{5}-720\bigg] 
+ \chi_A \delta  \left(-\frac{49 e^8}{60}+\frac{339871 e^6}{180}+\frac{4957 e^4}{30}-\frac{58736 e^2}{15}-720\right) 
\Bigg\} , \\
\mathcal{X}^{\text{SS}} &= \frac{x^2 \epsilon ^4}{\left(1-e^2\right)^2} \Bigg\{
\chi_A^2 \left[e^6 \left(\frac{312 \nu }{5}-\frac{597}{40}\right)+e^4 \left(\frac{2176 \nu }{5}-\frac{1969}{20}\right)+e^2 \left(173-\frac{3232 \nu }{5}\right)-384 \nu +\frac{486}{5}\right]
\nonumber\\
&\qquad
+\chi_S^2 \left[e^6 \left(-\frac{27 \nu }{10}-\frac{597}{40}\right)+e^4 \left(-\frac{207 \nu }{5}-\frac{1969}{20}\right)+e^2 \left(173-\frac{228 \nu }{5}\right)-\frac{24 \nu }{5}+\frac{486}{5}\right]
\nonumber\\
&\qquad
+ \chi_A \chi_S \delta  \left(-\frac{597 e^6}{20}-\frac{1969 e^4}{10}+346 e^2+\frac{972}{5}\right) 
+ \kappa_S \bigg[e^6 \left(\frac{33}{5}-\frac{66 \nu }{5}\right)+e^4 \left(\frac{1064}{5}-\frac{2128 \nu }{5}\right)
\nonumber\\
&\qquad \quad
+e^2 \left(\frac{2064}{5}-\frac{4128 \nu }{5}\right)-192 \nu +96\bigg]
+ \kappa_A \delta \left(\frac{33 e^6}{5}+\frac{1064 e^4}{5}+\frac{2064 e^2}{5}+96\right)  
\Bigg\} 
\nonumber\\
&\quad
+ \frac{x^3 \epsilon ^6}{\left(1-e^2\right)^3} \Bigg\{
\chi_S^2 \bigg[ e^8 \left(\frac{129 \nu ^2}{5}-\frac{4043 \nu }{40}+\frac{503031}{2240}\right)+e^6 \left(\frac{9497 \nu ^2}{20}-\frac{172211 \nu }{240}+\frac{164001}{112}\right)
\nonumber\\
&\qquad \quad
+e^4 \left(\frac{39311 \nu ^2}{18}-\frac{1239619 \nu }{360}-\frac{8736761}{2520}\right)+e^2 \left(\frac{141998 \nu ^2}{45}-\frac{827251 \nu }{90}-\frac{593801}{630}\right)+956 \nu ^2-\frac{16743 \nu }{5}+\frac{55817}{35}\bigg]
\nonumber\\
&\qquad
+\chi_A^2 \bigg[e^8 \left(50 \nu ^2-\frac{129319 \nu }{140}+\frac{503031}{2240}\right)+e^6 \left(\frac{5261 \nu ^2}{15}-\frac{10429817 \nu }{1680}+\frac{164001}{112}\right)
\nonumber\\
&\qquad \quad
+e^4 \left(\frac{49232 \nu ^2}{15}+\frac{31182647 \nu }{2520}-\frac{8736761}{2520}\right)+e^2 \left(\frac{115136 \nu ^2}{15}+\frac{851591 \nu }{630}-\frac{593801}{630}\right)+\frac{8448 \nu ^2}{5}-\frac{243029 \nu }{35}+\frac{55817}{35}\bigg]
\nonumber\\
&\qquad
+\delta  \chi_A \chi_S \bigg[e^8 \left(\frac{503031}{1120}-\frac{10121 \nu }{80}\right)+e^6 \left(\frac{164001}{56}-\frac{128231 \nu }{120}\right)+e^4 \left(-\frac{177739 \nu }{36}-\frac{8736761}{1260}\right)
\nonumber\\
&\qquad \quad
+e^2 \left(-\frac{104491 \nu }{9}-\frac{593801}{315}\right)-\frac{19566 \nu }{5}+\frac{111634}{35}\bigg]
+\delta  \kappa_A \bigg[e^8 \left(-\frac{689 \nu }{32}-\frac{10317}{280}\right)+e^6 \left(-\frac{68021 \nu }{60}-\frac{78391}{56}\right)
\nonumber\\
&\qquad \quad
+e^4 \left(-\frac{80302 \nu }{15}-\frac{573723}{140}\right)+e^2 \left(-4688 \nu -\frac{55623}{35}\right)-\frac{3916 \nu }{5}+\frac{17926}{35}\bigg]
+\kappa_S \bigg[e^8 \left(\frac{199 \nu ^2}{5}+\frac{58421 \nu }{1120}-\frac{10317}{280}\right)
\nonumber\\
&\qquad \quad
+e^6 \left(\frac{51299 \nu ^2}{30}+\frac{349859 \nu }{210}-\frac{78391}{56}\right)+e^4 \left(\frac{115112 \nu ^2}{15}+\frac{596941 \nu }{210}-\frac{573723}{140}\right)
\nonumber\\
&\qquad \quad
+e^2 \left(\frac{31992 \nu ^2}{5}-\frac{52834 \nu }{35}-\frac{55623}{35}\right)+\frac{4224 \nu ^2}{5}-\frac{63264 \nu }{35}+\frac{17926}{35}\bigg]
\nonumber\\
&\qquad
+ \sqrt{1-e^2} \bigg\{
\chi_S^2 \bigg[e^6 \left(-\frac{1304 \nu ^2}{15}+\frac{5216 \nu }{15}-\frac{3586}{15}\right)+e^4 \left(-\frac{392 \nu ^2}{5}+\frac{1568 \nu }{5}-\frac{1078}{5}\right)
\nonumber\\
&\qquad \quad
+e^2 \left(\frac{2864 \nu ^2}{15}-\frac{11456 \nu }{15}+\frac{7876}{15}\right)-\frac{128 \nu ^2}{5}+\frac{512 \nu }{5}-\frac{352}{5}\bigg]
+\delta \chi_A \chi_S \bigg[e^6 \left(\frac{1956 \nu }{5}-\frac{7172}{15}\right)+e^4 \left(\frac{1764 \nu }{5}-\frac{2156}{5}\right)
\nonumber\\
&\qquad \qquad
+e^2 \left(\frac{15752}{15}-\frac{4296 \nu }{5}\right)+\frac{576 \nu }{5}-\frac{704}{5}\bigg]
+\chi_A^2 \bigg[e^6 \left(-\frac{652 \nu ^2}{5}+\frac{14996 \nu }{15}-\frac{3586}{15}\right)
\nonumber\\
&\qquad \qquad
+e^4 \left(-\frac{588 \nu ^2}{5}+\frac{4508 \nu }{5}-\frac{1078}{5}\right)+e^2 \left(\frac{1432 \nu ^2}{5}-\frac{32936 \nu }{15}+\frac{7876}{15}\right)-\frac{192 \nu ^2}{5}+\frac{1472 \nu }{5}-\frac{352}{5}\bigg]
\nonumber\\
&\qquad \quad
+ \delta  \kappa_A \bigg[e^6 \left(\frac{163 \nu }{3}-\frac{2282}{15}\right)+e^4 \left(49 \nu -\frac{686}{5}\right)+e^2 \left(\frac{5012}{15}-\frac{358 \nu }{3}\right)+16 \nu -\frac{224}{5}\bigg]
\nonumber\\
&\qquad \quad
+\kappa_S \bigg[e^6 \left(-\frac{326 \nu ^2}{5}+\frac{1793 \nu }{5}-\frac{2282}{15}\right)+e^4 \left(-\frac{294 \nu ^2}{5}+\frac{1617 \nu }{5}-\frac{686}{5}\right)+e^2 \left(\frac{716 \nu ^2}{5}-\frac{3938 \nu }{5}+\frac{5012}{15}\right)
\nonumber\\
&\qquad \qquad
-\frac{96 \nu ^2}{5}+\frac{528 \nu }{5}-\frac{224}{5}\bigg]
\bigg\}
\Bigg\}.
\end{align}
\end{subequations}
\normalsize

The 3PN evolution equation for the relativistic anomaly is given by
%
\small
\begin{subequations}
 \label{eq:chidotFull}
\begin{align}
\dot \zeta &= \frac{x^{3/2} (1 + e \cos\zeta)^2}{M\left(1-e^2\right)^{3/2}} \Big[1 + Z^{1\text{PN}} + Z^{2\text{PN}} + Z^{2.5\text{PN}} + Z^{3\text{PN}} + Z^{\text{SO}} + Z^{\text{SS}}\Big], \\
Z^{1\text{PN}} &= \frac{x \epsilon ^2}{1-e^2} \left[-3 e^2-3 e \cos\zeta-3\right], \\
Z^{2\text{PN}} &= \frac{x^2 \epsilon ^4}{\left(1-e^2\right)^2} \bigg\{
e^4 (6-\nu )+e^3 (15-\nu ) \cos\zeta+e^2 \left(\frac{3 \nu }{2}+\frac{3}{4}\right) \cos (2 \zeta )+e^2 (4 \nu +20)+e (8 \nu -1) \cos\zeta+10 \nu -12 \nonumber\\
&\qquad
+\sqrt{1-e^2} \left[e^2 \left(3 \nu -\frac{15}{2}\right)-3 \nu +\frac{15}{2}\right]
\bigg\}, \\
Z^{2.5\text{PN}} &= \frac{\nu e x^{5/2} \epsilon ^5}{30 (1+e \cos\zeta)^2}  \sqrt{1-e^2} \left(5635 e^6-3265 e^4-1762 e^2-608\right) (2+e \cos\zeta) \sin\zeta, \\
Z^{3\text{PN}} &= \frac{x^3 \epsilon ^6}{\left(1-e^2\right)^3} \bigg\{
e^6 \left(\frac{21 \nu }{4}-9\right)+e^5 \left(\frac{45 \nu }{4}-45\right) \cos\zeta+e^4 \left(\nu ^2-10 \nu -\frac{21}{4}\right) \cos (2 \zeta )+e^4 \left(\frac{9 \nu ^2}{8}-\frac{3 \nu }{2}\right) \cos (4 \zeta ) \nonumber\\
&\quad
+e^4 \left(\frac{5 \nu ^2}{12}-\frac{215 \nu }{12}-\frac{139}{2}\right)+e^3 \left(-\frac{25 \nu ^2}{12}-\frac{158 \nu }{3}-\frac{79}{8}\right) \cos\zeta+e^3 \left(\frac{15 \nu ^2}{4}-\nu +\frac{1}{8}\right) \cos (3 \zeta ) \nonumber\\
&\quad
+e^2 \left[-\nu ^2+\left(\frac{191}{6}-\frac{41 \pi ^2}{128}\right) \nu +\frac{11}{4}\right] \cos (2 \zeta )+e^2 \left[-\frac{44 \nu ^2}{3}+\left(\frac{565}{12}-\frac{41 \pi ^2}{16}\right) \nu +\frac{113}{2}\right]\nonumber\\
&\quad
+\sqrt{1-e^2} \bigg[e^4 \left(-\frac{5 \nu ^2}{2}-17 \nu +45\right)+e^3 \left(\frac{75}{2}-15 \nu \right) \cos\zeta+e^2 \left(-\frac{5 \nu ^2}{2}+\left(\frac{176}{3}-\frac{41 \pi ^2}{64}\right) \nu -\frac{45}{2}\right)\nonumber\\
&\qquad
+e \left(15 \nu -\frac{75}{2}\right) \cos\zeta+5 \nu ^2+\left(\frac{41 \pi ^2}{64}-\frac{125}{3}\right) \nu -\frac{45}{2}\bigg]
+e \left[-\frac{41 \nu ^2}{3}+\left(\frac{1303}{12}-\frac{41 \pi ^2}{16}\right) \nu +37\right] \cos\zeta\nonumber\\
&\quad
-12 \nu ^2+\left(\frac{1943}{12}-\frac{287 \pi ^2}{64}\right) \nu +9
\bigg\}, \\
Z^{\text{SO}} &= \frac{x^{3/2} \epsilon ^3}{\left(1-e^2\right)^{3/2}} \left[\left(e^2+e \cos\zeta+2\right) \left(2 \delta  \chi _A-(\nu -2) \chi _S\right)\right] \nonumber\\
&\quad
+\frac{x^{5/2} \epsilon ^5}{\left(1-e^2\right)^{5/2}} \bigg\{
\delta \chi _A  \left(22-\frac{25 \nu }{2}\right)+\left(4 \nu ^2-\frac{89 \nu }{2}+22\right) \chi _S
+e^4 \left[ \left(-\frac{11 \nu }{16}-12\right) \delta \chi _A+\left(\frac{5 \nu ^2}{8}+\frac{55 \nu }{16}-12\right) \chi _S\right]\nonumber\\
&\qquad
+e^3 \cos (3 \zeta ) \left[\delta \chi _A  \left(\frac{1}{2}-\frac{27 \nu }{64}\right)+\left(\frac{9 \nu ^2}{32}-\frac{73 \nu }{64}+\frac{1}{2}\right) \chi _S\right]
+e^2 \left[\left(\frac{17 \nu ^2}{4}-\frac{71 \nu }{8}-36\right) \chi _S-\delta \chi _A  \left(\frac{77 \nu }{8}+36\right)\right]\nonumber\\
&\qquad
+e^3 \cos\zeta \left[\delta \chi _A \left(-\frac{125 \nu }{64}-\frac{45}{2}\right)+\left(\frac{47 \nu ^2}{32}+\frac{385 \nu }{64}-\frac{45}{2}\right) \chi _S\right] 
+e^2 \cos (2 \zeta ) \bigg[\delta \chi _A \left(2-\frac{59 \nu }{16}\right)\nonumber\\
&\qquad\quad
+\left(\frac{17 \nu ^2}{8}-\frac{153 \nu }{16}+2\right) \chi _S\bigg]
+e \cos\zeta \left[\delta \chi _A  \left(6-\frac{109 \nu }{8}\right)+\left(\frac{25 \nu ^2}{4}-\frac{295 \nu }{8}+6\right) \chi _S\right]
\nonumber\\
&\qquad
+\sqrt{1-e^2} \left[\delta \chi _A (4 \nu -12) +e^2 \left(\delta  (12-4 \nu ) \chi _A+\left(2 \nu ^2-16 \nu +12\right) \chi _S\right)+\left(-2 \nu ^2+16 \nu -12\right) \chi _S\right]
\bigg\}, \\
Z^{\text{SS}} &=  \frac{x^2 \epsilon ^4}{\left(1-e^2\right)^2} \bigg\{
\left(6 \nu -\frac{3}{2}\right) \chi _A^2-3 \delta  \chi _A \chi _S-\frac{3 \chi _S^2}{2}+e^2 \left[(4 \nu -1) \chi _A^2-2 \delta  \chi _A \chi _S-\chi _S^2+\left(\nu -\frac{1}{2}\right) \kappa _S-\frac{\delta  \kappa _A}{2}\right] \nonumber\\
&\qquad
+e \cos\zeta \left[(4 \nu -1) \chi _A^2-2 \delta  \chi _A \chi _S-\chi _S^2+\left(\nu -\frac{1}{2}\right) \kappa _S-\frac{\delta  \kappa _A}{2}\right]
+\left(3 \nu -\frac{3}{2}\right) \kappa _S-\frac{3 \delta  \kappa _A}{2}
\bigg\} \nonumber\\
&\quad
+ \frac{x^3 \epsilon ^6}{\left(1-e^2\right)^3}
\Bigg\{
e^4 \bigg[\left(\frac{8 \nu ^2}{3}-\frac{170 \nu }{3}+14\right) \chi _A^2+\left(28-\frac{28 \nu }{3}\right) \delta  \chi _S \chi _A+\left(2 \nu ^2-\frac{26 \nu }{3}+14\right) \chi _S^2+\left(\frac{7}{2}-\frac{\nu }{3}\right) \delta  \kappa _A\nonumber\\
&\qquad\quad
+\left(\frac{2 \nu ^2}{3}-\frac{22 \nu }{3}+\frac{7}{2}\right) \kappa _S\bigg] 
+ e^3 \cos\zeta \bigg[\left(-\frac{11 \nu ^2}{6}-\frac{2293 \nu }{24}+24\right) \chi _A^2+\left(48-\frac{79 \nu }{12}\right) \delta  \chi _S \chi _A\nonumber\\
&\qquad\quad
+\left(\frac{15 \nu ^2}{8}-\frac{169 \nu }{24}+24\right) \chi _S^2
+\left(\frac{19 \nu }{24}+\frac{37}{8}\right) \delta  \kappa _A+\left(-\frac{19 \nu ^2}{12}-\frac{203 \nu }{24}+\frac{37}{8}\right) \kappa _S\bigg]\nonumber\\
&\qquad
+e^3 \cos (3 \zeta ) \left[\left(\frac{3 \nu }{8}-\frac{3 \nu ^2}{2}\right) \chi _A^2+\frac{9}{4} \delta  \nu  \chi _S \chi _A+\left(\frac{15 \nu }{8}-\frac{3 \nu ^2}{8}\right) \chi _S^2+\left(\frac{3 \nu }{8}-\frac{5}{8}\right) \delta  \kappa _A-\left(\frac{3 \nu ^2}{4}-\frac{13 \nu }{8}+\frac{5}{8}\right) \kappa _S\right] \nonumber\\
&\qquad
+ e^2\cos (2 \zeta ) \bigg[\left(-9 \nu ^2+\frac{9 \nu }{2}-\frac{1}{2}\right) \chi _A^2+\delta  (16 \nu -1) \chi _S \chi _A+\left(-3 \nu ^2+\frac{27 \nu }{2}-\frac{1}{2}\right) \chi _S^2+\left(\frac{11 \nu }{4}-\frac{9}{2}\right) \delta  \kappa _A\nonumber\\
&\qquad\quad
-\left(\frac{9 \nu ^2}{2}-\frac{47 \nu }{4}+\frac{9}{2}\right) \kappa _S\bigg]
+ e^2\bigg[\left(\frac{79}{2}-\frac{50 \nu ^2}{3}-\frac{911 \nu }{6}\right) \chi _A^2+\left(\frac{61 \nu }{3}+79\right) \delta  \chi _S \chi _A+\left(\frac{79}{2}-4 \nu ^2+\frac{85 \nu }{6}\right) \chi _S^2\nonumber\\
&\qquad\quad
+\left(\frac{47 \nu }{6}+\frac{3}{2}\right) \delta  \kappa _A+\left(\frac{3}{2}-\frac{23 \nu ^2}{3}+\frac{29 \nu }{6}\right) \kappa _S\bigg]
+ e\cos\zeta \bigg[\left(\frac{145 \nu }{3}-15\right) \delta  \chi _S \chi _A - \left(\frac{62 \nu ^2}{3}-\frac{223 \nu }{6}+\frac{15}{2}\right) \chi _A^2\nonumber\\
&\qquad\quad
+\left(-\frac{21 \nu ^2}{2}+\frac{247 \nu }{6}-\frac{15}{2}\right) \chi _S^2+\left(\frac{53 \nu }{6}-12\right) \delta  \kappa _A+\left(-\frac{29 \nu ^2}{3}+\frac{197 \nu }{6}-12\right) \kappa _S\bigg]
+\delta  (57 \nu -39) \chi _A \chi _S\nonumber\\
&\qquad
+\left(-16 \nu ^2+\frac{171 \nu }{2}-\frac{39}{2}\right) \chi _A^2+\left(-14 \nu ^2+\frac{99 \nu }{2}-\frac{39}{2}\right) \chi _S^2+\delta  \left(11 \nu -\frac{35}{2}\right) \kappa _A+\left(-8 \nu ^2+46 \nu -\frac{35}{2}\right) \kappa _S\nonumber\\
&\qquad
+\sqrt{1-e^2} \bigg\{
e^2 \bigg[\left(-6 \nu ^2+46 \nu -11\right) \chi _A^2+\delta  (18 \nu -22) \chi _S \chi _A+\left(-4 \nu ^2+16 \nu -11\right) \chi _S^2+\left(\frac{5 \nu }{2}-7\right) \delta  \kappa _A\nonumber\\
&\qquad\quad
+\left(-3 \nu ^2+\frac{33 \nu }{2}-7\right) \kappa _S\bigg]
+\left(6 \nu ^2-46 \nu +11\right) \chi _A^2+\left(4 \nu ^2-16 \nu +11\right) \chi _S^2+\left(7-\frac{5 \nu }{2}\right) \delta  \kappa _A\nonumber\\
&\qquad\quad
+\left(3 \nu ^2-\frac{33 \nu }{2}+7\right) \kappa _S+\delta  (22-18 \nu ) \chi _A \chi _S\bigg\}
\Bigg\}.
\end{align}
\end{subequations}
\normalsize

The 3PN relation between the orbit-averaged and instantaneous frequencies is given by
%
\small
\begin{subequations}
\begin{align} \label{eq:xxinstFull}
x &\equiv \frac{(M \Omega)^{2/3} \left(1-e^2\right)}{(1 + e \cos \zeta)^{4/3}} \Big[1 + X^{1\text{PN}} + X^{2\text{PN}} + X^{3\text{PN}} + X^{\text{SO}} + X^{\text{SS}}\Big], \\
X^{1\text{PN}} &= \frac{2 e \epsilon ^2 (M \Omega )^{2/3} [3 e+2 \cos\zeta]}{3 (1+e \cos\zeta)^{4/3}}, \\
X^{2\text{PN}} &= \frac{\epsilon ^4 (M \Omega )^{4/3}}{(1+e \cos\zeta)^{8/3}} \bigg\{
e^4 \left(\frac{2 \nu }{3}+5\right)+\left[e^3 \left(\frac{4 \nu }{9}+\frac{16}{3}\right)+e \left(\frac{8}{3}-\frac{4 \nu }{9}\right)\right] \cos\zeta \nonumber\\
&\qquad
+2 e^2 \cos (2 \zeta )+e^2 \left(-\frac{7 \nu }{3}-\frac{5}{6}\right)-2 \nu +5
+\sqrt{1-e^2} \left[e^2 (5-2 \nu )+2 \nu -5\right]
\bigg\}, \\
X^{3\text{PN}} &=\frac{M^2 \Omega ^2 \epsilon ^6}{(1+e \cos\zeta)^4} \bigg\{
e^6 \left(\frac{23 \nu }{6}+\frac{40}{3}\right)
+e^5 \left(\frac{43 \nu }{9}+20\right) \cos\zeta
+e^4 \left(\frac{44 \nu }{27}+\frac{88}{9}\right) \cos (2 \zeta )
+e^3 \left(\frac{242}{27}-\frac{95 \nu }{9}\right) \cos\zeta \nonumber\\
&\quad
+e^4 \left(-\frac{13 \nu ^2}{36}-\frac{277 \nu }{27}-\frac{77}{9}\right)
+e^3 \left(\frac{308}{81}-\frac{\nu }{3}\right) \cos (3 \zeta )
+e^2 \left[\frac{52 \nu ^2}{9}+\left(\frac{205 \pi ^2}{192}-\frac{3349}{54}\right) \nu +\frac{346}{9}\right]  \nonumber\\
&\quad
+e^2 \left(\frac{88}{9}-\frac{98 \nu }{27}\right) \cos (2 \zeta )
+e \left(\frac{76}{3}-\frac{109 \nu }{9}\right) \cos\zeta
+\frac{10 \nu ^2}{3}+\left(\frac{41 \pi ^2}{96}-\frac{340}{9}\right) \nu +10 \nonumber\\
&\quad
+\sqrt{1-e^2} \bigg[e^4 \left(\frac{5 \nu ^2}{3}-\frac{32 \nu }{3}+25\right)+e^3 (20-8 \nu ) \cos\zeta+e^2 \left(\frac{5 \nu ^2}{3}+\left(\frac{41 \pi ^2}{96}-\frac{244}{9}\right) \nu -15\right) \nonumber\\
&\qquad
+e (8 \nu -20) \cos\zeta-\frac{10 \nu ^2}{3}+\left(\frac{340}{9}-\frac{41 \pi ^2}{96}\right) \nu -10\bigg]
\bigg\} , \\
X^{\text{SO}} &= -\frac{2 e M \Omega  \epsilon ^3 (e+\cos\zeta)}{3 (1+e \cos\zeta)^2}  \left[2 \delta  \chi _A-(\nu -2) \chi _S\right] \nonumber\\
&\quad
+ \frac{\epsilon ^5 (M \Omega )^{5/3}}{(1+e \cos\zeta)^{10/3}} \bigg\{
\delta \chi _A \left(\frac{8 \nu }{3}-8\right) 
+e^4 \left[ \left(\frac{11 \nu }{24}-\frac{16}{3}\right) \delta \chi _A+\left(-\frac{5 \nu ^2}{12}+\frac{35 \nu }{8}-\frac{16}{3}\right) \chi _S\right] \nonumber\\
&\qquad
+e^3 \cos (3 \zeta ) \left[ \left(\frac{27 \nu }{32}-\frac{1}{3}\right) \delta \chi _A+\left(-\frac{9 \nu ^2}{16}+\frac{187 \nu }{96}-\frac{1}{3}\right) \chi _S\right]
+e^3 \cos\zeta \bigg[\delta \chi _A  \left(\frac{179 \nu }{96}-\frac{89}{9}\right) \nonumber\\
&\qquad\quad
+\left(-\frac{65 \nu ^2}{48}+\frac{2771 \nu }{288}-\frac{89}{9}\right) \chi _S\bigg] 
+e^2 \cos (2 \zeta ) \left[ \left(\frac{113 \nu }{24}-\frac{58}{9}\right) \delta \chi _A+\left(-\frac{35 \nu ^2}{12}+\frac{985 \nu }{72}-\frac{58}{9}\right) \chi _S\right]\nonumber\\
&\qquad
+e^2 \left[ \left(\frac{61 \nu }{12}-\frac{46}{9}\right) \delta \chi _A+\left(-\frac{13 \nu ^2}{6}+\frac{545 \nu }{36}-\frac{46}{9}\right) \chi _S\right] \nonumber\\
&\qquad
+\sqrt{1-e^2} \left[ \left(8-\frac{8 \nu }{3}\right) \delta \chi _A+e^2 \left( \left(\frac{8 \nu }{3}-8\right) \delta \chi _A+\left(-\frac{4 \nu ^2}{3}+\frac{32 \nu }{3}-8\right) \chi _S\right)+\left(\frac{4 \nu ^2}{3}-\frac{32 \nu }{3}+8\right) \chi _S\right] \nonumber\\
&\qquad
+e \cos\zeta \left[ \left(\frac{75 \nu }{8}-\frac{52}{9}\right) \delta \chi _A+\left(-\frac{21 \nu ^2}{4}+\frac{1729 \nu }{72}-\frac{52}{9}\right) \chi _S\right]
+ \chi _S \left(-\frac{4 \nu ^2}{3}+\frac{32 \nu }{3}-8\right)
\bigg\}, \\
X^{\text{SS}} &= \frac{e \epsilon ^4 (M \Omega )^{4/3}}{(1+e \cos\zeta)^{8/3}} \bigg\{
e \left[\frac{\delta  \kappa _A}{3}+\left(\frac{2}{3}-\frac{8 \nu }{3}\right) \chi _A^2+\frac{4}{3} \delta  \chi _A \chi _S+\left(\frac{1}{3}-\frac{2 \nu }{3}\right) \kappa _S+\frac{2 \chi _S^2}{3}\right] \nonumber\\
&\qquad
+e \cos (2 \zeta ) \left[\left(\frac{1}{3}-\frac{4 \nu }{3}\right) \chi _A^2+\frac{2}{3} \delta  \chi _A \chi _S+\frac{\chi _S^2}{3}\right]
+\cos\zeta \left[\left(\frac{4}{3}-\frac{16 \nu }{3}\right) \chi _A^2+\frac{8}{3} \delta  \chi _A \chi _S+\frac{4 \chi _S^2}{3}\right]
\bigg\} \nonumber\\
&\quad
+ \frac{M^2 \Omega ^2 \epsilon ^6}{(1+e \cos\zeta)^4} \Bigg\{
e^4 \cos (2 \zeta ) \left[\left(-\frac{8 \nu ^2}{9}-\frac{46 \nu }{9}+\frac{4}{3}\right) \chi _A^2+\left(\frac{4 \nu }{9}+\frac{8}{3}\right) \delta  \chi _S \chi _A+\left(\frac{2 \nu }{9}+\frac{4}{3}\right) \chi _S^2\right] \nonumber\\
&\qquad
+  e^4\bigg[\left(-\frac{16 \nu ^2}{9}-\frac{100 \nu }{9}+\frac{26}{9}\right) \chi _A^2+\left(\frac{4 \nu }{3}+\frac{52}{9}\right) \delta  \chi _S \chi _A+\left(-\frac{\nu ^2}{9}+\frac{8 \nu }{9}+\frac{26}{9}\right) \chi _S^2+\left(\frac{2 \nu }{9}+\frac{4}{3}\right) \delta  \kappa _A \nonumber\\
&\qquad\quad
+\left(-\frac{4 \nu ^2}{9}-\frac{22 \nu }{9}+\frac{4}{3}\right) \kappa _S\bigg]
+ \cos (3 \zeta ) e^3\left[\left(\frac{11}{9}-\frac{44 \nu }{9}\right) \chi _A^2+\frac{22}{9} \delta  \chi _S \chi _A+\frac{11 \chi _S^2}{9}+\frac{\delta  \kappa _A}{6}+\left(\frac{1}{6}-\frac{\nu }{3}\right) \kappa _S\right]\nonumber\\
&\qquad
+e^3 \cos\zeta \left[\left(\frac{4 \nu ^2}{9}-\frac{8 \nu }{9}+\frac{101}{9}\right) \chi _S^2+\chi _A^2\left(\frac{101}{9}-\frac{32 \nu ^2}{9}-44 \nu \right) +\frac{202}{9} \delta  \chi _S \chi _A+\frac{29 \delta  \kappa _A}{18}+\left(\frac{29}{18}-\frac{29 \nu }{9}\right) \kappa _S\right]\nonumber\\
&\qquad
+ e^2\cos (2 \zeta ) \left[\left(\frac{8 \nu ^2}{9}-\frac{290 \nu }{9}+8\right) \chi _A^2+\left(16-\frac{26 \nu }{9}\right) \delta  \chi _S \chi _A+\left(\frac{11 \nu ^2}{18}-\frac{8 \nu }{3}+8\right) \chi _S^2+\delta  \kappa _A+(1-2 \nu ) \kappa _S\right]\nonumber\\
&\qquad
+e\cos\zeta \left[\left(\frac{32 \nu ^2}{9}-\frac{56 \nu }{9}+\frac{4}{3}\right) \chi _A^2+\left(\frac{8 \nu }{9}+\frac{8}{3}\right) \delta  \chi _S \chi _A+\left(-\frac{2 \nu ^2}{3}+\frac{16 \nu }{9}+\frac{4}{3}\right) \chi _S^2+\frac{4 \delta  \kappa _A}{3}+\left(\frac{4}{3}-\frac{8 \nu }{3}\right) \kappa _S\right] \nonumber\\
&\qquad
+ e^2\bigg[\left(\frac{70 \nu ^2}{9}-\frac{301 \nu }{9}+\frac{23}{3}\right) \chi _A^2+\left(\frac{46}{3}-\frac{62 \nu }{3}\right) \delta  \chi _S \chi _A+\left(\frac{83 \nu ^2}{18}-\frac{161 \nu }{9}+\frac{23}{3}\right) \chi _S^2+\left(\frac{14}{3}-\frac{61 \nu }{18}\right) \delta  \kappa _A \nonumber\\
&\qquad\quad
+\left(\frac{31 \nu ^2}{9}-\frac{229 \nu }{18}+\frac{14}{3}\right) \kappa _S\bigg]
+\left(4 \nu ^2-\frac{92 \nu }{3}+\frac{22}{3}\right) \chi _A^2+\left(\frac{8 \nu ^2}{3}-\frac{32 \nu }{3}+\frac{22}{3}\right) \chi _S^2+\left(\frac{14}{3}-\frac{5 \nu }{3}\right) \delta  \kappa _A\nonumber\\
&\qquad
+\left(2 \nu ^2-11 \nu +\frac{14}{3}\right) \kappa _S+\delta \chi _A \chi _S \left(\frac{44}{3}-12 \nu \right) 
+\sqrt{1-e^2} \bigg\{
e^2 \bigg[ \chi _A^2\left(4 \nu ^2-\frac{92 \nu }{3}+\frac{22}{3}\right)+\delta \chi _S \chi _A  \left(\frac{44}{3}-12 \nu \right) \nonumber\\
&\qquad\quad
+\left(\frac{8 \nu ^2}{3}-\frac{32 \nu }{3}+\frac{22}{3}\right) \chi _S^2+\left(\frac{14}{3}-\frac{5 \nu }{3}\right) \delta  \kappa _A+\left(2 \nu ^2-11 \nu +\frac{14}{3}\right) \kappa _S\bigg]
+\left(-4 \nu ^2+\frac{92 \nu }{3}-\frac{22}{3}\right) \chi _A^2 \nonumber\\
&\qquad\quad
+\left(-\frac{8 \nu ^2}{3}+\frac{32 \nu }{3}-\frac{22}{3}\right) \chi _S^2+\left(\frac{5 \nu }{3}-\frac{14}{3}\right) \delta  \kappa _A+\left(-2 \nu ^2+11 \nu -\frac{14}{3}\right) \kappa _S
+\delta  \chi _A \chi _S \left(12 \nu -\frac{44}{3}\right)\bigg\}
\Bigg\}.
\end{align}
\end{subequations}
\normalsize

\section{Waveform modes}

All the expressions for the PN-expanded waveform modes are provided in the Supplemental Material in terms of the EOB variables $ (r, p_r, p_\phi) $ (only for the instantaneous contributions), and also in terms of the Keplerian parameters $ \left( x, e, \zeta \right) $ for which we perform a resummation in eccentricity as described in Secs.~\ref{sec:tail_modes} and \ref{sec:memory_modes}.
In this appendix, we display the complete expressions for the different contributions to the $ (2,2) $ mode, except for the tail part which is only shown up to $ \mathcal O(e^2) $.

The 3PN instantaneous contribution to the $(2,2)$ mode in EOB coordinates and in terms of the Keplerian parameters $ \left( x, e, \zeta \right) $ is given by
\small
\begin{subequations}
\label{eq:app:instModes}
\begin{align}
H^\text{inst}_{22} &= H_\text{inst, N}^{22} + H^\text{inst, 1PN}_{22} + H^\text{inst, 2PN}_{22} + H^\text{inst, 2.5PN}_{22} + H^\text{inst, 3PN}_{22} + H^\text{inst, SO}_{22} + H^\text{inst, SS}_{22}, \\
H^\text{inst, N}_{22} &=  \frac{4+ (\text{e}^{-i \zeta } +5 \text{e}^{i \zeta }) e+2 \text{e}^{2 i \zeta } e^2}{4 (1- e^2)}, \\
H^\text{inst, 1PN}_{22} &= \frac{x \epsilon ^2}{42 \left(1-e^2\right)^2} \bigg\{
e^3 \e^{-3 i \zeta } \left(\frac{27 \nu }{4}+\frac{3}{8}\right)+e^3 \e^{3 i \zeta } \left(\frac{21 \nu }{4}-\frac{343}{8}\right)+\e^{-i \zeta } \left[e^3 \left(23 \nu -\frac{199}{8}\right)+e \left(\frac{379 \nu }{4}-\frac{383}{4}\right)\right] \nonumber\\
&\quad
+\e^{i \zeta } \left[e^3 \left(\frac{77 \nu }{2}-\frac{1001}{8}\right)+e \left(\frac{287 \nu }{4}-\frac{847}{4}\right)\right]
+\e^{2 i \zeta } \left[e^4 \left(\frac{17 \nu }{2}-\frac{57}{2}\right)+e^2 \left(\frac{61 \nu }{2}-\frac{673}{4}\right)\right]\nonumber\\
&\quad
+e^2 \e^{-2 i \zeta } \left(45 \nu -\frac{95}{4}\right)+e^2 \left(101 \nu -\frac{345}{2}\right)+55 \nu -107
\bigg\},\\
H^\text{inst, 2PN}_{22} &= \frac{x^2 \epsilon ^4}{84 \left(1-e^2\right)^4} \bigg\{
\left(8 \nu ^2+\frac{5103 \nu }{4}-\frac{14923}{16}\right) e^6+\left(-283 \nu ^2+\frac{14969 \nu }{12}+\frac{3879}{16}\right) e^4+\left(\frac{2903 \nu ^2}{18}-\frac{20479 \nu }{9}+\frac{44315}{36}\right) e^2 \nonumber\\
&\quad
+\bigg[\left(\frac{25 \nu ^2}{6}+\frac{3271 \nu }{16}-\frac{18875}{192}\right) e^7+\left(-\frac{551 \nu ^2}{3}+\frac{31579 \nu }{16}-\frac{107713}{192}\right) e^5+\left(-\frac{7555 \nu ^2}{72}-\frac{30697 \nu }{36}+\frac{130445}{144}\right) e^3\nonumber\\
&\qquad+\left(\frac{20479 \nu ^2}{72}-\frac{95431 \nu }{72}-\frac{2219}{9}\right) e\bigg] \e^{-i \zeta }
+\left[\left(-\frac{205 \nu ^2}{48}+\frac{425 \nu }{96}-\frac{655}{384}\right) e^7+\left(\frac{205 \nu ^2}{48}-\frac{425 \nu }{96}+\frac{655}{384}\right) e^5\right] \e^{-5 i \zeta } \nonumber\\
&\quad
+\left[\left(-\frac{353 \nu ^2}{48}+\frac{6349 \nu }{96}-\frac{767}{384}\right) e^7+\left(-\frac{20545 \nu ^2}{144}+\frac{139781 \nu }{288}+\frac{72677}{1152}\right) e^5+\left(\frac{5401 \nu ^2}{36}-\frac{39707 \nu }{72}-\frac{8797}{144}\right) e^3\right] \e^{-3 i \zeta } \nonumber\\
&\quad
+\left[\left(-\frac{1435 \nu ^2}{36}+\frac{7025 \nu }{72}+\frac{1355}{288}\right) e^6+\left(\frac{1435 \nu ^2}{36}-\frac{7025 \nu }{72}-\frac{1355}{288}\right) e^4\right] \e^{-4 i \zeta } 
+\bigg[\left(16 \nu ^2+\frac{17845 \nu }{48}-\frac{29717}{64}\right) e^7\nonumber\\
&\qquad
+\left(-\frac{202 \nu ^2}{3}+\frac{20879 \nu }{16}-\frac{252989}{192}\right) e^5+\left(-\frac{835 \nu ^2}{8}-\frac{3403 \nu }{3}+\frac{47397}{16}\right) e^3+\left(\frac{3737 \nu ^2}{24}-\frac{4339 \nu }{8}-\frac{3541}{3}\right) e\bigg] \e^{i \zeta } \nonumber\\
&\quad
+\bigg[\left(\frac{9 \nu ^2}{4}+\frac{259 \nu }{4}-\frac{155}{4}\right) e^8+\left(\frac{401 \nu ^2}{36}+\frac{4750 \nu }{9}-\frac{12131}{9}\right) e^6+\left(-\frac{3017 \nu ^2}{36}-\frac{8989 \nu }{36}+\frac{162337}{72}\right) e^4 \nonumber\\
&\qquad
+\left(\frac{845 \nu ^2}{12}-\frac{2057 \nu }{6}-\frac{20833}{24}\right) e^2\bigg] \e^{2 i \zeta } 
+\left[\left(-\frac{11 \nu ^2}{4}+\frac{185 \nu }{8}+\frac{131}{32}\right) e^6+\left(\frac{11 \nu ^2}{4}-\frac{185 \nu }{8}-\frac{131}{32}\right) e^4\right] \e^{4 i \zeta }\nonumber\\
&\quad
+\left[\left(\frac{57 \nu ^2}{16}+\frac{9553 \nu }{96}-\frac{36617}{128}\right) e^7+\left(-\frac{2461 \nu ^2}{144}+\frac{11105 \nu }{288}+\frac{512825}{1152}\right) e^5+\left(\frac{487 \nu ^2}{36}-\frac{9941 \nu }{72}-\frac{22909}{144}\right) e^3\right] \e^{3 i \zeta } \nonumber\\
&\quad
+\left[\left(-\frac{11 \nu ^2}{16}-\frac{\nu }{32}+\frac{351}{128}\right) e^7+\left(\frac{11 \nu ^2}{16}+\frac{\nu }{32}-\frac{351}{128}\right) e^5\right] \e^{5 i \zeta }
+\frac{2047 \nu ^2}{18}-\frac{4459 \nu }{18}-\frac{9733}{18} \nonumber\\
&\quad
+\left[\left(-57 \nu ^2+\frac{8303 \nu }{12}-\frac{721}{4}\right) e^6+\left(-231 \nu ^2+\frac{3643 \nu }{6}+\frac{2779}{8}\right) e^4+\left(288 \nu ^2-\frac{15589 \nu }{12}-\frac{1337}{8}\right) e^2\right] \e^{-2 i \zeta } \nonumber\\
&\quad
+ \sqrt{1-e^2} \Big\{e^4 (420-168 \nu )+e^2 (336 \nu -840)+\e^{-i \zeta } \left[e^5 (105-42 \nu )+e^3 (84 \nu -210)+e (105-42 \nu )\right] \nonumber\\
&\qquad
+\e^{i \zeta } \left[e^5 (525-210 \nu )+e^3 (420 \nu -1050)+e (525-210 \nu )\right] -168 \nu +420\nonumber\\
&\qquad
+\e^{2 i \zeta } \left[e^6 (210-84 \nu )+e^4 (168 \nu -420)+e^2 (210-84 \nu )\right]
\Big\}
\bigg\},\\
H^\text{inst, 2.5PN}_{22} &= \frac{i \nu x^{5/2} \epsilon ^5}{10 \left(1-e^2\right)^{7/2}} \bigg\{
\frac{16}{3} (6 \alpha -4 \beta -39)
+e \left[\e^{-i \zeta } \left(56 \alpha -\frac{112 \beta }{3}-\frac{229}{14}\right)+\e^{i \zeta } \left(88 \alpha -\frac{176 \beta }{3}-\frac{39961}{42}\right)\right] \nonumber\\
&\quad
+e^2 \left[\e^{2 i \zeta } \left(100 \alpha -\frac{200 \beta }{3}-\frac{178039}{126}\right)+\e^{-2 i \zeta } \left(36 \alpha -24 \beta -\frac{21361}{126}\right)+120 \alpha -80 \beta -\frac{30964}{21}\right] \nonumber\\
&\quad
+e^3 \bigg[\e^{-3 i \zeta } \left(10 \alpha -\frac{20 \beta }{3}-\frac{9223}{72}\right)+\e^{3 i \zeta } \left(58 \alpha -\frac{116 \beta }{3}-\frac{439627}{504}\right)+\e^{-i \zeta } \left(54 \alpha -36 \beta +\frac{45979}{84}\right)\nonumber\\
&\qquad
+\e^{i \zeta } \left(\!102 \alpha -68 \beta -\frac{1477649}{252} \!\right)\!\bigg] 
+e^4 \bigg[30 \alpha -20 \beta -\frac{234155}{63}+\e^{-4 i \zeta } \left(\!\alpha -\frac{2 \beta }{3}-\frac{13913}{252} \!\right)+\e^{4 i \zeta } \left(\! 17 \alpha -\frac{34 \beta }{3}-\frac{12763}{84} \!\right)\nonumber\\
&\qquad
+\e^{-2 i \zeta } \left(8 \alpha -\frac{16 \beta }{3}+\frac{332387}{504}\right)+\e^{2 i \zeta } \left(40 \alpha -\frac{80 \beta }{3}-\frac{4048747}{504}\right)\bigg]
+e^5 \bigg[\e^{-i \zeta } \left(2 \alpha -\frac{4 \beta }{3}+\frac{288297}{224}\right)\nonumber\\
&\qquad
+\e^{5 i \zeta } \left(\! 2 \alpha -\frac{4 \beta }{3}+\frac{4139}{6720} \!\right)+\e^{i \zeta } \left(\! 6 \alpha -4 \beta -\frac{4406435}{288} \!\right)+\e^{3 i \zeta } \left(\! 6 \alpha -4 \beta -\frac{18462635}{4032} \!\right)+\frac{936623 \e^{-3 i \zeta }}{4032}-\frac{8851 \e^{-5 i \zeta }}{2880}\bigg] \nonumber\\
&\quad
+e^6 \!\left[\frac{1113067}{576} \e^{-2 i \zeta }-\frac{3707689}{192} \e^{2 i \zeta }-\frac{6383}{360} \e^{-4 i \zeta }-\frac{511249}{600} \e^{4 i \zeta }-\frac{16811 \e^{-6 i \zeta }}{14400}+\frac{359}{320} \e^{6 i \zeta }-\frac{622097}{72}\right]\!
+ \Order(e^7)\!
\bigg\} ,\\
H^\text{inst, 3PN}_{22} &= \frac{x^3 \epsilon ^6}{168 \left(1-e^2\right)^4} \Bigg\{
\left(-\frac{875 \nu ^3}{1408}-\frac{48265 \nu ^2}{4224}+\frac{7735 \nu }{2112}+\frac{245}{1056}\right) \e^{-7 i \zeta } e^7
+\left(-\frac{57 \nu ^3}{128}+\frac{871 \nu ^2}{128}+\frac{45 \nu }{16}-\frac{9}{128}\right) \e^{7 i \zeta } e^7 \nonumber\\
&\quad
+\left(-\frac{2625 \nu ^3}{352}-\frac{192425 \nu ^2}{1056}+\frac{62591 \nu }{1056}-\frac{242831}{21120}\right) \e^{-6 i \zeta } e^6
+\left(-\frac{95 \nu ^3}{32}+\frac{2585 \nu ^2}{96}+\frac{1041 \nu }{32}+\frac{12081}{640}\right) \e^{6 i \zeta } e^6 \nonumber\\
&\quad
+\left(\! \frac{5185655 \nu }{528}-\frac{487643 \nu ^3}{2376}-\frac{286673 \nu ^2}{264}-\frac{4026923}{1056} \!\right)\! e^6
+\left(\!\frac{74577563}{4400}-\frac{71906 \nu ^3}{99}-\frac{4044037 \nu ^2}{396}-\frac{9471 \pi ^2 \nu }{32}+\frac{6998291 \nu }{176} \!\right)\! e^4 \nonumber\\
&\quad
+\left(\! \frac{76231 \nu ^3}{198}-\frac{421551 \nu ^2}{44}-\frac{4305 \pi ^2 \nu }{16}+\frac{283739 \nu }{22}+\frac{63527301}{1100} \!\right)\! e^2
+\bigg[\!\left(\! \frac{2442317 \nu }{2112} -\frac{1485443 \nu ^3}{38016}-\frac{1525841 \nu ^2}{12672}-\frac{1448701}{4224} \!\right)\! e^7 \nonumber\\
&\qquad
+\left(\frac{270067757}{158400}-\frac{393323 \nu ^3}{528}-\frac{2451829 \nu ^2}{396}-\frac{26691 \pi ^2 \nu }{256}+\frac{17934849 \nu }{704}\right) e^5 
+\bigg(\frac{167254997}{3300}-\frac{173747 \nu ^3}{792}-\frac{26446709 \nu ^2}{1584}\nonumber\\
&\qquad\quad
-\frac{50799 \pi ^2 \nu }{64}+\frac{11302525 \nu }{264}\bigg) e^3
+\left(\frac{1270235 \nu ^3}{2376}-\frac{854797 \nu ^2}{198}-\frac{23325 \nu }{22}-\frac{2583 \pi ^2 \nu }{32}+\frac{89359568}{2475}\right) e\bigg] \e^{-i \zeta }\nonumber\\
&\quad
+\bigg[\left(-\frac{27725 \nu ^3}{1056}-\frac{1908089 \nu ^2}{1584}-\frac{12915 \pi ^2 \nu }{512}+\frac{4123627 \nu }{4224}+\frac{13384643}{21120}\right) e^5 
+\bigg(-\frac{20245 \nu ^3}{1408}-\frac{119505 \nu ^2}{1408}\nonumber\\
&\qquad\quad
+\frac{88875 \nu }{1408}-\frac{6535}{352}\bigg) e^7\bigg] \e^{-5 i \zeta } 
+\bigg[\left(\frac{1225 \nu ^3}{396}-\frac{3385981 \nu ^2}{792}-\frac{24969 \pi ^2 \nu }{128}+\frac{6247435 \nu }{1056}+\frac{99480817}{15840}\right) e^4 \nonumber\\
&\qquad
+\left(-\frac{71645 \nu ^3}{528}-\frac{549149 \nu ^2}{528}+\frac{1395313 \nu }{1056}+\frac{103437}{3520}\right) e^6\bigg] \e^{-4 i \zeta } 
+\bigg[\left(-\frac{55655 \nu ^3}{1408}-\frac{708265 \nu ^2}{4224}+\frac{599461 \nu }{1408}-\frac{57383}{1408}\right) e^7\nonumber\\
&\qquad
+\left(-\frac{166483 \nu ^3}{352}-\frac{8090617 \nu ^2}{1584}-\frac{43911 \pi ^2 \nu }{512}+\frac{42090875 \nu }{4224}+\frac{123132257}{35200}\right) e^5
+\bigg(\frac{88945 \nu ^3}{396}-\frac{13671529 \nu ^2}{1584}-\frac{35301 \pi ^2 \nu }{64} \nonumber\\
&\qquad\quad
+\frac{1531340 \nu }{99}+\frac{74820017}{3300}\bigg) e^3\bigg] \e^{-3 i \zeta } 
+\bigg[\left(-\frac{102131 \nu ^3}{352}-\frac{5346941 \nu ^2}{3168}+\frac{2106845 \nu }{352}-\frac{3605645}{4224}\right) e^6\nonumber\\
&\quad
+\left(-\frac{68488 \nu ^3}{99}-\frac{5051657 \nu ^2}{396}-\frac{30135 \pi ^2 \nu }{64}+\frac{24986869 \nu }{792}+\frac{425040593}{19800}\right) e^4+\bigg(\frac{18010 \nu ^3}{33}-\frac{7402627 \nu ^2}{792}-\frac{19803 \pi ^2 \nu }{32} \nonumber\\
&\qquad\quad
+\frac{686371 \nu }{44}+\frac{88338663}{2200}\bigg) e^2\bigg] \e^{-2 i \zeta }
+\bigg[\left(-\frac{60593 \nu ^3}{3456}-\frac{69107 \nu ^2}{1152}+\frac{207923 \nu }{96}-\frac{88291}{64}\right) e^7 \nonumber\\
&\quad
+\left(-\frac{45107 \nu ^3}{144}-\frac{30889 \nu ^2}{12}+\frac{1265919 \nu }{64}-\frac{12915 \pi ^2 \nu }{256}-\frac{30195011}{4800}\right) e^5
+\bigg(-\frac{8855 \nu ^3}{72}-\frac{1081577 \nu ^2}{144}+\frac{214145 \nu }{12} \nonumber\\
&\qquad\quad
-\frac{18081 \pi ^2 \nu }{64}+\frac{11145947}{300}\bigg) e^3
+\left(\frac{59261 \nu ^3}{216}-\frac{16340 \nu ^2}{9}+\frac{40109 \nu }{9}-\frac{2583 \pi ^2 \nu }{32}+\frac{4901998}{225}\right) e\bigg] \e^{i \zeta } \nonumber\\
&\quad
+\bigg[\left(-\frac{2195 \nu ^3}{1188}-\frac{458 \nu ^2}{33}+\frac{9725 \nu }{44}+\frac{2005}{44}\right) e^8+\left(-\frac{121751 \nu ^3}{3168}-\frac{775945 \nu ^2}{3168}+\frac{2248627 \nu }{352}-\frac{134452609}{21120}\right) e^6 \nonumber\\
&\qquad
+\left(-\frac{58147 \nu ^3}{396}-\frac{999205 \nu ^2}{396}-\frac{4305 \pi ^2 \nu }{64}+\frac{1596785 \nu }{264}+\frac{218238563}{19800}\right) e^4+\bigg(\frac{153983 \nu ^3}{1188}-\frac{316229 \nu ^2}{264}\nonumber\\
&\qquad\quad
+\frac{61174 \nu }{11}-\frac{6027 \pi ^2 \nu }{32}+\frac{291333017}{19800}\bigg) e^2\bigg] \e^{2 i \zeta }
+\bigg[\left(-\frac{679 \nu ^3}{384}-\frac{9217 \nu ^2}{1152}+\frac{356119 \nu }{384}-\frac{71827}{64}\right) e^7 \nonumber\\
&\qquad
+\left(-\frac{3371 \nu ^3}{96}-\frac{65905 \nu ^2}{144}+\frac{572653 \nu }{384}-\frac{2583 \pi ^2 \nu }{512}+\frac{2985907}{3200}\right) e^5
+\bigg(\frac{391 \nu ^3}{18}-\frac{22261 \nu ^2}{144}-\frac{2583 \pi ^2 \nu }{64}+\frac{18953 \nu }{72} \nonumber\\
&\qquad\quad
+\frac{1083707}{300}\bigg) e^3\bigg] \e^{3 i \zeta }
+\frac{114635 \nu ^3}{594}+\frac{203879 \nu }{198}-\frac{29962 \nu ^2}{33}+\frac{6758911}{550} \nonumber\\
&\quad
+\left[\left(-\frac{117 \nu ^3}{16}-\frac{2575 \nu ^2}{48}+\frac{10473 \nu }{32}+\frac{22711}{320}\right) e^6+\left(-\frac{113 \nu ^3}{36}+\frac{281 \nu ^2}{72}+\frac{861 \pi ^2 \nu }{128}-\frac{78157 \nu }{288}-\frac{277621}{1440}\right) e^4\right] \e^{4 i \zeta } \nonumber\\
&\quad
+\left[\left(-\frac{215 \nu ^3}{128}+\frac{621 \nu ^2}{128}-\frac{449 \nu }{128}+\frac{3495}{128}\right) e^7+\left(-\frac{1765 \nu ^3}{288}+\frac{3383 \nu ^2}{144}+\frac{861 \pi ^2 \nu }{512}+\frac{4359 \nu }{128}-\frac{350837}{5760}\right) e^5\right] \e^{5 i \zeta } \nonumber\\
&\quad
+ \sqrt{1-e^2} \bigg\{ e^4 \left[1336 \nu ^2-7024 \nu +7740\right]
+e^2 \left[-1016 \nu ^2+\frac{28376 \nu }{3}-\frac{287 \pi ^2 \nu }{4}-5140\right] \nonumber\\
&\qquad
+\e^{3 i \zeta } \left[e^5 \left(84 \nu ^2-896 \nu +1715\right)+e^3 \left(896 \nu-84 \nu ^2 -1715\right)\right]
+\e^{-3 i \zeta } \left[e^5 \left(108 \nu ^2-264 \nu -15\right)+e^3 \left(264 \nu -108 \nu ^2+15\right)\right] \nonumber\\
&\qquad
+\e^{i \zeta } \!\left[e^5 \!\left(266 \nu ^2-3822 \nu +6055\right)+e^3 \!\left(182 \nu ^2+\frac{16492 \nu }{3}-\frac{1435 \pi ^2 \nu }{16}+315\right)+e \!\left(\frac{1435 \pi ^2 \nu }{16}-448 \nu ^2-\frac{5026 \nu }{3}-6370\right)\!\right] \nonumber\\
&\qquad
+\e^{-i \zeta } \!\left[e^5 \!\left(298 \nu ^2-1374 \nu +1205\right)\!+e^3 \!\left(1078 \nu ^2-\frac{7084 \nu }{3}-\frac{287 \pi ^2 \nu }{16}+2205\right)\!+e \!\left(\frac{287 \pi ^2 \nu }{16}-1376 \nu ^2+\frac{11206 \nu }{3}-3410\right)\!\right] \nonumber\\
&\qquad
+\e^{-2 i \zeta } \left[e^4 \left(720 \nu ^2-2180 \nu +950\right)+e^2 \left(-720 \nu ^2+2180 \nu -950\right)\right] 
+\e^{2 i \zeta } \bigg[e^2 \left(-208 \nu ^2+\frac{287 \pi ^2 \nu }{8}+\frac{2216 \nu }{3}-5890\right)\nonumber\\
&\qquad\quad
+e^6 \left(-4 \nu ^2-908 \nu +1560\right) 
+e^4 \left(212 \nu ^2+\frac{508 \nu }{3}-\frac{287 \pi ^2 \nu }{8}+4330\right)\bigg]
-320 \nu ^2+\frac{287 \pi ^2 \nu }{4}-\frac{7304 \nu }{3}-2600 \bigg\} \nonumber\\
&\quad
+\bigg[\frac{321}{4} e^5 \e^{-5 i \zeta }-\frac{107}{4} e^5 \e^{5 i \zeta }+749 e^4 \e^{-4 i \zeta }-107 e^4 \e^{4 i \zeta }+\frac{12198 e^4}{5}+\frac{33384 e^2}{5}+\left(\frac{5029 e^5}{10}+\frac{30816 e^3}{5}+\frac{20972 e}{5}\right) \e^{-i \zeta } \nonumber\\
&\qquad
+\left(\frac{6741 e^5}{20}+\frac{13696 e^3}{5}\right) \e^{-3 i \zeta }+\left(\frac{2889 e^5}{10}+\frac{17976 e^3}{5}+\frac{12412 e}{5}\right) \e^{i \zeta }+\left(\frac{321 e^5}{20}+\frac{856 e^3}{5}\right) \e^{3 i \zeta }\nonumber\\
&\qquad
+\left(\frac{11984 e^4}{5}+\frac{24396 e^2}{5}\right) \e^{-2 i \zeta }+\left(\frac{3424 e^4}{5}+\frac{7276 e^2}{5}\right) \e^{2 i \zeta }+\frac{6848}{5}\bigg] \ln \left[\frac{x \left(1+\frac{1}{2} e \left(\e^{-i \zeta }+\e^{i \zeta }\right)\right)}{\left(1-e^2\right) x_0}\right]
\Bigg\}, \\
H^\text{inst, SO}_{22} &= \frac{x^{3/2} \epsilon ^3}{\left(1-e^2\right)^{5/2}}  \Bigg\lbrace
\delta\chi_A \left[\frac{3}{4} e^3 \e^{3 i \zeta }+\left(\frac{e^3}{4}-\frac{11 e}{6}\right) \e^{-i \zeta }+\left(e^3-\frac{e}{6}\right) \e^{i \zeta }-\frac{1}{4} e^2 \e^{-2 i \zeta }+\frac{25}{12} e^2 \e^{2 i \zeta }-\frac{e^2}{2}-\frac{4}{3}\right] \nonumber\\
&\qquad
+ \chi_S \bigg\{
\frac{4 \nu }{3}-\frac{4}{3} + e^3 \e^{3 i \zeta } \left(\frac{3}{4}-\frac{13 \nu }{48}\right)+\frac{1}{16} e^3 \e^{-3 i \zeta } \nu +\e^{-i \zeta } \left[e^3 \left(\frac{5 \nu }{48}+\frac{1}{4}\right)+e \left(\frac{11 \nu }{6}-\frac{11}{6}\right)\right] \nonumber\\
&\qquad\quad
+\e^{i \zeta } \left[e^3 \left(1-\frac{11 \nu }{48}\right)+e \left(\frac{7 \nu }{6}-\frac{1}{6}\right)\right]
+e^2 \e^{-2 i \zeta } \left(\frac{13 \nu }{24}-\frac{1}{4}\right)
+e^2 \e^{2 i \zeta } \left(\frac{25}{12}-\frac{11 \nu }{24}\right)+e^2 \left(\frac{5 \nu }{4}-\frac{1}{2}\right)
\bigg\}
\Bigg\rbrace \nonumber\\
&\quad
+ \frac{x^{5/2} \epsilon ^5}{\left(1-e^2\right)^{7/2}}  \Bigg\lbrace
\delta\chi_A \bigg\{
e^5 \e^{5 i \zeta } \!\left(\frac{81}{448}-\frac{89 \nu }{672}\right)-e^5 \e^{-5 i \zeta } \!\left(\frac{25 \nu }{448}+\frac{15}{448}\right)-e^4 \e^{-4 i \zeta } \!\left(\frac{641 \nu }{1536}+\frac{19}{48}\right)+e^4 \e^{4 i \zeta } \!\left(\frac{295}{336}-\frac{16295 \nu }{10752}\right) \nonumber\\
&\qquad\quad
+e^4 \!\left(\frac{9211 \nu }{5376}-\frac{221}{56}\right)+e^2 \!\left(\frac{37}{28}-\frac{2545 \nu }{504}\right)
+\e^{-i \zeta } \!\left[e^5 \!\left(\frac{13 \nu }{128}-\frac{1009}{672}\right)+e^3 \!\left(\frac{943}{504}-\frac{2125 \nu }{8064}\right)+e \!\left(-\frac{14281 \nu }{4032}-\frac{563}{126}\right)\right] \nonumber\\
&\qquad\quad
+\e^{i \zeta } \left[e^5 \left(\frac{365 \nu }{2688}-\frac{4301}{672}\right)+e^3 \left(-\frac{22265 \nu }{8064}-\frac{2657}{252}\right)+e \left(\frac{1390}{63}-\frac{59909 \nu }{4032}\right)\right] -\frac{40 \nu }{9}+\frac{344}{63}\nonumber\\
&\qquad\quad
+\e^{-3 i \zeta } \left[e^5 \left(-\frac{11 \nu }{128}-\frac{5}{64}\right)+e^3 \left(-\frac{4409 \nu }{2688}-\frac{19}{14}\right)\right]
+\e^{3 i \zeta } \left[e^5 \left(-\frac{115 \nu }{896}-\frac{6401}{1344}\right)+e^3 \left(\frac{385}{72}-\frac{20917 \nu }{2688}\right)\right]\nonumber\\
&\qquad\quad
+\e^{-2 i \zeta } \left[e^4 \left(\frac{341 \nu }{2688}-\frac{233}{168}\right)+e^2 \left(-\frac{4967 \nu }{1344}-\frac{1475}{504}\right)\right]
+\e^{2 i \zeta } \left[e^4 \left(\frac{1027 \nu }{8064}-\frac{1067}{56}\right)+e^2 \left(\frac{10265}{504}-\frac{71999 \nu }{4032}\right)\right]
\bigg\} \nonumber\\
&\qquad
+ \chi_S \bigg\{
e^5 \e^{-5 i \zeta } \left(\frac{95 \nu ^2}{448}+\frac{15 \nu }{112}-\frac{15}{448}\right)+e^5 \e^{5 i \zeta } \left(\frac{33 \nu ^2}{448}-\frac{535 \nu }{1344}+\frac{81}{448}\right)+e^4 \e^{-4 i \zeta } \left(\frac{1451 \nu ^2}{768}+\frac{1525 \nu }{1536}-\frac{19}{48}\right)\nonumber\\
&\qquad\quad
+e^4 \e^{4 i \zeta } \left(\frac{1679 \nu ^2}{1792}-\frac{45197 \nu }{10752}+\frac{295}{336}\right)+e^4 \left(\frac{12391 \nu ^2}{2688}+\frac{3673 \nu }{5376}-\frac{221}{56}\right)+e^2 \left(\frac{374 \nu ^2}{21}-\frac{1943 \nu }{252}+\frac{37}{28}\right)\nonumber\\
&\qquad\quad
+\e^{-i \zeta } \left[e^5 \left(\frac{4229 \nu ^2}{4032}-\frac{349 \nu }{2688}-\frac{1009}{672}\right)+e^3 \left(\frac{53923 \nu ^2}{4032}-\frac{307 \nu }{896}+\frac{943}{504}\right)+e \left(\frac{8041 \nu ^2}{672}+\frac{21757 \nu }{4032}-\frac{563}{126}\right)\right]\nonumber\\
&\qquad\quad
+\e^{i \zeta } \left[e^5 \left(\frac{439 \nu ^2}{576}+\frac{789 \nu }{896}-\frac{4301}{672}\right)+e^3 \left(\frac{39047 \nu ^2}{4032}-\frac{26261 \nu }{2688}-\frac{2657}{252}\right)+e \left(\frac{2935 \nu ^2}{224}-\frac{151255 \nu }{4032}+\frac{1390}{63}\right)\right]\nonumber\\
&\qquad\quad
+\e^{-3 i \zeta } \left[e^5 \left(\frac{37 \nu ^2}{48}+\frac{17 \nu }{128}-\frac{5}{64}\right)+e^3 \left(\frac{9271 \nu ^2}{1344}+\frac{6353 \nu }{2688}-\frac{19}{14}\right)\right]
+\e^{3 i \zeta } \bigg[e^5 \left(\frac{697 \nu ^2}{2016}+\frac{1643 \nu }{2688}-\frac{6401}{1344}\right)\nonumber\\
&\qquad\qquad
+e^3 \!\left(\frac{20065 \nu ^2}{4032}-\frac{169625 \nu }{8064}+\frac{385}{72}\right)\!\bigg]\!
+\e^{-2 i \zeta } \!\left[e^4 \!\left(\frac{20459 \nu ^2}{4032}+\frac{529 \nu }{384}-\frac{233}{168}\right)+e^2 \!\left(\frac{25883 \nu ^2}{2016}+\frac{12605 \nu }{4032}-\frac{1475}{504}\right)\!\right]\nonumber\\
&\qquad\quad
+\e^{2 i \zeta } \left[e^4 \left(\frac{145 \nu ^2}{64}+\frac{17089 \nu }{8064}-\frac{1067}{56}\right)+e^2 \left(\frac{8243 \nu ^2}{672}-\frac{27991 \nu }{576}+\frac{10265}{504}\right)\right]+\frac{116 \nu ^2}{21}-\frac{470 \nu }{63}+\frac{344}{63}
\bigg\} \nonumber\\
&\qquad
+ \chi_S\sqrt{1-e^2} \bigg\{
\e^{-i \zeta } \!\left[e^3 \left(\frac{\nu ^2}{3}-\frac{8 \nu }{3}+2\right)+e \!\left(\frac{8 \nu }{3}-\frac{\nu ^2}{3}-2\right)\right]\!
+\e^{i \zeta } \!\left[e^3 \!\left(\frac{5 \nu ^2}{3}-\frac{40 \nu }{3}+10\right)+e \!\left(\frac{40 \nu }{3}-\frac{5 \nu ^2}{3}-10\right)\right]\nonumber\\
&\qquad\quad
+e^2 \left(\frac{4 \nu ^2}{3}-\frac{32 \nu }{3}+8\right)
+\e^{2 i \zeta } \left[e^4 \left(\frac{2 \nu ^2}{3}-\frac{16 \nu }{3}+4\right)+e^2 \left(-\frac{2 \nu ^2}{3}+\frac{16 \nu }{3}-4\right)\right]
-\frac{4 \nu ^2}{3}+\frac{32 \nu }{3}-8
\bigg\} \nonumber\\
&\qquad
+ \delta\chi_A\sqrt{1-e^2} \bigg\{
\e^{-i \zeta } \left[e^3 \left(2-\frac{2 \nu }{3}\right)+e \left(\frac{2 \nu }{3}-2\right)\right]
+\e^{i \zeta } \left[e^3 \left(10-\frac{10 \nu }{3}\right)+e \left(\frac{10 \nu }{3}-10\right)\right]\nonumber\\
&\qquad\quad
+\e^{2 i \zeta } \left[e^4 \left(4-\frac{4 \nu }{3}\right)+e^2 \left(\frac{4 \nu }{3}-4\right)\right]+e^2 \left(8-\frac{8 \nu }{3}\right)+\frac{8 \nu }{3}-8
\bigg\} 
\Bigg\rbrace \nonumber\\
&\quad
+ \frac{i x^3 \epsilon ^6}{\left(1-e^2\right)^4} \left[\delta  \chi _A+(1-2 \nu ) \chi _S\right] \bigg\lbrace
-\frac{4}{3}
-\frac{5}{64}  e^5 \e^{-5 i \zeta }+\frac{5}{192} e^5 \e^{5 i \zeta }-\frac{35}{48} e^4 \e^{-4 i \zeta }+\frac{5}{48} e^4 \e^{4 i \zeta }-\frac{19 e^4}{8}-\frac{13 e^2}{2}\nonumber\\
&\qquad
+\left(-\frac{47 e^5}{96}-6 e^3-\frac{49 e}{12}\right) \e^{-i \zeta }
-\left(\frac{21 e^5}{64}+\frac{8 e^3}{3}\right) \e^{-3 i \zeta }
+\left(-\frac{9 e^5}{32}-\frac{7 e^3}{2}-\frac{29 e}{12}\right) \e^{i \zeta }+\left(-\frac{e^5}{64}-\frac{e^3}{6}\right) \e^{3 i \zeta }\nonumber\\
&\qquad
+\left(-\frac{7 e^4}{3}-\frac{19 e^2}{4}\right) \e^{-2 i \zeta }+\left(-\frac{2 e^4}{3}-\frac{17 e^2}{12}\right) \e^{2 i \zeta }
\bigg\rbrace, \\
H^\text{inst, SS}_{22} &= \frac{x^2 \epsilon ^4}{\left(1-e^2\right)^3} \Bigg\lbrace
\delta\kappa_A \left[
1+\frac{3}{32} e^3 \e^{-3 i \zeta }-\frac{1}{32} e^3 \e^{3 i \zeta }+\left(\frac{9 e^3}{32}+\frac{7 e}{4}\right) \e^{-i \zeta }+\left(\frac{5 e^3}{32}+\frac{5 e}{4}\right) \e^{i \zeta }+\frac{11}{16} e^2 \e^{-2 i \zeta }+\frac{3}{16} e^2 \e^{2 i \zeta }+\frac{13 e^2}{8}
\right] \nonumber\\
&\qquad
+ \chi_S^2 \left[
1-\frac{1}{8} e^4 \e^{4 i \zeta }+\frac{e^4}{8}+\frac{1}{16} e^3 \e^{-3 i \zeta }-\frac{5}{16} e^3 \e^{i \zeta }-\frac{15}{16} e^3 \e^{3 i \zeta }+\left(\frac{3 e^3}{16}+\frac{3 e}{2}\right) \e^{-i \zeta }+\frac{1}{2} e^2 \e^{-2 i \zeta }-\frac{7}{4} e^2 \e^{2 i \zeta }+\frac{3 e^2}{4}
\right]\nonumber\\
&\qquad
+ \delta\chi_S\chi_A \left[
2-\frac{1}{4} e^4 \e^{4 i \zeta }+\frac{e^4}{4}+\frac{1}{8} e^3 \e^{-3 i \zeta }-\frac{5}{8} e^3 \e^{i \zeta }-\frac{15}{8} e^3 \e^{3 i \zeta }+\left(\frac{3 e^3}{8}+3 e\right) \e^{-i \zeta }+e^2 \e^{-2 i \zeta }-\frac{7}{2} e^2 \e^{2 i \zeta }+\frac{3 e^2}{2}
\right] \nonumber\\
&\qquad
+ \chi_A^2 \bigg[
1-4 \nu+e^4 \e^{4 i \zeta } \left(\frac{\nu }{2}-\frac{1}{8}\right)+e^4 \left(\frac{1}{8}-\frac{\nu }{2}\right)+e^3 \e^{-3 i \zeta } \left(\frac{1}{16}-\frac{\nu }{4}\right)+e^3 \e^{i \zeta } \left(\frac{5 \nu }{4}-\frac{5}{16}\right)+e^3 \e^{3 i \zeta } \left(\frac{15 \nu }{4}-\frac{15}{16}\right)\nonumber\\
&\qquad\quad
+\e^{-i \zeta } \left(e^3 \left(\frac{3}{16}-\frac{3 \nu }{4}\right)+e \left(\frac{3}{2}-6 \nu \right)\right)+e^2 \e^{2 i \zeta } \left(7 \nu -\frac{7}{4}\right)+e^2 \e^{-2 i \zeta } \left(\frac{1}{2}-2 \nu \right)+e^2 \left(\frac{3}{4}-3 \nu \right)
\bigg] \nonumber\\
&\qquad
+ \kappa_S \bigg[
1-2 \nu +e^3 \e^{-3 i \zeta } \left(\frac{3}{32}-\frac{3 \nu }{16}\right)+e^3 \e^{3 i \zeta } \left(\frac{\nu }{16}-\frac{1}{32}\right)+\e^{-i \zeta } \left(e^3 \left(\frac{9}{32}-\frac{9 \nu }{16}\right)+e \left(\frac{7}{4}-\frac{7 \nu }{2}\right)\right)\nonumber\\
&\qquad\quad
+\e^{i \zeta } \left(e^3 \left(\frac{5}{32}-\frac{5 \nu }{16}\right)+e \left(\frac{5}{4}-\frac{5 \nu }{2}\right)\right)+e^2 \e^{-2 i \zeta } \left(\frac{11}{16}-\frac{11 \nu }{8}\right)+e^2 \e^{2 i \zeta } \left(\frac{3}{16}-\frac{3 \nu }{8}\right)+e^2 \left(\frac{13}{8}-\frac{13 \nu }{4}\right)
\bigg]
\Bigg\rbrace \nonumber\\
&\quad
+ \frac{x^3 \epsilon ^6}{\left(1-e^2\right)^4} \Bigg\lbrace
\delta\kappa_A \bigg\{
e^5 \e^{-5 i \zeta } \left(\frac{55 \nu }{896}+\frac{5}{128}\right)-e^5 \e^{5 i \zeta } \left(\frac{199 \nu }{2688}+\frac{19}{896}\right)+e^4 \e^{-4 i \zeta } \left(\frac{713 \nu }{1344}+\frac{103}{1344}\right)-e^4 \e^{4 i \zeta } \left(\frac{563 \nu }{1344}+\frac{703}{1344}\right)\nonumber\\
&\qquad\quad
+e^4 \left(\frac{109 \nu }{224}-\frac{911}{112}\right)+e^2 \left(-\frac{211 \nu }{336}-\frac{3721}{336}\right)+\e^{-i \zeta } \left[e^5 \left(\frac{97 \nu }{336}-\frac{34}{21}\right)+e^3 \left(\frac{3125 \nu }{1344}-\frac{16145}{1344}\right)+e \left(\frac{29}{168}-\frac{83 \nu }{42}\right)\right]\nonumber\\
&\qquad\quad
+\e^{-3 i \zeta } \left[e^5 \left(\frac{213 \nu }{896}-\frac{449}{896}\right)+e^3 \left(\frac{305 \nu }{192}-\frac{851}{1344}\right)\right]
+\e^{3 i \zeta } \left[e^5 \left(\frac{193}{384}-\frac{255 \nu }{896}\right)+e^3 \left(-\frac{277 \nu }{1344}-\frac{1765}{448}\right)\right]\nonumber\\
&\qquad\quad
+\e^{-2 i \zeta } \left[e^4 \left(\frac{893 \nu }{672}-\frac{1319}{336}\right)+e^2 \left(\frac{269 \nu }{224}-\frac{143}{96}\right)\right]
+\e^{2 i \zeta } \left[e^4 \left(-\frac{109 \nu }{224}-\frac{17}{24}\right)+e^2 \left(\frac{1535 \nu }{672}-\frac{8485}{672}\right)\right]\nonumber\\
&\qquad\quad
+\e^{i \zeta } \left[e^5 \left(-\frac{11 \nu }{112}-\frac{373}{672}\right)+e^3 \left(\frac{223 \nu }{192}-\frac{2155}{192}\right)+e \left(\frac{163 \nu }{84}-\frac{305}{24}\right)\right]
+\frac{2 \nu }{3}-\frac{86}{21}
\bigg\} \nonumber\\
&\qquad
+\kappa_S\bigg\{
e^5 \e^{-5 i \zeta } \left(-\frac{15 \nu ^2}{64}-\frac{15 \nu }{896}+\frac{5}{128}\right)+e^5 \e^{5 i \zeta } \left(\frac{55 \nu ^2}{448}-\frac{85 \nu }{2688}-\frac{19}{896}\right)+e^4 \e^{-4 i \zeta } \left(-\frac{471 \nu ^2}{224}+\frac{169 \nu }{448}+\frac{103}{1344}\right) \nonumber\\
&\qquad\quad
+e^4 \e^{4 i \zeta } \left(\frac{165 \nu ^2}{224}+\frac{281 \nu }{448}-\frac{703}{1344}\right)+e^4 \left(-\frac{461 \nu ^2}{112}+\frac{3753 \nu }{224}-\frac{911}{112}\right)+e^2 \left(-\frac{2525 \nu ^2}{168}+\frac{1033 \nu }{48}-\frac{3721}{336}\right)\nonumber\\
&\qquad\quad
+\e^{-i \zeta } \left[e^5 \left(-\frac{87 \nu ^2}{112}+\frac{395 \nu }{112}-\frac{34}{21}\right)+e^3 \left(-\frac{9557 \nu ^2}{672}+\frac{11805 \nu }{448}-\frac{16145}{1344}\right)+e \left(-\frac{415 \nu ^2}{42}-\frac{65 \nu }{28}+\frac{29}{168}\right)\right]\nonumber\\
&\qquad\quad
+\e^{i \zeta } \left[e^5 \left(\frac{19 \nu ^2}{112}+\frac{85 \nu }{84}-\frac{373}{672}\right)+e^3 \left(-\frac{703 \nu ^2}{96}+\frac{1511 \nu }{64}-\frac{2155}{192}\right)+e \left(-\frac{25 \nu ^2}{3}+\frac{383 \nu }{14}-\frac{305}{24}\right)\right]\nonumber\\
&\qquad\quad
+\e^{-3 i \zeta } \left[e^5 \left(-\frac{339 \nu ^2}{448}+\frac{1111 \nu }{896}-\frac{449}{896}\right)+e^3 \left(-\frac{1677 \nu ^2}{224}+\frac{1279 \nu }{448}-\frac{851}{1344}\right)\right]
+ e^5 \e^{3 i \zeta } \left(\frac{35 \nu ^2}{64}-\frac{3467 \nu }{2688}+\frac{193}{384}\right)\nonumber\\
&\qquad\quad
+e^3 \e^{3 i \zeta } \left(\frac{151 \nu ^2}{224}+\frac{10313 \nu }{1344}-\frac{1765}{448}\right)
+\e^{-2 i \zeta } \left[e^4 \left(-\frac{587 \nu ^2}{112}+\frac{6169 \nu }{672}-\frac{1319}{336}\right)+e^2 \left(-\frac{203 \nu ^2}{16}+\frac{2809 \nu }{672}-\frac{143}{96}\right)\right]\nonumber\\
&\qquad\quad
+\e^{2 i \zeta } \left[e^4 \left(\frac{13 \nu ^2}{48}+\frac{625 \nu }{672}-\frac{17}{24}\right)+e^2 \left(-\frac{1375 \nu ^2}{336}+\frac{18505 \nu }{672}-\frac{8485}{672}\right)\right]-\frac{110 \nu ^2}{21}+\frac{62 \nu }{7}-\frac{86}{21}
\bigg\} \nonumber\\
&\qquad
+\chi_A^2 \bigg\{
e^6 \left(-\frac{15 \nu ^2}{28}+\frac{45 \nu }{16}-\frac{75}{112}\right)+e^5 \e^{-5 i \zeta } \left(-\frac{51 \nu ^2}{224}+\frac{3 \nu }{56}+\frac{1}{28}\right)+e^5 \e^{5 i \zeta } \left(\frac{97 \nu ^2}{224}-\frac{719 \nu }{672}+\frac{53}{224}\right)\nonumber\\
&\qquad\quad
+e^4 \e^{-4 i \zeta } \left(-\frac{27 \nu ^2}{16}+\frac{167 \nu }{192}+\frac{1}{6}\right)+e^4 \left(\frac{419 \nu ^2}{56}+\frac{3225 \nu }{224}-\frac{199}{56}\right)+e^2 \left(-\frac{86 \nu ^2}{21}+\frac{1123 \nu }{28}-\frac{1511}{168}\right)\nonumber\\
&\qquad\quad
+\e^{-i \zeta } \left[e^5 \left(\frac{3 \nu ^2}{16}+\frac{65 \nu }{56}-\frac{7}{48}\right)+e^3 \left(-\frac{5 \nu ^2}{3}+\frac{6345 \nu }{112}-\frac{305}{24}\right)+e \left(-\frac{479 \nu ^2}{42}-\frac{407 \nu }{36}+\frac{989}{252}\right)\right]\nonumber\\
&\qquad\quad
+\e^{4 i \zeta } \left[e^6 \left(\frac{15 \nu ^2}{28}-\frac{45 \nu }{16}+\frac{75}{112}\right)+e^4 \left(\frac{335 \nu ^2}{112}-\frac{11135 \nu }{1344}+\frac{157}{84}\right)\right]
+ e^5 \e^{-3 i \zeta }\left(-\frac{115 \nu ^2}{224}+\frac{179 \nu }{96}-\frac{209}{672}\right)\nonumber\\
&\qquad\quad
+e^3 \e^{-3 i \zeta }\left(-\frac{39 \nu ^2}{7}+\frac{949 \nu }{336}+\frac{53}{112}\right)
+\e^{3 i \zeta } \left[e^5 \left(\frac{1273 \nu ^2}{224}-\frac{778 \nu }{21}+\frac{2995}{336}\right)+e^3 \left(\frac{79 \nu ^2}{14}-\frac{1145 \nu }{112}+\frac{95}{42}\right)\right] \nonumber\\
&\qquad\quad
+\e^{i \zeta } \left[e^5 \left(\frac{401 \nu ^2}{112}-\frac{8527 \nu }{336}+\frac{2083}{336}\right)+e^3 \left(\frac{565 \nu ^2}{42}-\frac{1435 \nu }{48}+\frac{2423}{336}\right)+e \left(-\frac{79 \nu ^2}{6}+\frac{10783 \nu }{126}-\frac{731}{36}\right)\right]\nonumber\\
&\qquad\quad
+\e^{-2 i \zeta } \left[e^4 \left(-\frac{17 \nu ^2}{14}+\frac{695 \nu }{48}-\frac{317}{112}\right)+e^2 \left(-\frac{295 \nu ^2}{28}-\frac{1637 \nu }{336}+\frac{493}{168}\right)\right] -\frac{220 \nu ^2}{21}+\frac{2005 \nu }{63}-\frac{454}{63}\nonumber\\
&\qquad\quad
+\e^{2 i \zeta } \left[e^4 \left(\frac{344 \nu ^2}{21}-\frac{33739 \nu }{336}+\frac{2711}{112}\right)+e^2 \left(-\frac{29 \nu ^2}{12}+\frac{45467 \nu }{1008}-\frac{389}{36}\right)\right]
\bigg\} \nonumber\\
&\qquad
+ \delta\chi_A\chi_S \bigg\{
e^6 \left(\frac{15 \nu }{56}-\frac{75}{56}\right)+e^5 \e^{-5 i \zeta } \left(\frac{71 \nu }{448}+\frac{1}{14}\right)+e^5 \e^{5 i \zeta } \left(\frac{53}{112}-\frac{563 \nu }{1344}\right)+e^4 \e^{-4 i \zeta } \left(\frac{89 \nu }{96}+\frac{1}{3}\right)\nonumber\\
&\qquad\quad
-e^4 \left(\frac{351 \nu }{112}+\frac{199}{28}\right)-e^2 \left(\frac{127 \nu }{14}+\frac{1511}{84}\right)
+\e^{-i \zeta } \left[e^3 \left(\frac{1385 \nu }{504}-\frac{305}{12}\right)-e^5 \left(\frac{61 \nu }{224}+\frac{7}{24}\right)+e \left(\frac{989}{126}-\frac{4765 \nu }{252}\right)\right] \nonumber\\
&\qquad\quad
+\e^{4 i \zeta } \left[e^6 \left(\frac{75}{56}-\frac{15 \nu }{56}\right)+e^4 \left(\frac{157}{42}-\frac{1997 \nu }{672}\right)\right]
+\e^{-3 i \zeta } \left[e^5 \left(\frac{283 \nu }{448}-\frac{209}{336}\right)+e^3 \left(\frac{29 \nu }{56}+\frac{53}{56}\right)\right]\nonumber\\
&\qquad\quad
+\e^{3 i \zeta } \left[e^5 \left(\frac{2995}{168}-\frac{8119 \nu }{1344}\right)+e^3 \left(\frac{95}{21}-\frac{2473 \nu }{504}\right)\right]
+\e^{-2 i \zeta } \left[e^4 \left(\frac{46 \nu }{21}-\frac{317}{56}\right)+e^2 \left(\frac{493}{84}-\frac{4405 \nu }{504}\right)\right]\nonumber\\
&\qquad\quad
+\e^{i \zeta } \left[e^5 \left(\frac{2083}{168}-\frac{3559 \nu }{672}\right)+e^3 \left(\frac{2423}{168}-\frac{1997 \nu }{504}\right)+e \left(\frac{3505 \nu }{252}-\frac{731}{18}\right)\right]\nonumber\\
&\qquad\quad
+\e^{2 i \zeta } \left[e^4 \left(\frac{2711}{56}-\frac{1265 \nu }{84}\right)+e^2 \left(\frac{3929 \nu }{504}-\frac{389}{18}\right)\right]+\frac{56 \nu }{9}-\frac{908}{63}
\bigg\} \nonumber\\
&\qquad
+\chi_S^2 \bigg\{
e^6 \!\left(\frac{15 \nu }{112}-\frac{75}{112}\right)+e^5 \e^{-5 i \zeta } \!\left(\frac{1}{28}-\frac{5 \nu ^2}{128}-\frac{17 \nu }{448}\right)+e^5 \e^{5 i \zeta } \!\left(\frac{29 \nu ^2}{384}-\frac{397 \nu }{1344}+\frac{53}{224}\right)+e^4 \e^{-4 i \zeta } \!\left(\frac{1}{6}-\frac{47 \nu ^2}{192}-\frac{39 \nu }{64}\right)\nonumber\\
&\qquad\quad
-e^4 \left(\frac{41 \nu ^2}{32}+\frac{743 \nu }{224}+\frac{199}{56}\right)+e^2 \left(\frac{21 \nu ^2}{8}-\frac{1109 \nu }{84}-\frac{1511}{168}\right)
+e^4 \e^{4 i \zeta }\left(\frac{101 \nu ^2}{192}-\frac{969 \nu }{448}+\frac{157}{84}\right) \nonumber\\
&\qquad\quad
+ e^6 \e^{4 i \zeta }\left(\frac{75}{112}-\frac{15 \nu }{112}\right)
+\e^{-3 i \zeta } \left[e^5 \left(-\frac{33 \nu ^2}{128}+\frac{5 \nu }{448}-\frac{209}{672}\right)+e^3 \left(-\frac{\nu ^2}{32}-\frac{1411 \nu }{336}+\frac{53}{112}\right)\right] +\frac{29 \nu }{9}-\frac{454}{63}\nonumber\\
&\qquad\quad
+\e^{-i \zeta } \left[e^5 \left(-\frac{19 \nu ^2}{64}-\frac{571 \nu }{672}-\frac{7}{48}\right)+e^3 \left(-\frac{595 \nu ^2}{288}-\frac{3095 \nu }{1008}-\frac{305}{24}\right)+e \left(\frac{27 \nu ^2}{4}-\frac{1468 \nu }{63}+\frac{989}{252}\right)\right]\nonumber\\
&\qquad\quad
+\e^{i \zeta } \left[e^5 \left(\frac{115 \nu ^2}{192}-\frac{3169 \nu }{672}+\frac{2083}{336}\right)+e^3 \left(-\frac{455 \nu ^2}{288}-\frac{2935 \nu }{1008}+\frac{2423}{336}\right)+e \left(-\frac{19 \nu ^2}{12}+\frac{2407 \nu }{252}-\frac{731}{36}\right)\right]\nonumber\\
&\qquad\quad
+\e^{3 i \zeta } \left[e^5 \left(\frac{289 \nu ^2}{384}-\frac{6247 \nu }{1344}+\frac{2995}{336}\right)+e^3 \left(\frac{227 \nu ^2}{288}-\frac{3761 \nu }{1008}+\frac{95}{42}\right)\right]
+e^4\e^{-2 i \zeta }  \left(-\frac{65 \nu ^2}{48}-\frac{325 \nu }{336}-\frac{317}{112}\right)\nonumber\\
&\qquad\quad
+e^2 \e^{-2 i \zeta } \left(\frac{445 \nu ^2}{144}-\frac{15731 \nu }{1008}+\frac{493}{168}\right)
+\e^{2 i \zeta } \left[e^4 \left(\frac{65 \nu ^2}{48}-\frac{3853 \nu }{336}+\frac{2711}{112}\right)+e^2 \left(-\frac{77 \nu ^2}{48}+\frac{5959 \nu }{1008}-\frac{389}{36}\right)\right]
\bigg\} \nonumber\\
&\qquad
+\delta\kappa_A \sqrt{1-e^2} \bigg\{
\e^{-i \zeta } \left[e^3 \left(\frac{5 \nu }{12}-\frac{7}{6}\right)+e \left(\frac{7}{6}-\frac{5 \nu }{12}\right)\right]
+\e^{i \zeta } \left[e^3 \left(\frac{25 \nu }{12}-\frac{35}{6}\right)+e \left(\frac{35}{6}-\frac{25 \nu }{12}\right)\right]\nonumber\\
&\qquad\quad
+e^2 \left(\frac{5 \nu }{3}-\frac{14}{3}\right)+\e^{2 i \zeta } \left[e^4 \left(\frac{5 \nu }{6}-\frac{7}{3}\right)+e^2 \left(\frac{7}{3}-\frac{5 \nu }{6}\right)\right]-\frac{5 \nu }{3}+\frac{14}{3}
\bigg\}\nonumber\\
&\qquad
+ \kappa_S \sqrt{1-e^2}\bigg\{
\e^{-i \zeta } \left[e^3 \left(-\frac{\nu ^2}{2}+\frac{11 \nu }{4}-\frac{7}{6}\right)+e \left(\frac{\nu ^2}{2}-\frac{11 \nu }{4}+\frac{7}{6}\right)\right] +e^2 \left(-2 \nu ^2+11 \nu -\frac{14}{3}\right)  +2 \nu ^2-11 \nu +\frac{14}{3}\nonumber\\
&\qquad\quad
+\e^{i \zeta } \left[e^3 \left(-\frac{5 \nu ^2}{2}+\frac{55 \nu }{4}-\frac{35}{6}\right)+e \left(\frac{5 \nu ^2}{2}-\frac{55 \nu }{4}+\frac{35}{6}\right)\right]
+\e^{2 i \zeta } \left[e^4 \left(-\nu ^2+\frac{11 \nu }{2}-\frac{7}{3}\right)+e^2 \left(\nu ^2-\frac{11 \nu }{2}+\frac{7}{3}\right)\right]\!
\bigg\} \nonumber\\
&\qquad
+ \chi_A^2 \sqrt{1-e^2}\bigg\{
\e^{-i \zeta } \left[e^3 \left(-\nu ^2+\frac{23 \nu }{3}-\frac{11}{6}\right)+e \left(\nu ^2-\frac{23 \nu }{3}+\frac{11}{6}\right)\right] +4 \nu ^2-\frac{92 \nu }{3}+\frac{22}{3} +e^2 \left(\frac{92 \nu }{3}-4 \nu ^2-\frac{22}{3}\right)\nonumber\\
&\qquad\quad
+\e^{i \zeta } \!\left[e^3 \!\left(\frac{115 \nu }{3}-5 \nu ^2-\frac{55}{6}\right)+e \!\left(5 \nu ^2-\frac{115 \nu }{3}+\frac{55}{6}\right)\!\right]\!
+\e^{2 i \zeta } \!\left[e^4 \!\left(\frac{46 \nu }{3}-2 \nu ^2-\frac{11}{3}\right)\!+e^2 \!\left(2 \nu ^2-\frac{46 \nu }{3}+\frac{11}{3}\right)\!\right]\!
\bigg\} \nonumber\\
&\qquad
+ \delta\chi_A\chi_S \sqrt{1-e^2}\bigg\{
\e^{-i \zeta } \left[e^3 \left(3 \nu -\frac{11}{3}\right)+e \left(\frac{11}{3}-3 \nu \right)\right]
+e^2 \left(12 \nu -\frac{44}{3}\right) -12 \nu +\frac{44}{3}\nonumber\\
&\qquad\quad
+\e^{i \zeta } \left[e^3 \left(15 \nu -\frac{55}{3}\right)+e \left(\frac{55}{3}-15 \nu \right)\right]
+\e^{2 i \zeta } \left[e^4 \left(6 \nu -\frac{22}{3}\right)+e^2 \left(\frac{22}{3}-6 \nu \right)\right]
\bigg\} \nonumber\\
&\qquad
+ \chi_S^2 \sqrt{1-e^2}\bigg\{
\e^{-i \zeta } \left[e^3 \left(-\frac{2 \nu ^2}{3}+\frac{8 \nu }{3}-\frac{11}{6}\right)+e \left(\frac{2 \nu ^2}{3}-\frac{8 \nu }{3}+\frac{11}{6}\right)\right] +e^2 \left(-\frac{8 \nu ^2}{3}+\frac{32 \nu }{3}-\frac{22}{3}\right)\nonumber\\
&\qquad\quad
+\e^{i \zeta } \left[e^3 \left(-\frac{10 \nu ^2}{3}+\frac{40 \nu }{3}-\frac{55}{6}\right)+e \left(\frac{10 \nu ^2}{3}-\frac{40 \nu }{3}+\frac{55}{6}\right)\right] 
+\frac{8 \nu ^2}{3}-\frac{32 \nu }{3}+\frac{22}{3}
\nonumber\\
&\qquad\quad
+\e^{2 i \zeta } \left[e^4 \left(-\frac{4 \nu ^2}{3}+\frac{16 \nu }{3}-\frac{11}{3}\right)+e^2 \left(\frac{4 \nu ^2}{3}-\frac{16 \nu }{3}+\frac{11}{3}\right)\right]
\bigg\}
\Bigg\rbrace.
\end{align}
\end{subequations}
\normalsize

The 3PN tail contribution to the $(2,2)$ mode in the EOB coordinate system and in terms of the Keplerian parameters $ \left( x, e, \zeta \right) $ expanded to $\Order(e^2)$, is given by
\small
\begin{subequations}
\label{eq:app:tailModes}
\begin{align}
H^\text{tail}_{22} &= H^\text{tail, 1.5PN}_{22} + H^\text{tail, 2.5PN}_{22} + H^\text{tail, 3PN}_{22} + H^\text{tail, SO}_{22}, \\
H^\text{tail, 1.5PN}_{22} &= \frac{x^{3/2} \epsilon ^3}{\left(1-e^2\right)^{5/2}} \bigg\{
2 \pi +6 i \ln \left(\frac{x}{x_0'}\right)
+ e \bigg[\pi  \left(\frac{11 \e^{-i \zeta }}{4}+\frac{13 \e^{i \zeta }}{4}\right)+\frac{3}{2} i \e^{i \zeta } \ln 2+\e^{-i \zeta } \left(\frac{27}{2} i \ln 3-\frac{27}{2} i \ln 2\right) \nonumber\\
&\qquad
+\left(\frac{33}{4} i \e^{-i \zeta }+\frac{39}{4} i \e^{i \zeta }\right) \ln \left(\frac{x}{x_0'}\right)\bigg]
+ e^2 \bigg[\pi  \left(\frac{5}{4} \e^{-2 i \zeta }+\frac{7}{4} \e^{2 i \zeta }+3\right)+\e^{-2 i \zeta } \left(\frac{145}{2} i \ln 2 
-\frac{81}{2} i \ln 3\right)\nonumber\\
&\qquad
+\frac{1}{2} i \e^{2 i \zeta } \ln 8+\left(\frac{15}{4} i \e^{-2 i \zeta }+\frac{21}{4} i \e^{2 i \zeta }+9 i\right) \ln \left(\frac{x}{x_0'}\right)-42 i \ln 2+\frac{81}{2} i \ln 3\bigg] + \Order(e^3)
\bigg\}, \\
H^\text{tail, 2.5PN}_{22} &=  \frac{x^{5/2} \epsilon ^5}{\left(1-e^2\right)^{7/2}} \bigg\{
\pi  \left(\frac{34 \nu }{21}-\frac{107}{21}\right)+\left(\frac{34 i \nu }{7}-\frac{107 i}{7} \right) \ln \left(\frac{x}{x_0'}\right)
+ e \bigg[\pi  \left(\left(\frac{227 \nu }{84}-\frac{1409}{168}\right) \e^{i \zeta }+\left(\frac{673 \nu }{84}-\frac{2467}{168}\right) \e^{-i \zeta }\right)\nonumber\\
&\qquad\quad
+\e^{i \zeta } \left(\frac{15}{14} i \nu  \ln 2-\frac{9 i}{2}+\frac{347}{28} i \ln 2\right)+\e^{-i \zeta } \left(\frac{45}{2} i \nu  \ln \left(\frac{3}{2}\right)-\frac{27 i}{2}-\frac{27}{4} i \ln \left(\frac{243}{32}\right)\right)\nonumber\\
&\qquad\quad
+\left(\left(\frac{227 i \nu }{28}-\frac{1409 i}{56} \right) \e^{i \zeta }+\left(\frac{673 i \nu }{28}-\frac{1}{56} (2467 i)\right) \e^{-i \zeta }\right) \ln \left(\frac{x}{x_0'}\right)\bigg]\nonumber\\
&\qquad
+ e^2 \bigg[\pi  \left(\left(\frac{133 \nu }{12}-14\right) \e^{-2 i \zeta }+\left(\frac{125 \nu }{84}-\frac{46}{21}\right) \e^{2 i \zeta }+\frac{296 \nu }{21}-\frac{218}{7}\right)+\e^{2 i \zeta } \left(\frac{15}{14} i \nu  \ln 2-\frac{9 i}{2} +\frac{289}{14} i \ln 2\right)\nonumber\\
&\qquad\quad
+\e^{-2 i \zeta } \left(\nu  \left(\frac{6227}{42} i \ln 2-\frac{135}{2} i \ln 3\right)-\frac{15 i}{2} -\frac{7139}{42} i \ln 2+\frac{135}{2} i \ln 3\right)+\nu  \left(\frac{135}{2} i \ln 3-\frac{480}{7} i \ln 2\right)\nonumber\\
&\qquad\quad
+\left(\left(\frac{133 i \nu }{4}-42 i\right) \e^{-2 i \zeta }+\left(\frac{125 i \nu }{28}-\frac{46 i}{7} \right) \e^{2 i \zeta }+\frac{296 i \nu }{7}-\frac{654 i}{7}\right) \ln \left(\frac{x}{x_0'}\right)-36 i+\frac{328}{7} i \ln 2-\frac{135}{2} i \ln 3\bigg] \nonumber\\
&\qquad
+ \Order(e^3) \bigg\}, \\
H^\text{tail, 3PN}_{22} &= 
\frac{x^3 \epsilon ^6}{\left(1-e^2\right)^4} \bigg\{
-18 \ln^2\left(\frac{x}{x_0'}\right)+\left(-\frac{428}{35}+12 i \pi \right) \ln \left(\frac{x}{x_0'}\right)+\frac{2 \pi ^2}{3}-\frac{515063}{22050}+\frac{428 i \pi }{105} \nonumber\\
&\qquad
+ e \bigg[\pi ^2 \left(\frac{49 \e^{-i \zeta }}{24}+\frac{29 \e^{i \zeta }}{24}\right)+\e^{-i \zeta } \left(-\frac{515063}{7200}-\frac{81}{2} \ln^2 2-\frac{81 \ln^2 3}{2}+81 \ln 3 \ln 2+\frac{2889 \ln 2}{70}-\frac{2889 \ln 3}{70}\right) \nonumber\\
&\qquad\quad
+\e^{i \zeta } \left(-\frac{14936827}{352800}+\frac{3 \ln^2 2}{2}-\frac{107 \ln 2}{70}\right)+\pi  \left(\e^{i \zeta } \left(\frac{3103 i}{420}+\frac{3}{2} i \ln 2\right)+\e^{-i \zeta } \left(\frac{749 i}{60}-\frac{81}{2} i \ln 2+\frac{81}{2} i \ln 3\right)\right) \nonumber\\
&\qquad\quad
+\left(\pi  \left(\frac{147}{4} i \e^{-i \zeta }+\frac{87}{4} i \e^{i \zeta }\right)+\e^{i \zeta } \left(-\frac{3103}{140}-\frac{1}{2} 9 \ln 2\right)+\e^{-i \zeta } \left(-\frac{749}{20}+\frac{243 \ln 2}{2}-\frac{243 \ln 3}{2}\right)\right) \ln \left(\frac{x}{x_0'}\right) \nonumber\\
&\qquad\quad
+\left(-\frac{441}{8} \e^{-i \zeta }-\frac{261 \e^{i \zeta }}{8}\right) \ln^2\left(\frac{x}{x_0'}\right) \bigg] 
+e^2 \bigg[\pi ^2 \left(\frac{19}{8} \e^{-2 i \zeta }+\frac{17}{24} \e^{2 i \zeta }+\frac{13}{4}\right)
+\e^{2 i \zeta } \left(-\frac{8756071}{352800}+\frac{3 \ln^2 2}{2}-\frac{107 \ln 2}{70}\right)\nonumber\\
&\qquad\quad
+\e^{-2 i \zeta } \left(-\frac{9786197}{117600}-\frac{13}{2}  \ln^2 2+\frac{243 \ln^2 3}{2}-243 \ln 3 \ln 2-\frac{53393 \ln 2}{210}+\frac{8667 \ln 3}{70}\right)\nonumber\\
&\qquad\quad
+\pi  \left(\e^{2 i \zeta } \left(\frac{1819 i}{420}+\frac{3}{2} i \ln 2\right)+\e^{-2 i \zeta } \left(\frac{2033 i}{140}+\frac{499}{2} i \ln 2-\frac{243}{2} i \ln 3\right)+\frac{1391 i}{70}-123 i \ln 2+\frac{243}{2} i \ln 3\right)\nonumber\\
&\qquad\quad
+\bigg(\pi  \left(\frac{171}{4} i \e^{-2 i \zeta }+\frac{51}{4} i \e^{2 i \zeta }+\frac{117 i}{2}\right)+\e^{-2 i \zeta } \left(-\frac{6099}{140}-\frac{1497}{2}  \ln 2+\frac{729 \ln 3}{2}\right)+\e^{2 i \zeta } \left(-\frac{1819}{140}-\frac{ 9}{2} \ln 2\right)\nonumber\\
&\qquad\qquad
-\frac{4173}{70}+369 \ln 2-\frac{729 \ln 3}{2}\bigg) \ln \left(\frac{x}{x_0'}\right)
+\left(-\frac{513}{8} \e^{-2 i \zeta }-\frac{153}{8} \e^{2 i \zeta }-\frac{351}{4}\right) \ln^2\left(\frac{x}{x_0'}\right)\nonumber\\
&\qquad\quad
-\frac{6695819}{58800}-123 \ln^2 2-\frac{243 \ln^2 3}{2}+243 \ln 3 \ln 2+\frac{4387 \ln 2}{35}-\frac{8667 \ln 3}{70}\bigg]
+ \Order(e^3)\bigg\}, \\
H^\text{tail, SO}_{22} &= \frac{\epsilon^6 x^3}{\left(1-e^2\right)^4} \Bigg\lbrace
\delta \chi_A \bigg\{
-\frac{8 \pi }{3} -8 i \ln \left(\frac{x}{x_0'}\right)
+e \bigg[\pi  \left(-\frac{31}{6}  \e^{-i\zeta}-\frac{35 \e^{i \zeta }}{6}\right)+i (6-23 \ln 2) \e^{i \zeta }+i (18+45 \ln 2-45 \ln 3) \e^{-i\zeta}\nonumber\\
&\qquad\quad
-i\left(\frac{31}{2}  \e^{-i\zeta}+\frac{35}{2}  \e^{i \zeta }\right) \ln \left(\frac{x}{x_0'}\right)\bigg] 
+e^2 \bigg[\pi  \left(-\frac{41}{12} \e^{-2 i \zeta }-\frac{79}{12} \e^{2 i \zeta }-2\right)+i \left(10-\frac{2047}{6} \ln 2+\frac{405 \ln 3}{2}\right) \e^{-2 i \zeta } \nonumber\\
&\qquad\quad
+i \left(6-\frac{67 \ln 2}{2}\right) \e^{2 i \zeta }+\left(-\frac{1}{4} 41 i \e^{-2 i \zeta }-\frac{79}{4} i \e^{2 i \zeta }-6 i\right) \ln \left(\frac{x}{x_0'}\right)+i \left(48+236 \ln 2-\frac{405 \ln 3}{2}\right)\bigg]
\bigg\} \nonumber\\
&\quad
+ \chi_S \bigg\{
\pi \! \left(\frac{8 \nu }{3}-\frac{8}{3}\right)\! + i (8 \nu -8) \ln \left(\frac{x}{x_0'}\right)\!
+e \bigg[ \pi \e^{i \zeta } \! \left(\frac{16 \nu }{3}-\frac{35}{6}\right)+ \pi \e^{-i\zeta} \! \left(\frac{20 \nu }{3}-\frac{31}{6}\right)
+i \e^{i \zeta } (\nu  (12 \ln 2-3)+6-23 \ln 2) \nonumber\\
&\qquad\quad
+i \e^{-i\zeta} (\nu  (-9-36 \ln 2+36 \ln 3)+18+45 \ln 2-45 \ln 3) 
+\left(i \left(16 \nu -\frac{35}{2}\right) \e^{i \zeta}+i \left(20 \nu -\frac{31}{2}\right) \e^{-i\zeta}\right) \ln \left(\frac{x}{x_0'}\right)\bigg]\nonumber\\
&\qquad
+e^2 \bigg[\pi  \left(\left(\frac{113 \nu }{24}-\frac{79}{12}\right) \e^{2 i \zeta }+\left(\frac{155 \nu }{24}-\frac{41}{12}\right) \e^{-2 i \zeta }+\frac{15 \nu }{2}-2\right) 
+i \e^{2 i \zeta } \left(\nu  \left(\frac{69 \ln 2}{4}-3\right)+6-\frac{67 \ln 2}{2}\right) \nonumber\\
&\qquad\quad
+i \e^{-2 i \zeta } \left(\nu  \left(-5+\frac{1015 \ln 2}{4}-\frac{567 \ln 3}{4}\right)+10+\frac{405 \ln 3}{2}-\frac{2047 \ln 2}{6}\right)
+ i\nu  \left(-24-159 \ln 2+\frac{567 \ln 3}{4}\right)\nonumber\\
&\qquad\quad
+ i\left(48+236 \ln 2-\frac{405 \ln 3}{2}\right)
+\left(\left(\frac{155 \nu }{8}-\frac{41}{4}\right) \e^{-2 i \zeta }+ \left(\frac{113 \nu }{8}-\frac{79}{4}\right) \e^{2 i \zeta }+ \left(\frac{45 \nu }{2}-6\right)\right) i\ln \left(\frac{x}{x_0'}\right)\bigg]
\bigg\} \nonumber\\
&\quad
+ \Order(e^3)
\Bigg\rbrace.
\end{align}
\end{subequations}
\normalsize
The $\Order(e^6)$ expansion is provided in the Supplemental Material.

The 3PN oscillatory memory contribution to the $(2,2)$ mode in the EOB coordinate system and in terms of the Keplerian parameters $ \left( x, e, \zeta \right) $, expanded to $\Order(e^6)$, is given by
\small
\begin{subequations}
\label{eq:app:oscMemModes}
\begin{align}
H^\text{osc}_{22} &= H^{\text{osc, S}^0}_{22} + H^\text{osc, SO}_{22} + H^\text{osc, SS}_{22} + H^{\text{osc, S}^3}_{22}, \\
H^{\text{osc, S}^0}_{22} &= -\frac{i \left[13e^2 +2 e^4+ \Order(e^8)\right] \e^{2 i \zeta } \nu  x^{3/2} \epsilon ^3}{252 \left(1-e^2\right)^{5/2}}
+ \frac{i \e^{-4 i \zeta } \nu  x^{5/2} \epsilon ^5}{\left(1-e^2\right)^{7/2}} \bigg\{
e^6 \e^{6 i \zeta } \left(\frac{3097}{16128}-\frac{5281 \nu }{24192}\right)+e^5 \left(\frac{1}{126} \e^{5 i \zeta }-\frac{1}{126} \e^{7 i \zeta }\right) \nonumber\\
&\qquad
+e^4 \left[\e^{6 i \zeta } \left(\frac{4111}{1512}-\frac{5713 \nu }{1512}\right)+\frac{1}{84} \e^{2 i \zeta }+\frac{1}{168} \e^{4 i \zeta }+\frac{5}{336} \e^{8 i \zeta }+\frac{1}{336}\right]
+e^3 \left[\frac{\e^{i \zeta }}{42}+\frac{1}{18} \e^{3 i \zeta }-\frac{17}{252} \e^{5 i \zeta }+\frac{29}{252} \e^{7 i \zeta }\right] \nonumber\\
&\qquad
+e^2 \left[\e^{6 i \zeta } \left(\frac{865}{336}-\frac{6211 \nu }{1512}\right)+\frac{11}{168} \e^{2 i \zeta }+\frac{1}{42} \e^{4 i \zeta }\right]
+ e\left(\frac{4}{63} \e^{3 i \zeta }-\frac{4}{21} \e^{5 i \zeta }\right)
+ \Order(e^7)
\bigg\}
\nonumber\\
&\quad
-\frac{i \pi  \left[5568 e^2 + 7568 e^4+ 371 e^6+\Order(e^8)\right] \e^{2 i \zeta } \nu  x^3 \epsilon ^6}{24192 \left(1-e^2\right)^4}, \\
H^\text{osc, SO}_{22} &= -\frac{i \nu\epsilon ^4 }{378} x^2 e^2 \e^{2 i \zeta }  \left[13+41 e^2 +84 e^4+\Order(e^8)\right] \left[(\nu -2) \chi _S-2 \delta  \chi _A\right] 
+ \frac{i \nu  x^3 \epsilon ^6}{\left(1-e^2\right)^4} \bigg\lbrace
\delta\chi_A \bigg[
\left(\frac{2 e^5}{189}+\frac{13 e}{189}\right) \e^{i \chi }\nonumber\\
&\qquad\quad
-\left(\frac{2 e^5}{189}+\frac{13 e}{189}\right) \e^{3 i \chi }
+\e^{2 i \chi } \left(e^6 \left(\frac{24973}{96768}-\frac{1739 \nu }{6048}\right)+e^4 \left(\frac{69841}{18144}-\frac{3793 \nu }{756}\right)-\frac{8303 \nu }{1512}+\frac{21785}{6048}\right)
\bigg] \nonumber \\
&\qquad
+ \chi_S \bigg[
\e^{2 i \chi } \left(e^6 \left(\frac{1739 \nu ^2}{12096}-\frac{3331 \nu }{6048}+\frac{24973}{96768}\right)+e^4 \left(\frac{1901 \nu ^2}{756}-\frac{21661 \nu }{2268}+\frac{69841}{18144}\right)+e^2\left(\frac{2105 \nu ^2}{756}-\frac{1703 \nu }{168}+\frac{21785}{6048}\right)\right) \nonumber\\
&\qquad\quad
+\e^{i \chi } \left(e^5 \left(\frac{2}{189}-\frac{\nu }{189}\right)+e^3 \left(\frac{13}{189}-\frac{13 \nu }{378}\right)\right) 
+\e^{3 i \chi } \left(e^5 \left(\frac{\nu }{189}-\frac{2}{189}\right)+e^3 \left(\frac{13 \nu }{378}-\frac{13}{189}\right)\right)
\bigg] +\Order(e^7)\bigg\rbrace, \\
H^\text{osc, SS}_{22} &= \frac{i e^2 \nu \e^{2 i \chi } x^{5/2} \epsilon ^5}{\left(1-e^2\right)^{7/2}} \bigg\lbrace
\chi _S^2\left[e^2 \left(-\frac{2 \nu ^2}{567}+\frac{8 \nu }{567}-\frac{23}{2268}\right)-\frac{13 \nu ^2}{567}+\frac{52 \nu }{567}-\frac{299}{4536}\right]
+\chi _A^2 \left[e^2 \left(\frac{23 \nu }{567}-\frac{23}{2268}\right)+\frac{299 \nu }{1134}-\frac{299}{4536}\right] \nonumber\\
&\qquad
+\delta  \chi _A \chi _S\left[e^2 \left(\frac{8 \nu }{567}-\frac{23}{1134}\right)+\frac{52 \nu }{567}-\frac{299}{2268}\right] 
+\kappa _S\left[e^2 \left(\frac{1}{252}-\frac{\nu }{126}\right)-\frac{13 \nu }{252}+\frac{13}{504}\right]
+\delta \kappa _A \left[\frac{e^2}{252}+\frac{13}{504}\right] 
\bigg\rbrace, \\
H^{\text{osc, S}^3}_{22} &= \frac{i \nu \e^{2 i \chi } x^3 \epsilon ^6}{\left(1-e^2\right)^4} \bigg\lbrace
\delta  \chi _A^3 \left[e^4 \left(\frac{8 \nu }{243}-\frac{2}{243}\right)+e^2\left(+\frac{52 \nu }{243}-\frac{13}{243}\right)\right]
+\kappa _A \chi _A \left[e^4 \left(\frac{2}{189}-\frac{8 \nu }{189}\right)+e^2\left(\frac{13}{189}-\frac{52 \nu }{189}\right)\right] \nonumber\\
&\qquad
+\delta  \chi _A \kappa _S \left[e^4 \! \left(\frac{2}{189}-\frac{4 \nu }{189}\right)+e^2\!\left(-\frac{26 \nu }{189}+\frac{13}{189}\right)\right]
+\chi _A^2\chi _S \left[e^4 \! \left(-\frac{52 \nu ^2}{567}+\frac{23 \nu }{189}-\frac{2}{81}\right)+e^2 \!\left(\frac{299 \nu }{378}-\frac{338 \nu ^2}{567}-\frac{13}{81}\right)\right] \nonumber\\
&\qquad
+\delta  \kappa _A \chi _S \left[e^4\! \left(\frac{2}{189}-\frac{\nu }{189}\right)+e^2\!\left(\frac{13}{189}-\frac{13 \nu }{378}\right)\right]
+\delta  \chi _A \chi _S^2 \left[e^4\! \left(-\frac{8 \nu ^2}{567}+\frac{26 \nu }{567}-\frac{2}{81}\right)+e^2\!\left(\frac{169 \nu }{567}-\frac{52 \nu ^2}{567}-\frac{13}{81}\right)\right] \nonumber\\
&\qquad
+ \chi _S^3 \left[e^4 \left(\frac{4 \nu ^3}{1701}-\frac{8 \nu ^2}{567}+\frac{13 \nu }{567}-\frac{2}{243}\right)+e^2\left(\frac{26 \nu ^3}{1701}-\frac{52 \nu ^2}{567}+\frac{169 \nu }{1134}-\frac{13}{243}\right)\right]\nonumber\\
&\qquad
+ \kappa _S \chi _S \left[e^4 \left(\frac{2 \nu ^2}{189}-\frac{5 \nu }{189}+\frac{2}{189}\right)+e^2\left(\frac{13 \nu ^2}{189}-\frac{65 \nu }{378}+\frac{13}{189}\right)\right] + \Order(e^8)
\bigg\rbrace.
\end{align}
\end{subequations}
\normalsize
\end{widetext}


\bibliography{../references}


\end{document}